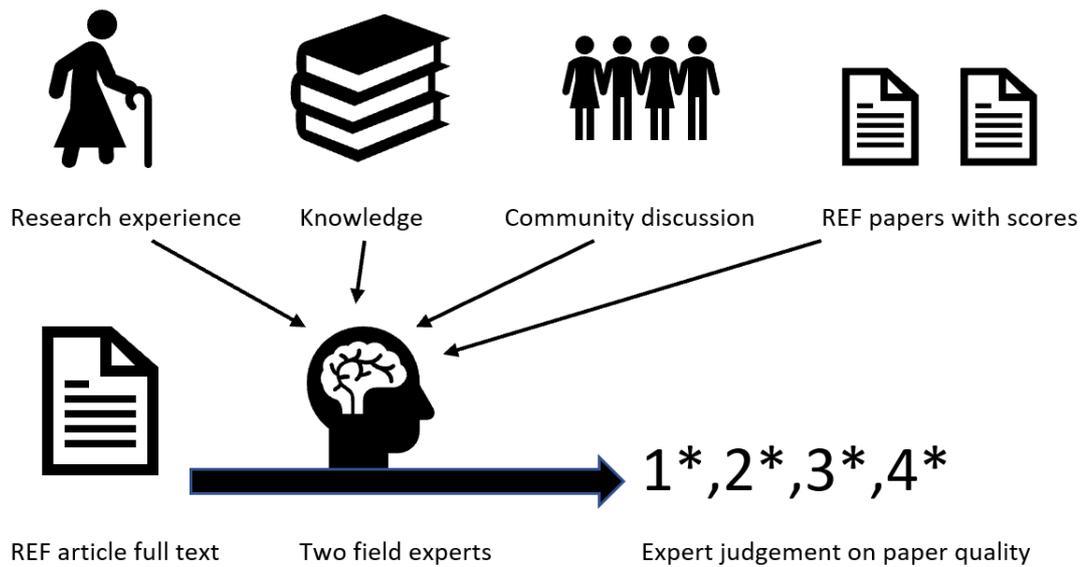
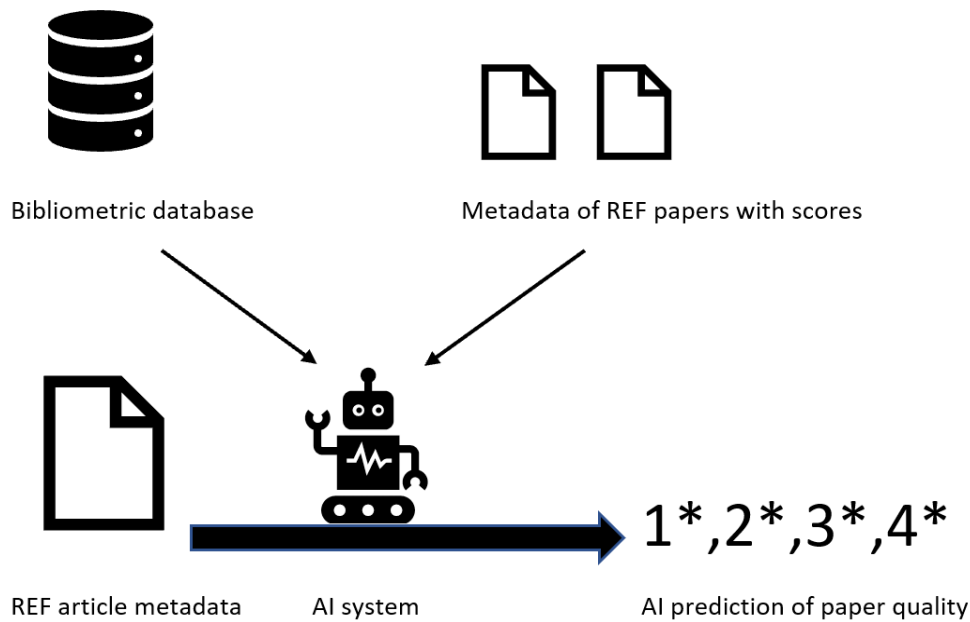

# Can REF output quality scores be assigned by AI? Experimental evidence

**13 October 2022**

A report for the four UK HE funding bodies by
Mike Thelwall, Kayvan Kousha, Paul Wilson, Mahshid Abdoli, Meiko Makita, Emma Stuart, Jonathan Levitt.
Statistical Cybermetrics and Research Evaluation Group, University of Wolverhampton, UK.

# Executive Summary

This document describes strategies for using Artificial Intelligence (AI) to predict some journal article scores in future research assessment exercises. Five strategies have been assessed. These are summarised here for completeness, but **we recommend that AI predictions are not used to help make scoring decisions yet but are further explored through pilot testing in the next REF or REF replacement.** The pilot testing should assess whether using AI predictions and prediction probabilities alongside, or instead of, bibliometric data would be helpful for any UoAs. For example, depending on UoA, AI predictions may be used to help mop up difficult scoring decisions near the end of the assessment period, to gain interdisciplinary input, as a tiebreaker in the way that bibliometrics are currently sometimes used, or to cross check the final scores.

**Background:** On behalf of the four UK HE funding bodies, as part of the Future Research Assessment Programme[1], we have developed an AI system to predict the scores assigned by sub-panels to journal articles submitted to the REF. The system uses machine learning to identify patterns in human-scored journal articles and leverages these patterns to predict the scores of new articles. The predictions are made based on 1000 properties extracted from each article. These 1000 properties include citations, journal impact, journal name, authorship team track record, and words and short phrases from the article title, abstract and keywords. The citation information is normalised for publication year and field so that high citation fields and older research have no advantage.

The accuracy of the system has been tested on provisional REF2021 scores from March 2021 and varies substantially between Units of Assessment (UoAs) and application strategies. In the worst case, the predictions are as poor as guessing but 3,688 score predictions can be made with 85% accuracy (although we are still not recommending that these are used). In particular:

| UoAs | Accuracy | Comments |
|---|---|---|
| 1, 2, 5, 7, 8, 9, 16 | High | Some article scores predicted with high accuracy are from these UoAs, depending on the strategy used. |
| 3, 4, 6, 10, 11, 12, 14, 17, 24 | Medium | Predictions can inform, but not replace, assessor scores. This could work similarly to the current use of bibliometrics in some UoAs or could be used at the end to mop up difficult cases. |
| 13, 15, 18-23, 25-34 | Low | Predictions probably have no value. |

As may be evident from the above, the predictions tend to be most accurate for the UoAs that currently use bibliometrics (1-9,11,16) except UoA 4 (uses bibliometrics but has weak AI predictions) as well as UoA 10 (reasonably accurate AI predictions but not using bibliometrics). It might be difficult from a policy perspective to have AI assistance in some sub-panels but not others, although bibliometrics were used in a fraction of panels in REF2021. Alternative strategies for implementing the AI system are summarised below.

**Strategy 1 (fixed fraction of output score prediction):** Sub-panel members in selected UoAs review 68.7% of the journal articles (all articles from the most recent two years and a random 50% from older years) and use AI predictions for the rest. This would work best for the UoAs with scores that can be predicted with 65%-75% accuracy (the longest blue bars[2] in Figure 1). For these UoAs, the AI predictions would avoid the need for human reviewing of 12,639 articles (effectively 9.3% of all REF2021 articles).

---

[1] https://www.jisc.ac.uk/future-research-assessment-programme/consultation

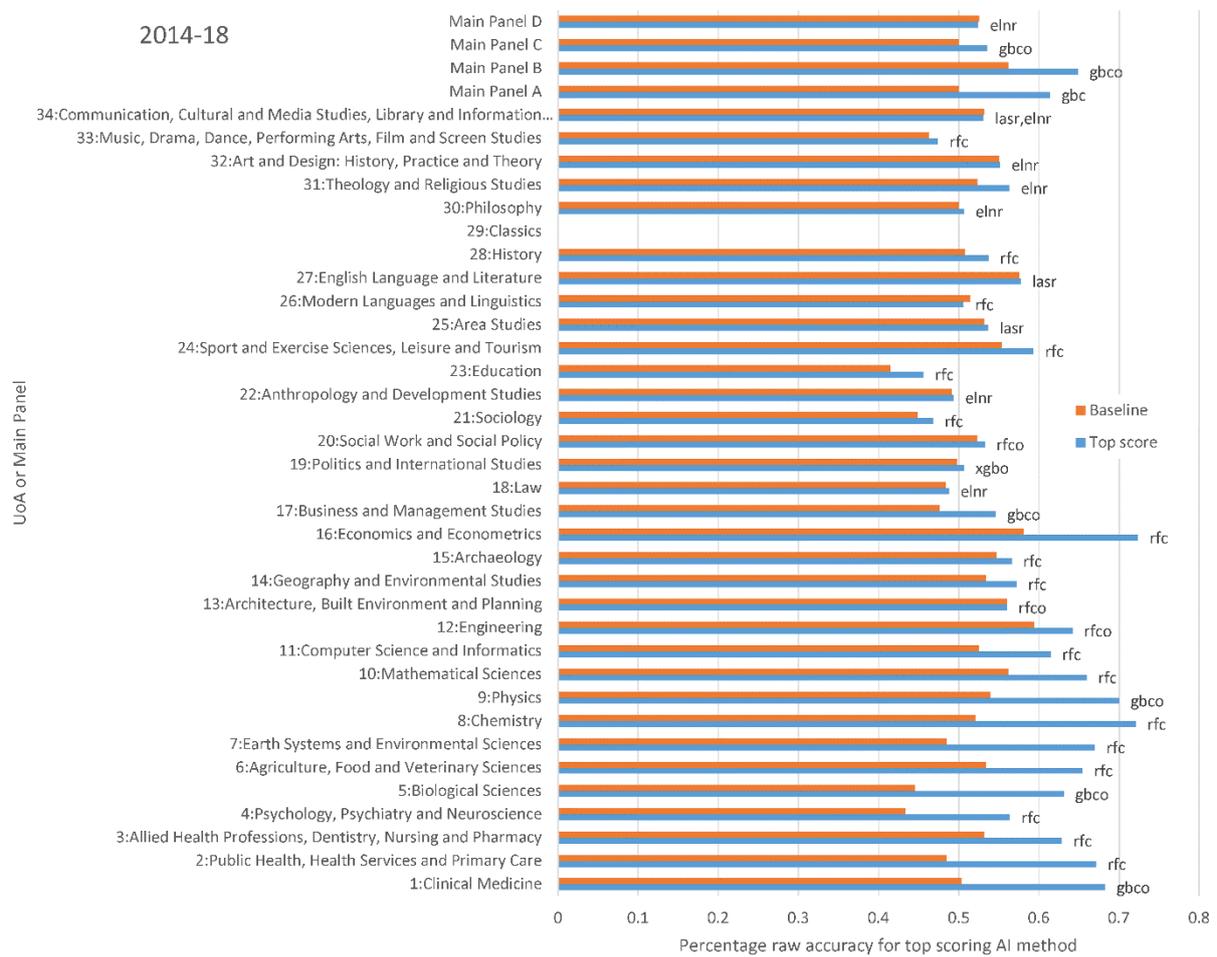

Figure 1. The percentage AI score accuracy for the most accurate machine learning method, trained on **50%** of the 2014-18 journal articles. The most accurate machine learning method is named, and the baseline score is for guessing. Articles from 2019-20 are not included due to lower prediction accuracies.

For the higher AI prediction accuracy UoAs, Pearson correlations between actual and AI predicted institutional scores range from 0.664 to 0.906 for average scores per article by institution, and from 0.945 to 0.998 after totalling all scores for each institution. Nevertheless, a switch from human scores to AI predictions substantially changes the results for smaller submissions, by up to 25% (8% overall, averaging with the 68.7% human assessed outputs), and for larger institutions (>250 outputs assessed) the score shift can be up to 6% (2% overall, averaging with the 68.7% human assessed outputs) (Figure 2). Most institutions can expect a ranking change if this strategy is adopted (Figure 3). **We think that the score shifts are much too large to be acceptable.**

---

[2] 1 Clinical Medicine; 2 Public Health, Health Services and Primary Care; 6 Agriculture and Veterinary Sciences; 7 Earth Systems and Environmental Sciences; 8 Chemistry; 9 Physics; 10 Mathematical Sciences; 16 Economics and Econometrics

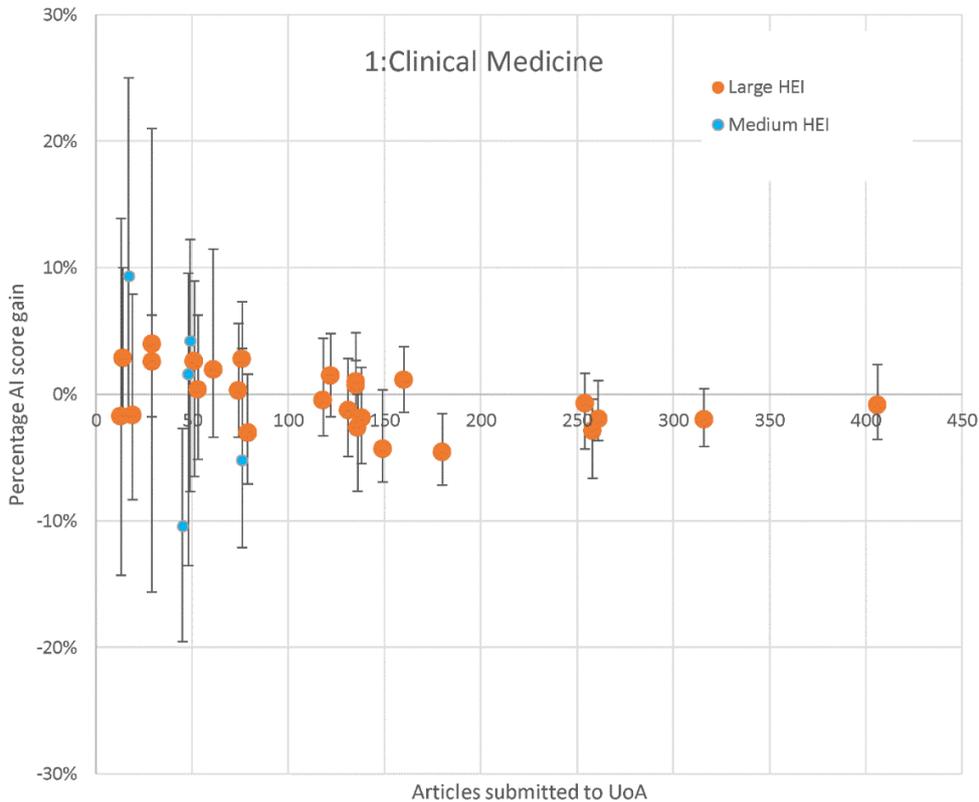

Figure 2. The average REF AI institutional score gain on UoA 1 for the most accurate machine learning method, trained on **50%** of the 2014-18 journal articles with abstracts. AI score gain is a financial calculation, sometimes called research power (4*=100% funding, 3*=25% funding, 0-2*=0% funding). The overall gain will be 31.3% of this, once the human classified articles are included. Error bars indicate the highest and lowest values from 10 random applications of the system.

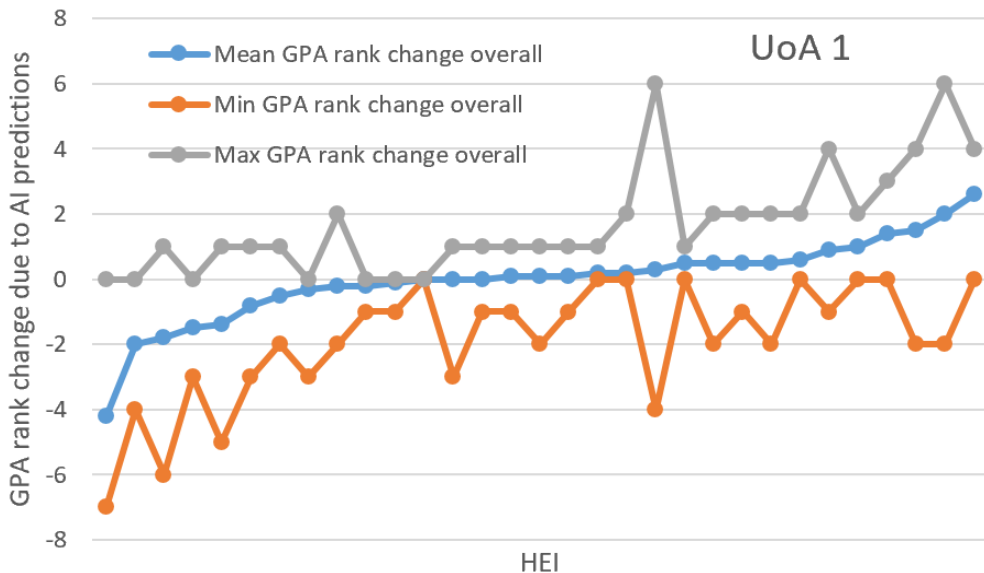

Figure 3. Average REF AI institutional grade point average (GPA) gain by replacing human scores with AI predictions for the predicted set, retaining the human scores for all non-journal outputs and for the non-predicted journal articles (i.e., the outcome if this strategy had been used in REF2021). Lower and upper lines are based on 10 iterations of the system.

- **Advantages**: The largest time saving of the proposed strategies.
- **Implementation**: The evaluation would need to be split into two phases. Sub-panels would need to ensure that a stable set of grades was entered for half of the journal articles at a suitable point, and then allow the system to remove some of the remaining articles from the pool to be assessed.
- **Disadvantages**: Inaccurate predictions for individual journal articles and substantial score shifts between institutions to a degree that seems unfair and unsupportable. Moderate incentives to publish in high citation areas, in high impact journals, with large successful teams, perhaps with prestigious methods.

**Strategy 2 (high probability output score prediction, two phases)**: Sub-panel members in selected UoAs review 68.7% of the journal articles (all articles from the most recent two years and a random 50% from older years) and use AI predictions for the remaining articles that can be predicted with above 85% accuracy; a second round of sub-panel member reviewing is used for the remaining articles. This would avoid the need for human classification of about 2,879 articles, mainly in UoAs 1-11 and 16 (effectively 2.1% of all REF2021 articles). This approach takes advantage of the AI system not only predicting article scores but also estimating the likelihood that each prediction is correct. Tests have shown that these prediction probabilities are relatively reliable. To illustrate the nature of these predictions, a high prediction probability output might be a highly cited article in field leading journal with a large international team of authors with a good research track record, mentioning a robust method (e.g., randomised control trial) (probably 4*). Alternatively, a high prediction probability output might also be a solo uncited article in rarely cited journal with a new author not mentioning a recognised research method (probably 1*/2*).

Strategy 2 will make little change to overall HEI scores, although it will probably tend to advantage smaller HEIs and lower scoring submissions (see below for Strategy 3, with similar effects).
- **Advantages**: Relatively accurate predictions would make relatively small shifts in overall institutional scores for UoAs.
- **Implementation**: A new procedure would be needed for sub-panels to evaluate the prediction probabilities, and the evaluation would need to be split into two phases.
- **Disadvantages**: System complexity. Larger score shifts for small institutions unless extra steps are taken for these. A minor degree of systematic bias. Mild incentives to publish in high citation areas, in high impact journals, with large successful teams, perhaps with prestigious methods. Relatively small saving in reviewer time.

**Strategy 3 (high probability output score prediction, multiple phases)**: After a sample of outputs is reviewed by the assessment panel (10%), the AI system can then predict the scores of the remaining outputs, but many of these will have a low degree of accuracy. The AI identifies the outputs with the lowest degree of accuracy, these are then reviewed by the panel in the next phase, and these scores are used to refine the AI system. This process continues until an appropriate level of accuracy for the AI predictions is achieved (i.e. over 85%). This would avoid the need for human classifications of 3,688 articles in UoAs 1, 2, 5, 7, 8, 9, and 16 (effectively 2.7% of all REF2021 articles). This approach will make little change to overall HEI scores (e.g., Figure 4), although it has a weak underlying tendency to advantage smaller HEIs and lower scoring submissions.

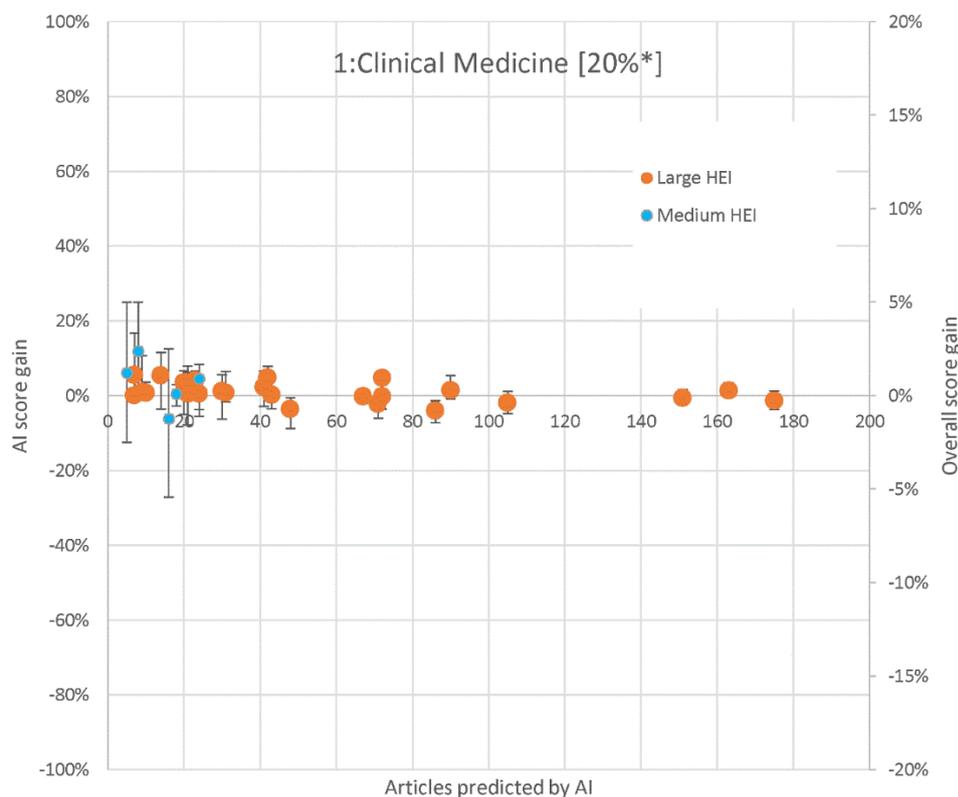

Figure 4. The average REF AI institutional score gain on UoA 1 for the most accurate machine learning method, trained on **80%** of the 2014-18 journal articles using active learning. AI score gain is a financial calculation (4*=100% funding, 3*=25% funding, 0-2*=0% funding). Error bars indicate the highest and lowest values from 10 random applications of the system. The right-hand axis shows the overall score gain for 2014-18 articles: the overall score gain for all articles 2014-20 is 75% of this.

- **Advantages**: Relatively accurate predictions would make little shift in institutional scores for UoAs.
- **Implementation**: The evaluation would need to be split into up to nine phases, with the system prioritizing articles for subpanel members to assess in each phase.
- **Disadvantages**: System complexity. The implementation system may conflict with existing procedures for allocating outputs to review by expertise, and the multiple phases may be undesirable and impractical. Larger score shifts for small institutions unless extra steps are taken for these. A minor degree of systematic bias. Mild incentives to publish in high citation areas, in high impact journals, with large successful teams, perhaps with prestigious methods.

**Strategy 4 (supporting decision making):** Sub-panels choose how to use the predictions to support their decision making. When at least half of the articles have been reliably scored, AI predictions and their associated probabilities are given to sub-panel reviewers to use how they see fit, in the way that the bibliometrics currently are in some sub-panels, to replace a third reviewer, as a final consistency check between reviewers, for support with decision making for interdisciplinary articles. For the interdisciplinary check, sub-panels would be given predictions based on the models built from their own panels as well the models built from other panels. This would give access to relevant disciplinary predictions for interdisciplinary articles that had been difficult to evaluate. For example, the reviewers might use the information as a tie breaker in difficult cases (as bibliometrics were in REF2021 for some UoAs). This would be a flexible approach, ensuring that each sub-panel oversees the way in which the predictions are used. This is safe in the sense that if the prediction system fails then the evaluation process could easily revert to the human review only approach. This is a natural evolution from the bibliometrics currently provided, by giving more accurate and specific quality estimates taking into account more factors.

- **Advantages**: Likely to improve score accuracy. Subpanels can choose the best way to use the information, and they may find extra uses for the prediction probabilities.
- **Implementation**: A procedure would also be needed for deciding what to do with the AI predictions in each UoA.
- **Disadvantages**: No time saving. Unavailability of predictions until halfway through the assessment period may delay and complicate scoring decisions. A very minor degree of systematic bias. Mild incentives to publish in high citation areas, in high impact journals, with large successful teams, perhaps with prestigious methods. Works against important changes in UK academic research culture to disregard publication venues.

# Technical Executive Summary

This summarises the technical findings of a series of investigations into the potential role of technology assisted assessment in the UK Research Excellence Framework (REF). It is based on a literature review and experiments with statistical regression and machine learning applied to an almost complete set of provisional REF2021 scores from early March 2022 for 148,977 journal articles from all 34 Units of Assessment (UoAs).

**AI system findings: Strategy 1 (classic AI)**

- Three different sets of inputs were tested for the AI methods: 1) article and author bibliometric data; 2) bibliometric data plus journal impact; 3) bibliometric data plus journal impact, plus title, abstract and keyword text, and journal names. Using all inputs combined produces more accurate predictions in most UoAs[3].
- Relying on information extracted from article full texts, as found online in open access repositories, reduced the number of predictions that could be made (due to incomplete full text data) and did not increase accuracy.
- Combining UoAs into main panels for prediction lowers accuracy in most cases. This is expected because an article evaluated in a different UoA has an approximately 54% chance of being allocated the same score, reflecting UoA-specific evaluation criteria.
- The experiments with AI suggest that in some cases scores for articles can be predicted with 65%-72% accuracy with the appropriate machine learning method (Random Forest Classifier standard or ordinal variant, Extreme Gradient Boost)[4], based on all the inputs combined. This would give AI predictions for 12,639 articles, which is less than 10% of REF outputs. Prediction accuracy is very low (below 10% above the baseline used) for articles in Main Panel D UoAs.
- Why is the accuracy not higher? AI predictions were made based on the relatively thin amount of information about the documents available and the software cannot learn enough from this to compete with the decades of experience of the 1000+ world leading field experts that allocate REF output scores.
- The highest accuracy predictions were achieved after removing articles not in Scopus (needed for bibliometric data), without abstracts (needed for text extraction), or published 2019-20 (with weaker citation data).
- At 65%-72% accuracy, scores for small submissions in most UoAs can change by up to 8% overall. Even the largest submissions in small UoAs can change by up to 7% overall, although overall shifts of up to 2% are more common for the largest submissions in UoAs. These changes occur despite high Pearson correlations between institutional scores and predicted institutional scores: from 0.664 to 0.906 for average scores, and from 0.945 to 0.998 for total scores.
- At 65%-72% accuracy, there is a moderate systematic score shift from AI predictions in favour of smaller submissions, smaller HEIs and lower scoring submissions.
- At 65%-72% accuracy, AI does not systematically work in favour or against Early Career Researchers (ECRs), women (compared to men), or research flagged as interdisciplinary[5]. There are shifts in these within some UoAs, however, such as an interdisciplinary research advantage for AI predictions in UoA 16.

---

[3] Key exceptions were Chemistry where set 1 is best and Physics, for which sets 1 and 3 give the best results.
[4] Articles published 2014-18 and in UoAs 1,2,6-10,16.
[5] Interdisciplinary flags were not reliable indicators of interdisciplinarity, however.

- The accuracy of the predictions decreases when applied to articles published in a different year, with more distant years being predicted with even lower accuracy.
- For the most predictable UoAs, AI predictions shift the results less than subpanel members classifying half the outputs for large institutions and doubling up scores.

**AI system findings: Strategy 2 (high probability predictions only)**
- Subsets of articles with higher prediction accuracy can be extracted by using the estimates provided by the AI methods. This allows some scores to be predicted highly accurately. For example, 2,879 articles could be predicted at 85% accuracy with this approach. These articles may be the "easy cases", such as highly cited large team research with high quality standard methods in high impact journals in some UoAs.

**AI system findings: Strategy 3 (active learning)**
- Higher prediction probabilities can also be obtained with an active learning strategy that uses iterative human classification, with the system selecting borderline cases for sub-panel members to evaluate. For example, this would allow 3,688 articles to be predicted by the AI with an accuracy of at least 85%[6]. The saving is concentrated in a few UoAs: 30% in UoA 8, 20% in UoA 1, and 10% in UoAs 2,5,7,9,16. The overall prediction accuracy, taking into account the vast majority of articles being human scored, would be at least 97.2% in all UoAs where it can be applied.
- For individual HEIs, active learning can induce overall score shifts of up to 14% in the worst case, but for larger submissions, the worst case is a score shift of 2.6%.
- With active learning, there is a weak systematic score shift from AI predictions in favour of smaller submissions, smaller HEIs and lower scoring submissions.
- Active learning does not systematically work in favour or against Early Career Researchers (ECRs), women (compared to men), or research flagged as interdisciplinary. There are small shifts in these within some UoAs, however.

**Supporting findings**
- Text inputs (titles, abstracts, keywords) seem to be mainly leveraged by the AI methods for journal style markers, learning whether an article is in a prestigious journal. In some fields the AI may also learn the names of highly regarded standard methods, such as randomised control trials. Qualitative methods seem to disproportionately attract low scores in some UoAs, which the AI may exploit.
- The reasons why the AI makes incorrect predictions may include: inability to judge the size of a contribution from its page length; inability to judge the significance or innovativeness of a finding; inability to evaluate methods details.
- Preliminary analysis of the data suggested that few journals are reliable indicators of the quality of their articles. No journal with more than 50 outputs submitted always hosted the same score (so articles in apparently prestigious high impact journals did not always score 4*). This supports the current REF instruction to ignore journal impact factors.
- A rough estimate of the degree of agreement for REF scores for the same articles submitted by different authors within the same UoA, when they are evaluated separately, is 86%. This can be thought of an estimate for the overall level of human agreement for the REF scores, although it is probably higher in UoAs with clearer research quality criteria.

---

[6] I.e., 2.5% of the REF journal articles, or 1 in 40, or saving the journal evaluation time of 25 of the 1000 REF sub-panel members.

**Contents**

# 1 Introduction

The UK Research Excellence Framework (REF) is used to quality assure UK academic research and to direct block grant funding to institutions based on their recent performance. It also influences reputations with students and researchers through league tables formed from, or partly from, the results and is used to generate institutional publicity, primarily to attract students. Although field-based reputations are primarily derived from international research communities (Langfeldt et al., 2020), they may also be affected by REF outcomes. The combination of peer-review research evaluation and performance-based funding is rare internationally (Sivertsen, 2017).

The REF is a labour-intensive process, both for institutions selecting content for submissions and writing their textual components, and for peer review of outputs and other documentation. Over a thousand experts in sub-panels review REF outputs and documentation over about a year. This presumably takes up a substantial amount of their time and it is therefore reasonable to assess whether it is possible to support or replace some or all their decisions with technology. Whilst citation-based indicators already play a minor role in UoAs 1-9, 11, and 16, most decisions seem to be made exclusively with peer review or with bibliometrics playing a minor role, such as to break ties between disagreeing sub-panel members. There is now a theoretical potential for technology to play a greater role by directly estimating scores for submitted journal articles.

This report contains a summary of experiments with machine learning applied to provisional REF2021 scores for journal articles from early March 2022. This is the first full scale evaluation of AI for a national research evaluation system. The results are combined to give suggestions and recommendations for the responsible use of technology to assist on assessment in future research exercises. The purpose is to clarify the potentials, limitations, and wider implications of a range of practical technology assisted assessment strategies. This report is published in parallel with a literature review of relevant technology assisted assessment topics and a set of supplementary analyses of REF data.

# 2 Ethical considerations for AI-assisted research assessment

A technically optimal or valid solution for AI-assisted research assessment may be unethical or undesirable so it is important to start with this issue.

## 2.1 Transparency

Transparency in research assessment is widely regarded as important. This is highlighted in The Metric Tide investigation of the responsible use of metrics for research assessment (Wilsdon et al., 2015a) and the Leiden Manifesto for research metrics (Hicks et al., 2015). Transparency safeguards researchers against being disadvantaged by deliberately biased or otherwise unfair decisions, including through mistakes.

The current REF has partial transparency, as summarised below. Full transparency is presumably impractical because (a) institutions may wish to challenge scores that they considered unfair and (b) publishing individual scores may have privacy and mental health implications for those assessed.

**Transparent aspects of current REF output scoring**
- The rules for submitting outputs to assess are fully transparent, published well in advance of key dates, and decided after widespread sector consultation.

- Evaluators (panel and sub-panel members and assessors) are selected after an open call to the sector for nominations and following a published procedure (REF, 2107).
- Evaluator names (i.e., sub-panel members and assessors) for each UoA are known.
- Evaluators are given public guidelines about how to evaluate outputs and how to interpret and apply the scoring system.
- Evaluators are instructed not to use Journal Impact Factors (JIFs) and cautioned about other unacceptable criteria (e.g., bias against protected characteristics).
- In UoAs where evaluators are allowed to consult citation-based indicators for articles to help them reach decisions[7], the nature of these indicators is transparent and the method to construct them is relatively transparent. Nevertheless, the data will be obtained partly using proprietary software from Clarivate (Web of Science), such as for citation extraction. In the remaining UoAs, evaluators are instructed to ignore all citation information.

**Opaque aspects of current REF output scoring**
- Individual output scores are only known to UoA panel members and the REF team. Only institutional score profiles for UoAs are published.
- Despite the evaluation guidelines, the thought processes used to score articles are unknown and presumably rely on complex knowledge of the academic research process in a field. In the absence of explanations for individual scores it is very difficult to know how influenced they are by conscious or unconscious biases (e.g., for or against topics, methodological approaches, people, or institutions).
- Individual or UoA output scores are not explained or justified, in contrast to many other peer review exercises, such as journal article and grant rejection letters, open peer review reports for journal articles. Institutions are given very general overall feedback on their outputs, such as a few sentences about strong or weak areas.
- There is no opportunity for those evaluated to challenge the results (there does not seem to have ever been any successful challenge to any aspect of the results).
- The names of the primary evaluators within a subpanel for individual outputs or institutions are unknown.

There are two types of AI: transparent and opaque, with opaque AI being by far the most powerful. Opaque AI would replace the currently invisible thought processes of current UoA assessors with an opaque algorithm, whereas transparent AI would reveal the decision-making process for the first time. Nevertheless, those evaluated might have greater concerns with opaque AI than hidden human expert thought processes based on a greater trust in the ability of the humans to be unbiased or avoid mistakes.

## 2.2 Dora and Journal Impact Factors

UKRI signed the San Francisco Declaration of Research Assessment (DORA) in 2020 (UKRI, 2020). This initiative was set up to counter the overuse of Journal Impact Factors (JIFs) in research evaluation, which many believed was unhelpful to scientific progress. The following general summary highlights the key principle:

---

[7] 1: Clinical Medicine; 2: Public Health, Health Services and Primary Care; 3: Allied Health Professions, Dentistry, Nursing and Pharmacy; 4: Psychology, Psychiatry and Neuroscience; 5: Biological Sciences; 6: Agriculture, Food and Veterinary Sciences; 7: Earth Systems and Environmental Sciences; 8: Chemistry; 9: Physics; 11: Computer Science and Informatics; 16: Economics and Econometrics

*Do not use journal-based metrics, such as Journal Impact Factors, as a surrogate measure of the quality of individual research articles, to assess an individual scientist's contributions, or in hiring, promotion, or funding decisions* (DORA, 2020).

The importance of DORA for the REF is highlighted in the advice to panel members:

*We encourage you to challenge research assessment practices that rely inappropriately on journal impact factors or conference rankings and promote and teach best practice that focuses on the value and influence of specific research outputs. If you are unsure about DORA, please speak to the panel convener or the panel chair.* (REF, 2017).

There are many aspects of DORA that seem uncontroversial in the REF context. The following advice for funding agencies is particularly relevant to the REF:

*For the purposes of research assessment, consider the value and impact of all research outputs (including datasets and software) in addition to research publications, and consider a broad range of impact measures including qualitative indicators of research impact, such as influence on policy and practice* (DORA, 2020).

In terms of the automatic assessment of research outputs, it is clear from DORA that decisions should not be based solely on JIFs. For example, a system that allocated scores to outputs based on JIFs or another journal or conference ranking system would not be appropriate for any UoA.

Nevertheless, DORA does not explicitly rule out the use of JIFs or other journal-based metrics, as part of the decision-making process. For example, some suggest that they can be used in conjunction with other approaches (Black et al., 2017). The argument in favour of taking into consideration the publication venue (journal or conference) is that journals in some fields are partly hierarchical gatekeepers of quality and so journal rankings may be rough indicators of article quality. In this context, ignoring journal or conference prestige risks losing valuable evidence from an established international system of journal editors and reviewers. This is most relevant in disparate fields where UoA panel members lack the expertise of specialist journal editors and reviewers, and so would not have the field knowledge to overrule the journal peer review decisions, despite having to for the REF.

There is some REF evidence about the value of journal impact metrics as evidence in different UoAs. An analysis of individual output scores for REF 2014 (Wilsdon et al., 2015b) found moderate statistical associations between article quality and journal citation rates in Economics and Econometrics (Spearman correlation: 0.67), Biological Sciences (0.55), Clinical Medicine (0.52), Chemistry (0.50), Public Health, Health Services and Primary Care (0.42), and Business and Management Studies (0.41). In the social sciences, arts and humanities, there were weak, negligible or even negative associations between article quality scores and journal citation rates, such as Social Work and Social Policy (0.29), Modern Languages and Linguistics (0.21), Area Studies (0.02), Art and Design: History, Practice and Theory (-0.07) and Classics (-0.79, based on only 5 articles) (Table A18 of: Wilsdon, et al., 2015b). This evaluation used Elsevier's Source Normalised Impact per Paper (SNIP), which is a field normalised variant of the JIF (and so is more appropriate in this context). A negative association in the humanities might occur, for example, if some humanities articles had been published in scientific journals, where the peer review for humanities contributions to the articles was less effective. Since the confidence intervals for negative values always contain zero (Figure 2.2.1), there is not necessarily an underlying negative association in any field, however. In the area where journal citation impact seems to be the strongest indicator of journal quality, Economics, a regression analysis has shown that JIFs may account for up to 89% of the variance in institutional REF score profiles (Stockhammer et al., 2021).

JIFs have been used in research assessment in Italy alongside citations in a dual system that runs alongside peer review, but an expert group has recommended that JIFs be discontinued based on a lack of transparency and a lack of alignment between the field normalisation practices used to calculate JIFs and the Italian context (Bonaccorsi, 2020).

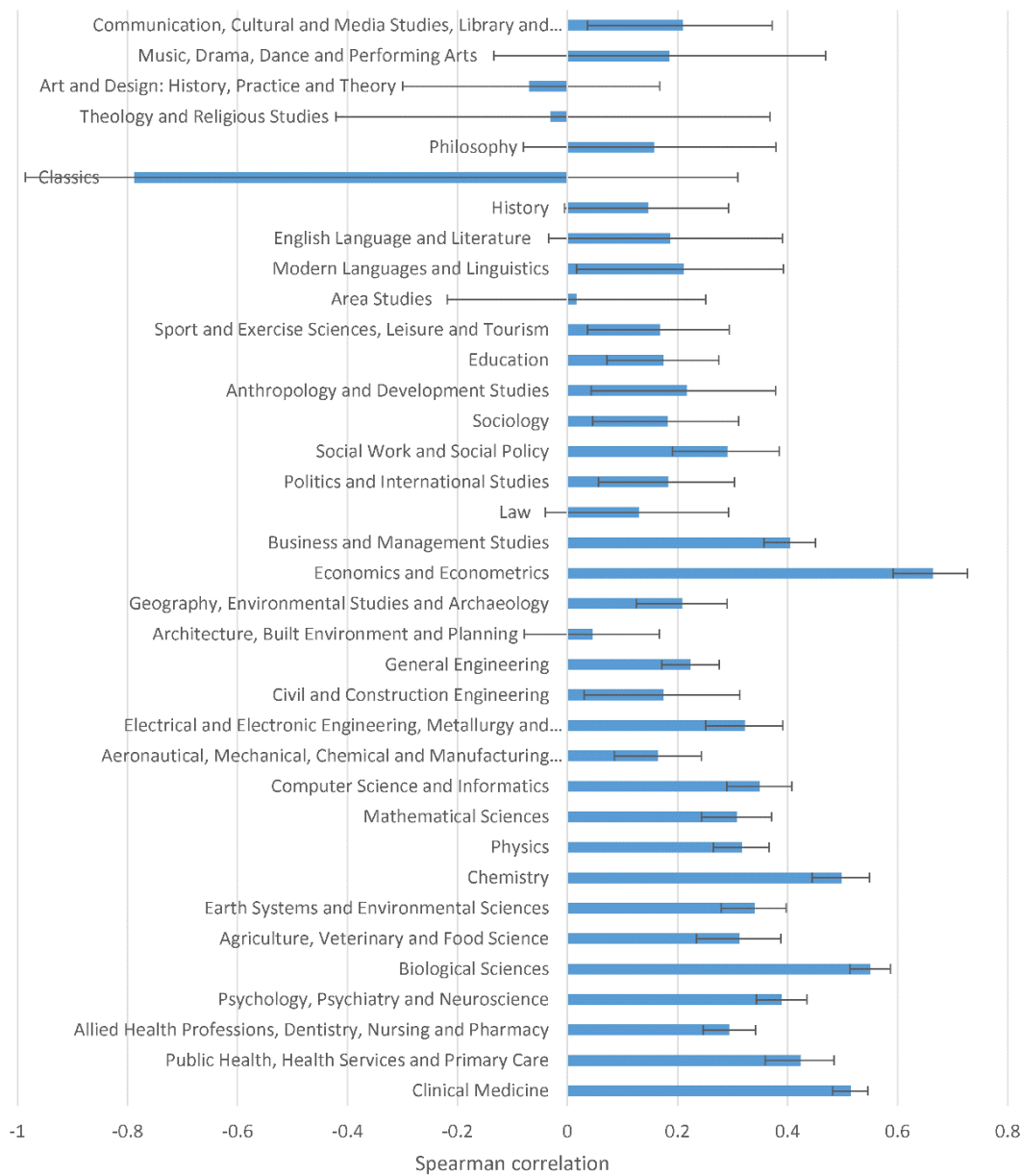

Figure 2.2.1. Spearman correlations between SNIP values and sub-panel member REF2014 scores for articles published in 2008 (Table A18 of: Wilsdon, et al., 2015b). Error bars illustrate 95% confidence intervals: wide confidence intervals are caused by small sample sizes.

## 2.3 Bias in human and AI-assisted research assessment

The potential for any type of bias in REF scores is an obvious concern. REF score bias is difficult to define precisely, however. Assuming that there is a correct score for a REF output, any deviation from this correct score would be an *error* and a systematic deviation from correct scores for a group of articles would be a (systematic) *bias*.

Although the accuracy of the REF2021 scores is unknown, any systematic difference between REF2021 scores and AI predictions (e.g., an overall increase in scores for one institution compared to another) suggests the presence of bias. At least small score differences between human and AI scores is inevitable, unless the AI was 100% accurate in mimicking the human scores. Unfortunately, it would not be possible to conclude from the existence of any size of score shift that the AI has introduced bias rather than corrected sub-panel member bias, however. For example, the AI may correct sub-panel member gender bias or institutional favouritism, or the AI may introduce a bias towards research with positive bibliometric characteristics, including more hierarchical specialties. Nevertheless, any systematic human-AI deviation needs to be identified and discussed.

### 2.3.1 Institutional bias

Since one of the core REF goals is to reward institutions for their prior performance with block grants, a system that disadvantaged some institutions would be undesirable, irrespective of its overall accuracy. This might occur, for example, for small specialist institutions that performed a type of research that was not well processed by a prediction algorithm. It is therefore important to check the institutional-level funding implications of switching from human scores to machine learning scores. Although institution-wide bias is important, institutional bias within UoAs seems more likely and is important because institutions presumably reward UoAs for their scores financially or through institutional kudos, and UoA level scores seem to be the most widely publicised.

### 2.3.2 Researcher characteristics bias

Society and academia have a long history of gender bias and therefore it is important to investigate whether AI scores seem to introduce a bias against any gender. Similarly, early career researchers are an important group within academia that may perform different types of research to other researchers, and it is therefore important to test whether AI scores would disadvantage them, given that they may have lower career statistics and some career data is an input for the AI model.

It is at least as important that the scores are not biased against researchers with other protected characteristics, including disability, LGBTQ+ status, and ethnic minority membership, but it is not practical to test for these due to a lack of personal data tied to REF outputs.

### 2.3.3 Interdisciplinarity, author contributions, and double-weighting biases

Interdisciplinary research is an issue of relevance for REF evaluations because outputs may draw upon different UoAs, despite being submitted to one. For a fairer analyses of such outputs, subpanel members and authors can flag outputs as interdisciplinary and request that they are also examined by other UoAs.

Although interdisciplinary research is recognised as being important, it may be undervalued by procedures designed primarily for single discipline research. It is therefore useful to assess whether AI scores would disadvantage it.

Another issue is that some articles are highly co-authored and HEIs in seven UoAs are asked to justify that the submitted author made a substantial contribution to a study with more than 15 authors.

> "Sub-panels 1-6 and 9 will only routinely refer to author contribution statements on outputs in cases where there are more than 15 authors and the submitted member of staff to whom the output is attributed is not identified as either the lead or corresponding author. If there are errors in an author contribution statement contained within the output, HEIs should flag this in the co-author contribution statement that they provide for the sub-panel to consider. If the sub-panel has any concerns about the information provided, the HEI may be asked to verify the co-author contribution through audit."[8]

The guidelines suggest that author contribution judgements will be binary in the sense of accepting or rejecting an output for evaluation. Nevertheless, it is theoretically possible that sub-panel members may allocate a lower grade to an output if they think that an author had not made a full contribution. In such a case, the AI predictions would tend to give higher scores these outputs because they would not have author contribution information from which to make judgements.

Double weighting outputs (i.e., a single output counting as two) is not expected to be an issue for AI because it is primarily for longer outputs than journal articles. For example,

> "Considering the patterns of publication across Main Panel A and B's areas of activity, the sub-panels expect that such requests will occur only exceptionally. In particular, the sub-panels anticipate that outputs published as journal articles and conference papers will not normally embody work of this nature, and they therefore do not normally expect to receive requests for double-weighting these types of outputs."[9]

## 3 Experimental setup for predicting REF journal article scores

The goal of the experimental work was to gain empirical evidence of the effectiveness of different AI approaches to predict scores for journal articles submitted to the REF. The work is supported by the literature review for the choice of inputs and analysis methods so that an informed decision can be made about the most promising approaches. The results can inform a decision about whether any gains, for example in time saving, outweigh any disadvantages, for example in inaccuracy, bias, transparency, or unintended consequences.

### 3.1 Background

Relatively few previous studies have systematically explored the possibility of predicting expert evaluator scores for academic journal articles because of the lack of public review scores. Binary peer review outcomes (i.e., publish/not publish) are a partial exception but limited because most publishers do not reveal peer review information and most that do only provide the information for accepted articles (exception: the F1000Research family of publishing platforms).

The largest previous attempt to predict peer review scores from document information (citations, altmetrics) was conducted by a HEFCE statistical team for REF 2014, available (unrefereed) as an annex to *The Metric Tide*[10]. It analysed 78% of REF2014 outputs (149,670 out of 191,080), reporting correlations against 17 bibliometric and altmetric indicators, as well as precision (% of correct 4*/non-4* predictions) and sensitivity/recall (% of 4* correctly predicted) with statistical (not AI) approaches. The 36 REF2014 UoAs were analysed separately, controlling for publication year. The results were best for Main Panels A and B, and weak for Main Panel D. There was some evidence of statistical prediction shifts (either adding bias or correcting existing bias) towards male researchers and people that were not early career researchers (ECRs) in some UoAs.

---

[8] https://www.ref.ac.uk/faqs/
[9] https://view.officeapps.live.com/op/view.aspx?src=https%3A%2F%2Fwww.ref.ac.uk%2Fmedia%2F1451%2Fref-2019_02-panel-criteria-and-working-methods.docx
[10] http://doi.org/10.13140/RG.2.1.3362.4162

The indicators analysed by HEFCE included three based on citation counts, two journal impact indicators, author count, country count, and altmetrics. The correlations between sub-panel member scores and indicator scores by publication year were low, and below 0.4 overall in the earliest year (with the best data), although higher in some UoAs, especially those in Main Panels A and B. The average precision and sensitivity for statistically predicting 4* outputs was low for most UoAs, and was below 70% in all cases, despite not using separate training and evaluation sets. This suggests that single indicators were insufficient to make reliable judgements for individual articles in any field. The fields most conforming to indicators were Clinical Medicine and Economics and Econometrics.

In addition to the correlations, HEFCE used statistical logistic regression to predict 4* REF scores from the bibliometric information (p. 69-70) using multiple inputs when available. This showed that a different subset of the metrics was significant in the regression but did not report the amount of improvement of the predictions from using multiple independent variables, so this factor is unknown. The regressions used dummy values for missing data and ran a single regression for all years, with publication year as an independent variable. Again, separate training and evaluation sets were not used.

The current investigation of machine learning for REF2021 learns from the HEFCE REF2014 exploration for inputs and attempts to build compact, powerful AI approach to predict the scores of individual outputs, rather than primarily investigating the value of inputs. It also uses separate training and evaluation sets. Other inputs suggested by the literature review, including a statistical analysis of Italian peer review scores, are also used.

## 3.2 REF 2022 journal article provisional scores

The provisional scores for journal articles submitted to REF2021 were made available for testing, excluding outputs from Wolverhampton and ineligible outputs, a total of 148,977 journal article scores (Figure 3.2.1). Based on REF2014[11], journal articles, this represents about 82% of REF outputs; and journal articles formed 82% of REF outputs again for REF2021. The number of articles per UoA varied from 227 (Classics) to 17,929 (Engineering) due to differing UoA sizes and differing proportions of journal articles in the submitted outputs (and some degree of double weighting). AI predictions for Main Panel D tended to be hampered by having too few articles to train the AI algorithms on.

---

[11] https://www.ref.ac.uk/2014/results/analysis/outputprofilesanddiversity/

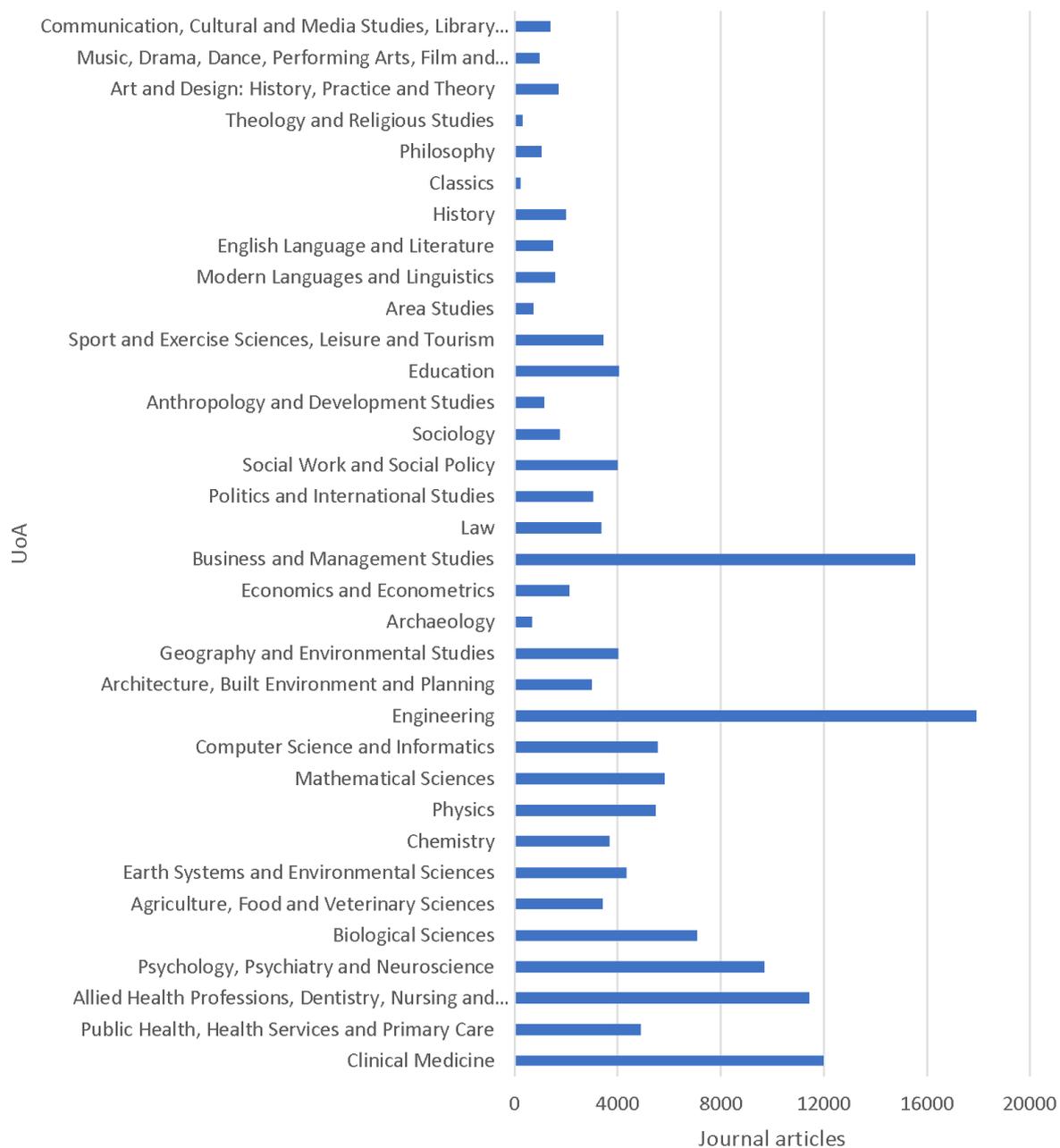

Figure 3.2.1. The number of journal articles with provisional REF2021 scores made available for analysis, by UoA.

Few journal articles received Scores of 0 or 1, with 3 being the modal score for journal articles in most UoAs. Given the small amounts of scores 0 and 1, and their financial equivalence to score 2, it seems reasonable to group them together into a single set (0-2) for analysis (Figure 3.2.2). The small number of articles with score 0 were subsequently removed, as discussed below, because some were ineligible rather than weak.

The proportion of the different scores varies significantly between UoAs (Figure 3.2.2). This is because the overall score profiles for all outputs differs between UoAs and/or because journal articles tended to get different scores from other outputs in some UoAs. For example, the low proportion of 4* scores in Main Panel D UoAs is presumably due to journal articles tending to score lower than books, performances, and/or exhibitions.

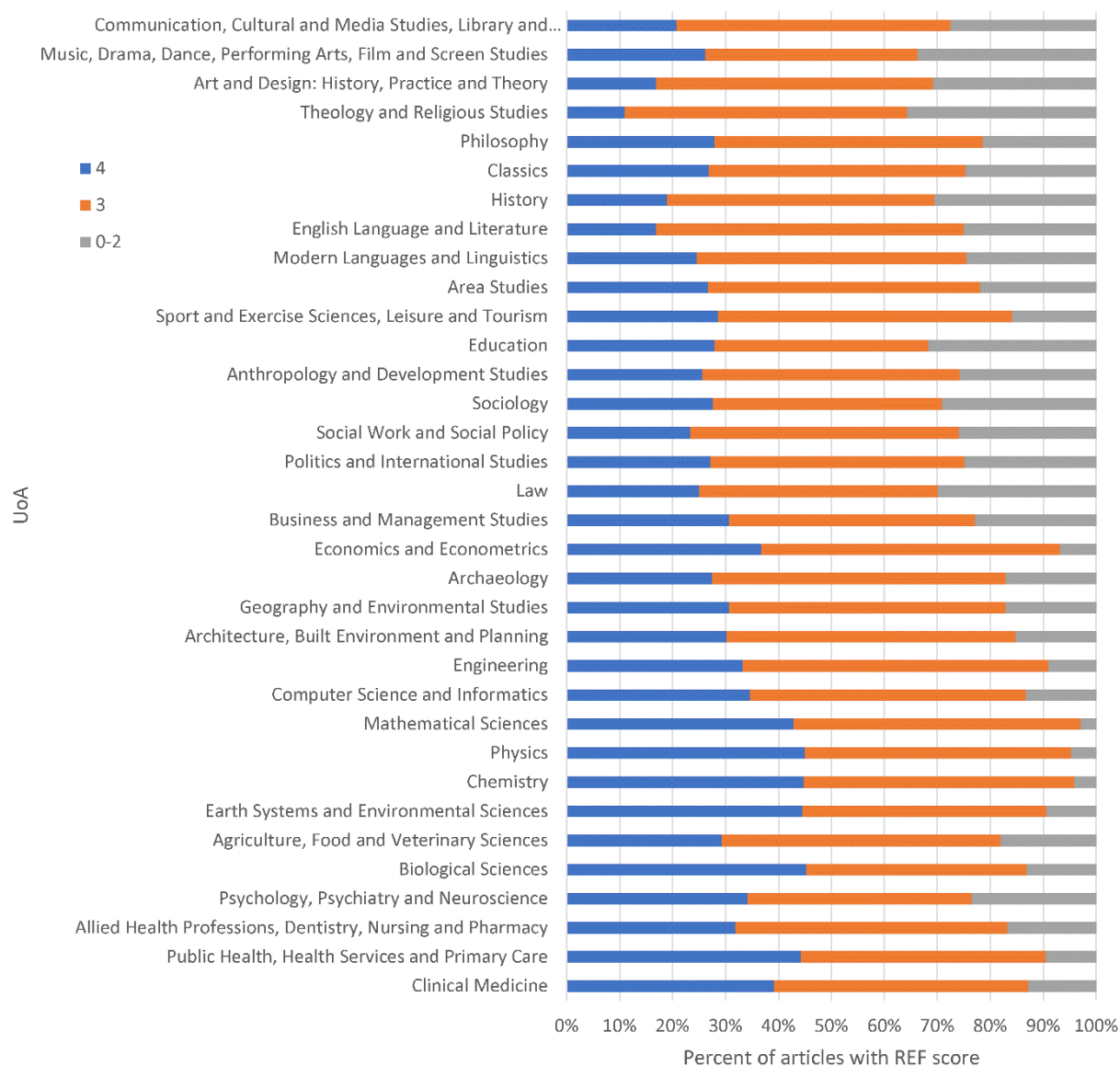

Figure 3.2.2. The provisional REF2021 scores of journal articles made available for analysis, by UoA.

### 3.2.1 Multiple provisional scores for the same article

Many articles were submitted multiple times to the REF for different authors, sometimes in different UoAs. This was common, with 25.1% of the outputs being duplicates in this sense. Such articles did not always have the same provisional REF scores. In a few cases, articles were given 4* for some authors but 0 for others, suggesting that the evaluators had not accepted a co-author's claim to have made a substantial contribution. Excluding all scores of 0, which could be disqualified authors, if two scores for an article were selected at random then the chance that they agreed on the four-point scale was 79.8%. If 1* and 2* are merged, then the overall agreement was only slightly larger at 80.0%. Disagreement might occur if different UoAs or sets of evaluators had different opinions of an article or it matched different UoA quality criteria.

The above figures include articles within the same UoA and articles submitted to different UoAs. Overall, if two scores were selected at random for the same article within a single UoA then there would be a 98.9% chance of the two scores being the same. Thus, almost all the disagreement between articles is discrepancies between UoAs. Agreement statistics for individual UoAs (Table 3.2.1.1) show that agreement levels within UoAs were almost universally high and were 100% in many. The 100% scores for large sets of articles are most likely due to these UoAs either systematically checking for discrepancies or ensuring that outputs are only assessed once (by the sub-panel member with the closest matching expertise) and using the same score for all copies of it. The very high agreement rates in other cases (above 95%) suggest that either cross-checking took place, but some differences were allowed to remain, or that the cross-checking was imperfect (e.g., due to late scores or late changes from a sub-panel member). It is also possible that some UoAs naturally agreed on the vast majority of outputs, but this seems unlikely because peer review disagreement is common for journal articles (Lee et al., 2013), even those winning Nobel Prozes (Campanario, 2009) and articles with marginal scores (e.g., could easily be either 3* or 4*) would cause problems.

Table 3.2.1.1. Agreement rates for scores for duplicate copies of the same article within the same UoA. Articles are considered duplicates if they have the same DOI (last two columns are the same except for UoA4).

| Sub-panel | Non-unique articles in UoA | Average agreement 1v2v3v4 | Average agreement 1&2v3v4 |
|---|---|---|---|
| 1:Clinical Medicine | 1249 | 98.8% | 98.8% |
| 2:Public Health, Health Services and Primary Care | 455 | 99.2% | 99.2% |
| 3:Allied Health Professions, Dentistry, Nursing and Pharmacy | 587 | 98.6% | 98.6% |
| 4:Psychology, Psychiatry and Neuroscience | 845 | 98.6% | 98.7% |
| 5:Biological Sciences | 472 | 98.7% | 98.7% |
| 6:Agriculture, Food and Veterinary Sciences | 119 | 100.0% | 100.0% |
| 7:Earth Systems and Environmental Sciences | 346 | 100.0% | 100.0% |
| 8:Chemistry | 306 | 99.8% | 99.8% |
| 9:Physics | 560 | 100.0% | 100.0% |
| 10:Mathematical Sciences | 384 | 100.0% | 100.0% |
| 11:Computer Science and Informatics | 372 | 100.0% | 100.0% |
| 12:Engineering | 864 | 97.1% | 97.1% |
| 13:Architecture, Built Environment and Planning | 95 | 100.0% | 100.0% |
| 14:Geography and Environmental Studies | 243 | 95.5% | 95.5% |
| 15:Archaeology | 44 | 86.4% | 86.4% |
| 16:Economics and Econometrics | 144 | 99.3% | 99.3% |
| 17:Business and Management Studies | 2112 | 100.0% | 100.0% |
| 18:Law | 116 | 100.0% | 100.0% |
| 19:Politics and International Studies | 152 | 98.7% | 98.7% |
| 20:Social Work and Social Policy | 107 | 97.2% | 97.2% |
| 21:Sociology | 31 | 100.0% | 100.0% |
| 22:Anthropology and Development Studies | 12 | 100.0% | 100.0% |
| 23:Education | 189 | 99.5% | 99.5% |
| 24:Sport and Exercise Sciences, Leisure and Tourism | 284 | 96.8% | 96.8% |
| 25:Area Studies | 1 | 100.0% | 100.0% |
| 26:Modern Languages and Linguistics | 20 | 100.0% | 100.0% |
| 27:English Language and Literature | 12 | 100.0% | 100.0% |
| 28:History | 40 | 100.0% | 100.0% |
| 30:Philosophy | 29 | 100.0% | 100.0% |
| 32:Art and Design: History, Practice and Theory | 24 | 66.7% | 66.7% |
| 33:Music, Drama, Dance, Performing Arts, Film and Screen Studies | 18 | 94.4% | 94.4% |
| 34:Communication, Cultural and Media Studies, Library & Info Man | 22 | 86.4% | 86.4% |

With the (unchecked) assumption that some or most within-UoA output score agreement rates are high due to cross checking or copied scores, calculating agreement rates only between UoAs would give a more reasonable estimate of the underlying human consensus level. The agreement rates for the same articles submitted to different UoAs are substantially lower, at 58.9% (or 59.1% if 1* and 2* are merged). Extrapolating to estimate the theoretical agreement rate for outputs submitted once, if they had accidentally been evaluated by a different but appropriate UoA, the between-UoA projected agreement rate for single submitted outputs is about 53% (Figure 3.2.2.1). This figure is not an estimate of the underlying REF evaluator agreement rate if it is accepted that different UoAs may legitimately give different scores to the same output because they are allowed to use different quality criteria. UoAs may also focus on the aspects of an output that most closely align with their subjects.

In the absence of strong evidence about human agreement rates within UoAs for the REF, the best estimate may be the agreement rate for the UoA with the most duplicate submissions that did not seem to systematically check scores for duplicate submissions. Very high agreement rates, for example above 95%, seem unlikely without systematic checking of discrepancies, even accounting for the fact that outputs are evaluated by multiple sub-panel members and ratified by the entire sub-panel. On this basis, the most likely within-UoA agreement level estimate would be for 15: Archaeology at 86.4%. **Thus, 86.4% seems to be the best available guess for inter-reviewer agreement within UoAs in the absence of checking.** This is a weak estimate because agreement rates seem likely to vary substantially for different UoAs and Archaeology may be atypical.

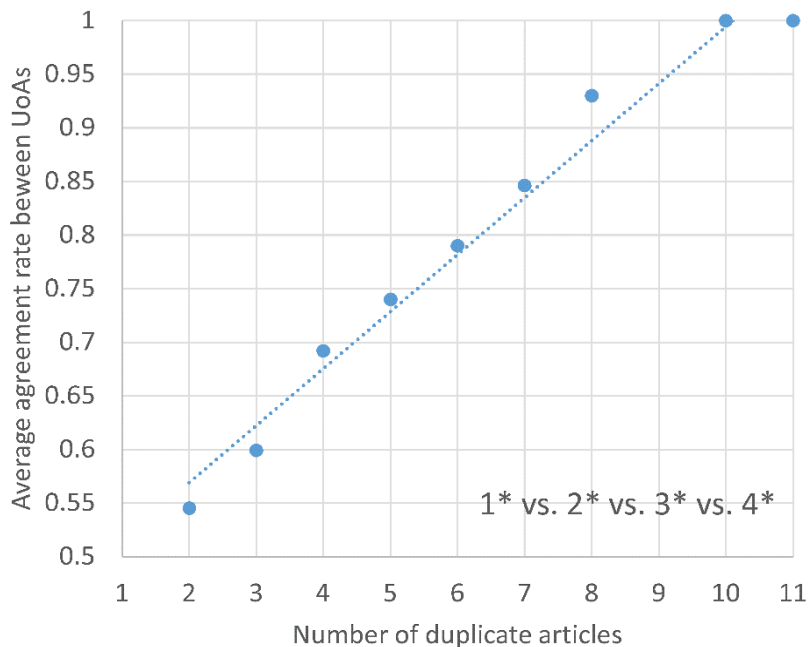

Figure 3.2.1.1. Agreement rate against number of duplicate submissions of the same article scoring at least 1*, only checking scores between different UoAs.

### 3.2.2 Similarity of scores for the same journal

The extent to which articles in the same journal tended to get the same score was assessed, since journals are widely used as quality indicators in some fields. Although this was explicitly banned in REF2021 guidelines, the data gives a good opportunity to check the validity of using journals as proxies for quality. Overall, if two articles were selected at random from the same journal, then the chance that they had received the same score was 48.7%. The extent to which journals tended to publish articles with the same score varied between UoA (Figure 3.2.2.1), being generally slightly higher than average in Main Panel A but highest in Economics and Econometrics (67%), and lowest in Main Panel D. Overall, then, with the partial exception of Economics and Econometrics, the publishing journal is not a reliable pointer to the REF score, although it is statistically helpful (greater journal homogeneity for article scores than expected by chance) in most UoAs.

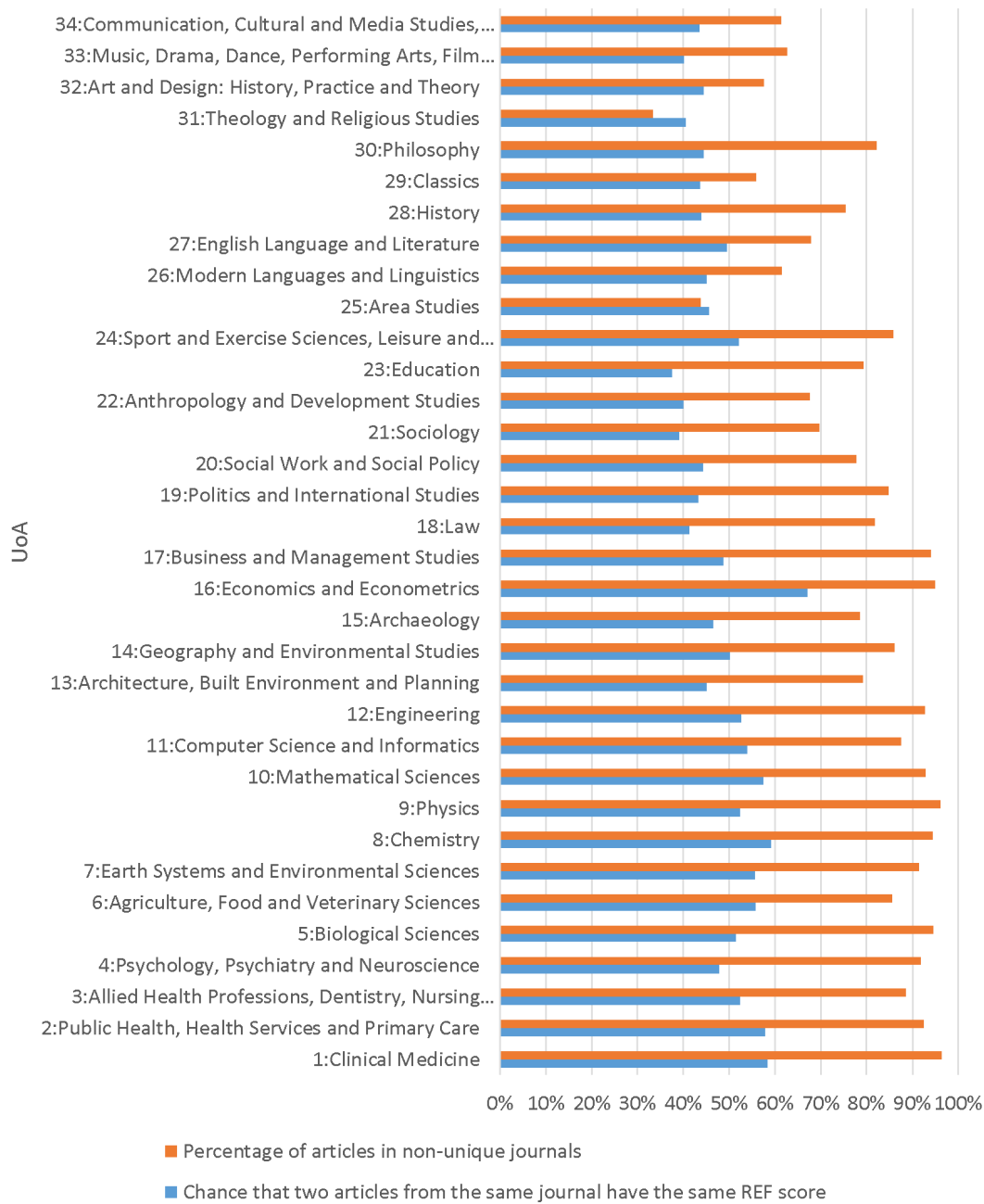

Figure 3.2.2.1. The percentage of articles in non-unique journals (i.e., more than one REF output from the journal) and the chance that two articles from the same journal have the same REF score.

The data were also checked for evidence that articles in prestigious journals reliably achieved 4* scores. Only limited evidence of this was found, however. For example, only ten journals with more than 10 REF articles were always awarded the same projected REF score (always 4* for six journals, always 3* for four journals). No journal publishing more than 50 REF articles had more than 90% of articles with the same provisional score. The fact that all apparently well-known high impact journals hosted articles with a variety of scores (less than 90% the same score each case) might caution against relying on journal prestige in the REF, although some outputs in prestigious journals might also be short form articles.

## *3.3 Statistical predictive power of the inputs*

A statistical investigation was conducted to investigate the relative predictive power of the proposed bibliometric inputs. The goal of this task is to investigate these inputs rather than to predict scores. Differently from the HEFCE REF2014 correlation analysis and regression study, separate models were built for each UoA rather than a combined model for all. This is because the importance of inputs varies substantially between fields, as shown by HEFCE REF2014 correlations. The preliminary statistical tests informed the choice of inputs for the machine learning. The final statistical results were available after the machine learning was complete and provide extra context for the report.

The main influence of the early statistical results was in emphasising the importance of journal citation rates. This, together with an error analysis of the initial machine learning models, led to the inclusion of journal names as inputs for the final machine learning models. Journal names were primarily added to capture journals with citation rates that did not reflect the average quality of their articles. The final statistical results are published as a supplementary file.

## *3.4 The predictive power of words and phrases in titles, abstracts, and keywords*

The words and phrases with high predictive power for each dataset were identified to generate insights into how text data inputs might be leveraged by AI to predict REF scores. For this, each title, keyword, and abstract word or phrase (two or three consecutive words) was tested for associating with a particular score using a chi-square test. This assesses the extent to which a word or phrase occurs in articles with REF scores (1* or 2*, 3* or 4*, using the above schema) in proportions similar to that of the category. For each category, the ten terms with the highest chi-squared score were extracted and the results are summarised below, by category.

When the chi-squared test was applied to the entire dataset (all articles with 500+ character abstracts), the terms with the highest chi-square bias scores were almost all stylistic and associated with higher scores. In particular, stylistic phrases like "here we show" and "we show that", and "the primary" associated with higher scores. The word "we" also occurred disproportionately in high scoring articles. It is almost always used in an exclusive sense (e.g., Doğan-Uçar & Akbasb, 2022). For example, "we" occurred in over 72% of 4* articles across the entire dataset but in under 44% of 1* or 2* articles. Some stylistic terms also associated with lower scores, such as "study", "this article", "paper", "was", and "research". Two topic-related terms associated with lower scores: "education" and "students". Several methods-related terms associated with higher scores: "primary outcome", "randomly assigned", and "adverse events". The results are discussed in more detail below.

### 3.4.1 Main Panel A: Words and phrases

For Main Panel A, most of the words and phrases most strongly associating with article scores were stylistic or methods-based. There were few top topic-related terms but mentions of funding associated with higher scores overall (Table 3.4.1.1). Overall, this suggests that higher scores tended to go to articles reporting specific types of (high quality) methods in their titles, abstracts, and keywords. There is a suggestion that qualitative methods tend not to score as highly, at least when they are mentioned in titles, abstracts, and keywords (e.g., interviews, thematic analyses – not necessarily Thematic Analysis). Funding was not a key term for any other Main panel, perhaps because journal styles do not include funding details in titles, abstracts, or keywords. *Lancet* family journals include funding information in abstracts, which is at least part of the underlying cause.

The stylistic terms, such as "we", "here we", and "show that" presumably reflect the styles of the abstracts of prestigious journals. Use of the past tense associates with lower scores, which may be a journal style issue.

Table 3.4.1.1. Examples of words and phrases with the strongest associations (chi-square test) with REF scores by UoA for Main Panel A. Bold terms associate with lower REF scores; other terms associate with higher REF scores.

| UoA | Style | Methods | Topic |
|---|---|---|---|
| **1:Clinical Medicine** | we | Funding, "randomly assigned", "the primary outcome" interpretation, masked, "is registered", "in the placebo" | [one minor topic] |
| **2:Public Health, Health Services and Primary Care** | "this is an" | Funding, "randomly assigned", "the primary outcome", interpretation, trial, "trial is registered", **interviews, participation** | |
| **3:Allied Health Professions, Dentistry, Nursing and Pharmacy** | We, "here we", "show that", **was, were** | Funding, CI, "is registered with", "adverse events", randomised, intention-to-treat, stratified, **thematic** | |
| **4:Psychology, Psychiatry and Neuroscience** | We, "we show", **were, "the aim", "current study"** | Funding, "randomly assigned", "Is registered", **online, web, "measures of"** | Neurons, gene, human, mouse |
| **5:Biological Sciences** | We, "here we", **be, were, was, investigated, some, study, significant** | **web, the effects** | [one minor topic] |
| **6:Agriculture, Food and Veterinary Sciences** | We, "here we", "we show that" **were, was, significantly, on, had** | Replication, **collected** | Evolution, genes, amino acid, genome |
| **Panel A** | We, "here we", interpretation, **were, study** | Funding, "randomly assigned", "is registered" | |

## 3.4.2 Main Panel B: words and phrases

For Main Panel B, most of the strongest words were stylistic or, in contrast to Main Panel A, topic-based. Most UoAs had words associating with topics that tended to receive high or low scores (Table 3.4.2.1). Thus, machine learning would predict scores partly from the topics of papers. For some UoAs, the language used seemed to be too diverse to identify words strongly associating with any score. This was particularly the case in UoA 10: Mathematical Sciences. For these UoAs, text features in machine learning may help little.

Table 3.4.2.1. Examples of words and phrases with the highest chi-square values by UoA for Main Panel B. Bold terms associate with lower scores; other terms associate with higher scores.

| UoA | Style | Methods | Topic |
|---|---|---|---|
| **7:Earth Systems and Environmental Sciences** | "Here we", we, find, **"was also"**, were, study, "this study" | | Global, warming, earth, climate, ocean, **reef, reproduction, conservation** |
| **8:Chemistry** | "Here we", we, **"were performed"**, was | **Reduce, tests** | Molecules, **distillery, wafers** |
| **9:Physics** | "here we", "we report", here, significance, **"it is shown"**, apparent, **"down to"**, thought | **formula** | Gravitational-wave, relativity, "the laser", **"laser pulses"** |
| **10:Mathematical Sciences** | We, **"the actual"** | "Related to" | **forms, waveguides** |
| **11:Computer Science and Informatics** | We, "we show that", our, **"this study", "the results"** | Experiments, **review, development** | Verification, complexity, **communities, people, national, video** |
| **12:Engineering** | We, "here we", "we show", demonstrate, **were, was, had, study** | Imaging, **qualitative** | **customer** |
| **Panel B** | We, "here we", "we show that", our, here, **was, were, paper, study, used** | | warming |

### 3.4.3 Main Panel C: words and phrases

For Main Panel C, there were strong words for style, methods, and topics, varying substantially between UoAs. Some qualitative approaches (interviews, thematic analyses – not necessarily Thematic Analysis) seemed to attract lower scores when mentioned in article titles, abstracts, and keywords. Work focused on teaching and learning in universities also seemed to attract lower scores (Table 3.4.3.1).

Table 3.4.3.1. Examples of words and phrases with the highest chi-square values by UoA for Main Panel C. Bold terms associate with lower scores; other terms associate with higher scores.

| UoA | Style | Methods | Topic |
|---|---|---|---|
| 13:Architecture, Built Environment and Planning | **Considered, "the purpose", "of this paper"** | Model, **literature, case** | Heating, **farm, "policies in", sustainable, cement, decision-makers** |
| 14:Geography and Environmental Studies | We, "here we", our, find, **study, "this study", were** | Assessment, applications, compared, **interviews** | Global, oceanic, sea, "Antarctic ice sheet", forests, **environment** |
| 15:Archaeology | "here we", during, latest, **article, "of archaeological interest", significant, "the results"** | experimental | Species, Asia, population, humans, "ancient genomes", **pottery, ceramic** |
| 16:Economics and Econometrics | We, "we develop", indicate, **"analysis of the", "this study,** | "Treatment effects", **literature** | **European, "in the UK", "women are", excellence, "asymmetric information"** |
| 17:Business and Management Studies | We, "we develop", because, "consistent with", when, outcomes, **paper, "this paper", been, significant** | Theory, **review** | **"the organization", "the firm", policy, "of management",** countries, business |
| 18:Law | We, argue, "we propose", why, about, focus, **involved, proposed, "it is"** |  | Role, rulings, redress, **legislation, "the European"** |
| 19:Politics and International Studies | We, our, find, "we show that", **also, "this article"** | "Data on", results, test, "evidence that", effects, theory, **reasons, interviews** | Preferences, "a new", American, voters, ethnic, **EU** |
| 20:Social Work and Social Policy | We, "find that", **would, "this paper is"** | Results, effects, longitudinal, data, evidence, CI, cohort, **thematic** | Life, parents, "of care" **students, learning, teaching** |
| 21:Sociology | first, however, **found, key** | Demonstrate, "factors that", **fieldwork, narratives, "case studies"** | Market, mobility, **public, international, violence, decision-making** |
| 22:Anthropology and Development Studies | Yet, nor, **article, "this article", "paper explores"** | "The model" | Political, social, "human capital", **countries, institutional** |
| 23:Education | We, **authors** | Multilevel, causal, longitudinal, effect, "sociology | British, ethnicity, governments, **learning, needs, university, physical, professional** |

| | | of", **semi-structured, qualitative** | |
|---|---|---|---|
| **24:Sport and Exercise Sciences, Leisure and Tourism** | **We, own, whereas,** "the study" | "Randomised controlled trial", "cluster randomised controlled" evidence, **correlated, investigation, observations** | Skeletal, "skeletal, muscle", human, care, **soccer, running, rugby** |
| **Panel C** | We, our, find, "purpose of this", "we show that, "find that", **"the findings", "of this paper"**, focuses, explored | Outcomes, "consistent with", theorize, "we develop", **interviews** | Organizational, skeletal, **"in higher education", education, teacher, undergraduate, "in the UK"** |

### 3.4.4 Main Panel D: words and phrases

For Main Panel D, the low numbers of articles to analyse gives weaker evidence of associations between words and scores. Nevertheless, there are words associated with writing style, methods, and topics, albeit differing between UoAs (Table 3.4.4.1).

Table 3.4.4.1. Examples of words and phrases with the highest chi-square values by UoA for Main Panel D. Bold terms associate with lower scores; other terms associate with higher scores.

| UoA | Style | Methods | Topic |
|---|---|---|---|
| **25:Area Studies** | Our, **"this article argues", study** | | European, "the global", **languages** |
| **26:Modern Languages and Linguistics** | Our, "show that", strongly, without, **I** | Results, "a model of", derive, lexically, influence, patterns | Semantics, syntactic, change, language, **narrative** |
| **27:English Language and Literature** | Whose, **"for this"** | "Example of", "implications for" | "Early modern", English, learners, "changes in" |
| **28:History** | "Would suggest" | | "the old", "the law", guilds, murder |
| **29:Classics** | | Approaches, implications, conventional, **relationship** | |
| **30:Philosophy** | **"In this paper", "I will"** | "Argued in", statistical, models, **assumptions** | |
| **31:Theology and Religious Studies** | First, "in some", **being** | **Associated** | Phenomenon, historical, Protestant, **culture** |
| **32:Art and Design: History, Practice and Theory** | | **Findings, values** | Large, artistic, political, **students** |
| **33:Music, Drama, Dance, Performing Arts, Film and Screen Studies** | "Show that" | Predicted | Music, voices, improvisations, **students, theatre** |
| **34:Communication, Cultural and Media Studies, Library and Information Management** | Most, "not only" | "in contrast" | Producers, government, **marketing, senior** |
| **Main Panel D** | "We show" | "Interpretation of", models, experiments, computational, measures, results, assumptions, **context** | Syntactic, **students, narrative** |

Overall, the results suggest that text inputs derived from titles, abstracts and keywords will leverage words and phrases that associate with writing styles, topics and methods that tend to get high or low REF scores (and in some cases average scores).

The strong style associations found presumably relate to the standard practice or guidelines of the publishing journals (e.g., *Lancet* journals for funding). The topic terms may reflect either important topics or the research orientations of high scoring groups within a UoA or panel.

Some of the methods terms associate with high quality research approaches (e.g., randomised controlled trials), which seem to be useful direct indicators of study quality. Other terms associate in some UoAs with methods (interviews, case studies, thematic approaches to analysis) that may have tended to be carried out less well than other submitted work, or that may have not been highly regarded by the UoA panel members evaluating them.

The topic terms varied between fields, but one recurring theme was that work focusing on higher education tended not to score as well. Purely speculatively, researchers with insufficient time or experience to conduct traditional studies in their own fields may have submitted small local investigations of their own teaching methods or students instead; alternatively, these may be high quality studies but not valued by UoA assessors.

### 3.5 Overview of the output score prediction testing

The overall approach for the machine learning was to extract all journal articles from the REF2021 data, and then develop a set of AI algorithms that would learn to predict the provisional scores given to these articles by the sub-panel members, assessing the accuracy of these predictions. The predictions would be based on metadata about the articles (e.g., citedness, journal impact, abstract text) fed as inputs to the algorithms.

A machine learning approach was used in all cases, with two types of algorithm: classification and regression. A classification algorithm predicts which class (e.g. 1*, 2*, 3*, 4*) an article should be in whereas a regression algorithm predicts a score for an article (e.g. any number; any whole number; any number in the range 1-4). Since the key score boundaries for the REF are between 2* (unfunded) and 3* (funded) and between 3* (funded) and 4* (maximum funding), and the 1* category was rare, the lower two categories were merged into one (1*, 2*) for prediction, giving a three-class problem (recall that the few 0 scores were removed). Merging was desirable because small categories create the technical problem of unbalanced training data for the classification algorithms. In fact, subsequent tests revealed that this did not affect overall accuracy.

Each algorithm was tested in a series of experiments. In each experiment, the algorithm was fed with a subset of the articles, the "training set" from which to learn how to predict article scores (1-2* or 3* or 4*) and then the accuracy of the algorithm was then evaluated on the "test set", a non-overlapping collection of the remaining articles from the set investigated (Figure 3.5.1). This is a standard approach for machine learning to avoid getting misleadingly high accuracy by training and testing on overlapping data. The experiments were repeated with different splits into training and testing data, with the average reported. Accuracy was measured in two ways: the percentage of computer predictions that agreed with the human scores and the average difference between the human and computer scores (i.e., 0 if they agree, 1 or 2 if they differ). Only agreement rates are reported here to reduce information overload.

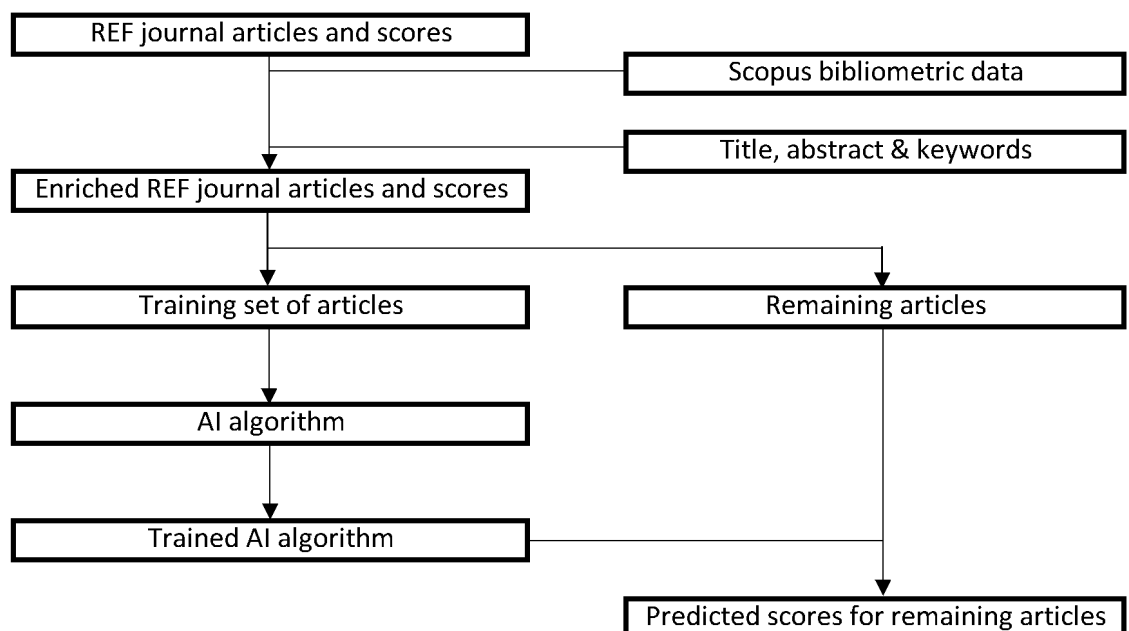

Figure 3.5.1. Overview of the process for predicting REF scores for articles, where the remaining articles form the "test set". Comparing the predicted scores for the remaining articles with their provisional REF scores gives a measure of the accuracy of the algorithm.

## 3.6 Inclusion criteria for the AI experiments

REF2021 outputs had to satisfy all the criteria below to be included in the testing (Figure 3.6.1). Essentially, they had to be non-Wolverhampton journal articles matching a Scopus record 2014-20. Note that multiple copies of the same output would be included if submitted in different UoAs, possibly with different provisional REF scores. Multiple copies of an article submitted to the same UoA were removed, leaving a single copy with the median score (if they differed; one of the two medians chosen with a random number generator when there were two medians).

- Registered as document type Journal Article when submitted to the REF (to analyse a standard type of document).
- Matching a journal article in Scopus with a registered publication date from 2014 to 2020. This excludes articles that were published online in 2020 but were allocated a 2021 publication year by Scopus when finally published in a journal issue. Matching was primarily achieved through DOIs: 89.4% had a matching DOI in Scopus. Papers without a DOI in Scopus were matched against Scopus by title, after removing non-alphabetic characters (including spaces) from the title (since there are variations in how whitespace was used in titles) and converting to lowercase. Title matches were manually checked for publication year, journal name, and author affiliations. In one case author information was checked online to see if they had worked at the Scopus affiliation institution even though it differed from their REF affiliation. This produced 997 extra matches. When there was a disagreement between the REF registered publication year and the Scopus publication year, the Scopus publication year was always used. Articles from 2021 were excluded (n=69) because comparable Scopus data would not be available until too late for score predictions.
- Not submitted by the University of Wolverhampton (for security reasons, Wolverhampton data was not supplied).

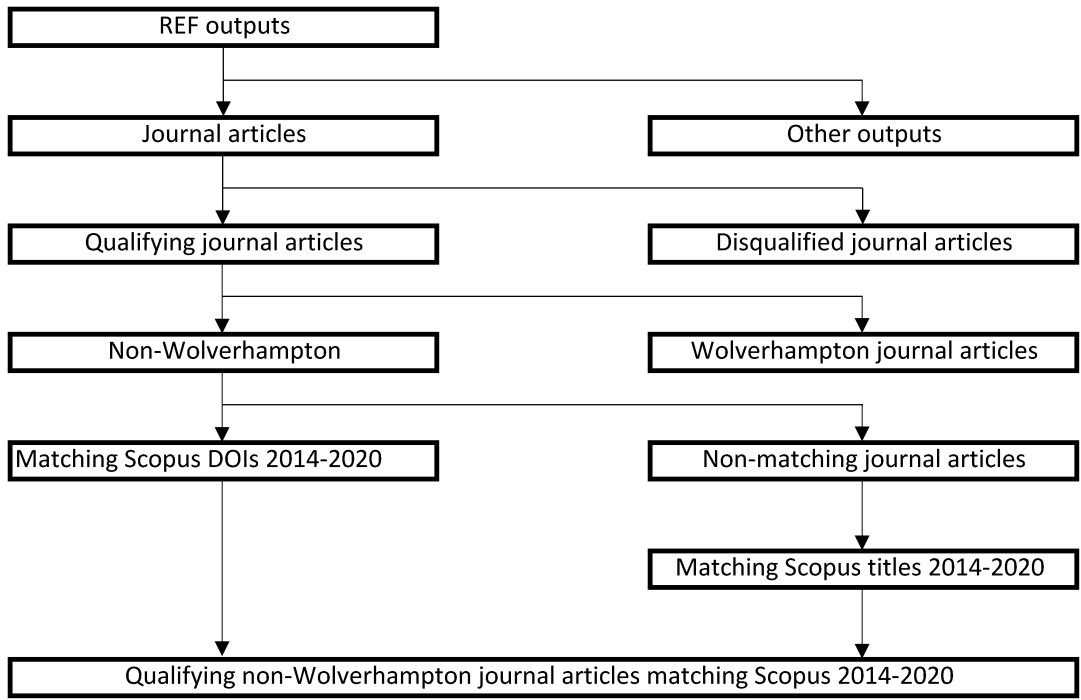

Figure 3.6.1. The process used to select REF2021 journal articles for the AI experiments.

The numbers of articles obtained after the various selection steps are summarised in Table 3.6.1. The table includes extra stages used to generate a more predictable set of articles from within the main set. This more predictable set, as justified by the experiments below, consists of articles with substantial abstracts published 2014-18.

Table 3.6.1. Descriptive statistics for creation of the experimental dataset. Effective percentages assume that predictions for duplicate articles will be applied to all copies. Duplicates were removed only within UoAs (for other output types, see: https://ref.ac.uk/results-analysis/output-profiles/).

| Set of articles | Journal articles |
| --- | --- |
| All REF2021 outputs of all types (e.g., 28,699 books or book parts). | [185,594] |
| All REF2021 journal articles. | 152,367 |
| REF2021 journal articles supplied. | 148,977 |
| With DOI. | 147,164 (98.8%) |
| With DOI and matching Scopus 2014-20 by DOI. | 133,218 (89.4%) |
| Not matching Scopus by DOI but matching with Scopus 2014-20 by title. | 997 (0.7%) |
| Not matched in Scopus and excluded from analysis. | 14,762 (9.9%) |
| All REF2021 journal articles matched in Scopus 2014-20. | 134,215 (90.1%) |
| All REF2021 journal articles matched in Scopus 2014-20 except score 0. | 134,031 (90.0%) |
| All REF2021 journal articles matched in Scopus 2014-20 except score 0 and except articles with less than 500 character cleaned abstracts. | 130,009 (87.3%) |
| All non-duplicate REF2021 journal articles matched in Scopus 2014-20 except score 0. | 122,331 [90.0% effective] |
| All non-duplicate REF2021 journal articles matched in Scopus 2014-20 except score 0 and except articles with less than 500 character cleaned abstracts. | 118,527 [87.3% effective] |
| All non-duplicate REF2021 journal articles matched in Scopus **2014-18** except score 0. These are the most accurate prediction years. | 87,739 [64.6% effective] |
| All non-duplicate REF2021 journal articles matched in Scopus **2014-18** except score 0 and except articles with less than 500 character cleaned abstracts. These are the most accurate prediction years. | 84,966 [62.6% effective] |

The 14,762 journal articles not matched in Scopus were logically either not in Scopus, in Scopus with a different publication year (e.g., 2021) or in Scopus with a different title and no matching DOI. A few missing articles had non-matching DOIs (e.g., for preprint servers) but most articles seemed to be not in Scopus. Articles matched in Scopus were more likely to be high scoring than articles not matched in Scopus (Table 3.6.2), but there were still substantial numbers of top scoring articles not matched in Scopus.

Checks of the 184 articles with scores of 0 suggested that these were sometimes due to author ineligibility (e.g., this was implied when other authors had high scores for the same article) or article type ineligibility (i.e., the article could reasonably be classified as not being research or being a review). Primarily to deal with the former issue, all 184 articles scoring 0 were excluded. This exclusion decision must be a human judgement in the future. The final dataset for the AI contained 134,031 (90.0%) of the qualifying non-Wolverhampton REF2021 journal articles. The analyses were repeated after excluding articles with short abstracts, and for years separately, with articles from 2014-18 merged eventually forming the main data set.

Table 3.6.2. Scores for articles matching and not matching a Scopus record 2014-20.

| Score | All articles | Matching Scopus | Not matching Scopus | Matching Scopus (%) |
|---|---|---|---|---|
| 0 | 318 | 184 | 134 | 57.9% |
| 1* | 2041 | 1407 | 634 | 68.9% |
| 2* | 22312 | 18855 | 3457 | 84.5% |
| 3* | 74067 | 67279 | 6788 | 90.8% |
| 4* | 50239 | 46490 | 3749 | 92.5% |
| All | 148977 | 134215 | 14762 | 90.1% |

For context, books and book parts had a much higher proportion of 4* ratings (48%) than did journal articles (43%) (Figure 3.6.2).

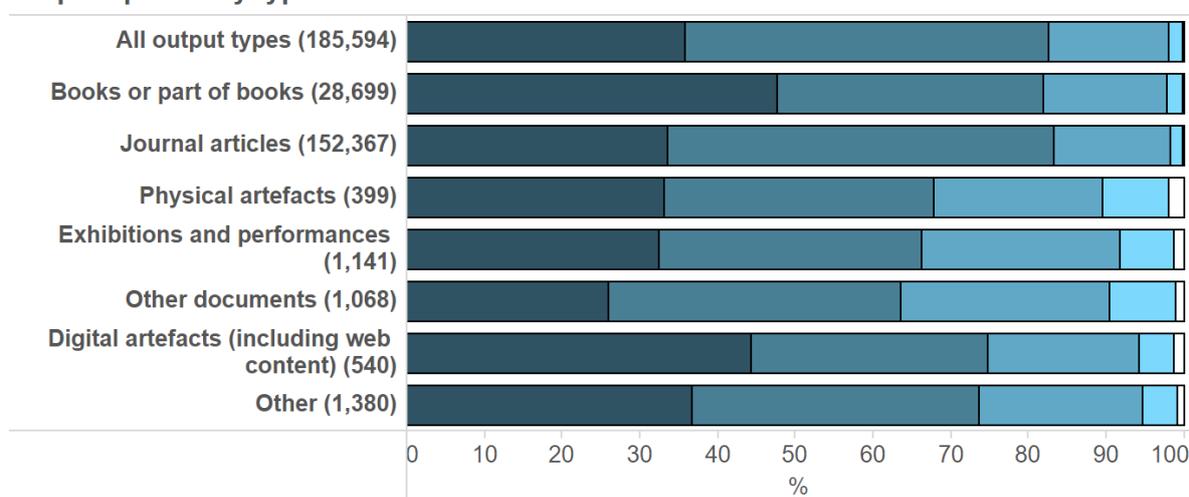

Note: Any outputs requested as double weighted are counted twice in this chart. The number of outputs in each category is shown in brackets.

Figure 3.6.2. Score profiles for different REF2021 output types (https://ref.ac.uk/results-analysis/output-profiles/). Only journal articles were available for the AI experiments.

## 3.7 Algorithms used to predict output scores

A wide range of classification and regression algorithms was tested (Table 3.7.1), based on preliminary tests with different data (Thelwall, 2022) and adding an additional classifier after subsequent extra testing. Since classification algorithms ignore the order of the scores, ordinal classification variants were created by combining two separate classification tasks (1*-3* vs. 4* and 1*-2* vs. 3*-4*) to deduce the likely overall classification (1*-2*, 3* or 4*). In theory, this should produce more accurate results from the classification algorithms although this was not the case in our preliminary tests with different data (Thelwall, 2022).

Gradient Boosting Classifier and Extreme Gradient Boosting Classifier were expected to provide the most accurate results, based on additional preliminary testing with different data, but all algorithms were assessed for completeness. All algorithms used their default parameters.

Most of the algorithms can be tuned by adjusting a range of input parameters to adjust the way in which they learn from data. These input parameters are known as hyperparameters. Different hyperparameter values were tested for the best algorithms to find the optimal configurations, in case this could improve accuracy. 5-fold cross validation hyperparameter tuning was used for this on the training set, to guard against overfitting. Preliminary testing on different data suggested that this would not work, but it was tried anyway and worked better than expected.

Table 3.7.1. Machine learning methods chosen for regression and classification. Those marked with /o have an ordinal version of the classification.

| Code | Method | Type |
|---|---|---|
| **bnb/o** | Bernoulli Naive Bayes | Classifier |
| **cnb/o** | Complement Naive Bayes | Classifier |
| **gbc/o** | Gradient Boosting Classifier | Classifier |
| **xgb/o** | Extreme Gradient Boosting Classifier | Classifier |
| **knn/o** | k Nearest Neighbours | Classifier |
| **lsvc/o** | Linear Support Vector Classification | Classifier |
| **log/o** | Logistic Regression | Classifier |
| **mnb/o** | Multinomial Naive Bayes | Classifier |
| **pac/o** | Passive Aggressive Classifier | Classifier |
| **per/o** | Perceptron | Classifier |
| **rfc/o** | Random Forest Classifier | Classifier |
| **rid/o** | Ridge classifier | Classifier |
| **sgd/o** | Stochastic Gradient Descent | Classifier |
| ~~**svc/o**~~ | ~~Support Vector Classification~~ | ~~Classifier *~~ |
| **elnr** | Elastic-net regression | Regression |
| **krr** | Kernel Ridge Regression | Regression |
| **lasr** | Lasso Regression | Regression |
| **lr** | Linear Regression | Regression |
| **ridr** | Ridge Regression | Regression |
| **sgdr** | Stochastic Gradient Descent Regressor | Regression |
| ~~**svr**~~ | ~~Support Vector Regression~~ | ~~Regression **~~ |

*Almost the same results as lsvc for the dummy data and so redundant for the real data.
**Inaccurate and slow in all tests with 1000 features for the dummy data and so not recommended for the real data.

Deep learning methods (e.g., convolutional neural networks) were *not* used because dataset sizes are not large enough to build the typical complex deep learning models, and there are no clearly effective deep learning network architectures for this type of task. Thus, developing an effective deep learning model would be time consuming and uncertain. Whilst there are some good results for citation prediction tasks using deep learning methods (Abrishami & Aliakbary, 2019; Xu et al., 2019), these work on narrow fields and rely on complete citation data, whereas REF scores are incomplete (a sample of the UK part of a field). A promising approach is SchuBERT (van Dongen et al., 2020) word embedding, but it seems unlikely to scale to the very broad fields of UoAs.

      Overall, the most accurate methods were usually rfc, gbc, and xbc and their ordinal variants rfco, gbco, and xbco. They had the highest accuracy above the baseline, on average across all datasets tested (Figure 3.7.1) and one of these was always the most accurate single method on the UoAs with the highest prediction accuracy. Thus, these six algorithms are the only serious candidates out of all tested. One other algorithm, xgbr, had a good overall

accuracy (Figure 3.7.1) but did not perform well on the most accurate UoAs and so is not included in the most accurate six algorithms.

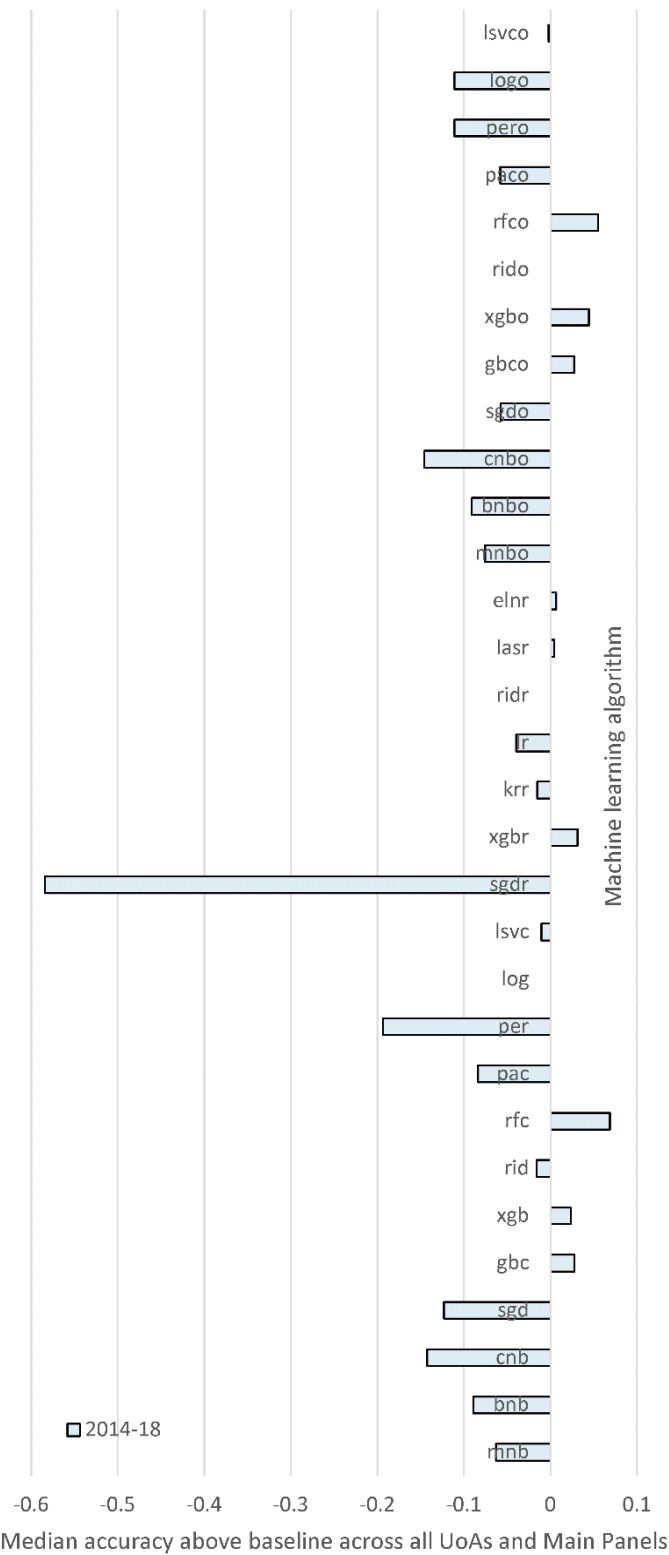

Figure 3.7.1. Median accuracy across the baseline (always predicting the modal score) across all UoAs and Main Panels for the algorithms tested.

## 3.8 Inputs for the predictions

Both text and bibliometric inputs were included as inputs for all the machine learning algorithms.

### 3.8.1 Bibliometric inputs for the predictions

The AI inputs were chosen for being relatively straightforward and proven to work on some fields, based on the literature review. They are split into three input sets for testing, listed below in order of increasing complexity. The purpose of testing different input sets is to allow decisions to be made by the sector about whether the additional accuracy given by the inclusion of more controversial or complex inputs is enough for them to be included.

Intuitively, longer documents may generate higher REF scores because they have more content. There may also be some short form articles submitted (letters, communications) that represent relatively minor contributions to scholarship. Thus, document length is a potential input for machine learning.

Document length is not a straightforward input for several reasons. Without article full text, page numbers were used as a proxy for article length, even though these are only approximate indicators of length due to differing journal page sizes (A4, A5) and font sizes. Another complicating factor is that some apparently prestigious journals have short maximum lengths, often with extensive supplementary materials containing essential details. This seems to be the reason why articles with 5 pages are the most likely to receive a 4* rating in UoA 1, Clinical Medicine. In UoAs 7, 8 and 9, shorter articles also tend to score higher. UoA 2 works in the opposite direction, with longer articles tending to score higher (UoAs 1,2,7,8,9 had the highest prediction accuracy so are illustrated in Figures 3.8.1.1 to 3.8.1.5). Article lengths from Scopus were included, with missing data (e.g., for electronic-only articles) replaced by the median for the UoA or panel (this slightly improved predictions compared to using zeros for articles with missing page numbers).

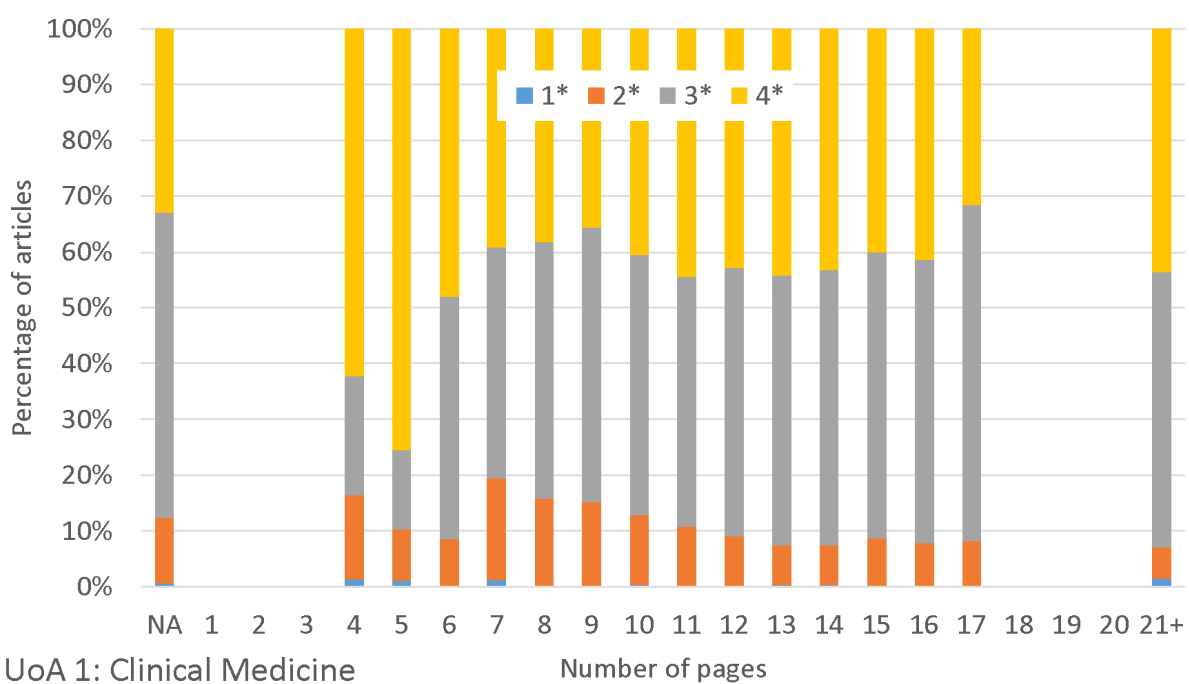

Figure 3.8.1.1. Distribution of REF 2021 provisional scores by number of pages, as reported in Scopus for UoA 1. Bars with fewer than 30 articles are not shown.

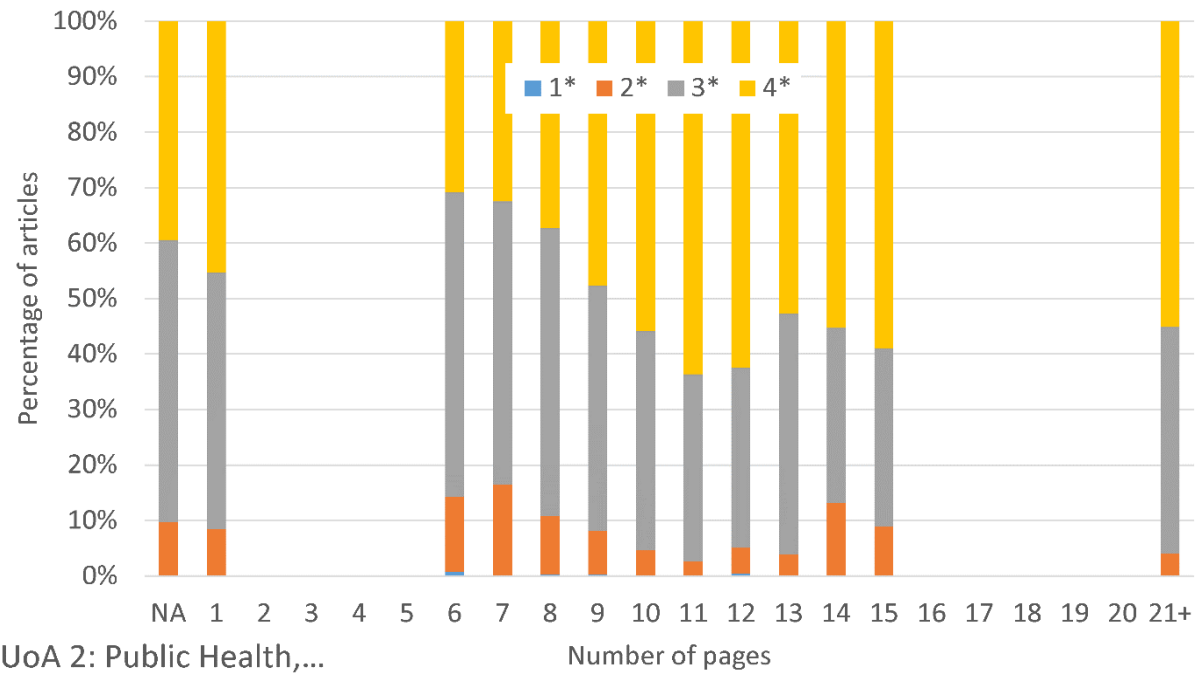

Figure 3.8.1.2. Distribution of REF 2021 provisional scores by number of pages, as reported in Scopus for UoA 2. Bars with fewer than 30 articles are not shown.

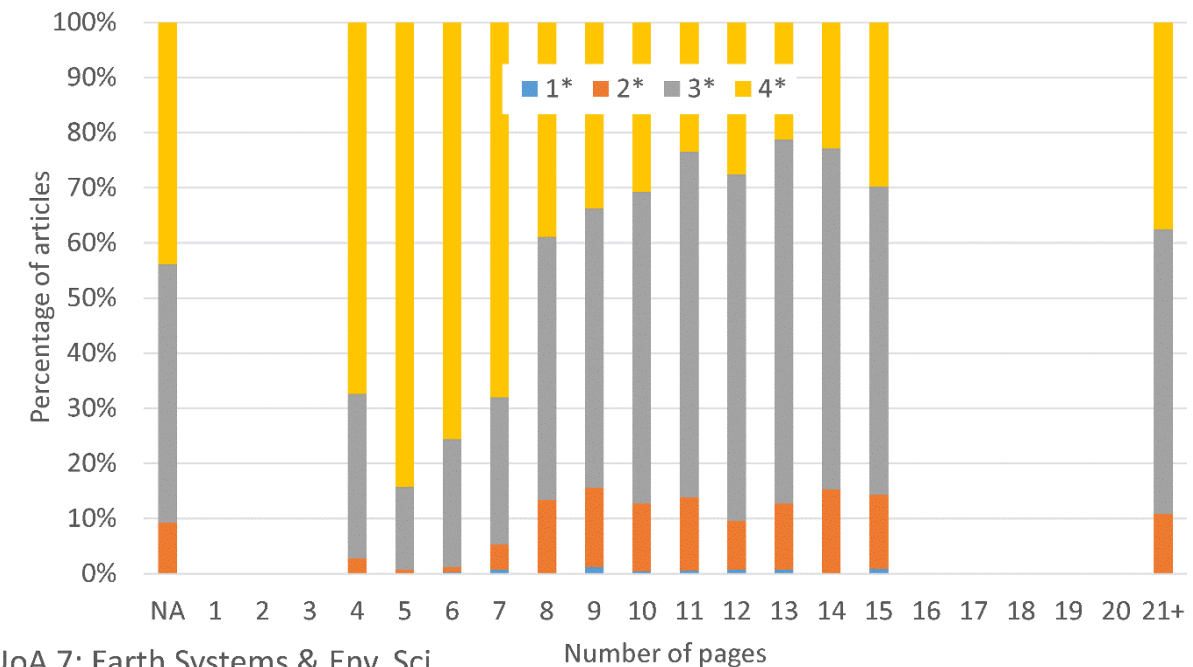

Figure 3.8.1.3. Distribution of REF 2021 provisional scores by number of pages, as reported in Scopus for UoA 7. Bars with fewer than 30 articles are not shown.

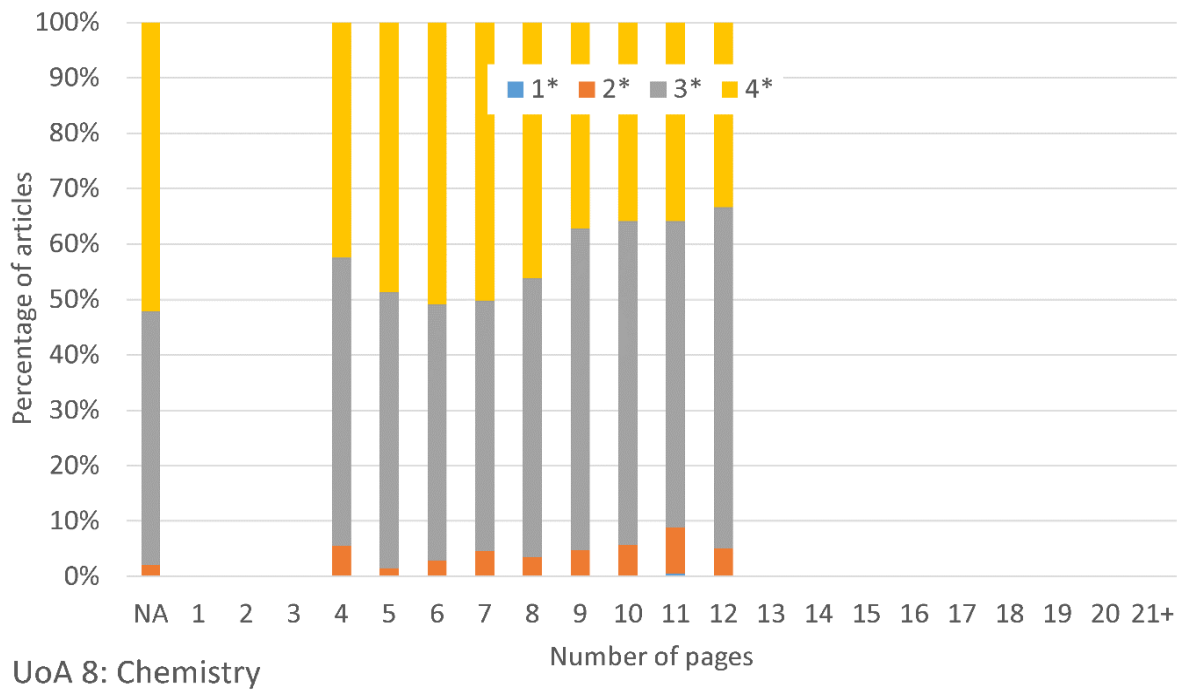

UoA 8: Chemistry

Figure 3.8.1.4. Distribution of REF 2021 provisional scores by number of pages, as reported in Scopus for UoA 8. Bars with fewer than 30 articles are not shown.

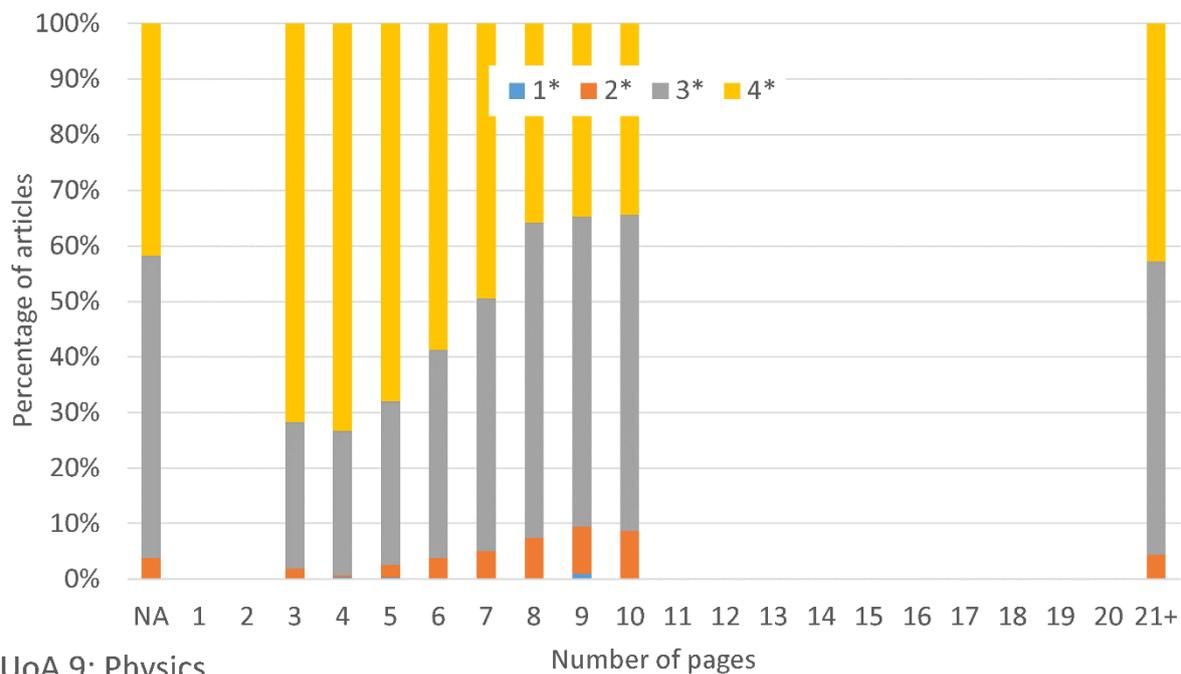

UoA 9: Physics

Figure 3.8.1.5. Distribution of REF 2021 provisional scores by number of pages, as reported in Scopus for UoA 9. Bars with fewer than 30 articles are not shown.

Journal ranking indicators were *excluded* from the first input set because they may be regarded as unacceptable by some interpretations of DORA[12] by partly evaluating researchers on their publishing journal and DORA is supported by UKRI. As argued above, however, it seems reasonable to include journal-level indicators on the basis that they are only one input and researchers are not being evaluated solely on them. Journal-level indicators might also penalise researchers publishing in new journals. Nevertheless, based on the substantial statistical score homogeneity for journals in some fields found above, and the preliminary statistical tests, journal indicators are expected to be powerful in some fields, especially for newer articles (e.g., Levitt & Thelwall, 2011), so two of the three input sets include them.

---

[12] https://www.ukri.org/our-work/supporting-healthy-research-and-innovation-culture/research-integrity/

**Input Set 1 (bibliometrics): Fully DORA compliant bibliometric input set, for all journal articles in Scopus.**
- **Citation counts** (field and year normalized to allow parity between fields and years, log transformed to reduce skewing to support linear-based algorithms). For the previous REF, the HEFCE analysis showed that in many UoAs, more cited articles tended to get higher REF scores. The Normalised Log-transformed Citation Score (NLCS) was used, field normalised by Scopus narrow field (e.g., 330 narrow fields in 2018). When averaged for institutions, it is known as the Mean Normalised Log-transformed Citation Score (MNLCS) (Thelwall, 2017).
- **Number of authors** (log transformed to reduce skewing). Articles with more authors tend to be more cited, so they are likely to also be more highly rated.
- **Number of institutions** (log transformed to reduce skewing). Articles with more institutional affiliations tend to be more cited, so they are likely to also be more highly rated.
- **Number of countries** (log transformed to reduce skewing). Articles with more country affiliations tend to be more cited, so they are likely to also be more highly rated.
- **Number of Scopus-indexed journal articles of the first author** during the REF period (log transformed to reduce skewing). More productive authors tend to be more cited, so this is a promising input.
- **Average citation rate of Scopus-indexed journal articles by the first author** during the REF period (field and year normalized, log transformed: the MNLCS). Authors with a track record of highly cited articles may also write higher quality articles. Note that the first author may not be the REF submitting author or from their institution because the goal is not to "reward" citations for the REF author but to predict the score of their article.
- **Average citation rate of Scopus-indexed journal articles by any author** during the REF period (maximum) (field and year normalized, log transformed: the MNLCS). Authors with a track record of highly cited articles may write higher quality articles.
- **Number of pages of article, as reported by Scopus, or the UoA/Main Panel median if missing from Scopus.** Longer papers may have more content, but short papers may be required by more prestigious journals.
- **Abstract readability**. There is evidence of a slight positive correlation between abstract readability and citation counts, but not all articles have abstracts (including some short form articles in medicine, plus some humanities style articles), so the likely slight gain in prediction accuracy for REF scores may not be worth the additional complexity for abstract-free articles. Nevertheless, abstract readability (Flesch-Kincaid grade level score) was included since it may have a small benefit.

**Input Set 2 (bibliometrics + journal impact): Bibliometric input set, for all journal articles in Scopus.**
All of the above factors plus the following measure of journal impact.

- **Journal citation rate** (field normalized, log transformed [MNLCS], based on the current year for older years, based on 3 years for 1-2 years' old articles).

Field normalised journal citation rates were used for Input set 2 instead of Journal Impact Factors (JIFs) because these align better with human journal rankings (Haddawy et al., 2016), probably because they are comparable between disciplines.

### 3.8.2 Text inputs for the predictions

The third input set adds text mining on the abstracts, keywords, and titles to the Input Set 2 information. Text mining for score prediction is likely to leverage hot topics in constituent fields (e.g., Hu et al., 2020; Thelwall & Sud, 2021), mainly predicting scores based on topic rather than quality. It might also identify the names of powerful statistical methods, however (Thelwall, 2015), suggesting higher quality. Hot topics in some fields tend to be highly cited and probably have higher quality articles, as judged by peers. Even the more stable arts and humanities-related UoAs are mixed with social sciences and other fields (e.g., computing technology for music), so text mining may still pick out hot topics within these UoAs. Whilst topics easily translate into obvious and common keywords, research quality has unknown and probably field dependant translation into research quality (e.g., "improved accuracy" [computing] vs. "surprising connection" [humanities]). Thus, text-based predictions of quality are likely to leverage topic-relevant keywords as indirect indicators of quality rather than more subtle textual expressions of quality. It is not clear whether input sets that include both citations and text information would leverage hot topics from the text, since the citations would point to the hot topics anyway, combining citations and text for machine learning may bypass the topic detection issue and will be tried. Similarly, AI applied to REF articles may identify the topics or methods of the best groups and learn to predict REF scores from them, which would be accurate but undesirable.

There are many ways of converting text into inputs for machine learning and counts of word unigrams, bigrams and trigrams were used: words, two word phrases and three word phrases. Since previous research has shown that text mining partly leverages journal styles, it has the same potential DORA issues as Input Set 2 and it therefore would not make sense to remove the journal impact indicator in an attempt to make it DORA compliant.

Since there are far more words, two word phrases and three word phrases than can be usefully exploited by machine learning with the training set sizes available, a method to select the most useful ones was included, with a default setting of 1000 inputs (features) overall, based on our preliminary testing with a different dataset. Different feature set sizes in the initial stages were also tested, however, and selected a different size when this substantially improved accuracy.

To get insights into what type of information was helping the score predictions, the apparently most powerful inputs were also checked. For example, in our preliminary tests of text-based citation prediction, in Materials Chemistry the term "graphene" was 5.8 times more likely to appear in a highly cited article than in a less cited article, suggesting that topic was partly driving the predictions. We examined the text features for each UoA to identify the overall patterns and anomalies (reported below). For example, perhaps metadata errors or omissions associate with incorrect predictions, or some topics or styles tend to generate bad predictions.

**Input Set 3 (bibliometrics + journal impact + text): Full input set, for all journal articles in Scopus or for all articles in Scopus with an abstract of at least 500 characters after text cleaning.**

This input set includes Input Set 2 plus title, keywords, and abstract text. This input set was expected to be the most accurate, based on the preliminary study.

- **Title, keywords, and abstract word unigrams, bigrams, and trigrams** within sentences. This means words and phrases of 2 or 3 words. From each set of articles, this is likely to generate hundreds of thousands of different inputs and feature selection was used (chi squared) but not feature compression (e.g., SVM) since our initial testing suggests that feature compression does not give an advantage and it makes the relative contributions of the inputs opaque. Different numbers of features were tested (e.g., 500, 1000), as selected by the chi squared method, but with 1000 as the initial default. More details are in the next sub-section.
- **Journal names**. On the basis that journals are key scientific gatekeepers and that a high average citation impact does not necessarily equate to publishing high quality articles, journal names were included as factors in input set 3. Testing with and without journal names suggested that their inclusion tended to slightly improve accuracy.

### 3.8.2.1 Number of text inputs to use

There can be hundreds of thousands of text inputs for the machine learning, but faster more accurate methods can be expected if they are not fed with all available inputs but with an optimal subset. This subset was chosen with the chi square feature selection method (applied to each training set generated), which selects inputs that differentiate best between article scores. It was not originally clear how many inputs were optimal, however, or even if the text inputs were beneficial. To test this, input set sizes of 9 (Bibliometrics only), 10 (Bibliometrics + Journal impact), 500 (with an additional 490 text features), 1000, 2000,…10,000 were tested with the six most accurate algorithms (see below) on the 12 UoAs with the highest accuracy rates (see below) to identify the optimal number of input features. The number 1000 was chosen after inspection of the results because it usually gave the optimal accuracy for rfc and rfco, and close to optimal accuracy for xgb, xgbo, gbc, and gbco. The graphs below illustrate this decision. The reason for choosing a universal feature set size rather than different sizes for each UoA or method is to avoid getting optimistic accuracy estimates through overfitting.

Figures Figure 3.8.2.1.1- Figure 3.8.2.1.4 illustrate the typical patterns found for the top six algorithms in terms of the relationship between the number of inputs and the size of the training set. For training set sizes of 5%, 10% and 15%, Figure 3.8.2.1.1 shows that for xgb (xgbo, gbc, and gbco are similar), there are clear increases in accuracy from 9 inputs (bibliometrics) to 10 inputs (bibliometrics + journal impact) and that the text inputs increase accuracy further. For xgb, increasing the number of input features tends to increase accuracy overall. In contrast, Figure 3.8.2.1.2 shows that increasing the number of input features can decrease overall accuracy for rfc after about 1000 features. Analysing the graphs for a wide range of training set proportions shows that similar patterns occur for different training set percentages, so 1000 is a reasonable overall input feature set size that performs well for all the main algorithms at low percentages of training data.

The larger graphs (Figure 3.8.2.1.3, Figure 3.8.2.1.4) include the same information as the smaller graphs, but for 19 different training set sizes, from 5% to 95%. These show the same pattern in terms of accuracy as a function of input feature set size, except that the accuracy difference between feature set sizes over 500 tends to be a bit smaller. For other UoAs, larger input feature sets sometimes associate with slightly lower accuracy. Overall, however, an input feature set size of 1000 seems reasonable for comparing between machine learning algorithms because it works well for all, but an input feature set size of 5000 tends to be 0.1% to 0.6% better for xgb, xgbo, gbc and gbco for most of the higher accuracy UoAs (Table Figure 3.8.2.1.1). Graphs for other UoAs and machine learning methods are available online. There isn't a clear advantage of using more than 5000 input features, however.

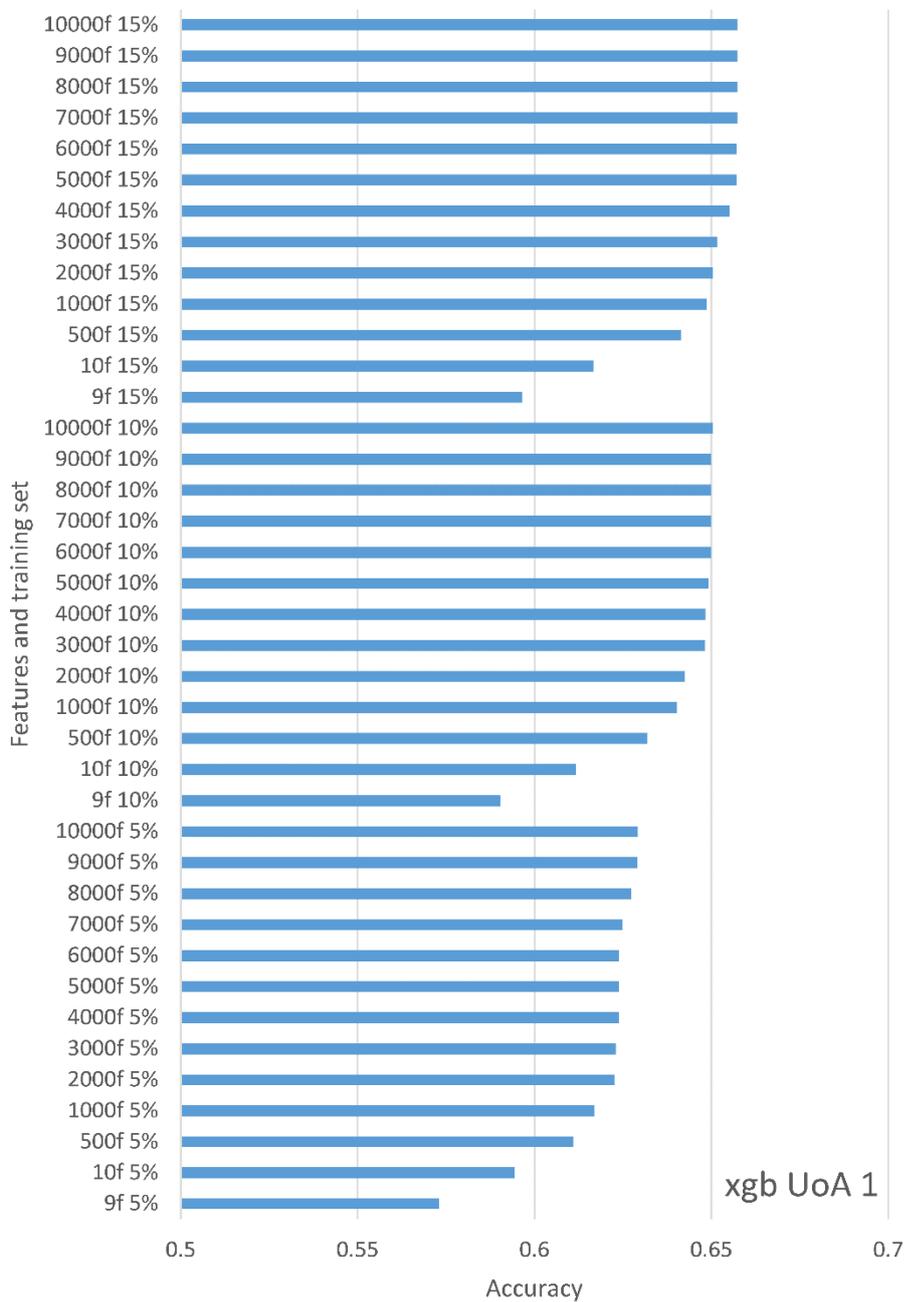

Figure 3.8.2.1.1. Accuracy of the xgb algorithm on predicting 3 levels of REF scores for articles with 500 character abstracts, with 13 different input feature set sizes (9 features to 10,000 features) and three different training set percentages (5%, 10% and 15%) on UoA1 2014-18. Each bar indicates the average accuracy over 10 attempts.

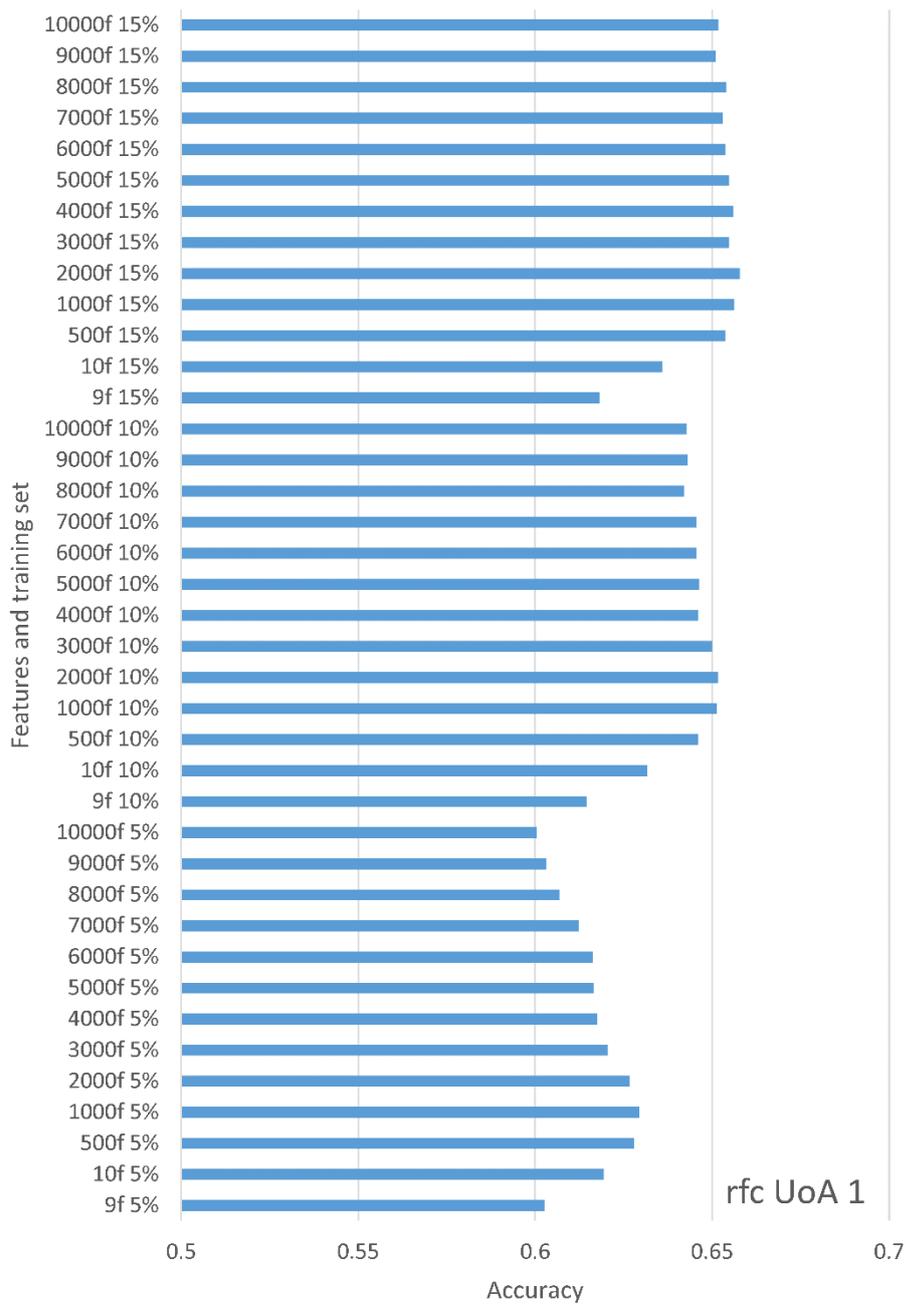

Figure 3.8.2.1.2. Accuracy of the rfc algorithm on predicting 3 levels of REF scores for articles with 500 character abstracts, with 13 different input feature set sizes (9 features to 10,000 features) and three different training set percentages (5%, 10% and 15%) on UoA1 2014-18. Each bar indicates the average accuracy over 10 attempts.

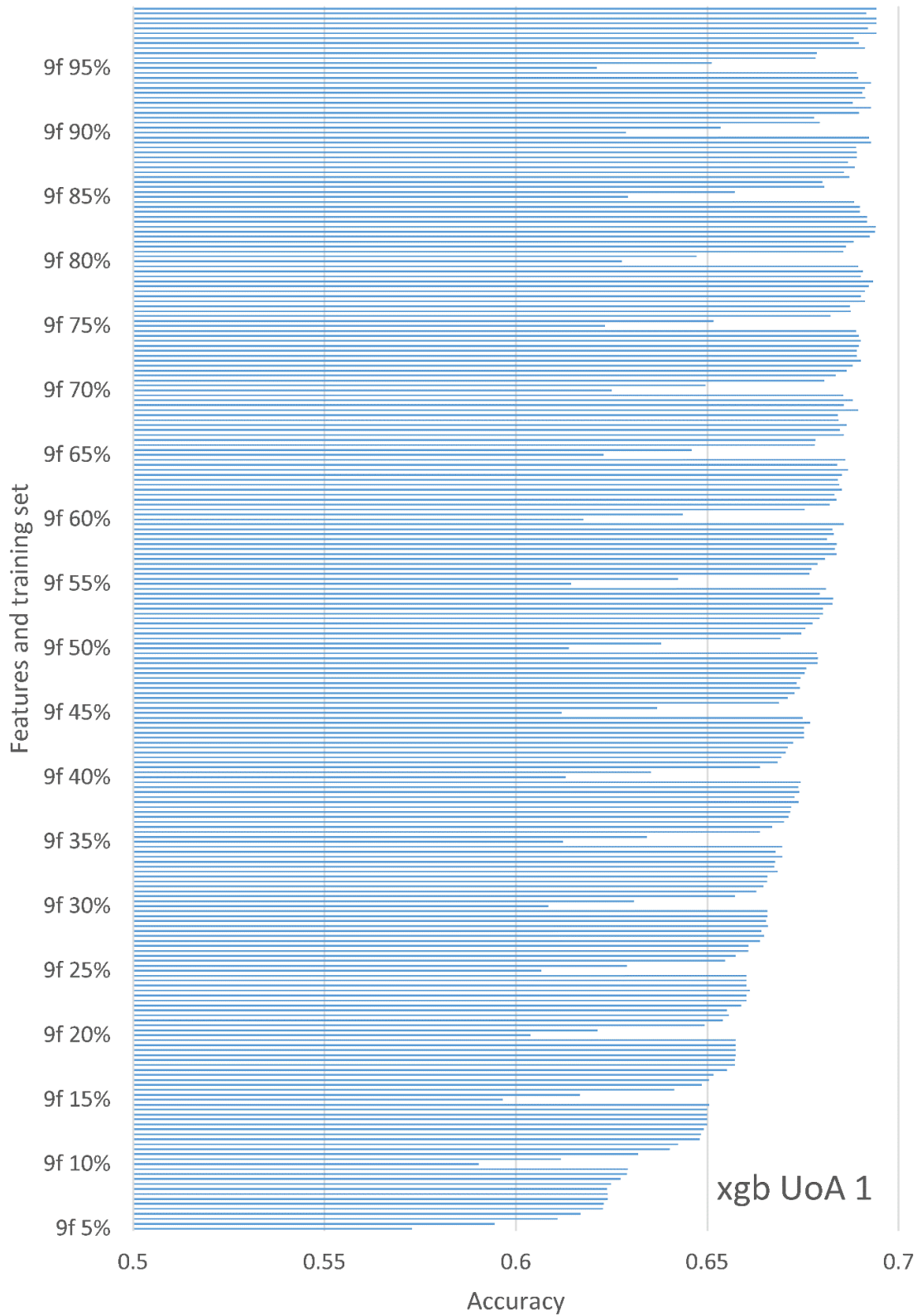

Figure 3.8.2.1.3. Accuracy of the xgb algorithm on predicting 3 levels of REF scores for articles with 500-character abstracts, with 13 different input feature set sizes (9 features to 10,000 features) and 19 different training set percentages (5%, to 95%) on UoA1 2014-18. Each bar indicates the average accuracy over 10 attempts. The patterns are more ragged for high percentages due to a smaller test set of articles to evaluate accuracy with.

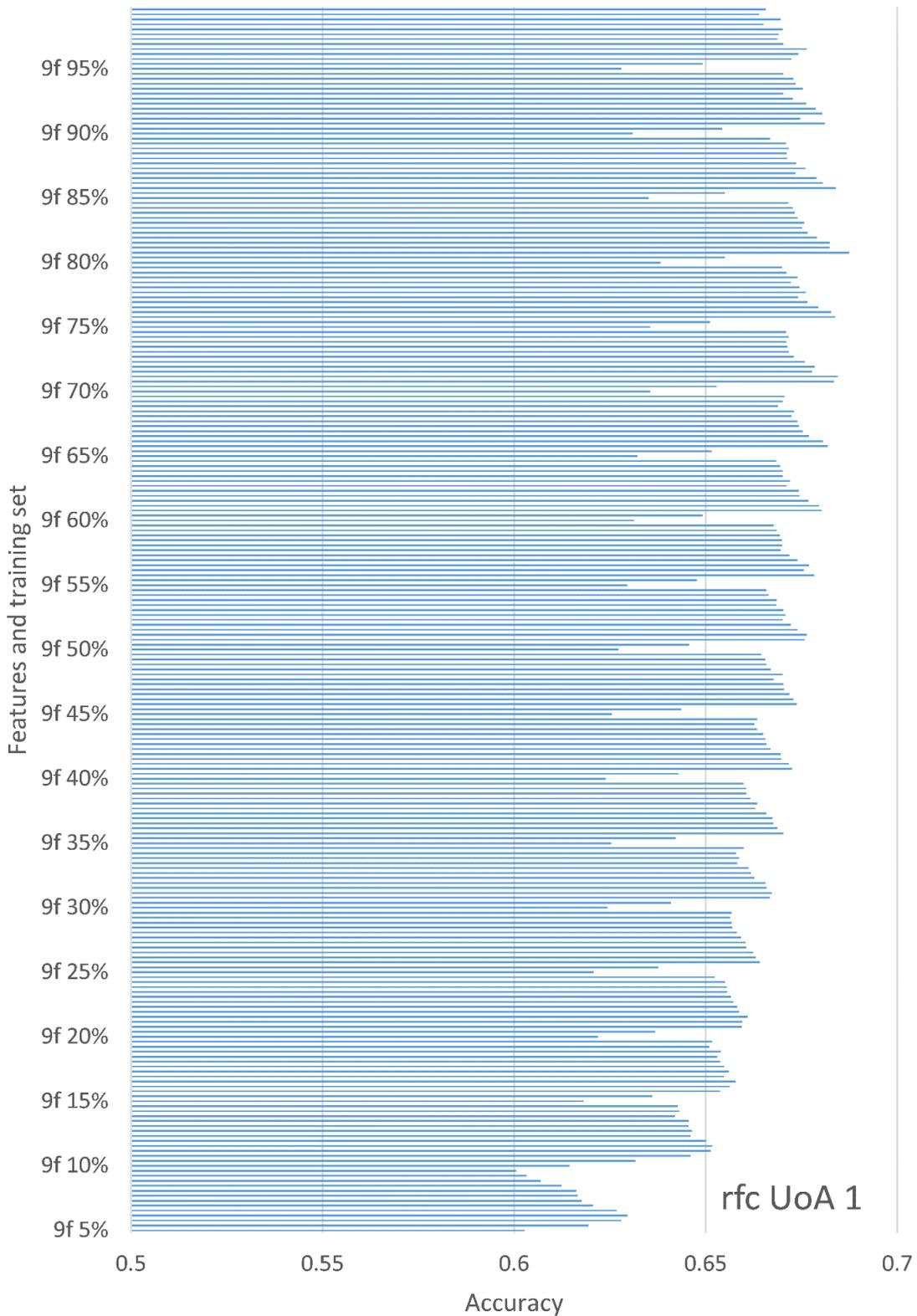

Figure 3.8.2.1.4. Accuracy of the rfc algorithm on predicting 3 levels of REF scores for articles with 500-character abstracts, with 13 different input feature set sizes (9 features to 10,000 features) and 19 different training set percentages (5%, to 95%) on UoA1 2014-18. Each bar indicates the average accuracy over 10 attempts.

Table 3.8.2.1.1. Average accuracy increases when using 5000 input features instead of 1000 for UoAs 1-11 and 16 by method and training set size, based on the average of 40 machine learning experiments.

| Method | Training set size | Minimum increase | Maximum increase | Average increase |
|---|---|---|---|---|
| **gbc** | 10% | 0.0% | 1.4% | 0.5% |
| **gbc** | 25% | 0.0% | 1.2% | 0.3% |
| **gbc** | 50% | -0.2% | 1.0% | 0.3% |
| **gbc** | 90% | -0.7% | 1.6% | 0.3% |
| **xgb** | 10% | 0.2% | 1.0% | 0.6% |
| **xgb** | 25% | 0.2% | 1.3% | 0.6% |
| **xgb** | 50% | 0.1% | 1.2% | 0.5% |
| **xgb** | 90% | -0.4% | 1.9% | 0.3% |
| **rfc** | 10% | -3.9% | -0.1% | -1.7% |
| **rfc** | 25% | -5.4% | 0.0% | -1.3% |
| **rfc** | 50% | -2.5% | -0.1% | -0.8% |
| **rfc** | 90% | -1.9% | 0.2% | -0.7% |
| **gbco** | 10% | -0.3% | 1.3% | 0.3% |
| **gbco** | 25% | -0.4% | 1.0% | 0.2% |
| **gbco** | 50% | -0.4% | 1.2% | 0.1% |
| **gbco** | 90% | -0.7% | 1.8% | 0.1% |
| **xgbo** | 10% | 0.0% | 1.0% | 0.5% |
| **xgbo** | 25% | 0.2% | 1.3% | 0.5% |
| **xgbo** | 50% | -0.2% | 1.1% | 0.4% |
| **xgbo** | 90% | -0.5% | 1.3% | 0.4% |
| **rfco** | 10% | -4.0% | -0.9% | -2.3% |
| **rfco** | 25% | -6.7% | -0.4% | -1.9% |
| **rfco** | 50% | -4.6% | -0.2% | -1.4% |
| **rfco** | 90% | -2.0% | 0.1% | -1.1% |

### 3.8.3 Inputs not used for predictions

Many other inputs were considered but not used. These include citation indicators that were considered to be suboptimal compared to variables already included, such as citation counts and Journal Impact Factor variants from different publishers. Altmetrics were also not included because of their mostly low correlations with REF2014 scores (limiting their usefulness) and their ease of gaming, which makes their use in this context irresponsible. The Mendeley Reader count had a substantial positive correlation with REF scores in the HEFCE correlation exercise so is the most logical to include but is easily gameable: universities could legitimately encourage academics to join Mendeley and share each other's REF work on it (e.g., routinely or for mock REFs).

**The number of Scopus-indexed journal articles of any author** (maximum) (log transformed to reduce skewing) had been initially proposed as an input because more productive authors tend to be more cited, so this is a promising input. This proposed input was withdrawn after the initial statistical analysis found that it did not contribute to the predictive power of any models.

Some full text mining methods were not used for various reasons. Article novelty, as measured by the difference between the words in the article and the words in its references (Yan et al., 2012; Zhu & Ban, 2018) seems to be a reasonable indicator of an aspect of research quality, but it was impractical to obtain text minable full text copies of all references of all articles in the dataset. Article topics have also been proposed for citation prediction based on helping to differentiate between more cited and less cited topics (Yan et al., 2012; Zhu & Ban, 2018) but this was partly covered by title, keyword, and abstract terms. Similarly, article breadth, in the sense of covering different topics and hence potentially being cited from a wide range of fields (Yan et al., 2012; Zhu & Ban, 2018), is also more relevant to citations than article quality. These last two topic-based inputs also require access to full text versions of articles from across science and beyond the REF submissions, and therefore require relatively comprehensive machine-readable full texts for articles (e.g., as the PubMed Central full text is). An alternative approach is to build a network of the relationships between each article and its references and citations, including those of its authors, and use various network features to predict its citations (Xu et al., 2019). This is more appropriate for predicting citations than article quality, however, and needs access to large collections of texts and citing-cited relationships. Finally, text mining peer review is promising (Li et al., 2019; Thelwall, et al., 2020), but open peer review reports seem to be available for too few articles to be a valuable input for a REF-based exercise yet.

The numbers of figures, tables and equations have also been proposed as machine learning inputs for citation or conference paper acceptance prediction (Elgendi, 2019; Joshi et al., 2021), presumably on the basis that each figure and table tends to represent a condensed summary of information, so a paper with more of these tend to report more work. More equations may also represent a more complex formulation of a problem, although this is less clear. There is limited evidence for these and presumably they will not be useful in less empirical fields. Moreover, some journals have strict limits on the number of figures or tables, leading to individual figures sometimes containing dozens of different images. Thus, it is not clear whether these will work at all as indicators of article quality. Some of these were tested, as discussed next.

### 3.8.4 Inputs extracted from article full texts

The full text of 59,194 ref-submitted articles was very kindly supplied from the CORE (https://core.ac.uk/) repository of open access papers (Knoth & Zdrahal, 2012) by Prof Petr Knoth, Maria Tarasiuk and Matteo Cancellieri (http://bsdtag.kmi.open.ac.uk/). These full texts were from various online institutional and subject open access repositories, such as White Rose Research Online (3356 articles) and arXiv (2763) as well as 1194 URLs from CrossRef. Through their DOIs, these matched 43.3% of the main set of articles analysed below, reducing the main sample size (articles with abstracts, excluding 0 scores) from 84,966 to 36,790.

The full texts were plain text files apparently typically extracted from PDF files in the repositories. They often contained publisher copyright statements, repository copyright statements, repository information, line numbers, and page numbers in addition to the article texts. This extra text substantially increased the word count document length metric and moderately increased the character count document length metric in affected cases. The inclusion of line numbers within many files, either at the start or end of lines, increased the error rate for identifying the number of tables and figures in each paper, as inputs for the machine learning. It was also not possible to accurately count the number of references in each document due to the varied positions and formats of references in the full text files. The full text files were also sometimes supplementary materials, acceptance letters from the editor, or partially scanned documents, rather than the full original papers. This presumably reflects various repositories not clearly labelling papers versions correctly, uploading incorrect documents or uploading multiple documents to the same page.

The experiments described below were repeated with four additional features: word count; character count; figure count; and table count. These were applied to a reduced set of 36,790 articles, which consists of the subset of the main dataset analysed that matched the full test. Machine learning accuracy was slightly worse with the extra full text features, however, so the results are not reported. For example, no accuracy exceeded 70%. The reason for the lower accuracy with full text features may be the smaller amount of training data offsetting the small advantage the extra information may have provided in some fields. Full text mining may be more powerful in some fields in the future if a solution is found to the need to extract simple and complete article full text for this task. We also tested whether the presence of data sharing/availability statements, appendices or supplements associated with quality, but none did in any UoA.

Figure 3.8.4.1 shows the low correlations between REF scores and arguably the most promising full text input: the number of figures in a paper. It is statistically significant in two UoAs (3, 16) and no main panel, but marginal positives are to be expected from the large number of tests even if there is no underlying correlation in any. After a Bonferroni correction to compensate for multiple tests, neither correlation is statistically significant. The correlation is weak in all UoAs and almost always below 0.05.

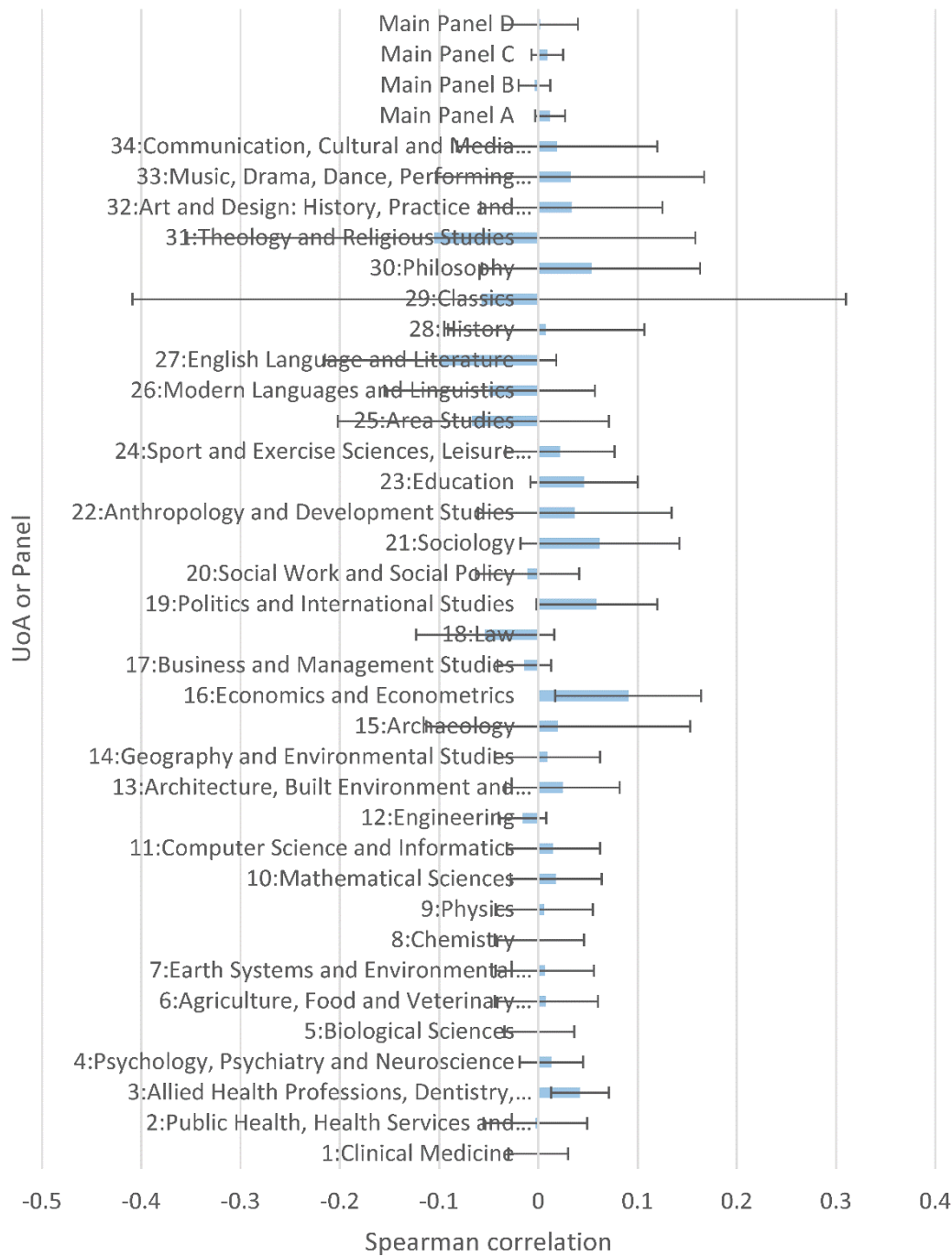

Figure 3.8.4.1. Spearman correlation between the number of figures and REF quality score by UoA or Main Panel.

### 3.8.5 Overfitting protection steps

Both statistical and machine learning approaches can lead to overfitting by making optimistic predictions that are more accurate than justified by the method. This means that the same method would be less accurate on a different but similar dataset. This occurs if data used to build the model is the same as the data used to evaluate the model, risking the model fitting the peculiarities of the specific data to some extent. Model overfitting was guarded against by testing model accuracy only on holdout data not used to train/build the models.

Model selection overfitting can also occur if many different models and parameters are tested, even if each is evaluated on hold-out data, because the best performing model might also vary between similar datasets. This is harder to control for given that this project aimed to compare a wide range of different strategies. This was guarded against by focusing on general solutions across UoAs and years rather than individualised solutions, and by reporting any identified sources of overfitting. We also guarded against this by starting with a set of methods and inputs identified by research on different data and only deviating from it when strong intuitive or empirical evidence supported its use. A small amount of overfitting is still likely, however, and particularly in reporting the accuracy of the most accurate machine learning method.

### *3.9 Training set sizes and merging UoAs and years*

Machine learning models for predicting REF output scores need to be trained on sub-panel member scores for a subset of REF outputs. AI models with more human-coded inputs (training data) tend to be more accurate so it is an advantage to have as much training data as possible. Nevertheless, a greater amount of training data means more human peer review and less time saving. The experimental work therefore assessed the relationship between training set size and accuracy. In other words, the accuracy of the models built was assessed on the basis that n of the outputs have been manually scored by panel members and the scores for the remainder were automatically assigned by an algorithm trained on the panel member scores, for different values of n. The report summarises the relationship between sample size and accuracy for the main models based on these results.

Based on our preliminary work, 1000 human classified articles were expected to be sufficient to train each algorithm to close to optimal accuracy. Nevertheless, to focus on the primary goal of the study the primary testing focused on using 10%, 25% and 50% of the sub-panel provisional REF scores for training the AI methods.

As a secondary step to give more context to the results, tests were run to analyse the effect of training set size on overall accuracy to make recommendations for whether this is acceptable or whether a different size would be optimal. Some UoAs had insufficient journal articles to make this possible. Because the goal of the experiment is to assess whether it is reasonable to replace many human scores by AI predictions, accuracy is also reported for smaller training set sizes to assess the trade-off between the number of human coded papers and overall accuracy.

Merging years and/or UoAs is likely to reduce machine learning accuracy by increasing data heterogeneity. Nonetheless, it may give substantial efficiency advantages by decreasing the total amount of human-coded training data needed, such as 1000 texts for five years combined rather than 1000 texts for each of five years separated. Based on the preliminary work, the first five years were merged as one training set in addition to analysing all years separately.

Merging different UoAs was not expected to be helpful since each one is already relatively broad. Merging all UoAs for each main panel (i.e., four in total for each year range) was tested anyway to assess whether it would give reasonable accuracy.

## 4  Predicting REF journal article scores: Results from three strategies

This section discusses and evaluates three ways in which machine learning could be used to score some REF outputs, focusing on the accuracy of the results and any score shifts induced (Figure 4.1).

- **Strategy 1** (**classic machine learning**: human -> AI): Sub-panel members classify a fraction of the REF outputs (e.g., 10%, 25%, 50%) and the scores are used to build a machine learning algorithm to predict the remaining scores.
- **Strategy 2 (three stage:** human -> AI -> human**)**: Sub-panel members classify a fraction of the REF outputs (e.g., 10%, 25%, 50%) and the scores are used to build a machine learning algorithm to predict the fraction of the remaining articles that matches a prediction accuracy threshold (e.g., 85%), then sub-panel members classify the rest.
- **Strategy 3 (active learning**: human -> AI -> human-> AI -> human …) Sub-panel members to classify a fraction of the REF outputs (e.g., 10%), the results are used to build a machine learning algorithm to identify a fraction (e.g., 10%) of the remaining articles that are difficult to predict with AI, sub-panel members classify this extra 10% and repeat this training-classifying cycle until the classification accuracy for the remaining articles exceeds a prediction accuracy threshold (e.g., 85%).

As mentioned above, initial experiments suggested that classification accuracy tended to be higher when articles with abstracts shorter than 500 characters were removed. Combining years 2014-20 gave more reliably accurate results than analysing years separately and combining 2014-18 gave slightly more accurate results than the complete set of years, so this year range was used for the main results. Graphs are available online for different versions.

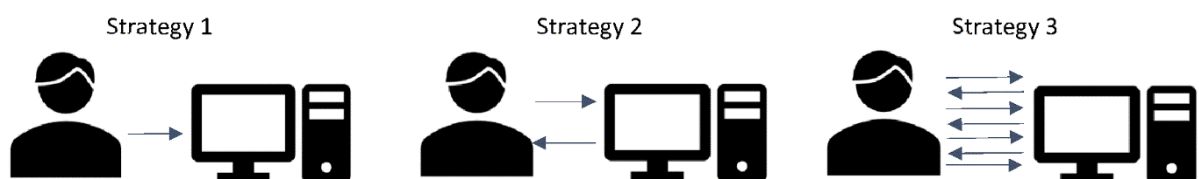

Figure 4.1. The three machine learning strategies.

## *4.1 Strategy 1: Predict scores for a fixed fraction of the articles*

Machine learning with the above parameters was used to investigate how accurately human scores for REF journal articles could be predicted from bibliometric information and text. More specifically the objectives were to estimate the accuracy of 1*-2*, 3*, 4* REF score prediction from text and bibliometrics for each UoA (or combination) and year range using the most accurate AI algorithm(s).

### 4.1.1 Overall accuracy (all articles, excluding score those with REF score 0)

The accuracy of all 31 machine learning methods mentioned above was evaluated on three different training set sizes (10%, 25% and 50%), on three different sets of inputs (bibliometrics, bibliometrics + journal impact, bibliometrics + journal impact + text), on eight different year ranges (2014, 2015, 2016, 2017, 2018, 2019, 2010, 2014-18; with 2014-2020 also sometimes reported) and on 38 different subject areas (UoA 1 to UoA 34, Main Panel A to Main Panel D). For each of the 62,496 combinations of these, the most accurate machine learning method was identified by training ten times on different splits of the data (so 624,960 models were built and tested for this phase).

Since there are too many results to report, a few graphs are included here to illustrate the main trends (all graphs are in a separate online Excel spreadsheet). The first graphs show accuracy above the baseline, where the baseline is giving each article the most common score (either 3* or 4* for most UoAs). The baseline accuracy varies between 42% and 59%, depending on UoA and year. Thus, an accuracy above the baseline of 0 means that the algorithm is as accurate as giving every article the most common score (e.g., 4*).

The attainable level of accuracy above the baseline (always predicting the most common score) varies considerably between UoAs but is not high for any (Figure 4.1.1.1). The accuracy for individual years is variable (due to low training set sizes) but is consistently high within each UoA for the grouped years 2014-18 and 2014-20 with the former tending to give slightly more accurate results than the latter (this is clearer in Figure 4.1.1.2). For the 2014-18 year range, the most accurate UoA/method combination gives 72% overall accuracy, with a large set of UoAs returning 60%-70% accuracy (Figure 4.1.1.3, Figure 4.1.1.4). The accuracy levels are not high, but this is unsurprising since a computer algorithm is attempting to replicate the judgments of world leading experienced field experts with limited information. Whilst the citation inputs may reflect an aspect of the scholarly significance of an output they poorly reflect originality, societal value or rigour (Aksnes et al, 2019).

The patterns are similar, but with generally lower accuracy, for the bibliometric and bibliometric (Figure 4.1.1.5) + journal impact data (Figure 4.1.1.6) sets.

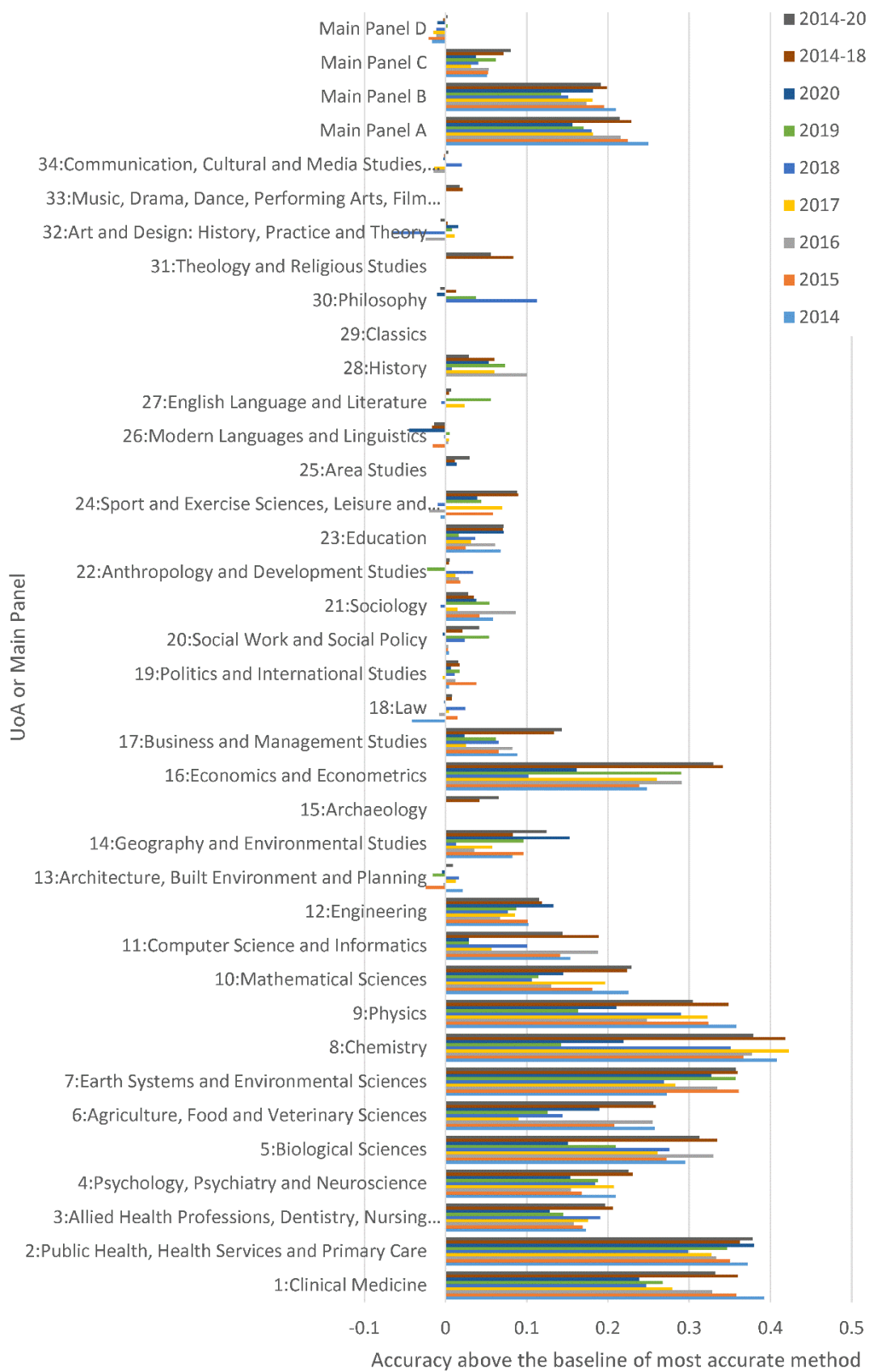

Figure 4.1.1.1. The percentage accuracy above the baseline for the most accurate machine learning method, trained on **50%** of the 2014-18 data and **bibliometric + journal impact + text inputs; 1000 features in total**. UoAs with less than 100 articles were ignored.

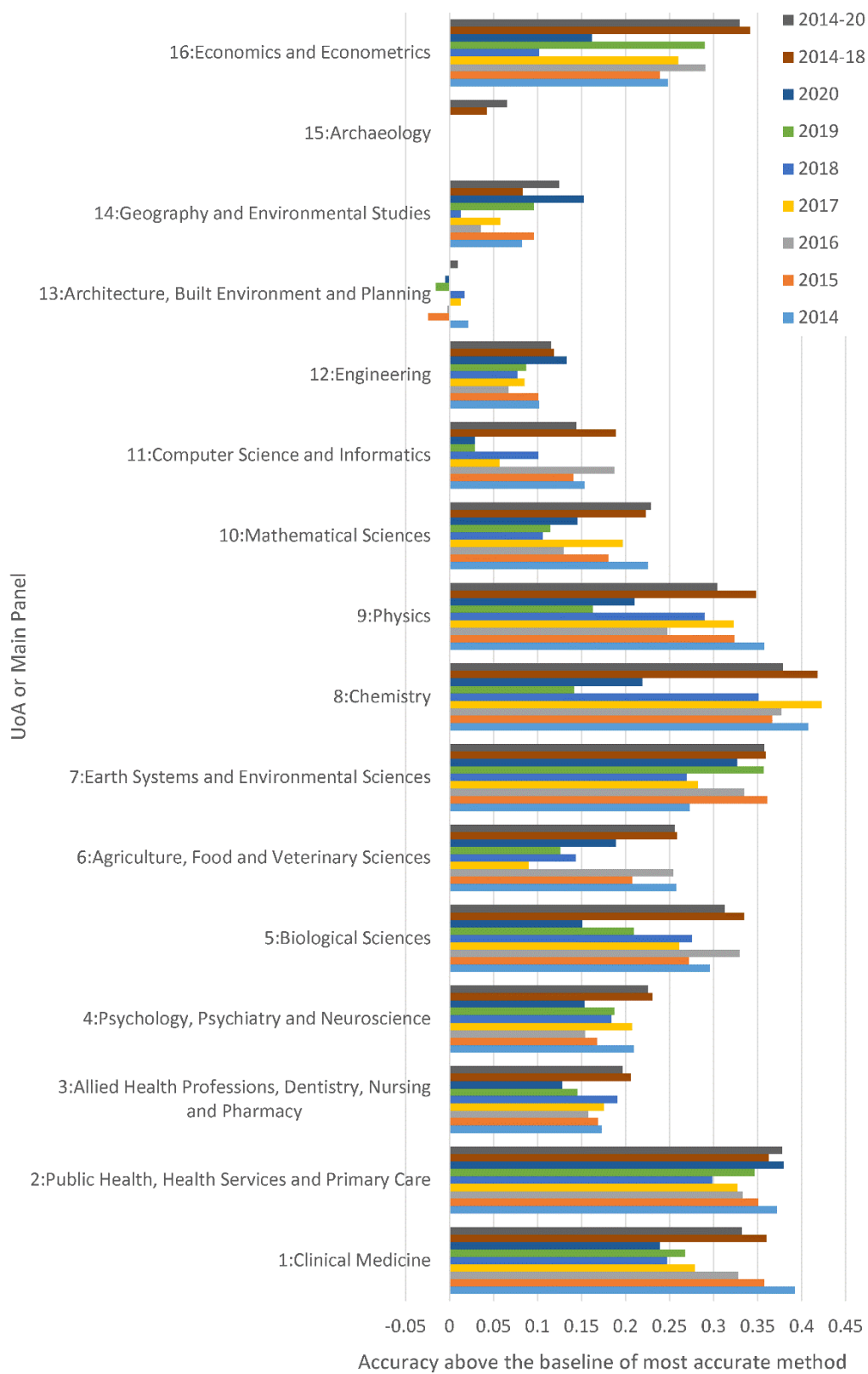

Figure 4.1.1.2. The percentage accuracy above the baseline for the most accurate machine learning method, trained on **50%** of articles and **bibliometric + journal impact + text inputs; 1000 features in total**. UoAs 1 to 16 only.

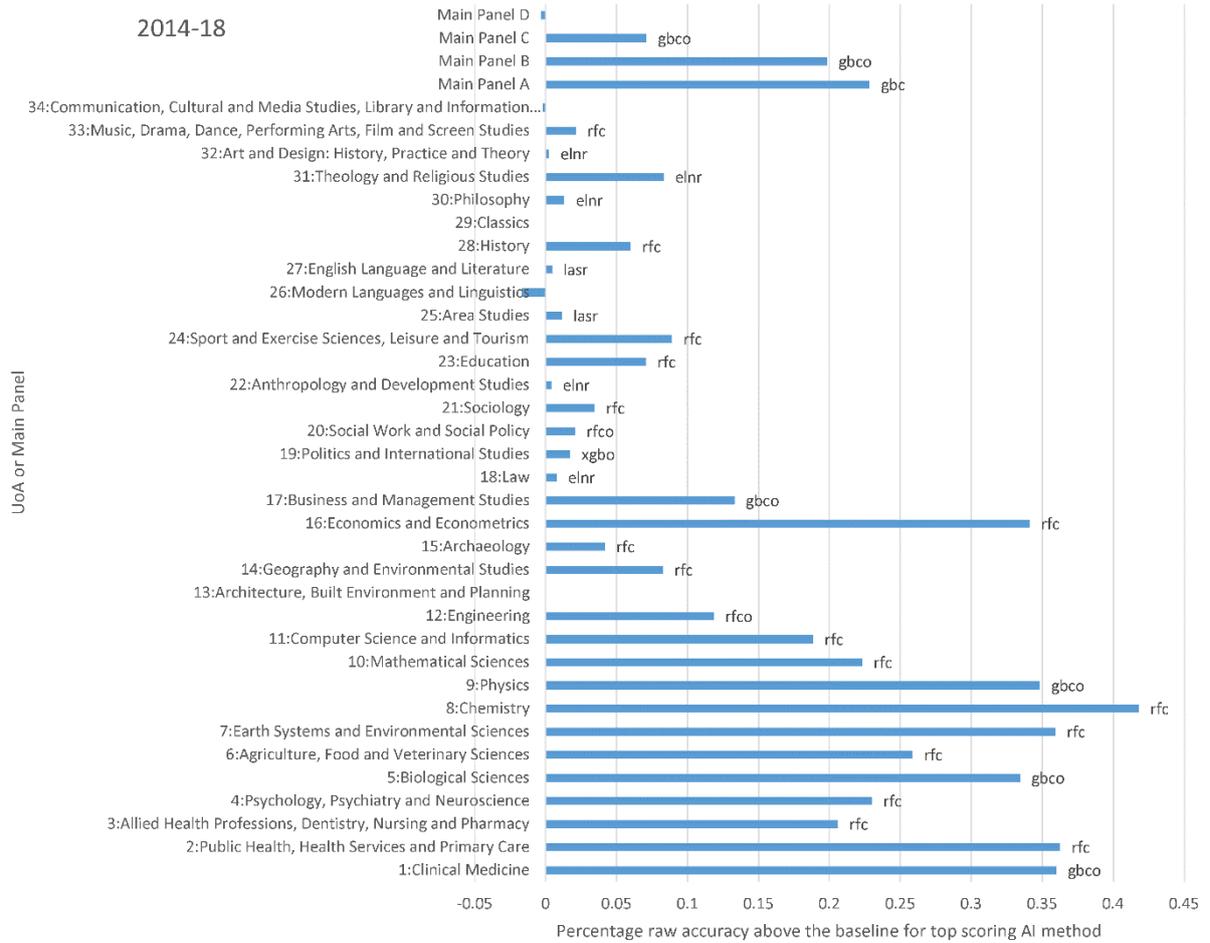

Figure 4.1.1.3. The percentage accuracy above the baseline for the most accurate machine learning method, trained on **50%** of the 2014-18 articles and **bibliometric + journal impact + text inputs; 1000 features in total**. The most accurate method is named.

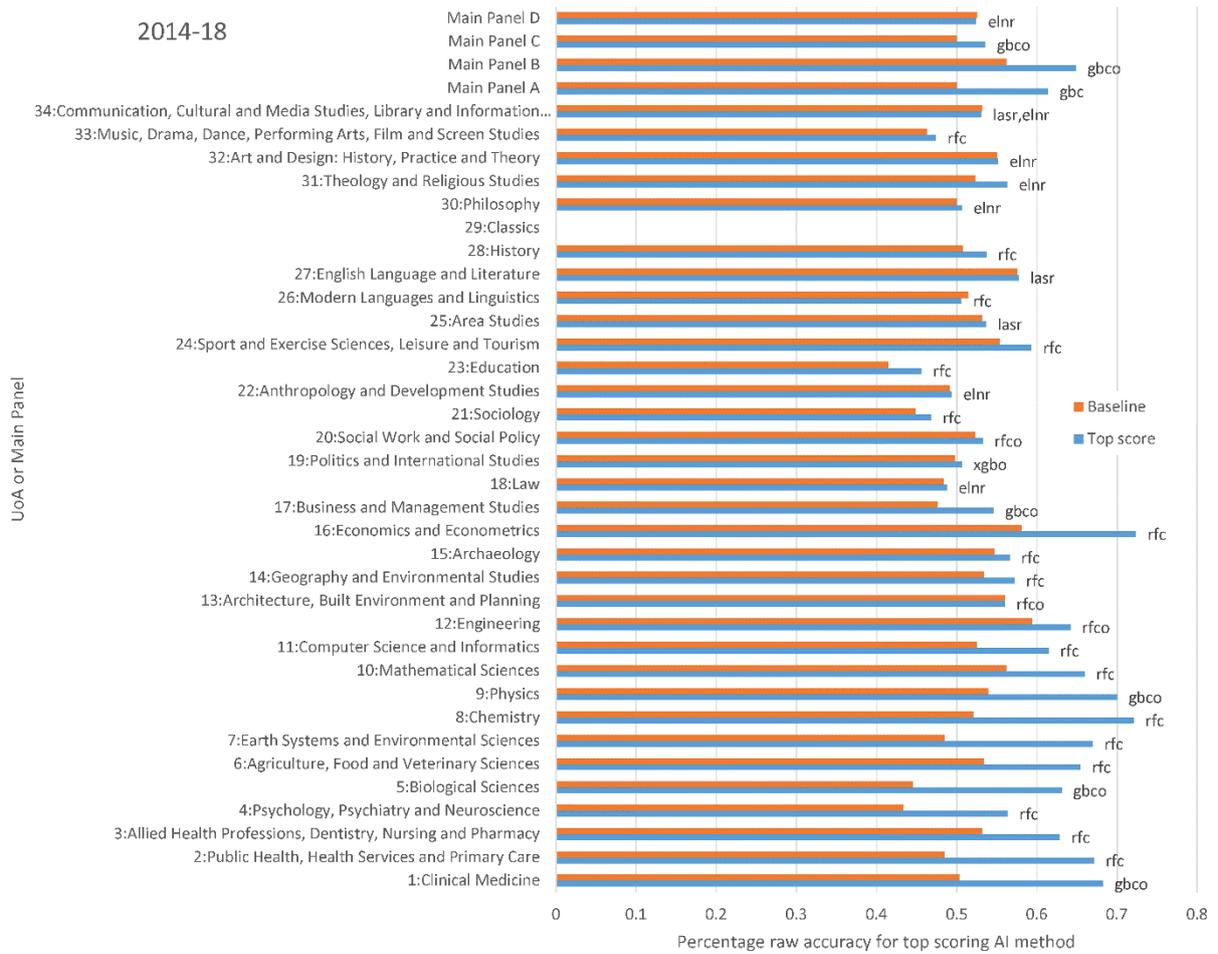

Figure 4.1.1.4. The percentage accuracy for the most accurate machine learning method, trained on **50%** of the 2014-18 articles and **bibliometric + journal impact + text inputs; 1000 features in total**. The most accurate method is named.

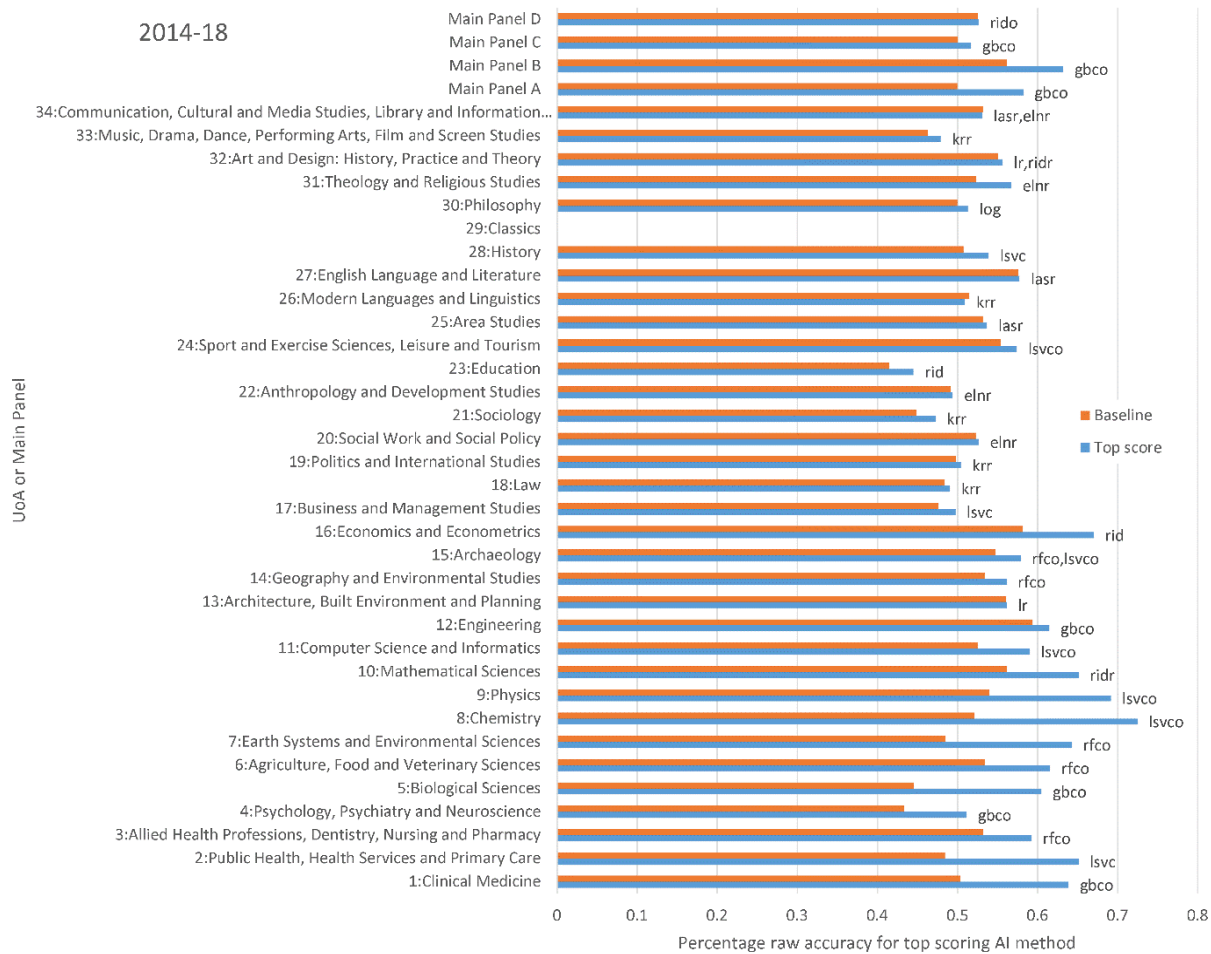

Figure 4.1.1.5. The percentage accuracy for the most accurate machine learning method, trained on **50%** of the 2014-18 articles and **bibliometric inputs; 9 features in total**. The most accurate method is named.

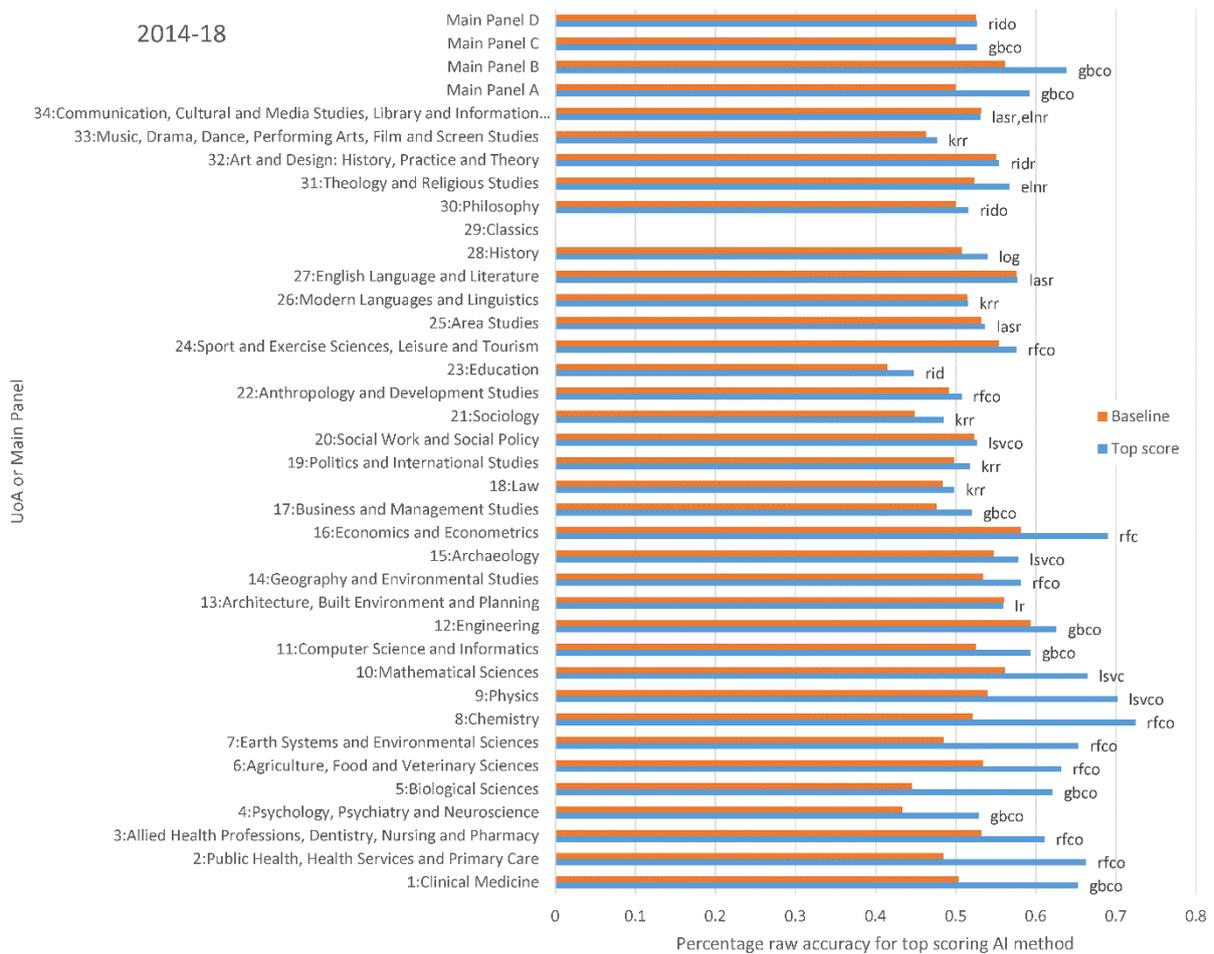

Figure 4.1.1.6. The percentage accuracy for the most accurate machine learning method, trained on **50%** of the 2014-18 articles and **bibliometric + journal impact inputs; 10 features in total**. The most accurate method is named.

With 50% training set size, 12,639 of the articles could be predicted by the AI in UoAs 1,2,6-10,16 with 65%-72% accuracy, or 42,446 if all UoAs are included irrespective of accuracy (Table 4.1.1.1).

      The correlations between AI predicted and actual scores are unsurprisingly positive for all UoAs but are weak for many (Table 4.1.1.1). The highest is only 0.536 (UoA 1), but correlations at the institutional level would much higher due to averaging out of the errors (see later).

Table 4.1.1.1. The number of articles 2014-18 predicted if 50% is used for training, following Strategy 1. Pearson correlations between AI predictions and actual scores are also reported (averaged across 10 iterations).

| Dataset | Articles 2014-18 | Predicted at 50% | Correlation |
|---|---|---|---|
| **1:Clinical Medicine** | 7274 | 3637 | 0.562 |
| **2:Public Health, Health Services and Primary Care** | 2855 | 1427 | 0.507 |
| **3:Allied Health Professions, Dentistry, Nursing and Pharmacy** | 6962 | 3481 | 0.406 |
| **4:Psychology, Psychiatry and Neuroscience** | 5845 | 2922 | 0.474 |
| **5:Biological Sciences** | 4728 | 2364 | 0.507 |
| **6:Agriculture, Food and Veterinary Sciences** | 2212 | 1106 | 0.452 |
| **7:Earth Systems and Environmental Sciences** | 2768 | 1384 | 0.491 |
| **8:Chemistry** | 2314 | 1157 | 0.505 |
| **9:Physics** | 3617 | 1808 | 0.472 |
| **10:Mathematical Sciences** | 3159 | 1579 | 0.328 |
| **11:Computer Science and Informatics** | 3292 | 1646 | 0.382 |
| **12:Engineering** | 12511 | 6255 | 0.271 |
| **13:Architecture, Built Environment and Planning** | 1697 | 848 | 0.125 |
| **14:Geography and Environmental Studies** | 2316 | 1158 | 0.277 |
| **15:Archaeology** | 371 | 185 | 0.283 |
| **16:Economics and Econometrics** | 1083 | 541 | 0.511 |
| **17:Business and Management Studies** | 7535 | 3767 | 0.353 |
| **18:Law** | 1166 | 583 | 0.101 |
| **19:Politics and International Studies** | 1595 | 797 | 0.181 |
| **20:Social Work and Social Policy** | 2045 | 1022 | 0.259 |
| **21:Sociology** | 949 | 474 | 0.180 |
| **22:Anthropology and Development Studies** | 618 | 309 | 0.040 |
| **23:Education** | 2081 | 1040 | 0.261 |
| **24:Sport and Exercise Sciences, Leisure and Tourism** | 1846 | 923 | 0.265 |
| **25:Area Studies** | 303 | 151 | 0.142 |
| **26:Modern Languages and Linguistics** | 630 | 315 | 0.066 |
| **27:English Language and Literature** | 424 | 212 | 0.064 |
| **28:History** | 583 | 291 | 0.141 |
| **29:Classics** | 0 | 0 | - |
| **30:Philosophy** | 426 | 213 | 0.070 |
| **31:Theology and Religious Studies** | 107 | 53 | 0.074 |
| **32:Art and Design: History, Practice and Theory** | 665 | 332 | 0.028 |
| **33:Music, Drama, Dance, Performing Arts, Film and Screen Studies** | 350 | 175 | 0.164 |
| **34:Communication, Cultural and Media Studies, Library and Information Management** | 583 | 291 | 0.084 |

The main results above used standard parameters for the machine learning algorithms. Hyperparameter tuning gives better results for some machine learning applications. Preliminary testing suggested that it would not work so it was tried for a limited set of UoAs with more accurate predictions and the main six methods (Figure 4.1.1.7). Whilst hyperparameter tuning marginally increases accuracy on some UoAs it marginally reduces it on others. The minor differences found could be due to random variations in the data, so hyperparameter tuning does not seem to provide clear enough evidence of an advantage to advocate its use.

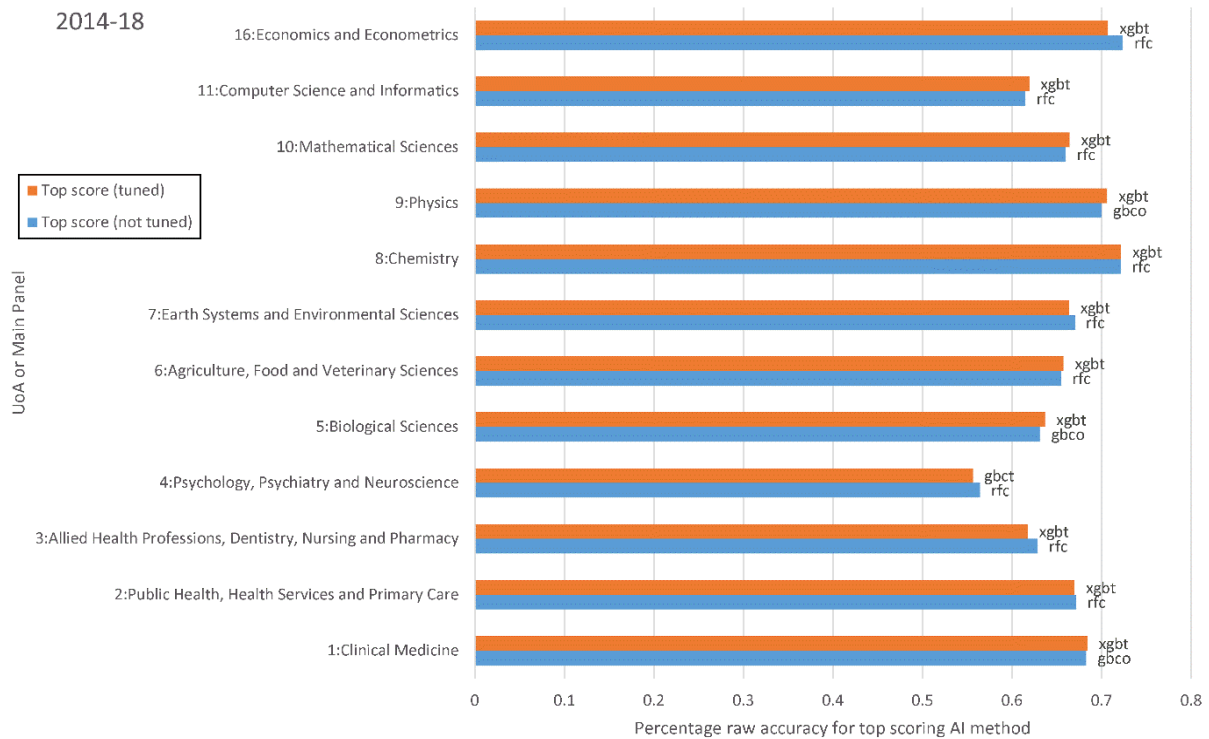

Figure 4.1.1.7. The percentage accuracy for the most accurate machine learning method with and without hyperparameter tuning (out of the main six), trained on **50%** of the 2014-18 articles and **bibliometric + journal impact + text inputs; 1000 features in total**. The most accurate method is named.

The main findings are as follows.
- The **maximum accuracy *above the baseline* for any case was 0.42** (Chemistry, Figure 4.1.1.3). In overall accuracy terms, **the maximum raw accuracy was 72%** (Chemistry, Economics and Econometrics, Figure 4.1.1.4): In other words, just under three quarters of the AI predictions are correct in the best case. Overall score shifts for institutions or researcher/research types can be tested to help make judgements about whether this level is high enough (see below).
- Since, from the above, the (very approximate) best guess level of agreement between provisional REF panel scores for the same article submitted by different authors was 86.4%, **the best-case AI predictions are 14% short of the agreement between separate groups of human experts**. The human experts have more information to analyse (e.g., full text, tacit knowledge of research fields), so the 72% accuracy attained in the best case may be close to the maximum achievable by AI, and the lower accuracy levels in most UoAs may reflect a greater degree of tacit knowledge needed for accurate scoring, or greater score uncertainty for human assessors in some UoAs.
- The UoAs for which the highest accuracy above the baseline is possible (at least 30%) are 1,2,5,7,8,9,16 (Figure 4.1.1.3). Reducing the threshold to 0.2 gives UoAs 1-10, 16 (Figure 4.1.1.3). Using raw accuracy instead changes the lists. The UoAs with raw accuracy above 65% are 1,2,6-10,16 and the UoAs with raw accuracy above 60% are 1-3,5-12,16 (Figure 4.1.1.4). **UoAs 1-3,5-11,16 will be called the most predictable UoAs** since this list includes all UoAs above the four thresholds, except UoA 12, which has high raw accuracy but marginal accuracy above the baseline. Much of the analysis below focuses on this set, with UoA 4 added since this UoA used bibliometrics in REF2021, as the most promising UoAs for AI predictions.

- The accuracy of the predictions increases as more inputs are added. In terms of accuracy: bibliometrics < bibliometrics + journal impact < bibliometrics + journal impact + text. The differences are relatively small, although it varies between UoAs. The main exceptions are Chemistry (bibliometrics alone is best) and Physics (bibliometrics and bibliometrics + journal impact + text both give the best results).
- The accuracy of the predictions increases as a higher percentage of training data is used. In terms of accuracy: 10% < 25% < 50%. The differences are again relatively small, although they vary between UoAs and input sets.
- Combining UoAs into main panels substantially decreases the maximum accuracy available from the predictions, even though the main panels have more training data. It is therefore better to predict separately for each UoA. This is also supported by the low agreement rate for articles classified in multiple different UoAs.
- The algorithms delivering the highest accuracy for the most predictable UoAs are rfc, rfco, gbco, xgbo (i.e., Random Forest Classifier; Random Forest Classifier, Ordinal; Gradient Boosting Classifier, Ordinal; Extreme Gradient Boosting Classifier; Extreme Gradient Boosting Classifier, Ordinal) on Input Set 3. Since the ordinal and standard variants of the algorithms give similar accuracies, it is reasonable to also consider the two missing methods, gbc and xgb.
- Combining years 2014 to 2018 into one set gives similar overall accuracy to classifying the years separately in the most predictable UoAs, although there is too little data to train a classifier in some UoAs for individual years. Individual year graphs for the most accurate dataset are available online, showing that the accuracy for individual years is sometimes greater than the aggregate 2014-18 accuracy, but combining years gives more consistent high accuracy. The higher accuracy rates for some years may be due to overfitting because there seems to be, in principle, little difference between years 2014-18 from an input strength perspective, although the bibliometric data is weaker for 2019-20. Graphs for all other datasets are available in the online spreadsheet. Combining years 2014 to 2020 gives a slightly lower overall accuracy than 2014-18, but the difference tends to be small so it would be possible to use 2014-20 instead.
- As mentioned above, classification accuracy is slightly lower for 2019 and 2020 than for the earlier years, presumably due to the weaker citation count indicator for recent years.
- Hyperparameter tuning does not seem worth using, although it may give marginal increases in accuracy for some UoAs.

In summary, the most promising avenue for AI score prediction is therefor for rfc, rfco, gbco, xgb, xgbo applied to 50% of the 2014-18 data with all inputs for UoAs 1,2,6-10,16 but the slightly expanded set of 1-11, 16 will be mainly analysed below. Score shifts (or bias) tests are needed to inform a decision about whether the achievable level of accuracy is acceptable.

### 4.1.1.1 Attainable levels of accuracy by UoA and Main Panel

The highest attainable **overall** level of accuracy for the set of articles evaluated, based on classifying 50% manually and 50% by machine learning is more than double the reported machine learning accuracy because the manually scored texts need to be considered. For the most accurate UoA/method combination, the machine learning accuracy is 72% on the 50% test set, which gives 86% accuracy on the eligible 2014-18 articles (i.e., 100% accuracy for the half that is hand classified and 72% accuracy for the half that is machine classified for the eligible articles 2014-18) and 91% accuracy on all REF-submitted journal articles. Increasing the percentage that is human classified and fed into the algorithms does not increase the accuracy for the remaining texts much (see Figure 3.8.2.1.1 to Figure 3.8.2.1.4) but would increase the overall accuracy (Table 4.1.1.1.1). For example, at 72% AI classification accuracy, 70% of the articles would need to be manually classified to give 95% accuracy overall (Table 4.1.1.1.1).

Table 4.1.1.1.1. Examples of overall accuracy levels with different proportions classified by hand and by machine, assuming 100% human accuracy and 72% machine accuracy. Eligible articles form 62.6% of all REF2021 journal articles (Table 3.6.1).

| Human scores | Human accuracy | AI scores | AI accuracy | Overall accuracy for eligible articles (62.6%) | Overall accuracy for all journal articles (100%) |
|---|---|---|---|---|---|
| **50%** | 100% | 50% | 72% | 86% | 91% |
| **60%** | 100% | 40% | 72% | 89% | 93% |
| **70%** | 100% | 30% | 72% | 92% | 95% |
| **80%** | 100% | 20% | 72% | 94% | 96% |
| **90%** | 100% | 10% | 72% | 97% | 98% |
| **100%** | 100% | 0% | 72% | 100% | 100% |

Thus, for all UoAs, it is possible to set an attainable level of overall accuracy and deduce from the machine learning accuracy data how many articles need to be human scored to achieve it. For reference, in UoAs where the machine classifications fall below the baseline, it would give more accurate results to award all the articles that have not been manually classified the majority score (e.g., 3* or 4*), although this is clearly outrageous and unacceptable.

### 4.1.2 Institutional score shifts, including institution size score shifts

The AI algorithms change some scores, and this may introduce bias through systematic mistakes, remove human bias in the original scores, or both. This section checks for differences between the average human scores and average AI scores for institutions without attributing bias as a cause.

Within UoAs 1,2,6-10,16 with the most accurate predictions (65%-72%), the predicted and actual results were compared for each institution using the optimal machine learning method and the maximum data (1000 features, 50% training set) using the 2014-18 dataset, which covers the most predictable years (Figure 4.1.2.1).

As could be expected statistically, the (financial) prediction gain varies most for units with few submitted articles (because each individual score change would alter the institutional average more). The prediction shifts are huge for small submissions, with gains or losses of 25% in most UoAs in extreme cases (Figure 4.1.2.1). This 25% translates into an overall change of 8%, however, considering the 50% human scores for eligible articles and the human scores for the 37.6% ineligible articles. This magnitude of shift might be avoided by using human scores for 100% of the small submissions.

Nontrivial (financial) score shifts are also evident for above average sized submissions for each UoA. All UoAs have changes of 5% or more for larger submissions (2% overall). For UoA 11, the largest submission has a score shift range (from 10 applications of the method) between 8% and 21% (2.5% to 6.6% overall).

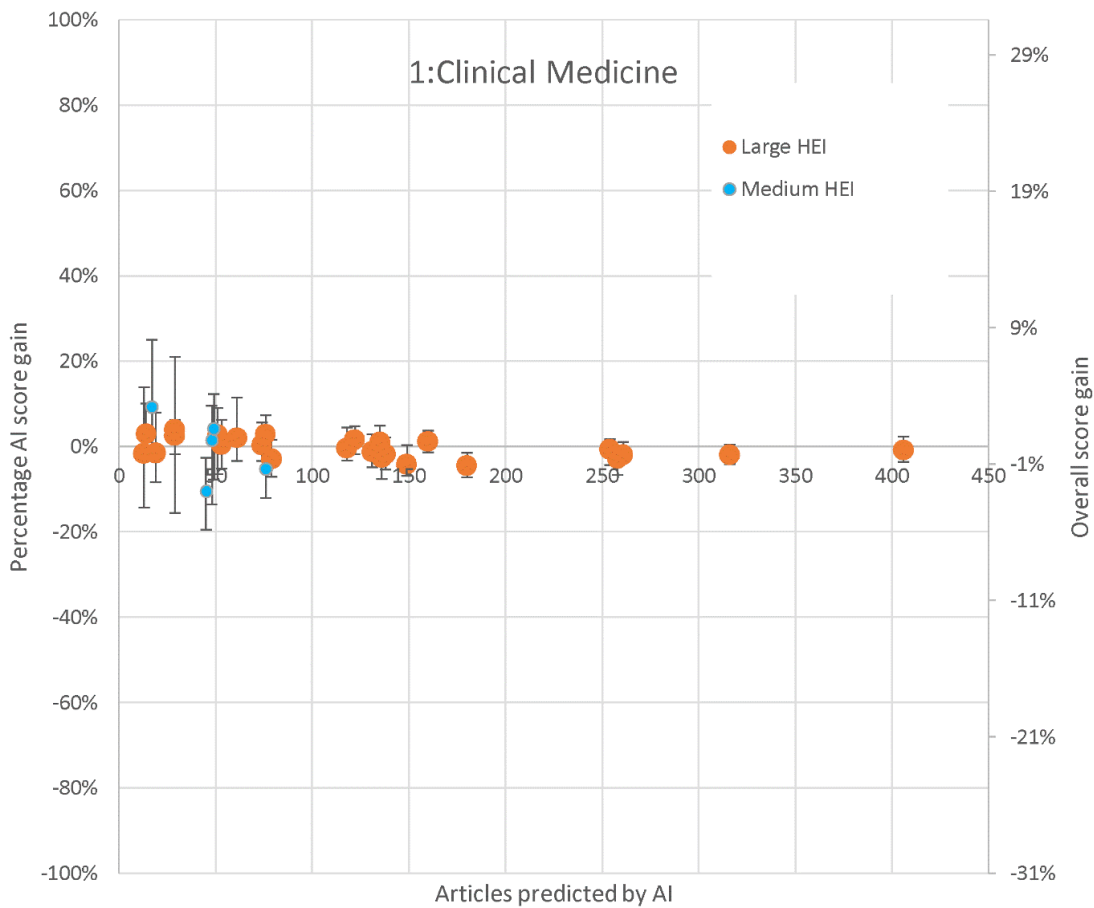

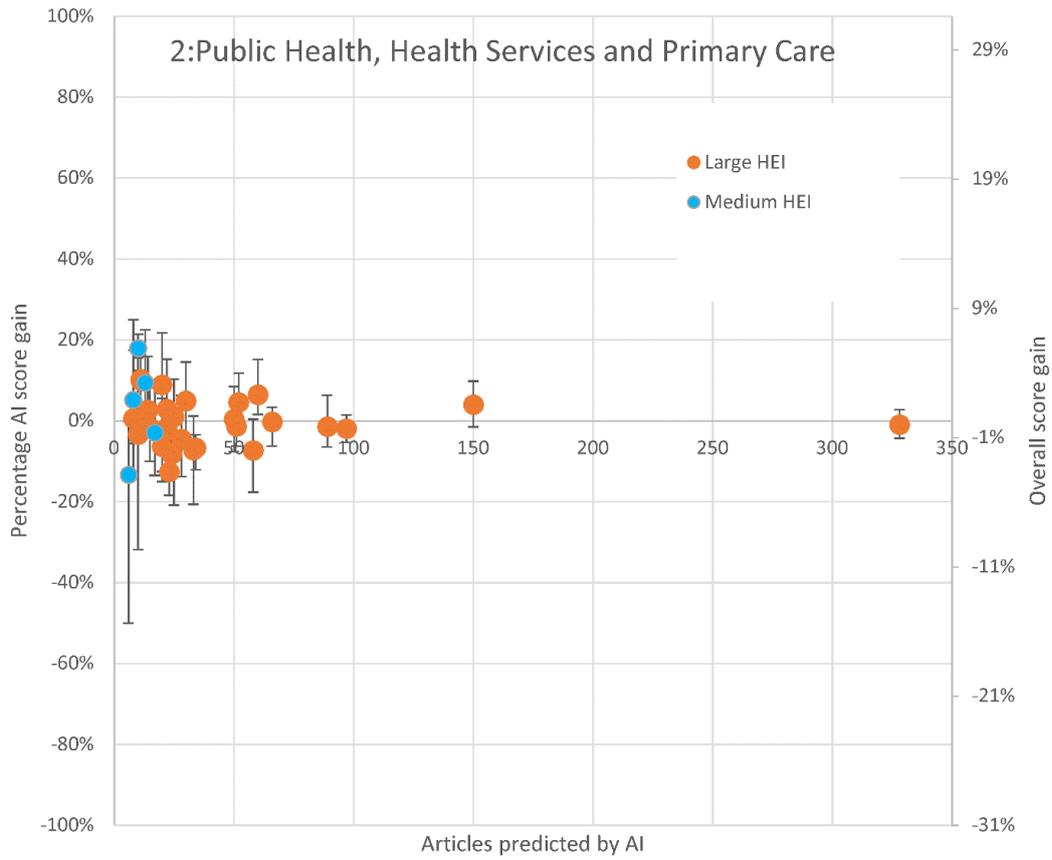
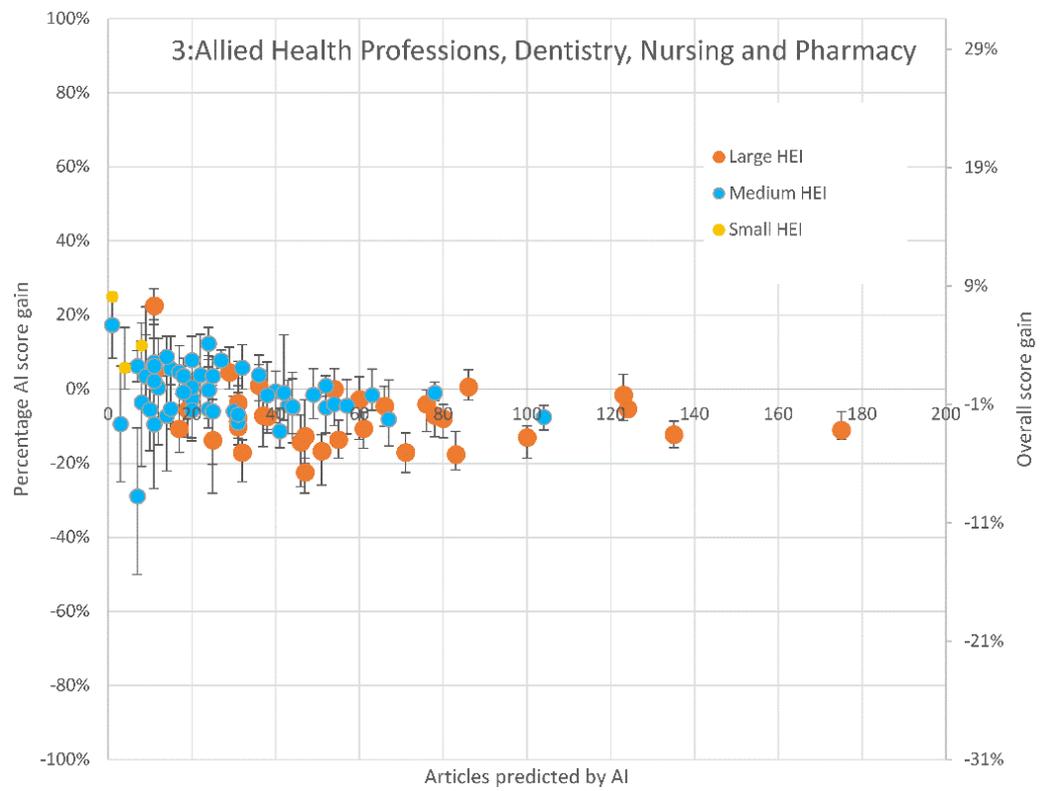

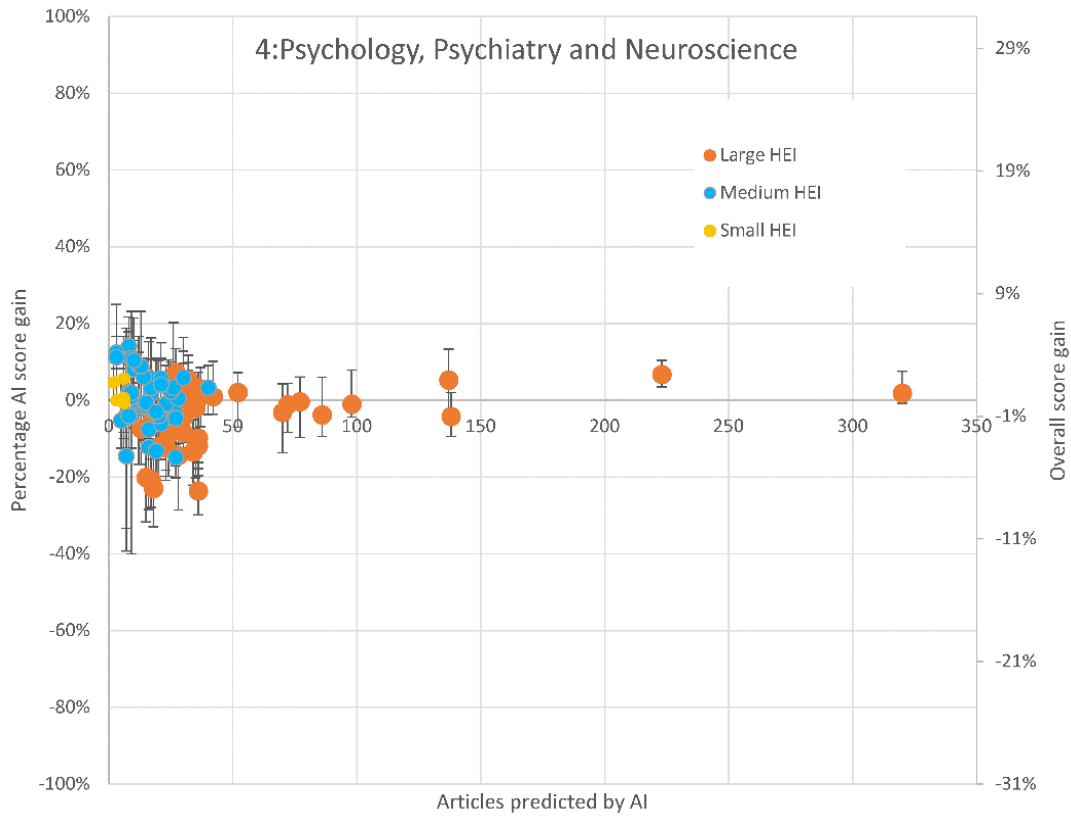

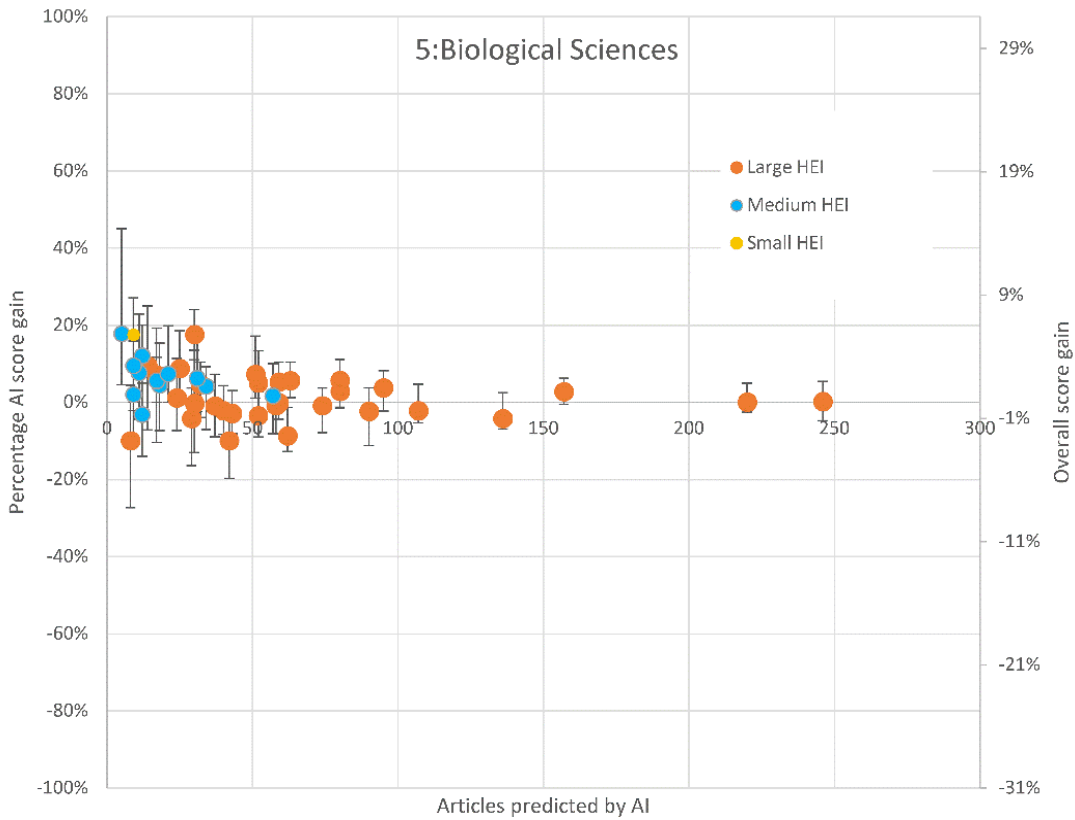

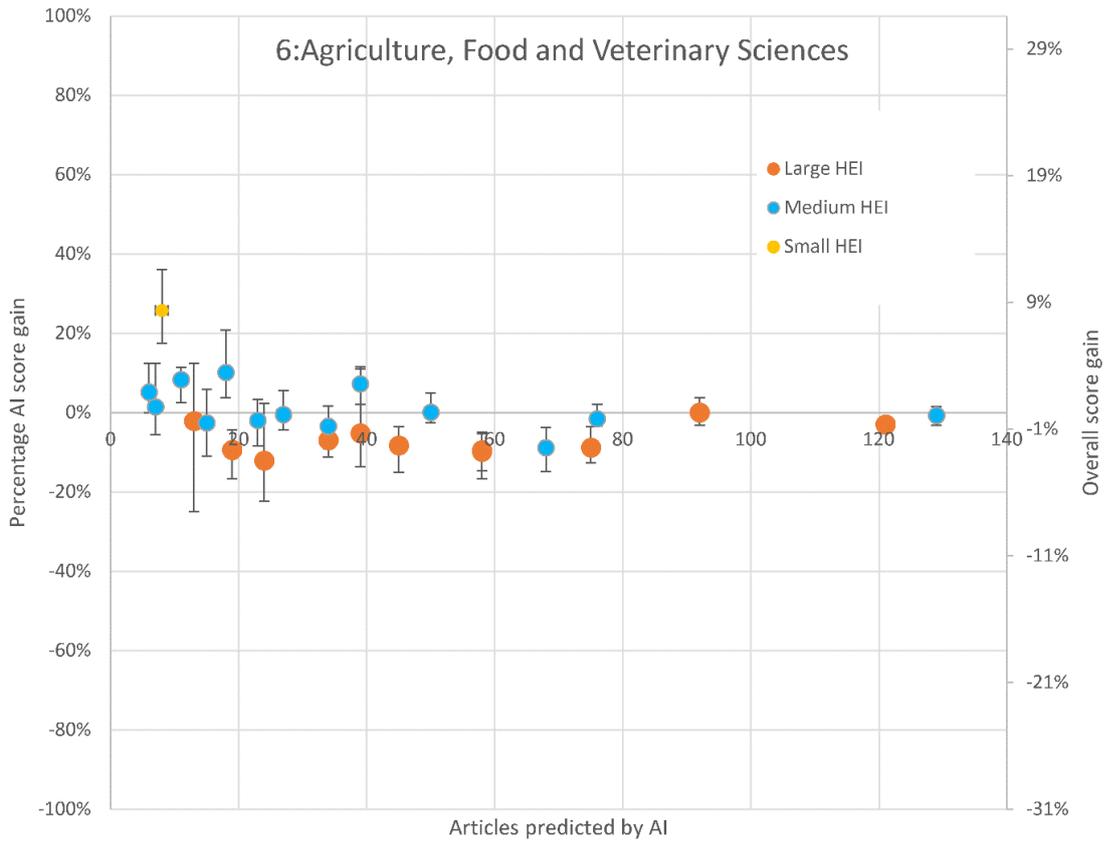
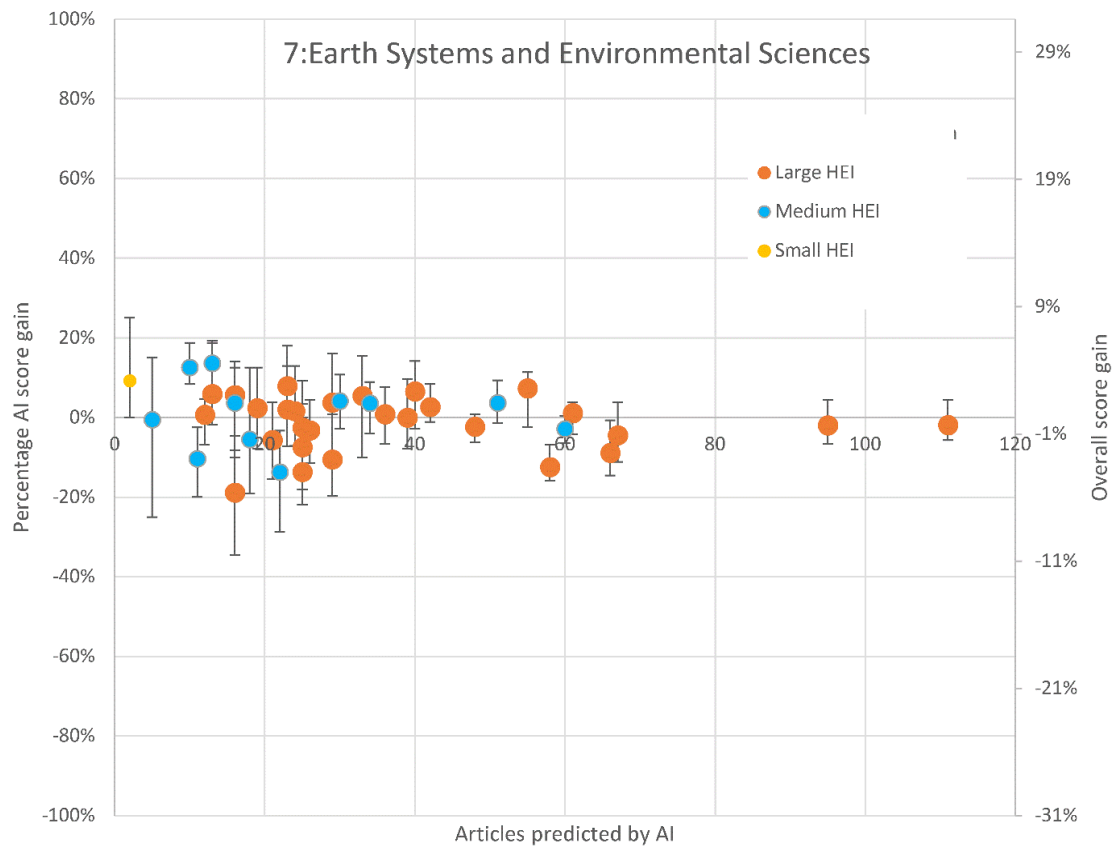

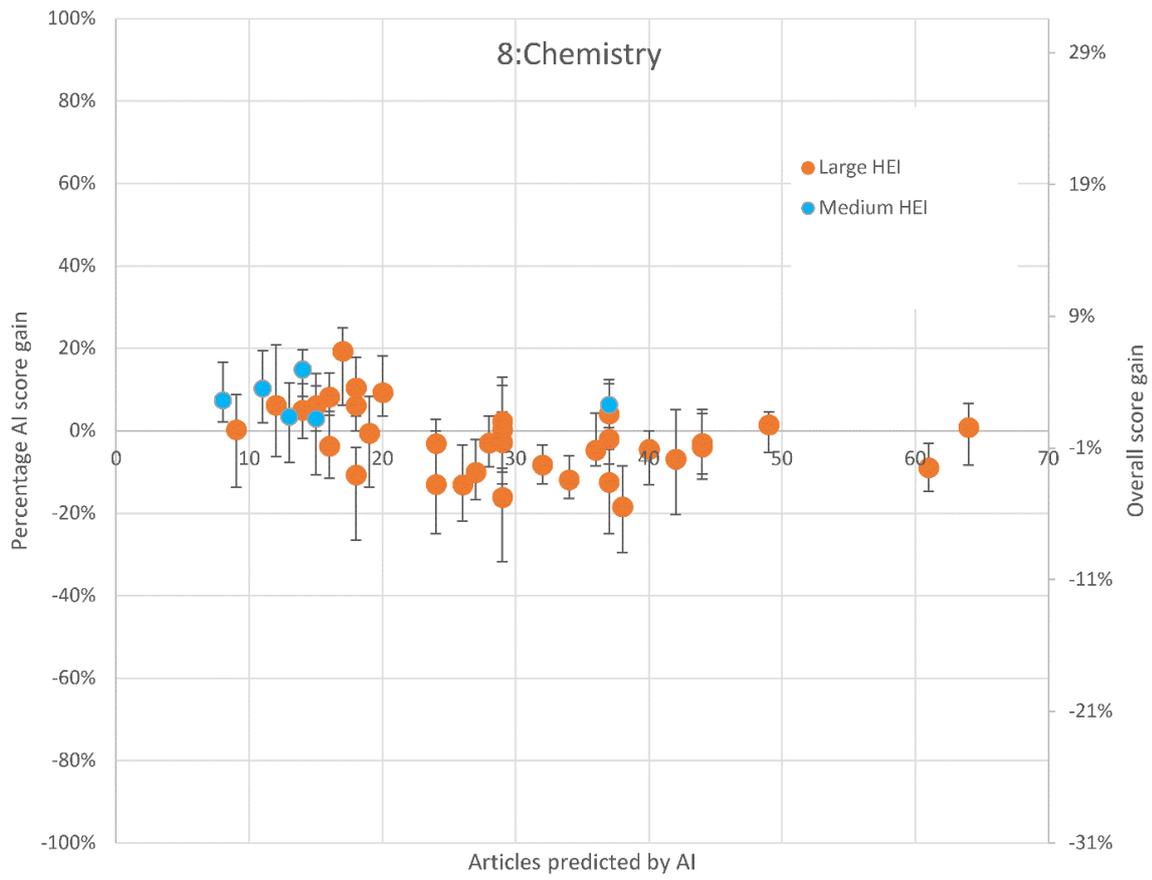
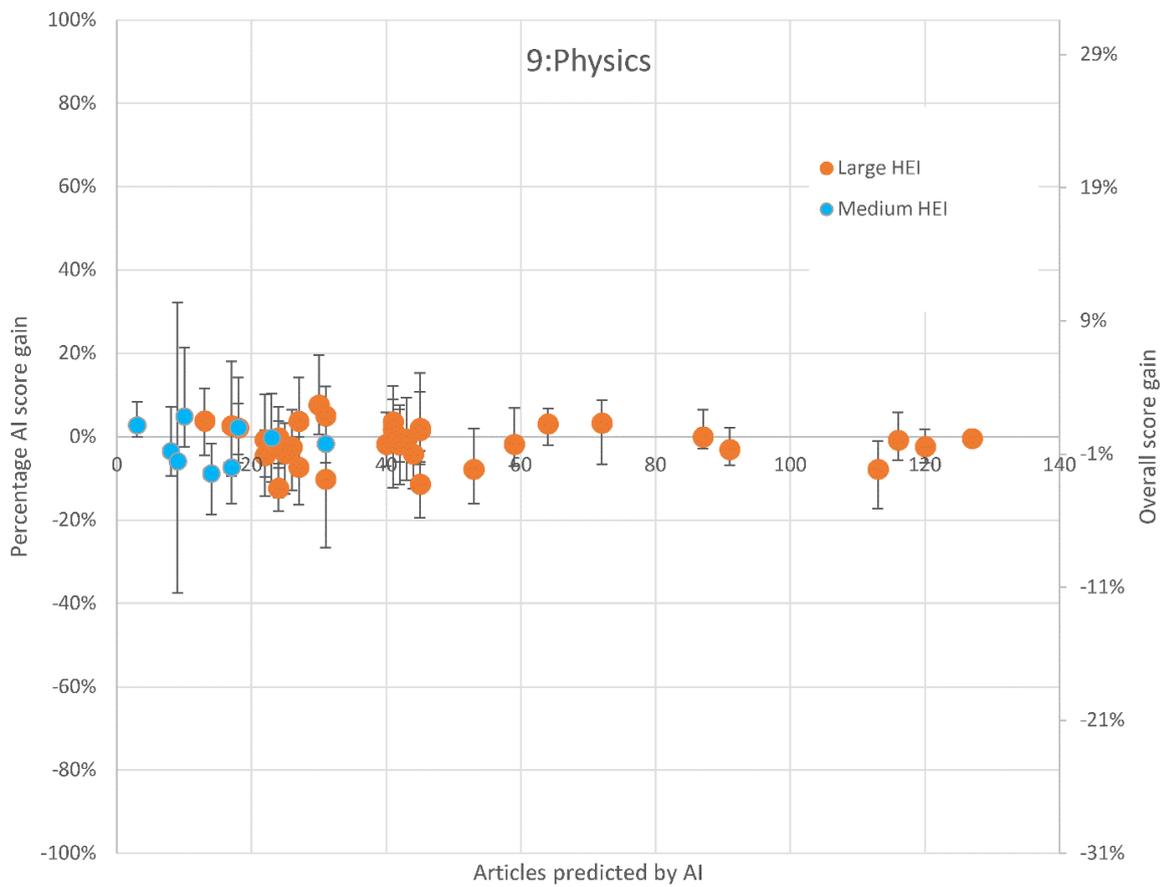

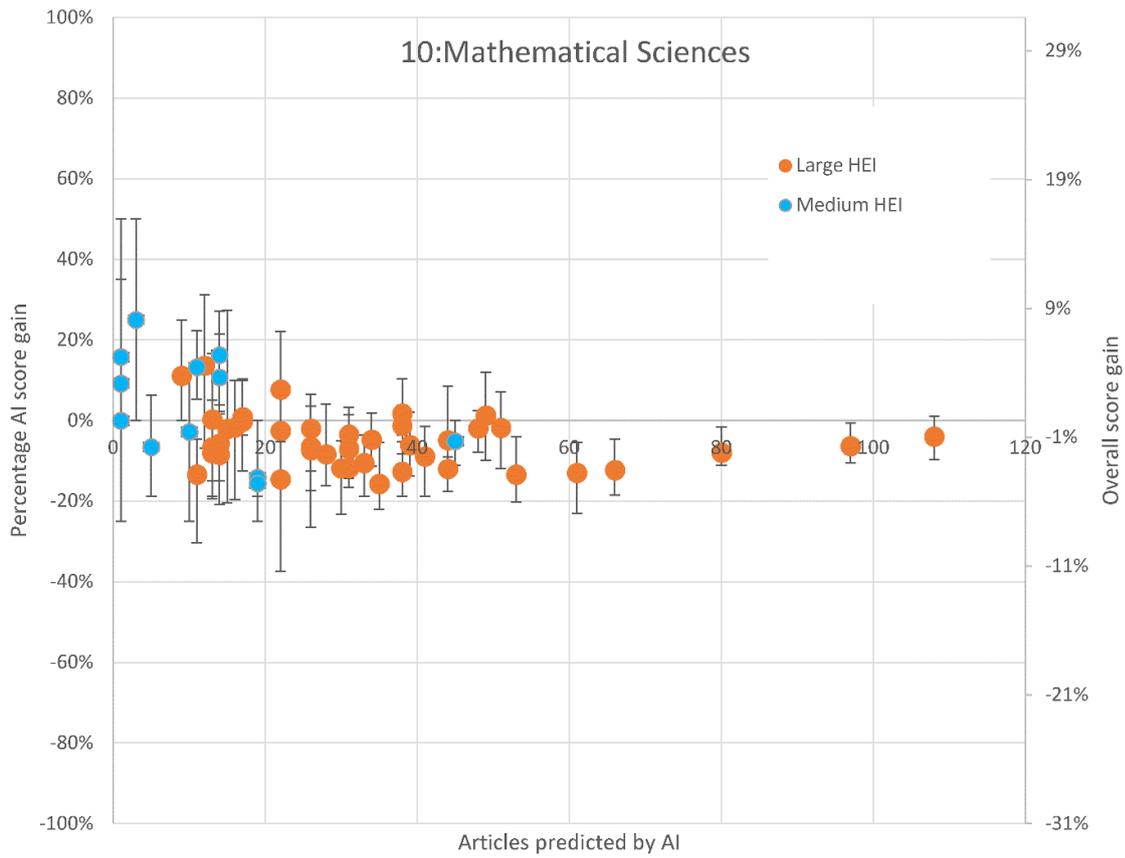
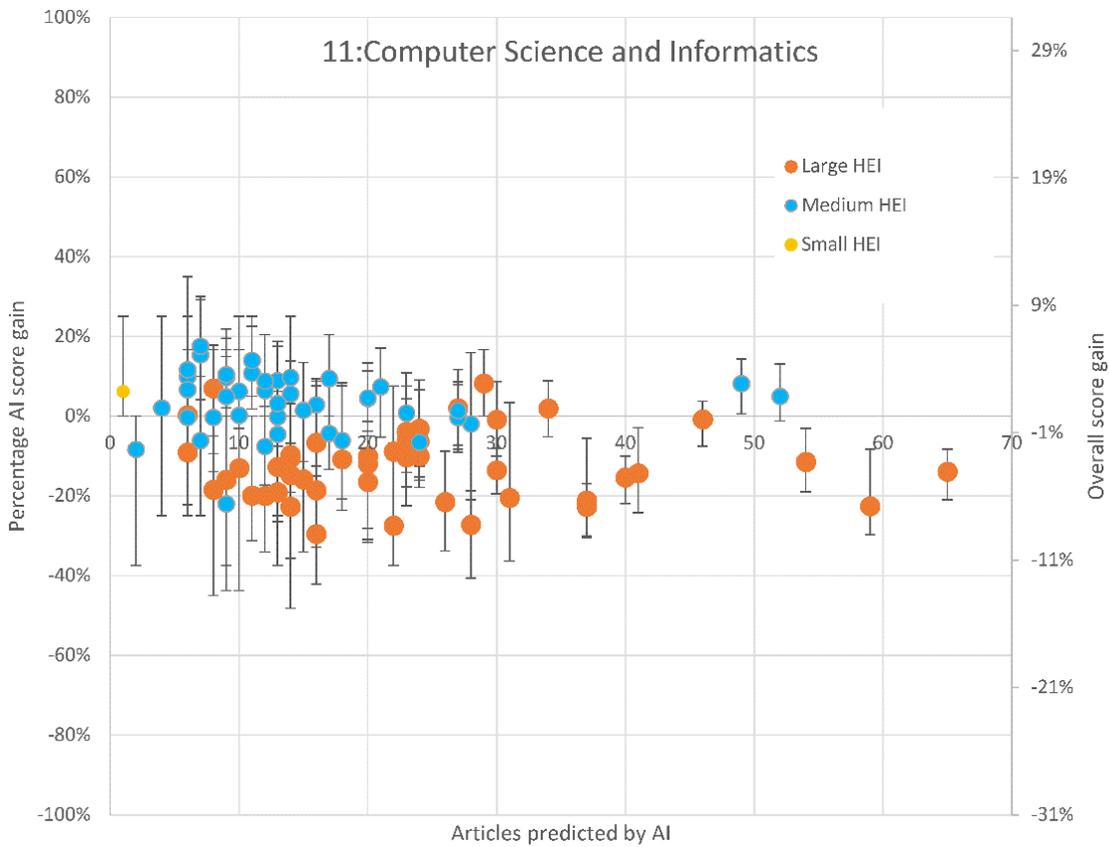

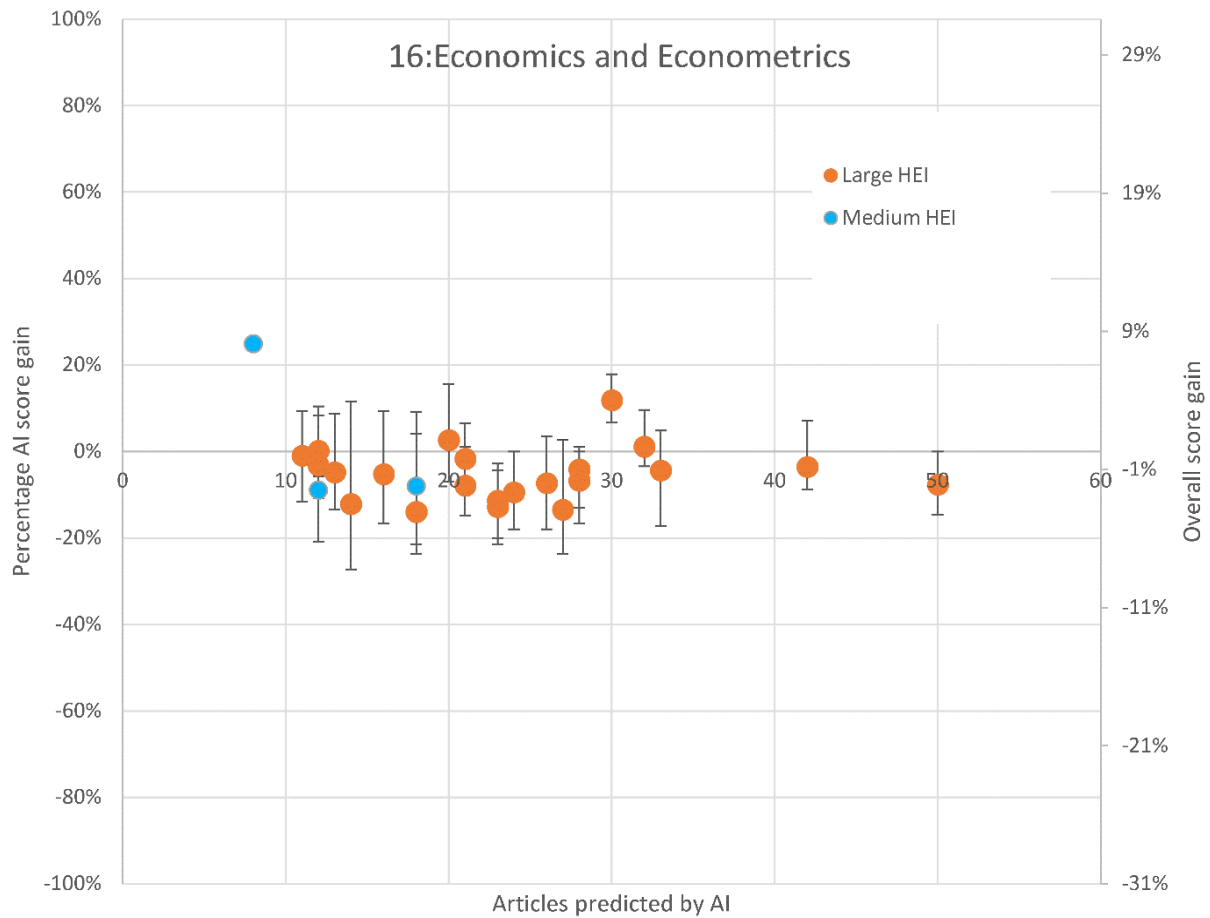

Figure 4.1.2.1. The average REF AI institutional score gain on UoA 1-11, 16 for the most accurate machine learning method, trained on **50%** of the 2014-18 data and **bibliometric + journal + text inputs, after excluding articles with shorter than 500 character abstracts**. UoAs 1,2,6-10,16 have 65%-72% raw accuracy. AI score is a financial calculation, sometimes called research power (4*=100% funding, 3*=25% funding, 0-2*=0% funding). Overall gain includes human classified articles (right hand axis for the same data) but not non-article outputs. Error bars indicate the highest and lowest values from 10 iterations.

Many intuitions focus on the grade point average scores of departments and some use rankings to benchmark against other departments or universities. Whilst this is not a recommended use of the data, it is widespread. Figure 4.1.2.2 shows the impact on overall rankings (including non-journal outputs) of replacing some journal scores with AI predictions for all 33 UoAs with AI predictions. Even in the UoAs with the most accurate predictions and the UoAs with the fewest journal articles, almost all institutions had at least one rank change in the ten experiments conducted. The HEI exceptions tend to be the large highly ranked institutions in UoAs without many small institutional submissions.

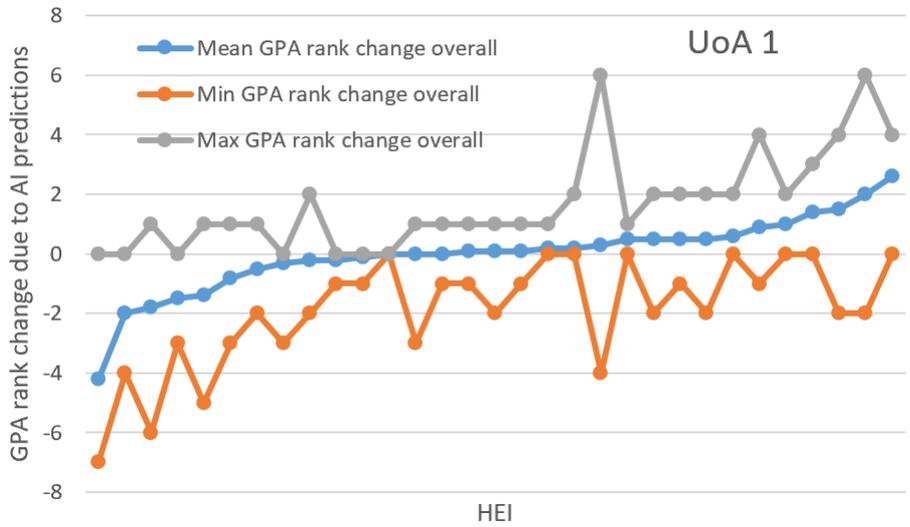
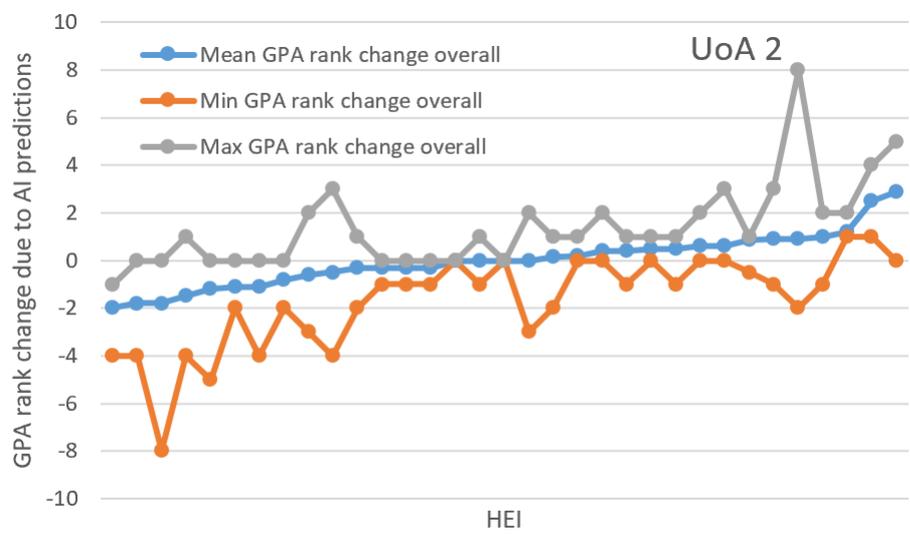
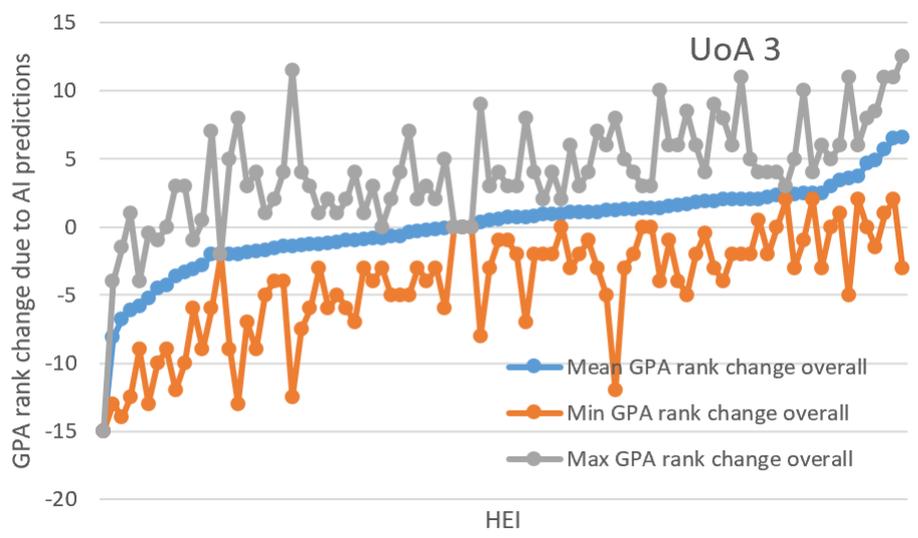

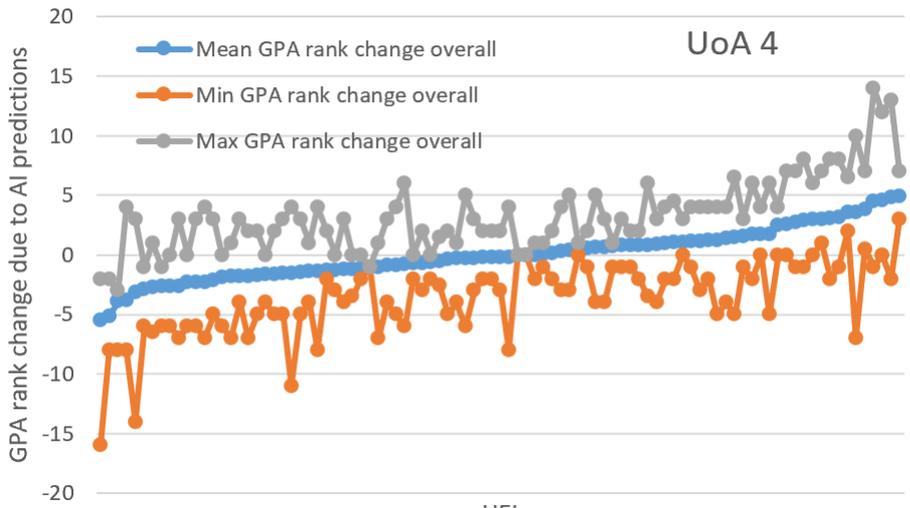
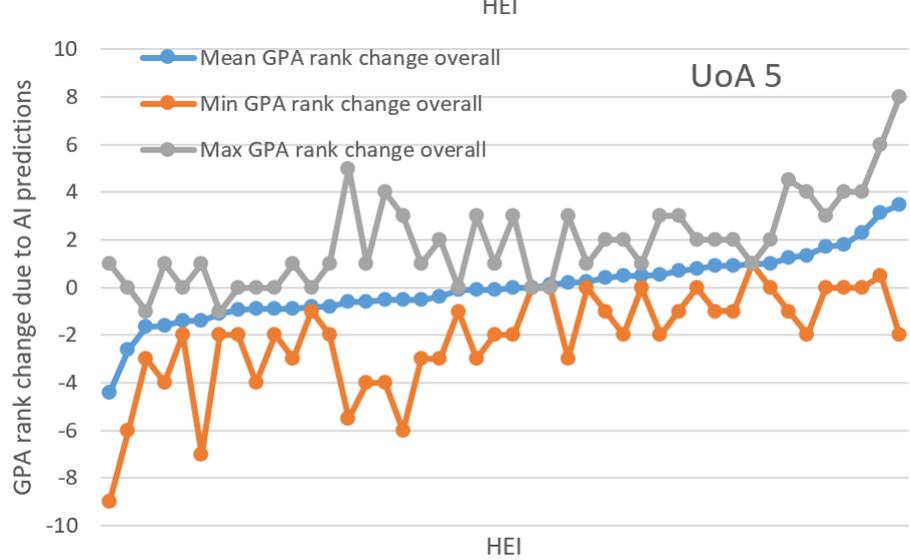
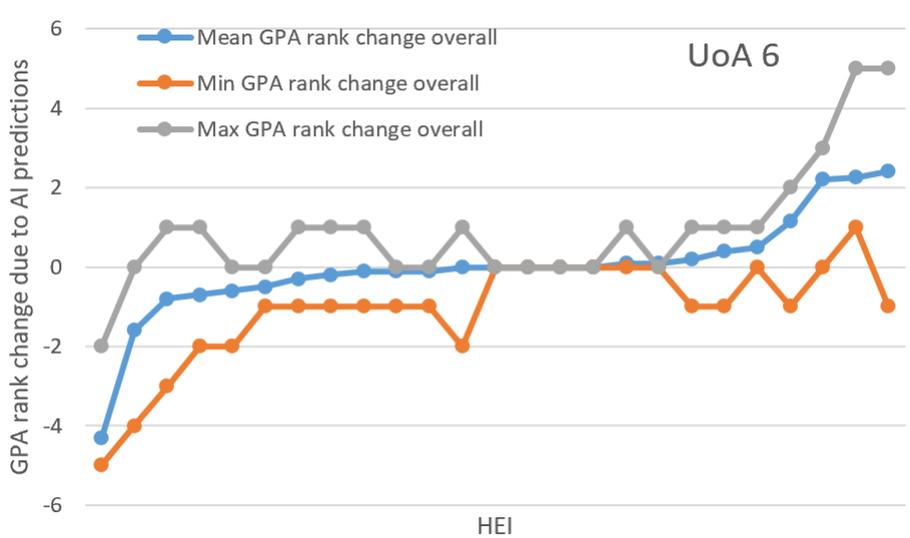

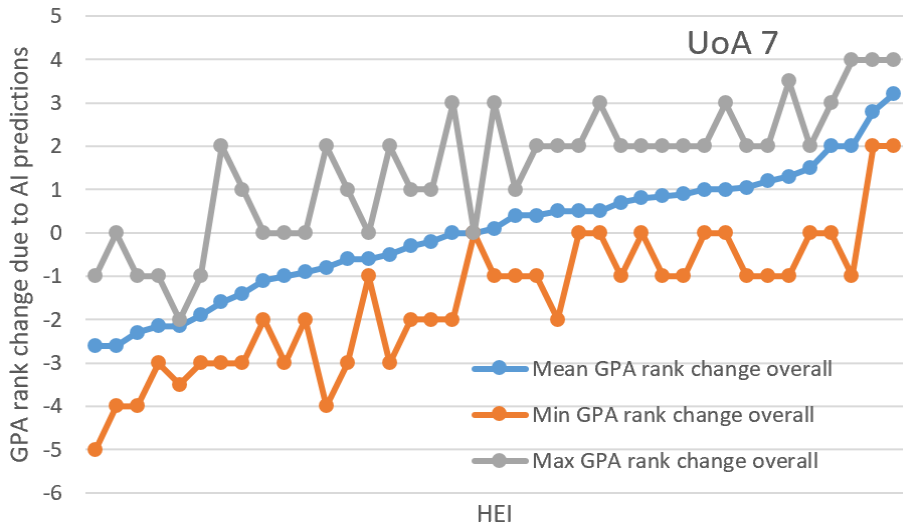

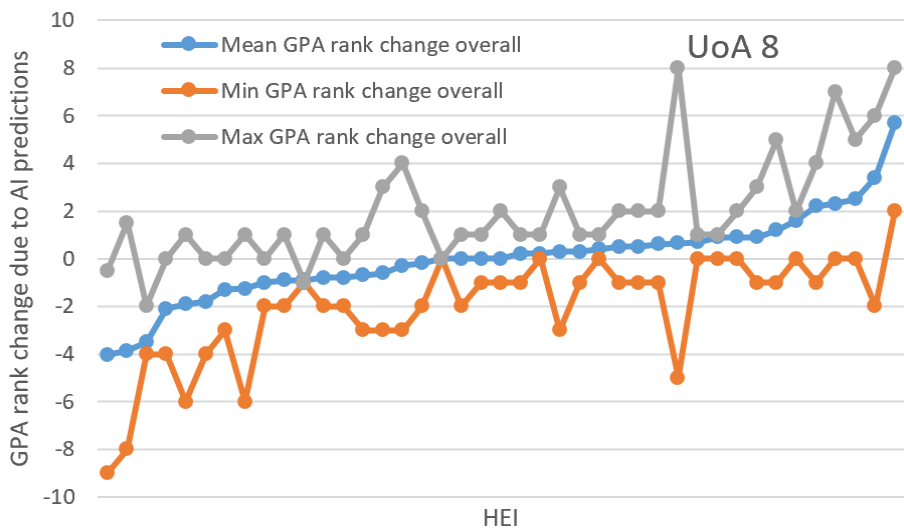

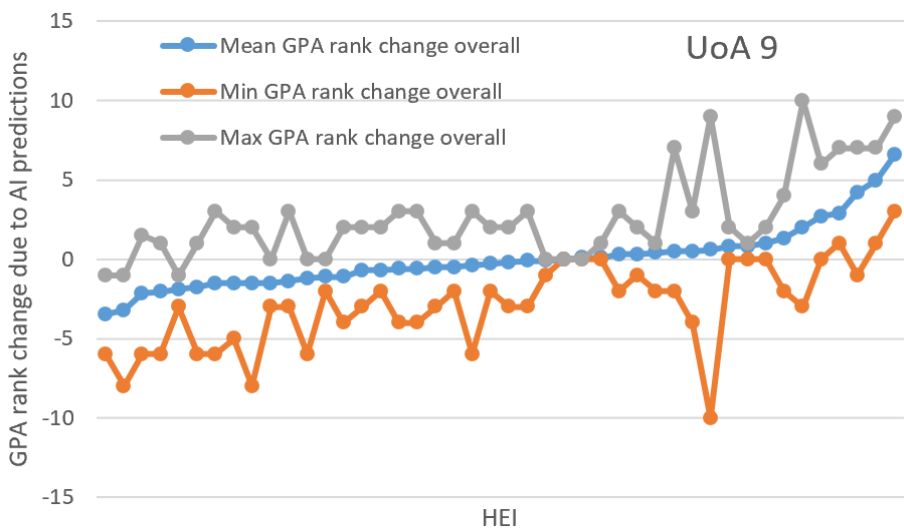

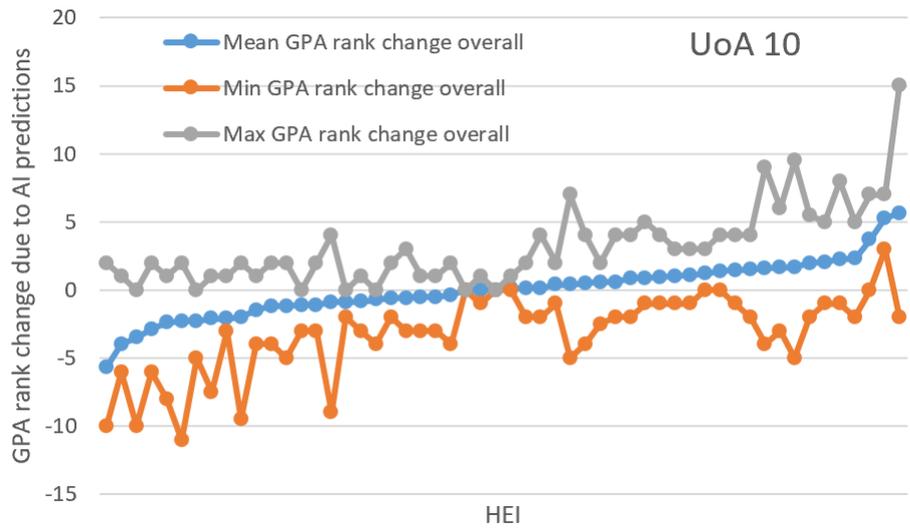
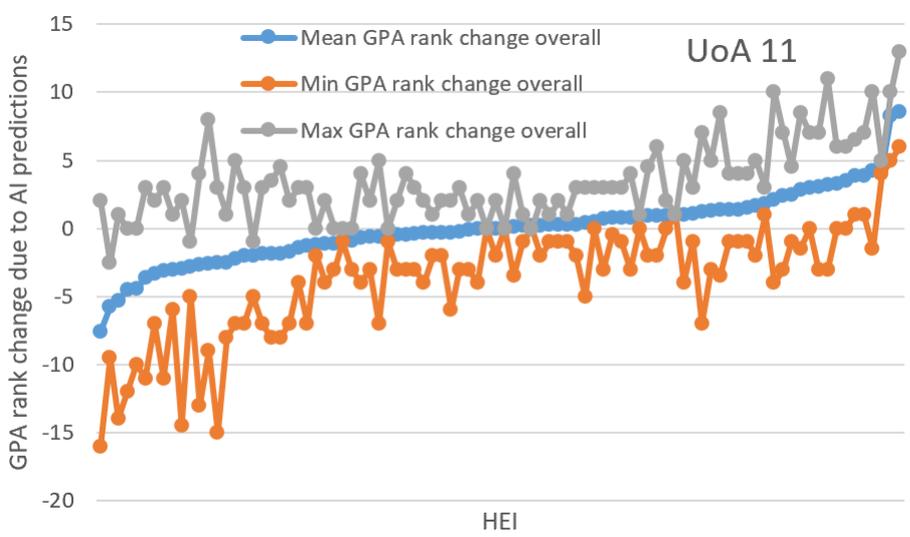
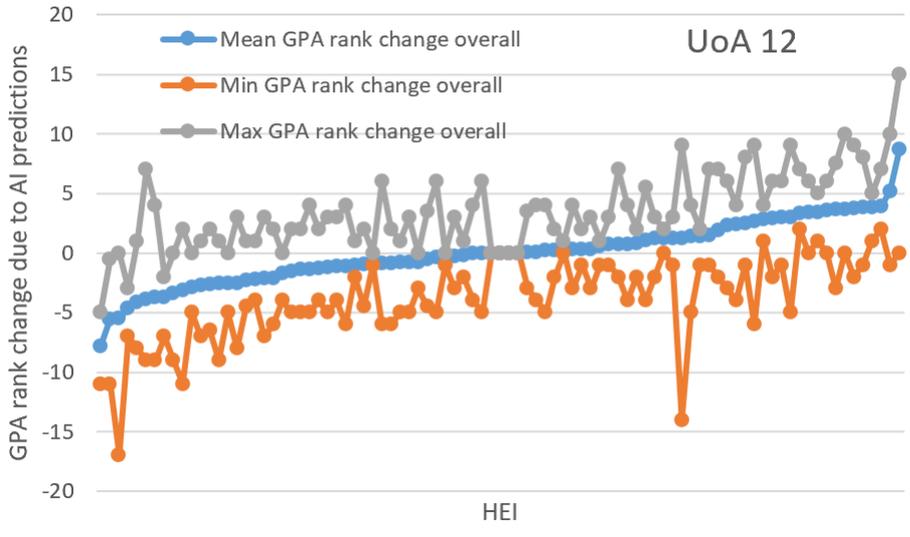

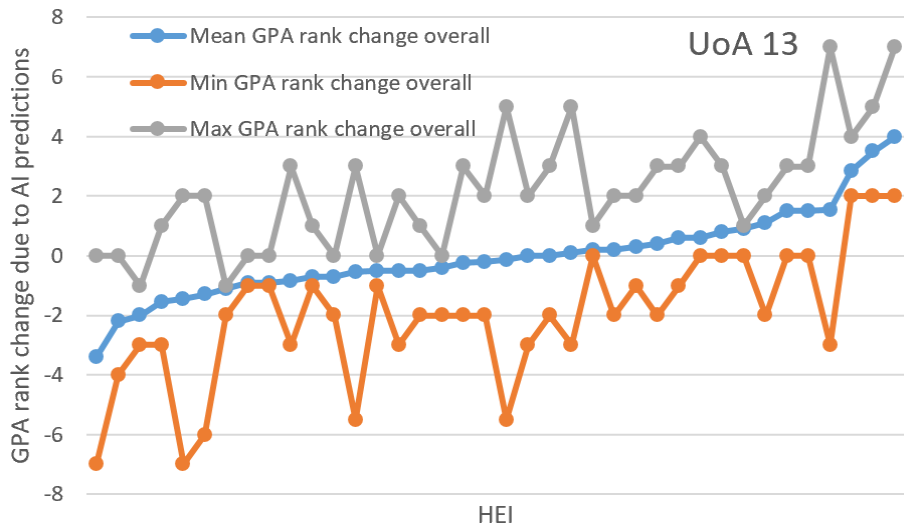

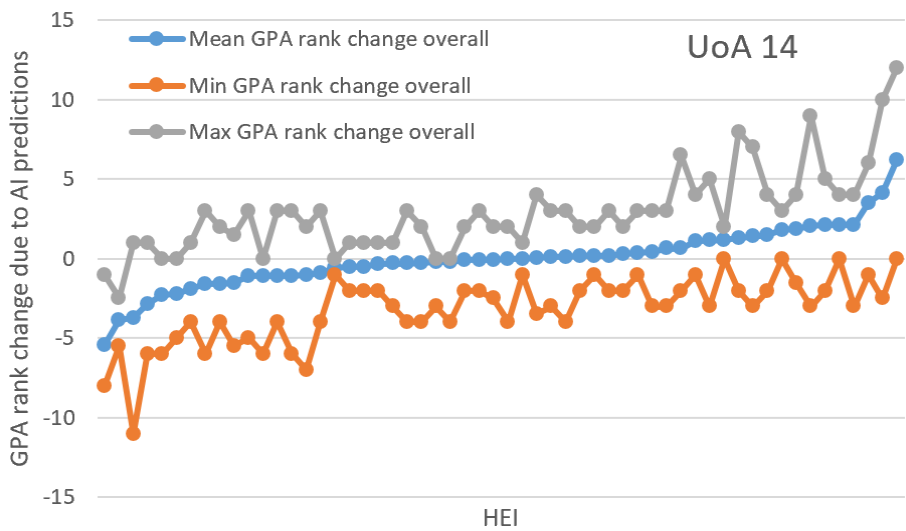

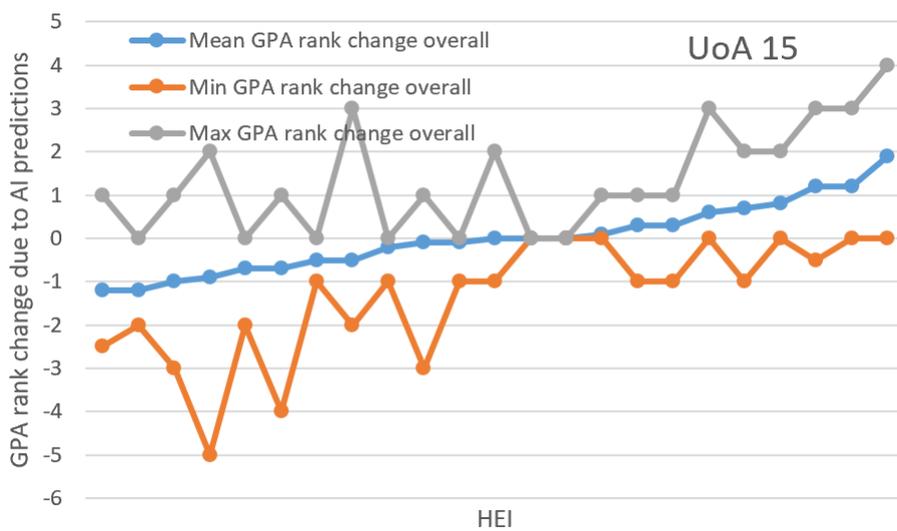

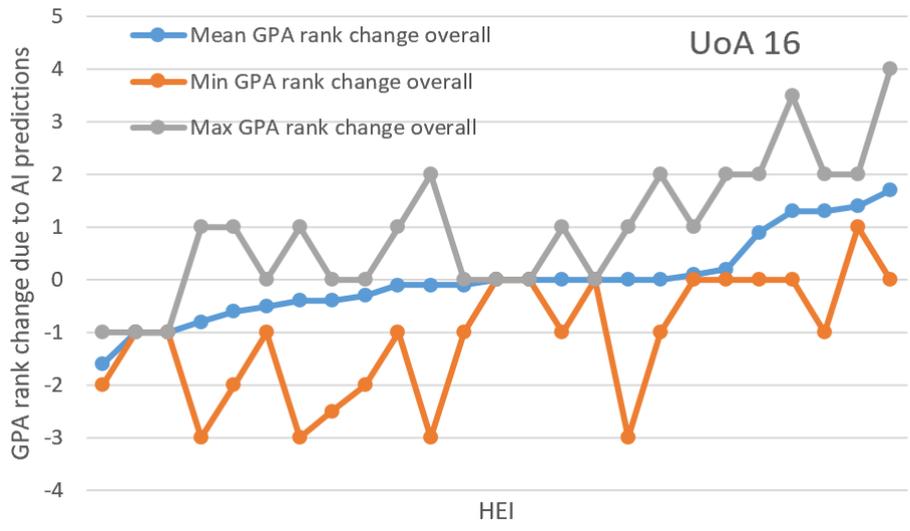
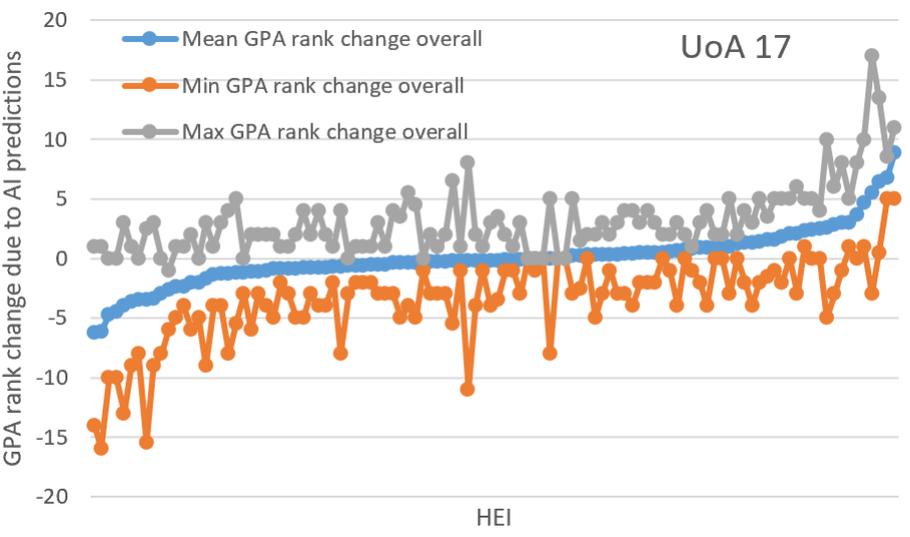
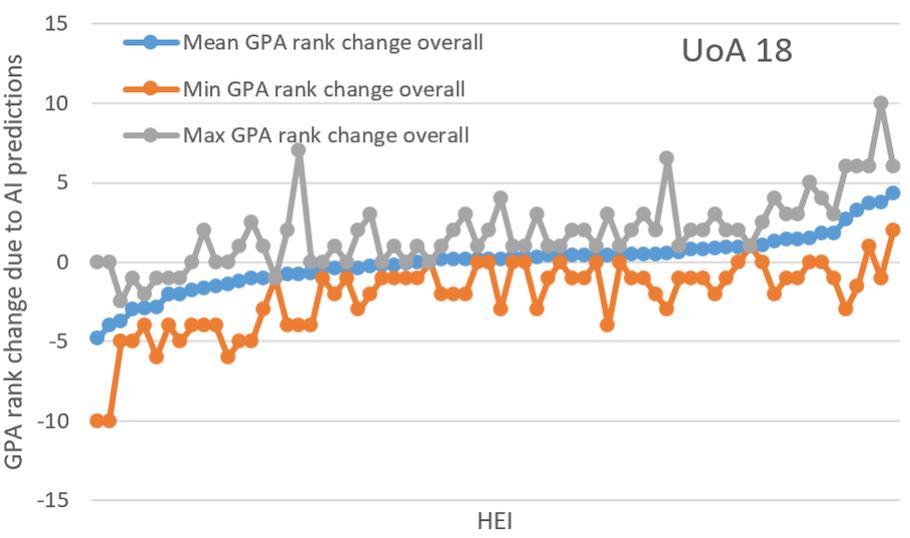

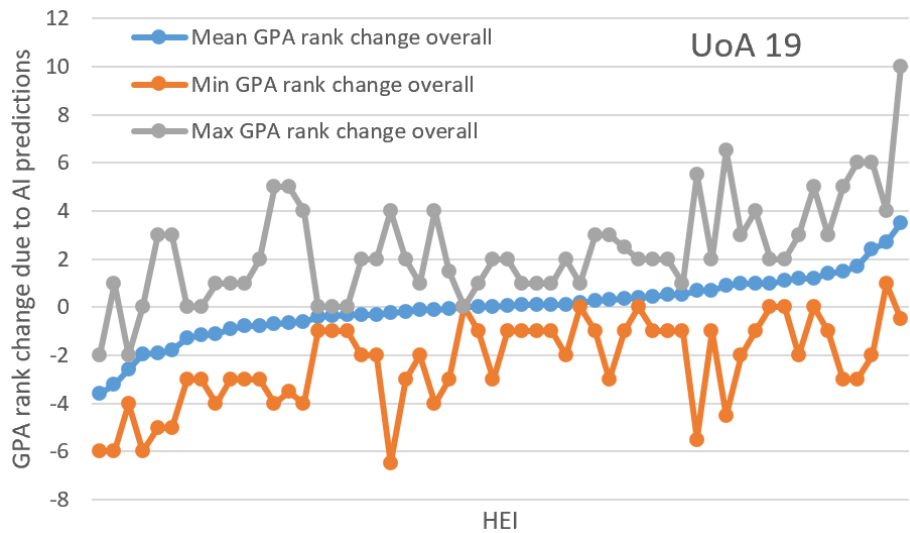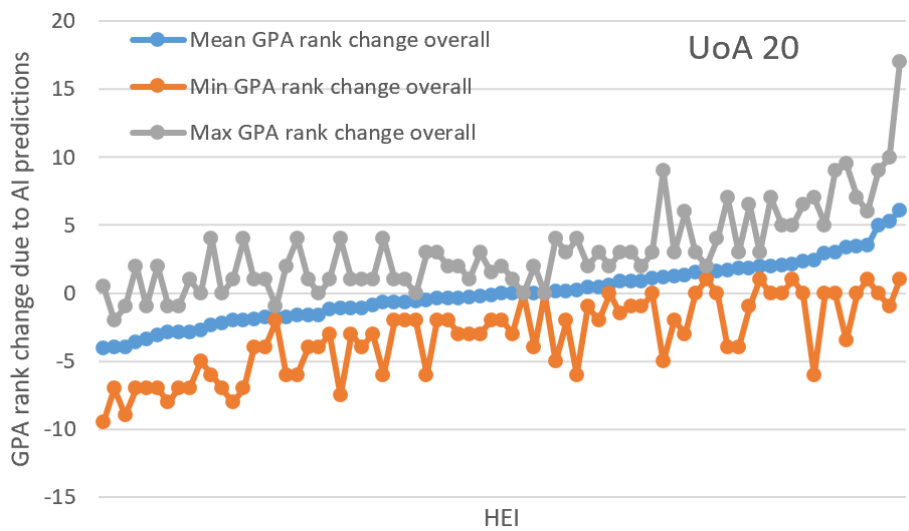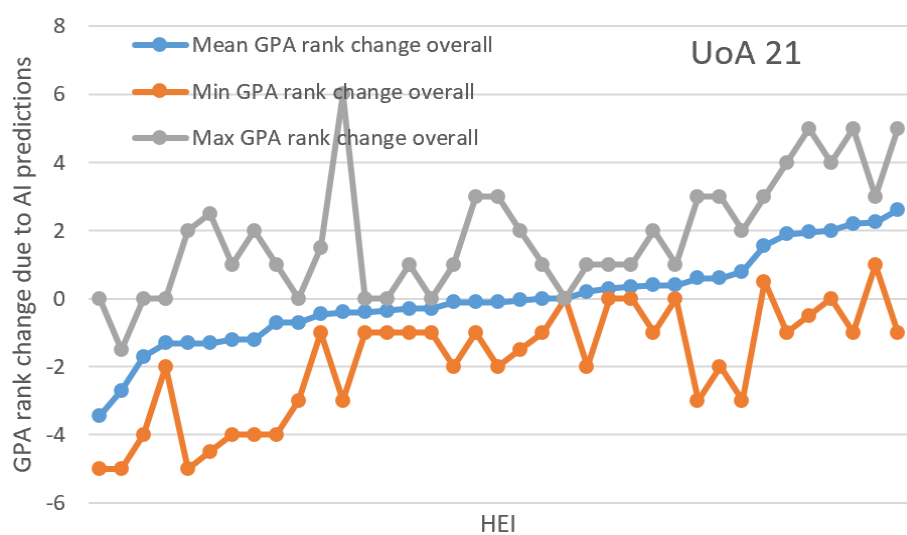

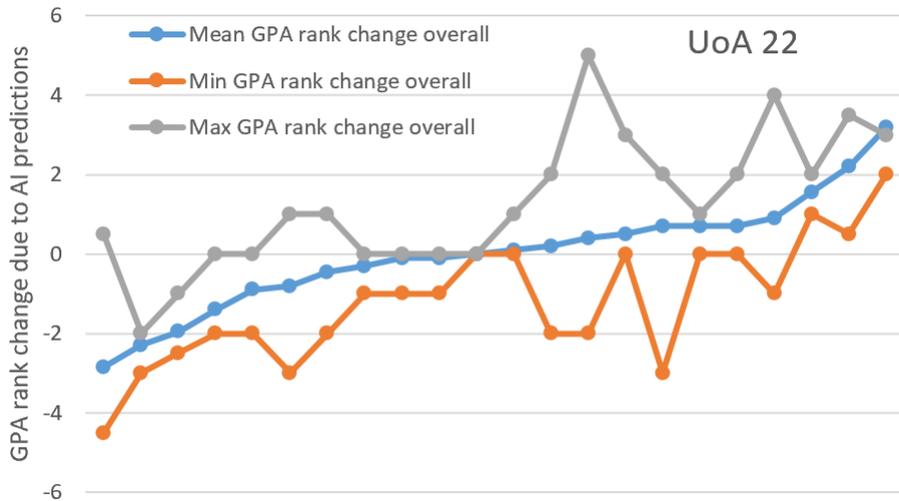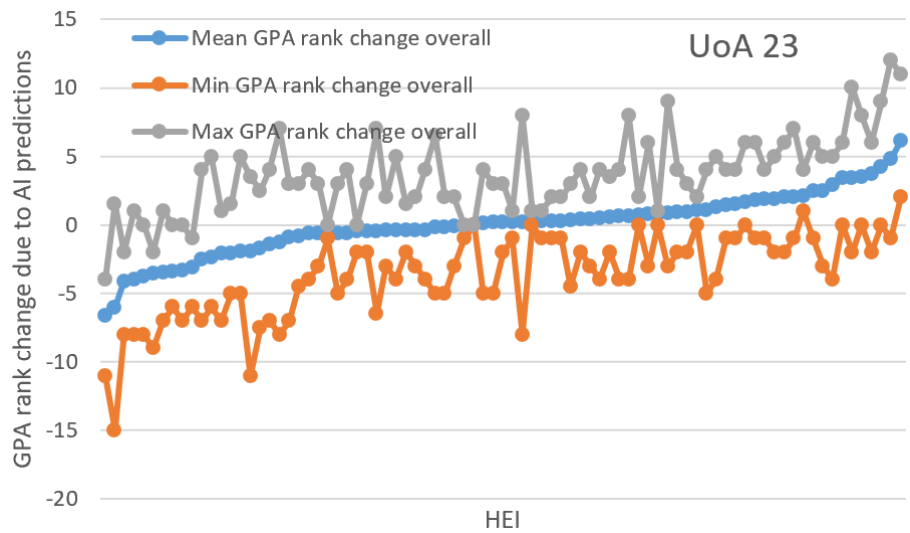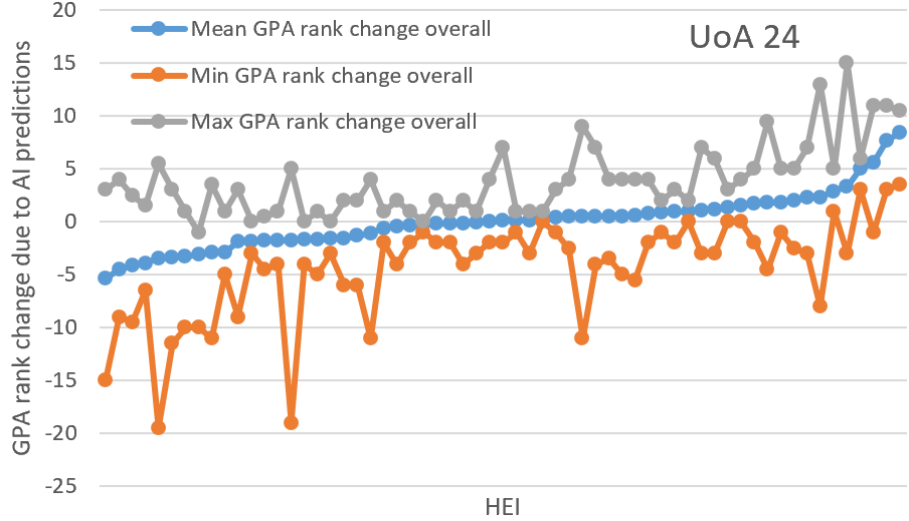

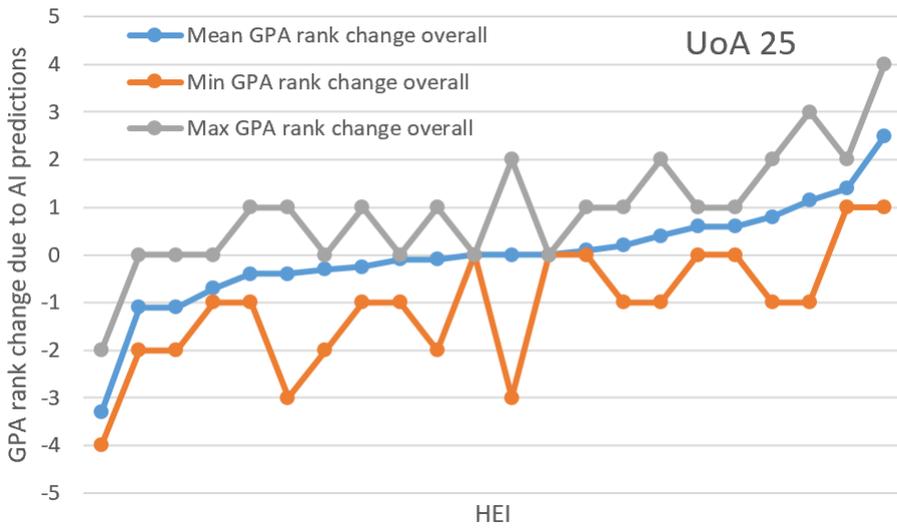
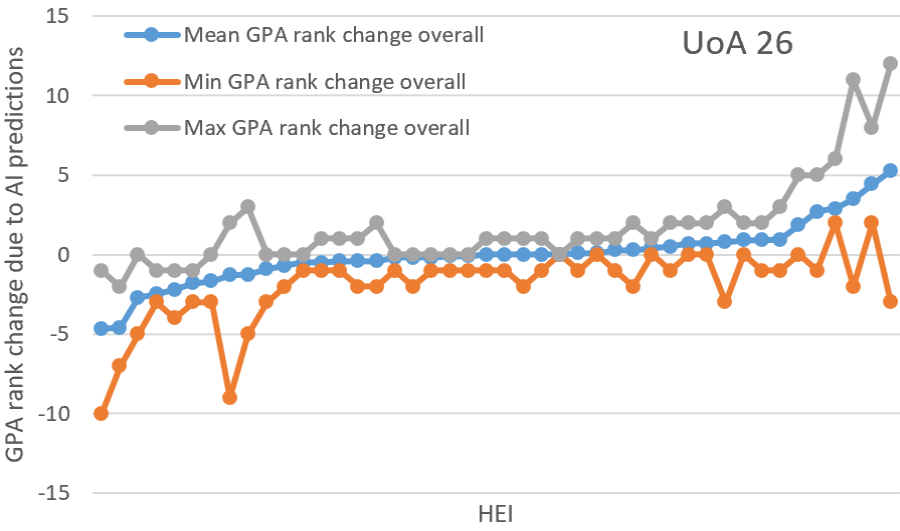
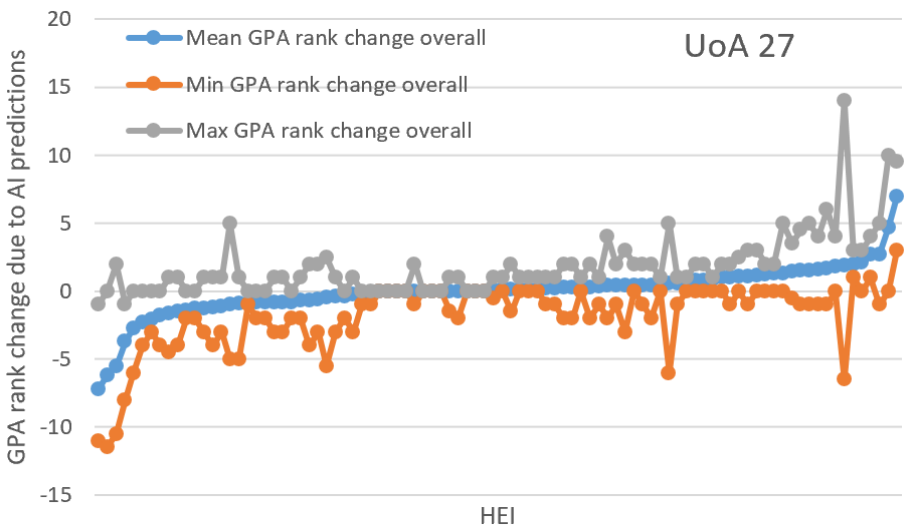

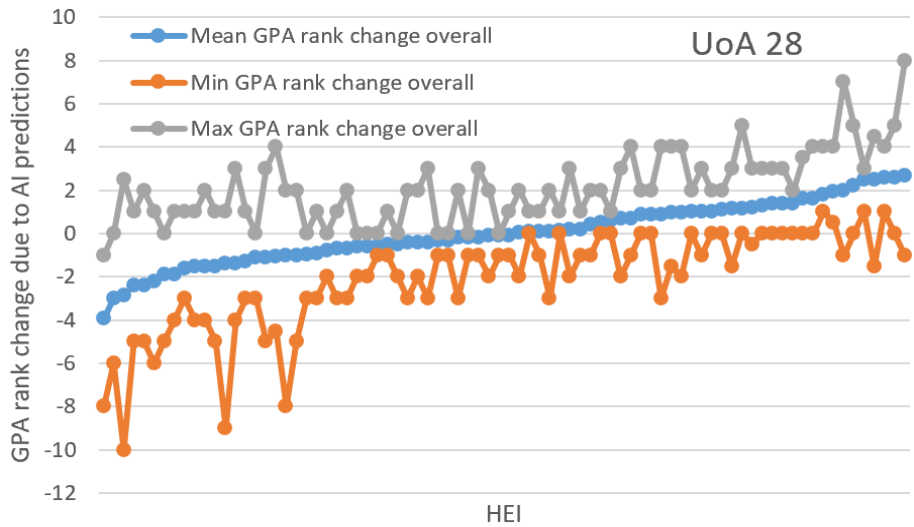

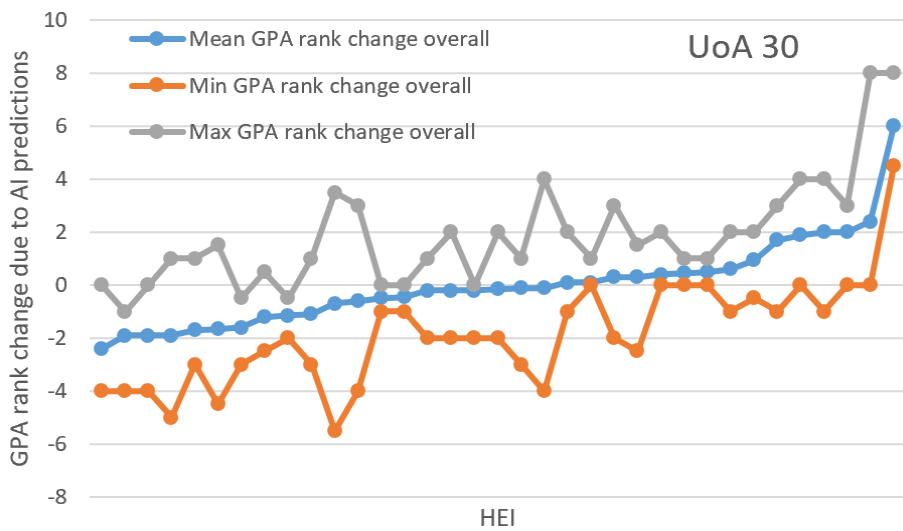

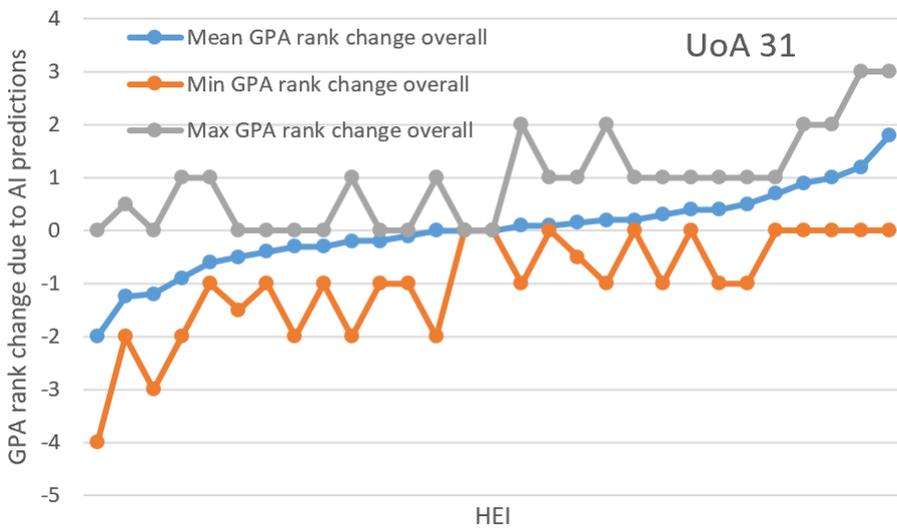

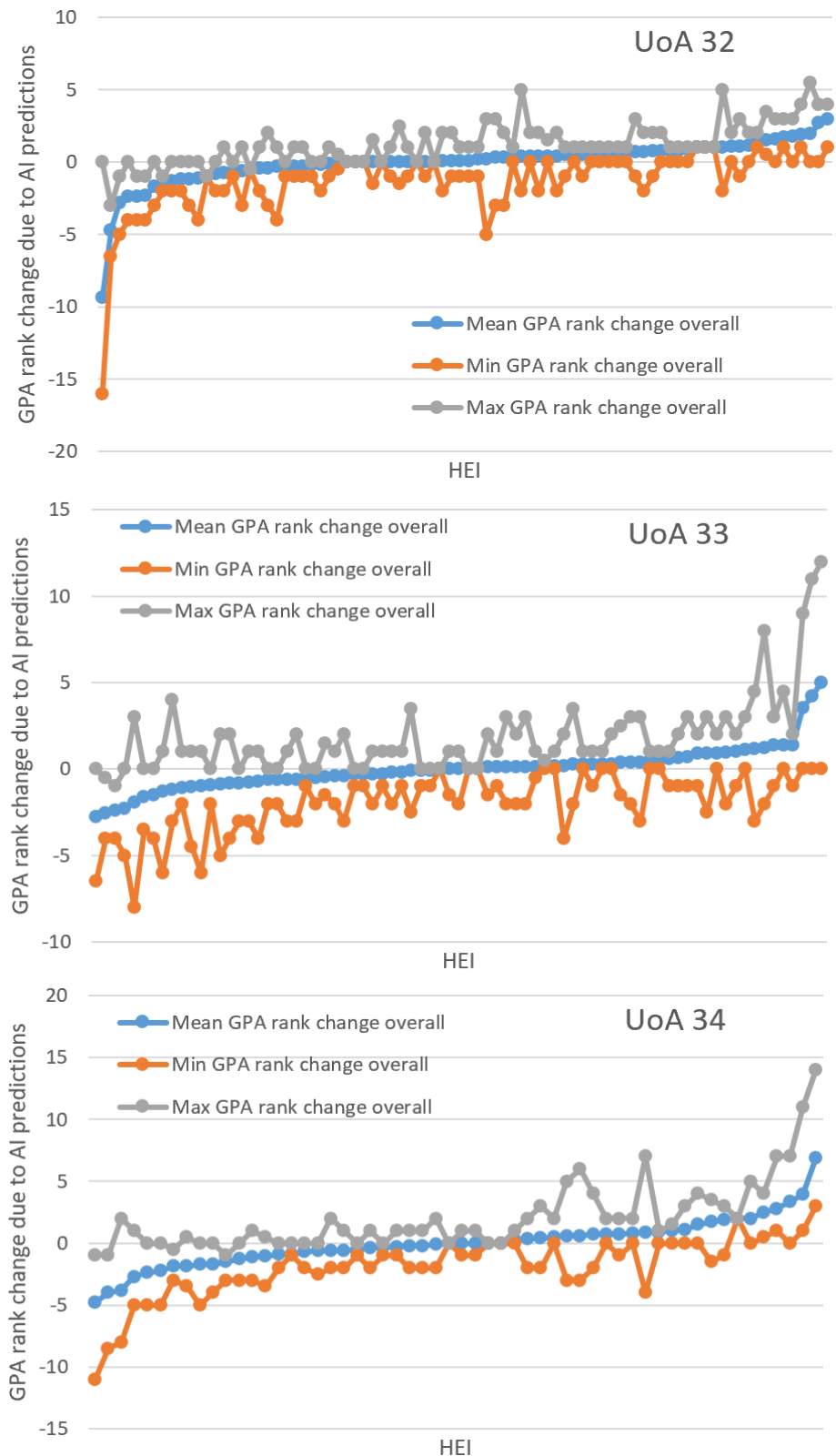

Figure 4.1.2.2. Average REF AI institutional grade point average (GPA) gain by replacing human scores with AI predictions for the predicted set, retaining the human scores for all non-journal outputs and for the non-predicted journal articles (i.e., the outcome if this strategy had been used in REF2021). Institutions are anonymised and presented in order of average rank gain. Rankings include the University of Wolverhampton without any prediction changes (its rank sometimes changes if other institutions' GPAs change around it). Scores of 1* are changed to 2* to align with the predictions. The min and max values are based on 10 iterations of the system. The results are for the most accurate ordinal machine learning method, trained on **50%** of the 2014-18 data and **bibliometric + journal + text inputs, after excluding articles with shorter than 500-character abstracts.**

Previous studies proposing methods to predict REF or RAE score profiles have often reported Spearman or Pearson correlations between predicted and actual (financial) scores. Using this approach (with Pearson correlations), the average AI predicted score correlates with the actual average score for UoAs 1-11,16 between 0.66 and 0.91 (Figure 4.1.2.3). If the scores are not size normalised but the correlations are between total scores then they are universally high for UoA 1-11, 16: between 0.945 and 9.998. Although these seem to be the highest correlations ever reported for attempts to predict REF score profiles (see the literature review), the detailed results above suggest that the AI accuracy is still insufficient.

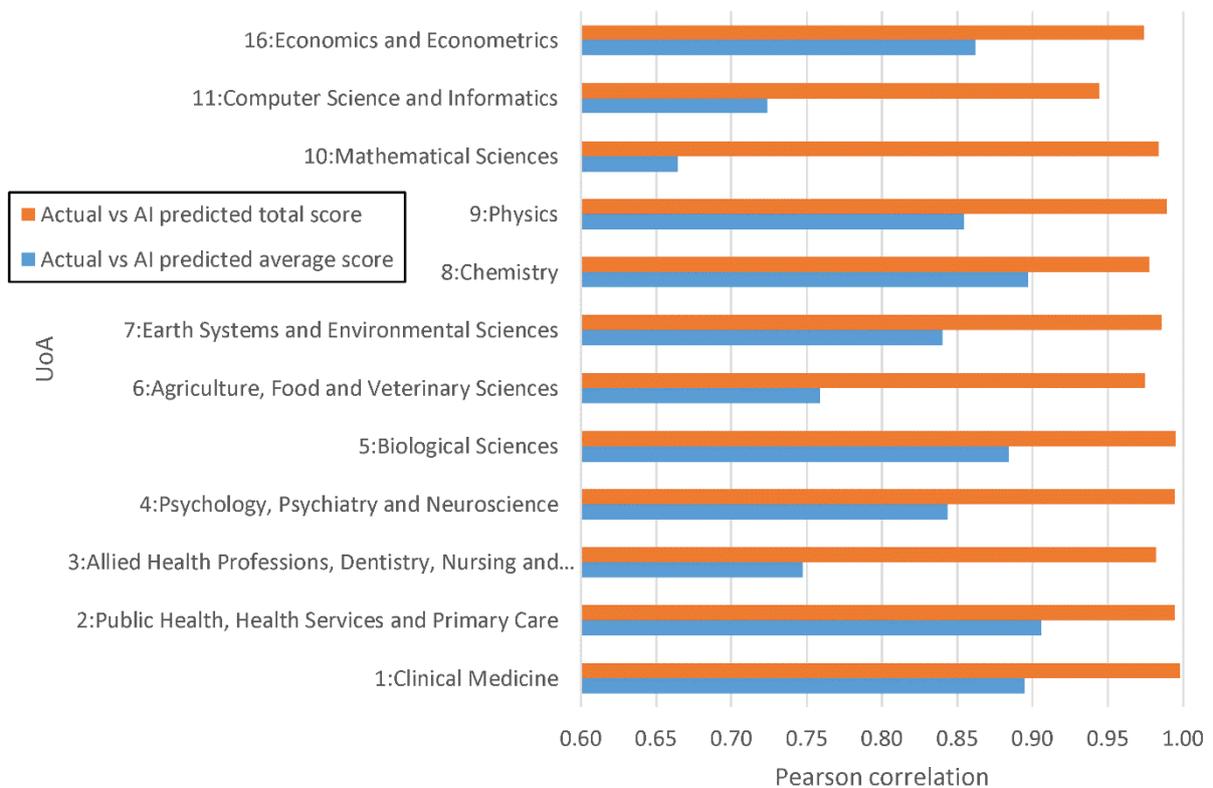

Figure 4.1.2.3. Pearson correlation between AI predicted and actual institutional score for UoA 1-11, 16 for the most accurate machine learning method, trained on **50%** of the 2014-18 data and **bibliometric + journal + text inputs, after excluding articles with shorter than 500 character abstracts**. UoAs 1,2,6-10,16 have 65%-72% raw accuracy. AI score is a financial calculation (4*=100% funding, 3*=25% funding, 0-2*=0% funding, and the average is taken for each institution).

Table 4.1.2.1. The numbers in Figure 4.1.2.2.

| UoA | Actual vs AI predicted average score | Actual vs AI predicted total score |
|---|---|---|
| 1:Clinical Medicine | 0.895 | 0.998 |
| 2:Public Health, Health Services and Primary Care | 0.906 | 0.995 |
| 3:Allied Health Professions, Dentistry, Nursing & Pharmacy | 0.747 | 0.982 |
| 4:Psychology, Psychiatry and Neuroscience | 0.844 | 0.995 |
| 5:Biological Sciences | 0.885 | 0.995 |
| 6:Agriculture, Food and Veterinary Sciences | 0.759 | 0.975 |
| 7:Earth Systems and Environmental Sciences | 0.840 | 0.986 |
| 8:Chemistry | 0.897 | 0.978 |
| 9:Physics | 0.855 | 0.989 |
| 10:Mathematical Sciences | 0.664 | 0.984 |
| 11:Computer Science and Informatics | 0.724 | 0.945 |
| 16:Economics and Econometrics | 0.862 | 0.974 |

### 4.1.3 Systematic score shifts by HEI size, submission size and submission quality

In many fields there is a substantial tendency for smaller HEIs, smaller submissions, and lower scoring submissions to have larger AI prediction gains (Figure 4.1.3.1). In all of the most predictable UoAs (1,2,6-10,16), smaller HEIs had an AI advantage, with the strength of the correlation usually being moderate (-0.2 to -0.5). Thus, smaller institutions would gain from a switch to AI with this strategy, and larger institutions would lose.

There is a strong tendency in most of the UoAs for lower scoring submissions to gain from the AI predictions, with the correlation strength being between -0.5 and -0.85 all except 2 (1,9) of the most predictable UoAs (1,2,6-10,16). Thus, replacing human scores with AI predictions using Strategy 1 would result in a substantial score shift in favour of weaker submissions. Presumably this is because, other factors being equal, an article with an AI prediction error is more likely to be a prediction loss if it is from a high scoring submission (because most scores would be 4* and could only decrease) and for the same reason it is more likely to be a prediction gain if it is from a low scoring submission.

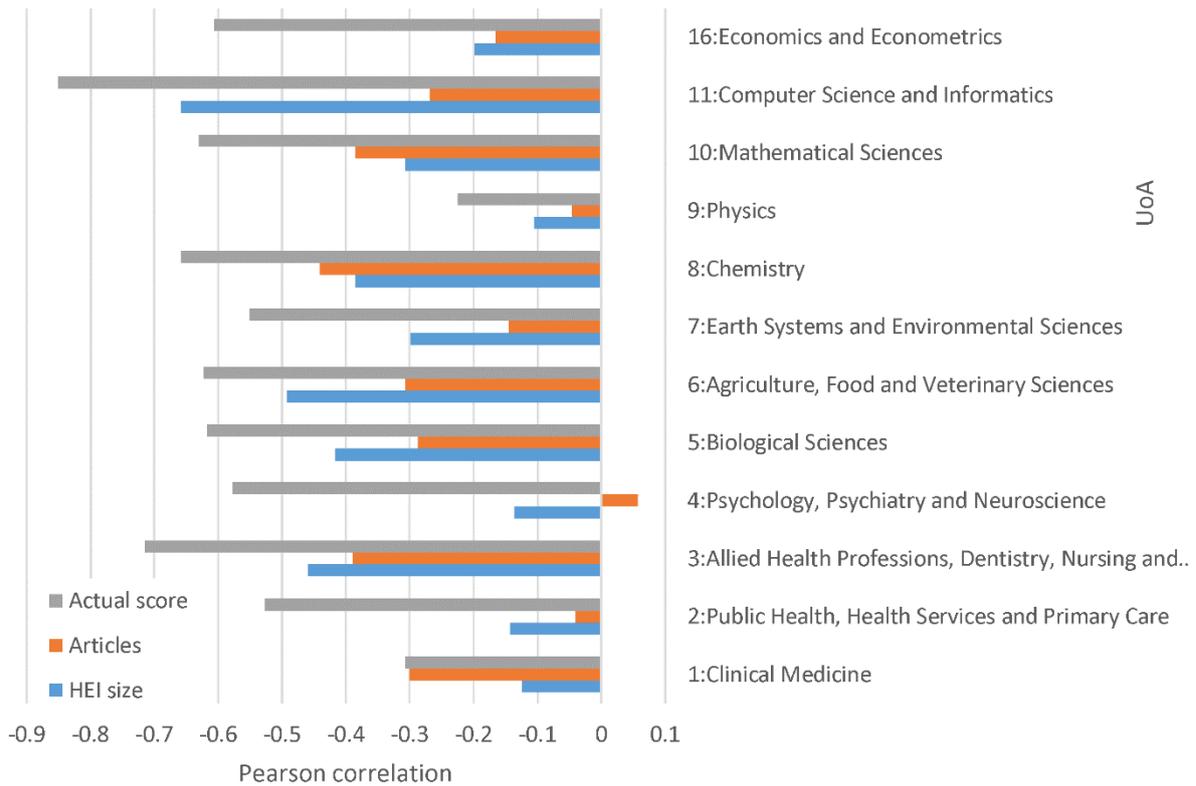

Figure 4.1.3.1. Pearson correlations between institutional size (number of articles submitted to REF), submission size (number of articles submitted to UoA) or average institutional REF score for the UoA and average REF AI institutional score gain on UoA 1-11, 16 for the most accurate machine learning method, trained on **50%** of the 2014-18 data and **bibliometric + journal + text inputs, after excluding articles with shorter than 500 character abstracts**. UoAs 1,2,6-10,16 have 65%-72% raw accuracy. Prediction gain is a financial calculation (4*=100% funding, 3*=25% funding, 0-2*=0% funding). Note: maximum score shift difference between institutions of 4% for institutions with at least 200 articles.

### 4.1.4 Institutional score shifts from avoiding AI and classifying 50% of the articles

An alternative time saving method to using AI would be to classify a randomly selected 50% of the articles from larger HEIs, doubling the scores to account for the remaining 50%. For example, with this approach an institution submitting 200 articles would have 100 human-classified and the remaining 100 would be assumed to receive the same scores as the first 100.

This approach would not induce an average score shift but would reduce the accuracy of the results. This approach was simulated 10 times for each UoA, reporting the highest and lowest averages out of 10 each time for the largest HEIs (Figure 4.1.4.1 to Figure 4.1.4.5). This would reduce accuracy, with a realistic chance than an institution's funding based on the revised scores would increase or decrease by 4% overall. This is larger than the apparent inaccuracy of the optimal AI method for UoA 1 but would be preferable to AI for UoAs where the AI is less accurate.

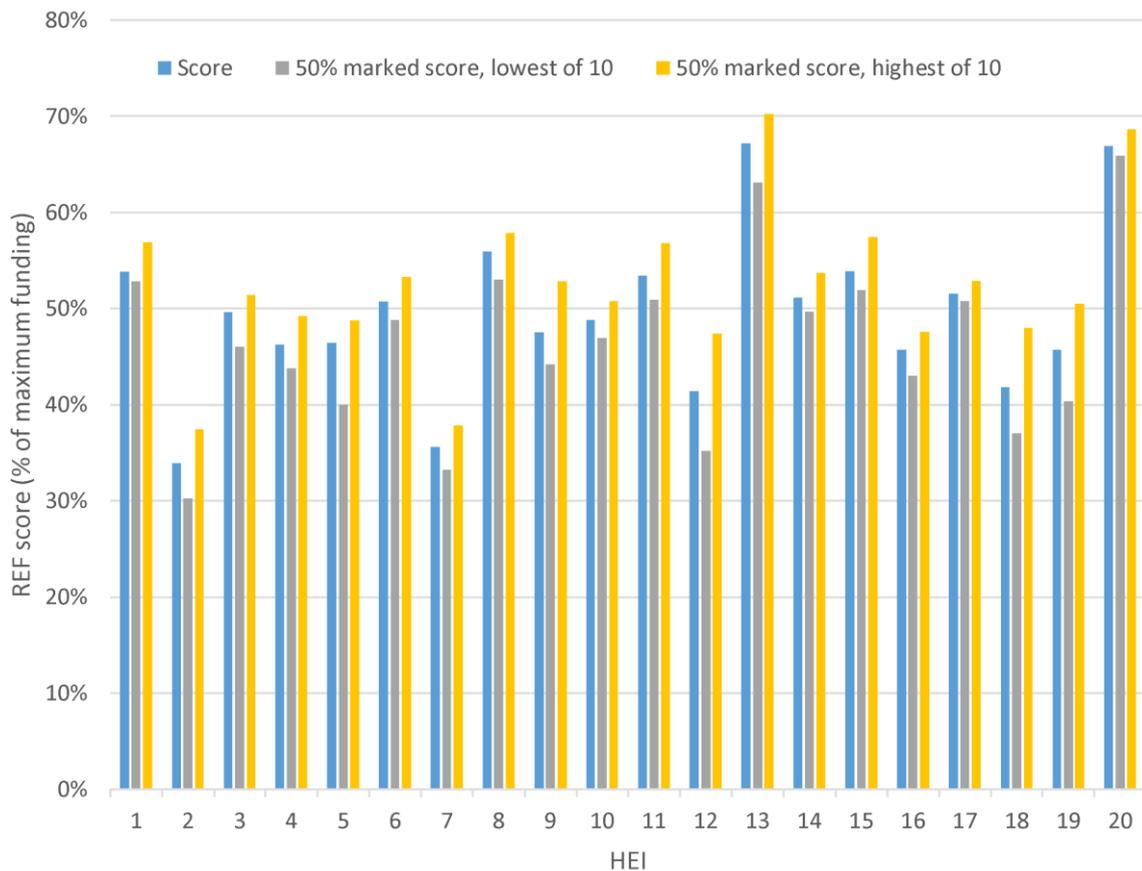

Figure 4.1.4.1. Average REF scores (4*=100% funding, 3*=25% funding, 1*-2*=0% funding) for institutions submitted to **UoA 1** and the lowest and highest scores out of 10 random samples of 50%. Qualification: at least 200 journal article outputs submitted to UoA. HEIs anonymised and listed in random order.

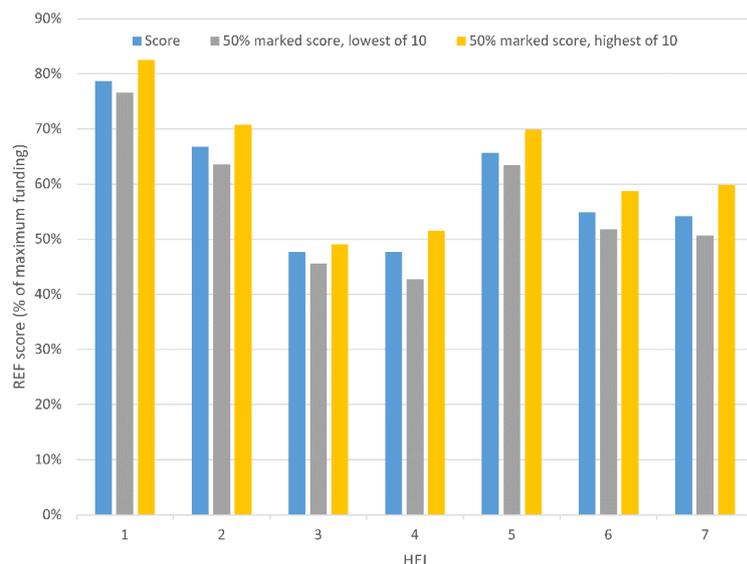

Figure 4.1.4.2. Average REF scores (4*=100% funding, 3*=25% funding, 1*-2*=0% funding) for institutions submitted to **UoA 2** and the lowest and highest scores out of 10 random samples of 50%. Qualification: at least 200 journal article outputs submitted to UoA. HEIs anonymised and listed in random order.

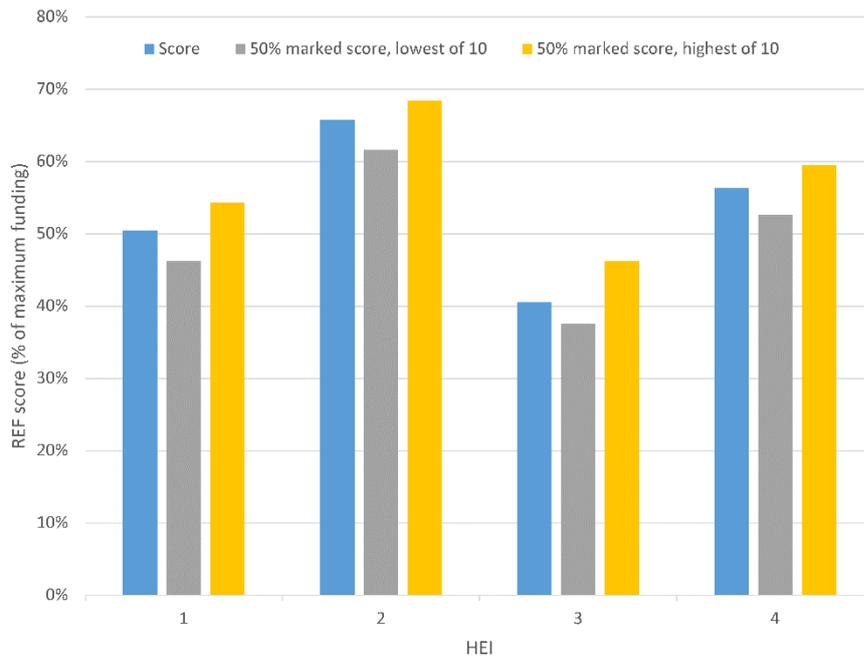

Figure 4.1.4.3. Average REF scores (4*=100% funding, 3*=25% funding, 1*-2*=0% funding) for institutions submitted to **UoA 7** and the lowest and highest scores out of 10 random samples of 50%. Qualification: at least 200 journal article outputs submitted to UoA. HEIs anonymised and listed in random order.

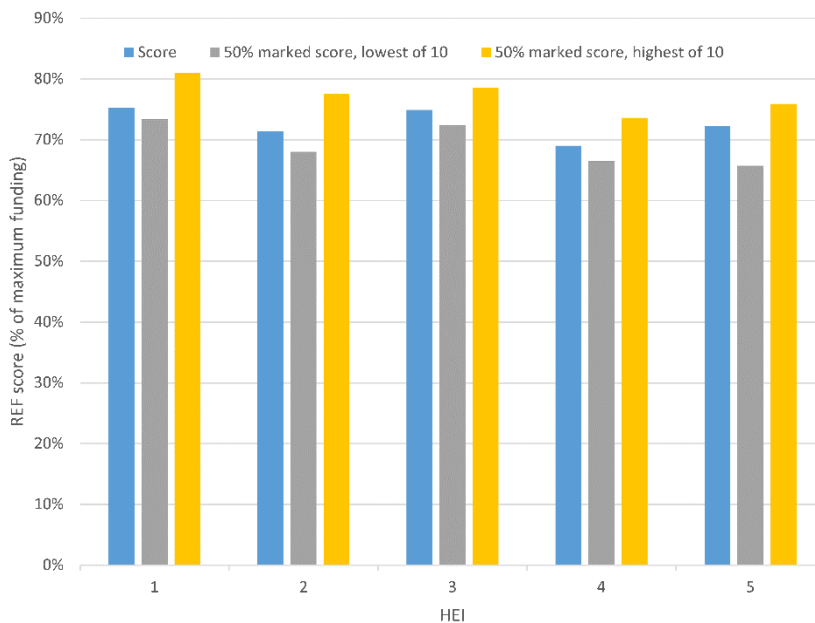

Figure 4.1.4.4. Average REF scores (4*=100% funding, 3*=25% funding, 1*-2*=0% funding) for institutions submitted to **UoA 8** and the lowest and highest scores out of 10 random samples of 50%. Qualification: at least **150** journal article outputs submitted to UoA. HEIs anonymised and listed in random order.

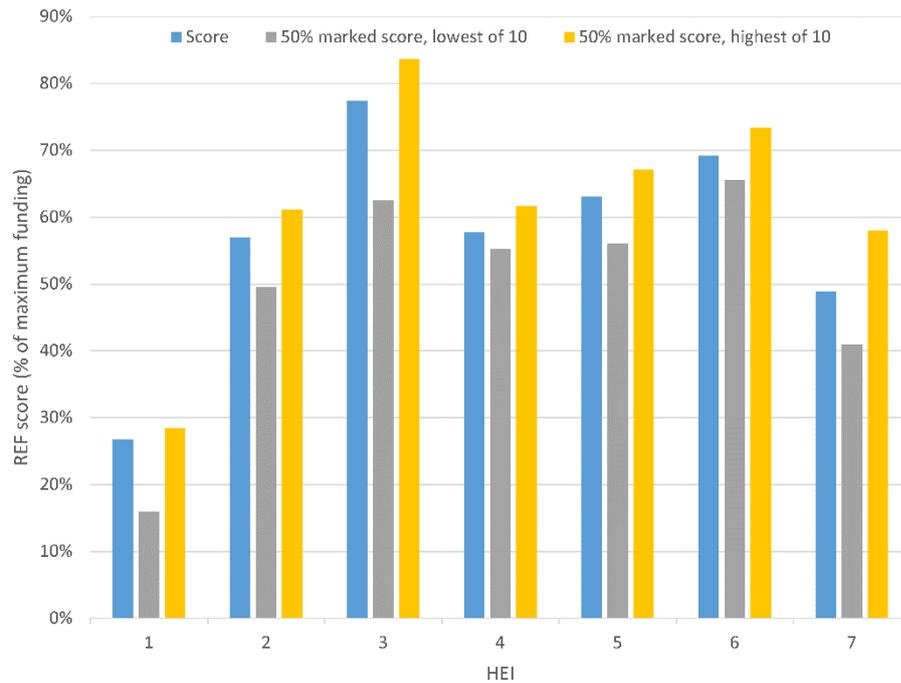

Figure 4.1.4.5. Average REF scores (4*=100% funding, 3*=25% funding, 1*-2*=0% funding) for institutions submitted to **UoA 9** and the lowest and highest scores out of 10 random samples of 50%. Qualification: at least 200 journal article outputs submitted to UoA. HEIs anonymised and listed in random order.

### 4.1.5 Gender and Early Career Researcher (ECR) status score shifts

Early Career Researchers (ECRs) are a special category of interest in the REF. Tests were run for the most accurate AI predictions to see whether they produced a shift in favour or against experienced researchers. Switching from human REF scores to AI predictions does not systematically work in favour or against ECRs in any of the top UoAs (Figure 4.1.5.1).

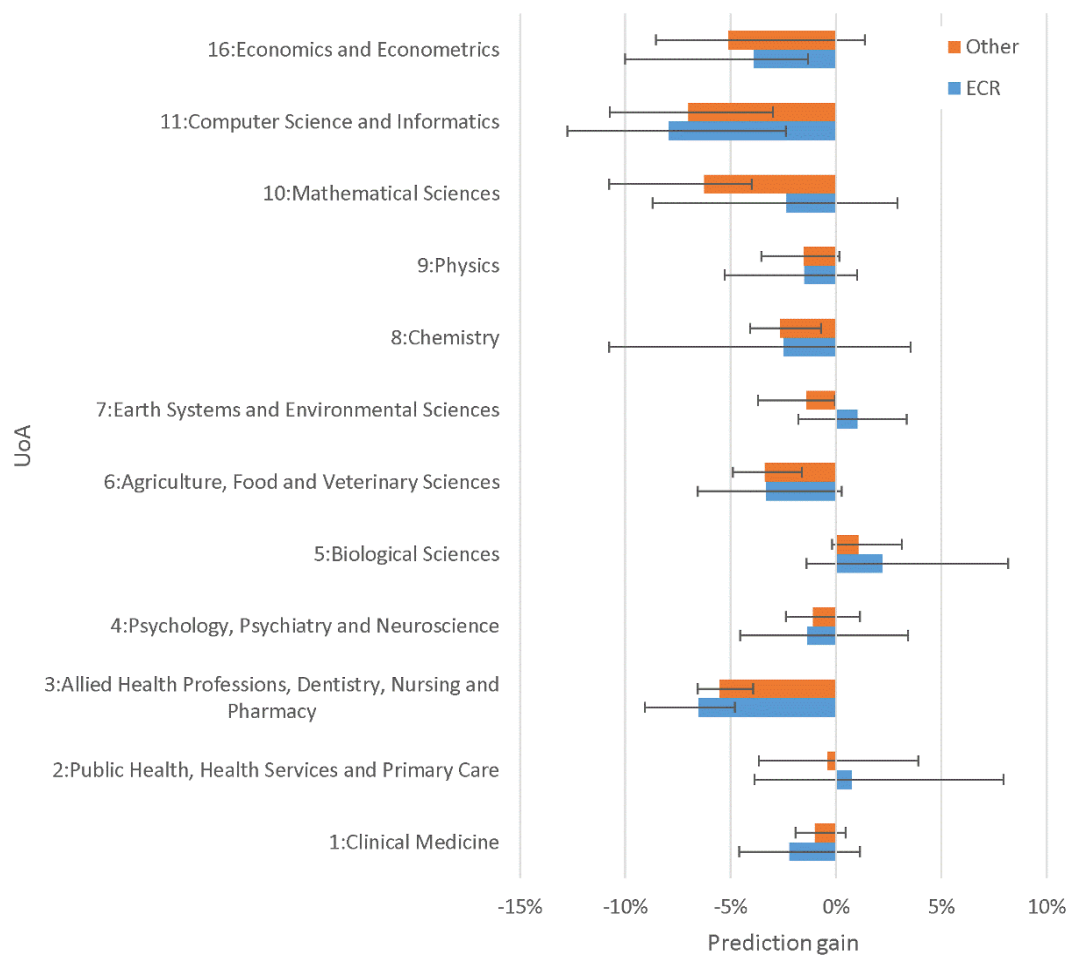

Figure 4.1.5.1. Average REF score AI prediction gains (AI score subtract reviewer score) for ECRs and experienced researchers in twelve UoAs for 2014-18 data and the most accurate machine learning method, trained on **50%** of the 2014-18 data and **bibliometric + journal + text inputs** ten times. REF score is a financial calculation (4*=100% funding, 3*=25% funding, 0-2*=0% funding). Error bars show the highest and lowest value from ten separate sets of AI predictions. UoAs 1,2,6-10,16 have 65%-72% raw accuracy.

Gender is a protected characteristic and, given historic extreme gender prejudice in HEIs and evidence of substantial ongoing gender problems (e.g., a shortfall of senior female researchers), it is natural to check for gender score shifts in AI REF score predictions. Tests were run for the most accurate data to see whether the AI produced a male or female gender score shift. Gender was inferred from first author first names and hence non-binary genders could not be analysed. The oversimplification was made that the first author was probably the submitting author (submitting author gender was not available). Switching from human REF scores to AI predictions does not systematically work in favour or against male or female first authored articles in any of the top UoAs (Figure 4.1.5.2).

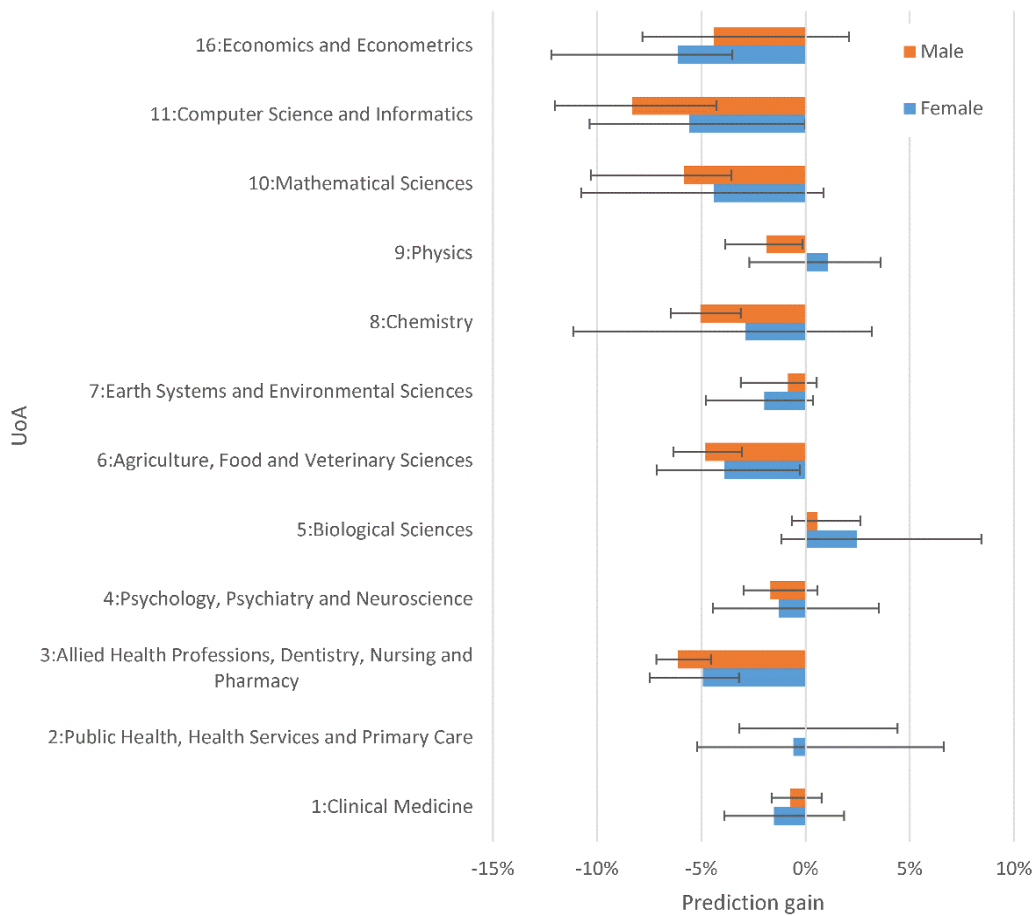

Figure 4.1.5.2. Average REF score AI prediction gains for male and female first authored articles in twelve UoAs for 2014-18 data and the most accurate machine learning method, trained on **50%** of the 2014-18 data and **bibliometric + journal + text inputs** ten times. REF score is a financial calculation (4*=100% funding, 3*=25% funding, 0-2*=0% funding). Error bars show the highest and lowest value from ten separate sets of AI predictions.

### 4.1.6 Interdisciplinarity score shift tests

The REF allows submitting HEIs to indicate that outputs are interdisciplinary such that the sub-panels are able to make appropriate allowances in their assessment. However, the 'flags' indicating interdisciplinary outputs were not reliable since they seem to have been applied differently between institutions, so the results here should be interpreted with caution. Scores and predictions for outputs flagged as interdisciplinary were compared with scores and predictions for the remaining outputs for the most accurate AI sets. AI score predictions did not systematically favour interdisciplinary or other research in any UoA, although interdisciplinary research may have an AI advantage in UoA 16 (Figure 4.1.6.1).

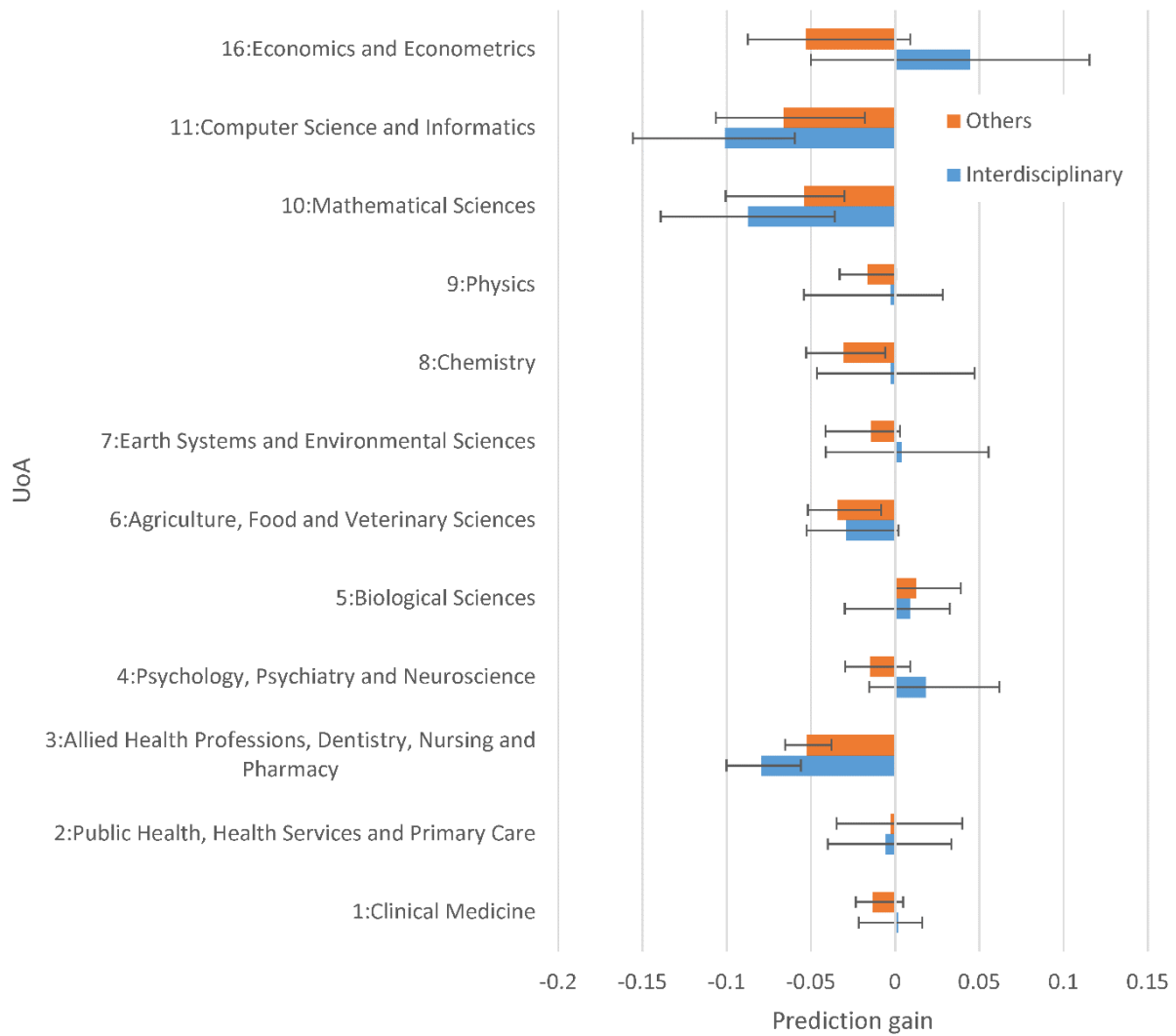

Figure 4.1.6.1. Average REF score AI prediction gains for articles flagged as interdisciplinary or not in twelve UoAs for 2014-18 data and the most accurate machine learning method, trained on **50%** of the 2014-18 data and **bibliometric + journal + text inputs** ten times. REF score is a financial calculation (4*=100% funding, 3*=25% funding, 0-2*=0% funding). Error bars show the highest and lowest value from ten separate sets of AI predictions.

### 4.1.7 Cross training between years

It would be useful to know whether models built on data from one year range would work as well on other year ranges. A positive answer might allow models from REF2021 outputs to predict scores in future exercises. To test the year sensitivity of the AI, models were built with 2014 data and then applied to articles from other years to see if and how the predictions decayed in accuracy over time.

Although the trend is sporadic, there is a general tendency in all main panels and the most predictable UoAs for the accuracy of the method to be similar or lower for the following two to four years, and then a clear decrease for years five, six and seven (Figure 4.1.7.1, Figure 4.1.7.2). The decrease for the three most recent years is probably due, at least in part, to the weaker citation count data (NLCS) for these years because of a shorter time to accrue citations. The tendency for the predictions to become less accurate over time up to five years is probably not due to citations but due to changes in the journals, topics, and standard methods in each field. In this context, it does not seem practical to use models built with REF2021 scores for future exercises. This is even though all inputs are normalised to allow comparability between years, as far as possible. There are known disciplinary differences in the extent to which issues change in importance over time (Thelwall & Sud, 2021), so it is expected that not all UoAs follow the same pattern.

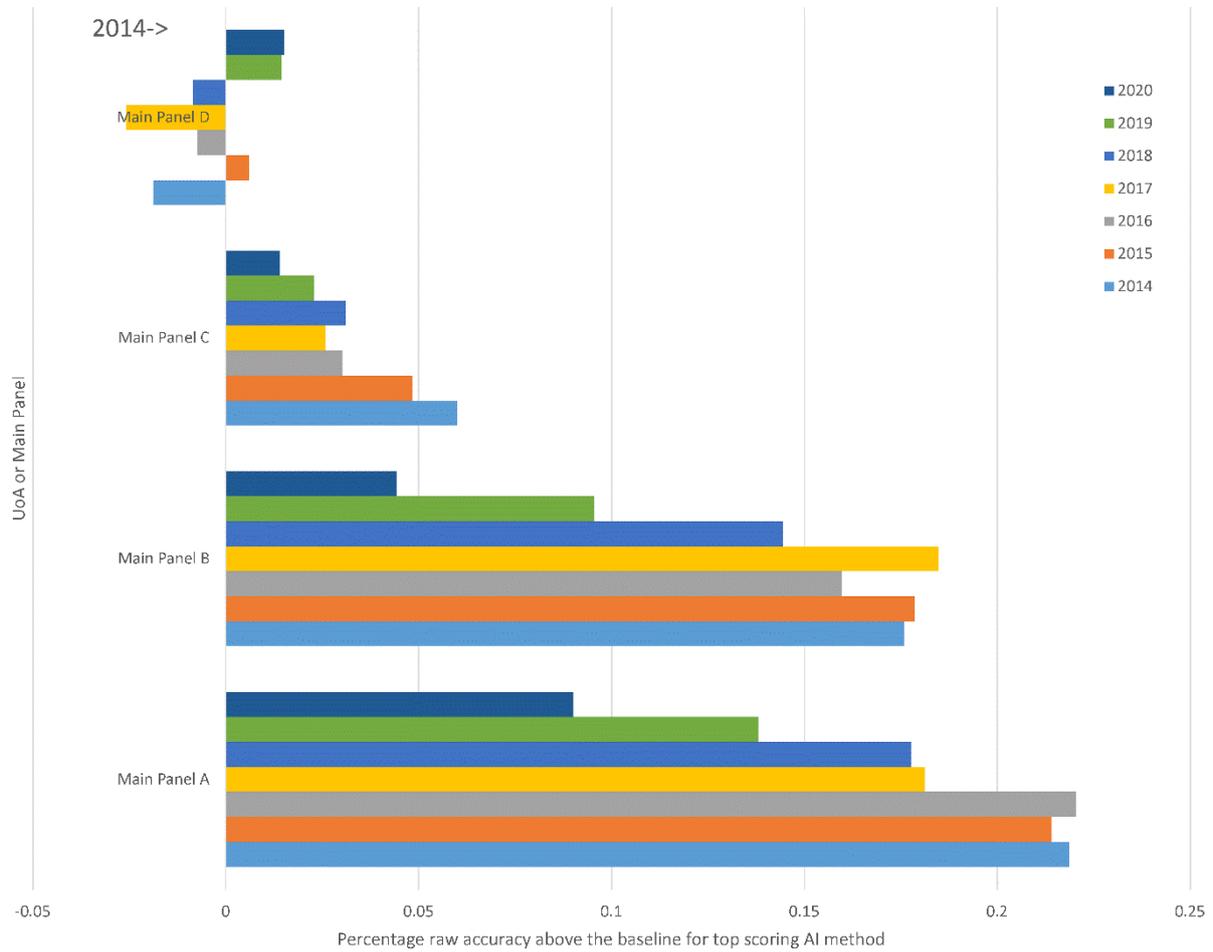

Figure 4.1.7.1. Average accuracy across 10 iterations for training on 50% of the 2014 articles from main panels and testing on either the remaining 50% of 2014 or 100% of another year.

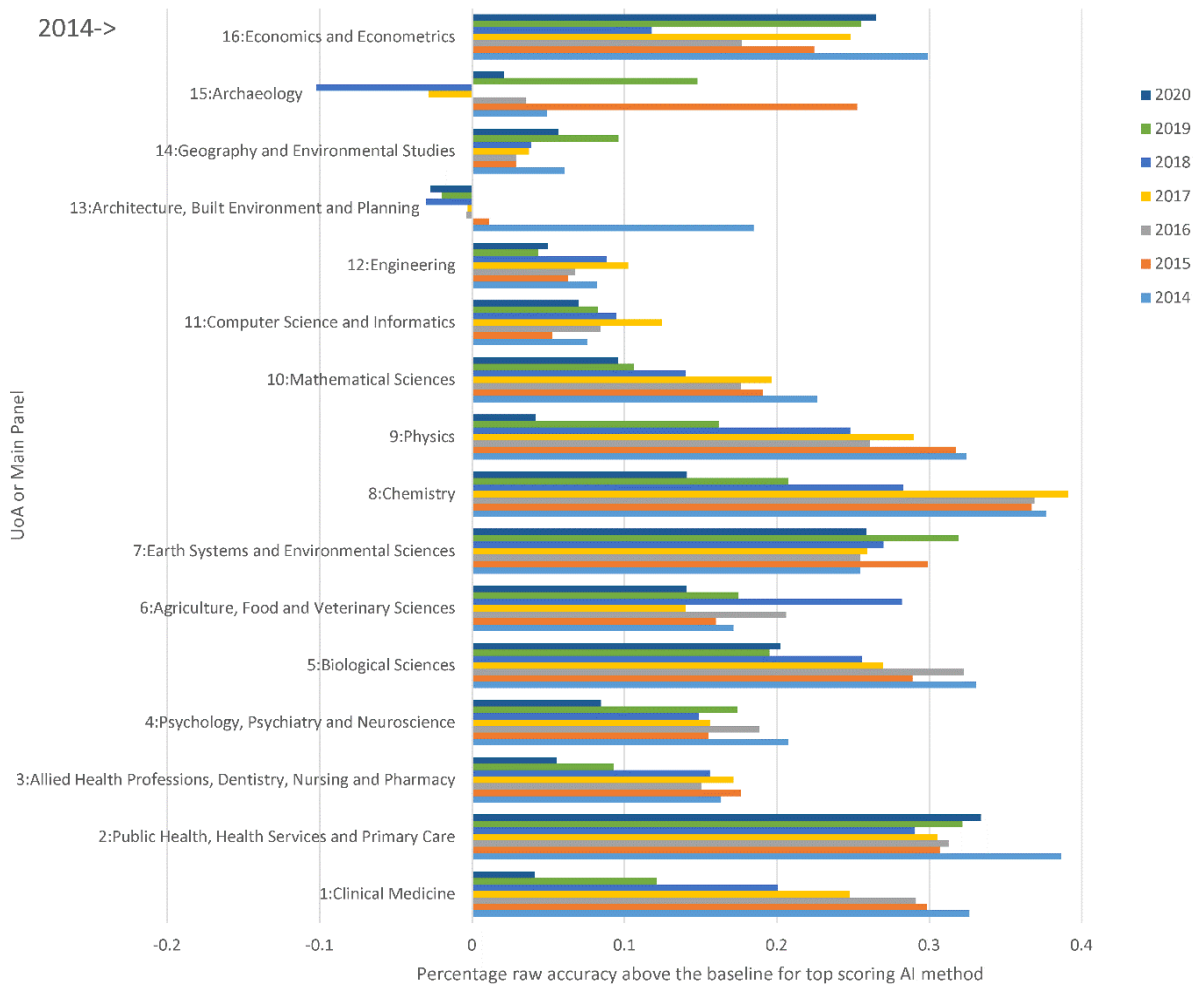

Figure 4.1.7.2. Average accuracy across 10 iterations for training on 50% of the 2014 articles from UoAs and testing on either the remaining 50% of 2014 or 100% of another year.

In more detail, accuracy might be lower when an AI algorithm is trained on data from a different year or REF than the one it is applied to, for the following reasons.

- **Changes over time in important words and phrases**. Because the text analysis component partly identifies topics and these change in importance over time, AI harnessing text analysis is likely to decrease in accuracy over time. For example, "Covid-19" might be a good indicator of highly valued research in 2020 but by the end of the next assessment exercise it might be an indicator of obsolete research in some fields.
- **Changes in time in the balance of importance for metadata**. The age of the citation data influences its importance, but the speed at which citations accumulate may be increasing (e.g., due preprints, online first, rapid reviewing online megajournals), so future models may need different bibliometrics weightings. Similarly, average collaboration rates are increasing over time, so author factors may also change in importance over time, and the values indicating likely higher quality research may also change over time. Similarly, the structure of science may change, altering the balance of other factors in unpredictable ways.

## 4.2 Strategy 2: High probability predictions only

This strategy identifies articles that can have their scores predicted by AI with a high degree of probability. The AI methods used to predict article scores report an estimate of the probability that these predictions are correct. By arranging the articles in descending order of these prediction probabilities, it is possible to identify a subset of the articles that can have their REF score estimated with a higher degree of accuracy than for the set overall. Strategy 2 entails accepting the highest probability predictions, running a second round of human scoring for the remaining articles with a lower probability of the AI prediction being correct. Examples of possible high prediction probability outputs might be a highly cited article in field leading journal with a large international team of authors with a good research track record, mentioning a robust method (e.g., randomised controlled trial) (probably 4*) or a solo uncited article in rarely cited journal with a new author not mentioning a recognised research method (probably 1*/2*).

The graphs below (Figure 4.2.1 to Figure 4.2.5) can be used to read the number of articles that can be predicted by the AI with any given degree of accuracy. For example, setting the accuracy threshold at 90%, 500 article scores could be predicted reliably by the AI for UoA 1, 160 for UoA 5, and less for the remainder. The graphs report the true accuracy (from comparing with the sub-panel provisional scores) rather than the AI accuracy estimate, which was only used to rank the articles.

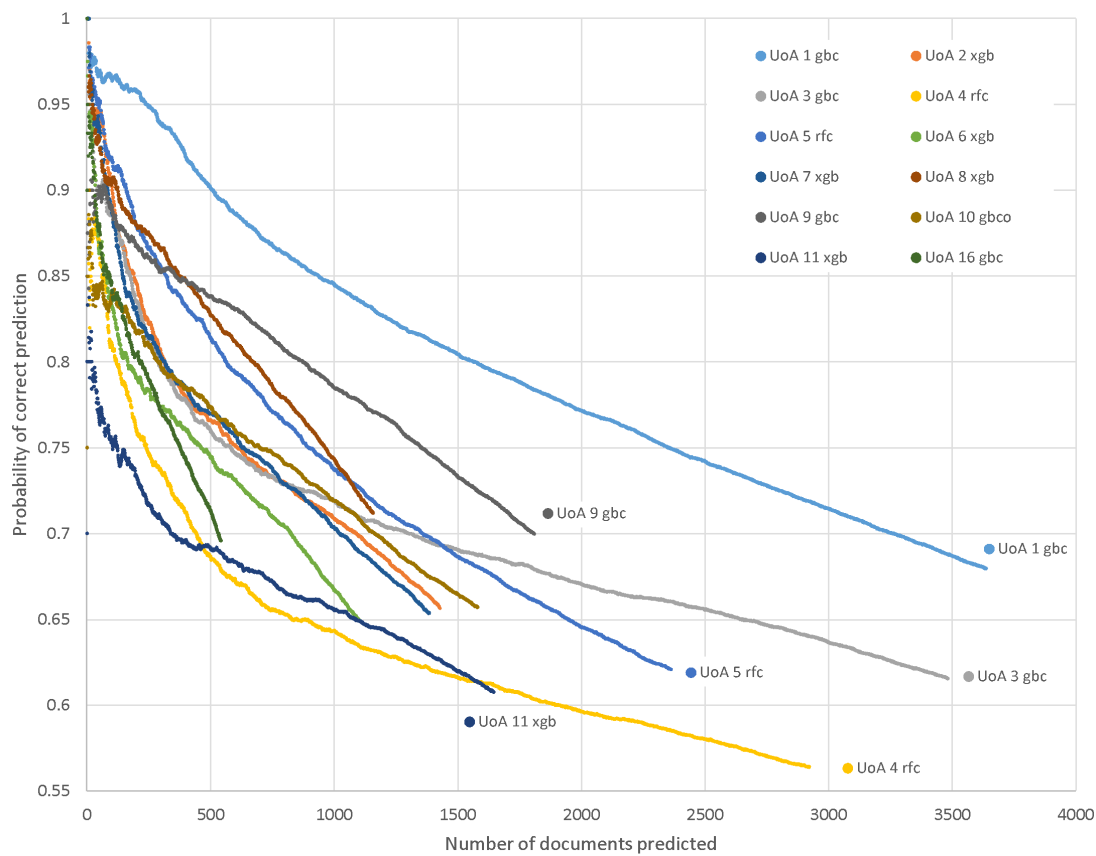

Figure 4.2.1. Probability of an AI prediction (best machine learning method at the 85% level, trained on 50% of the data 2014-18 with 1000 features) being correct against the number of predictions for twelve **UoAs**. The articles are arranged in order of the probability of the prediction being correct, as estimated by the AI. Each point is the average across 10 separate experiments.

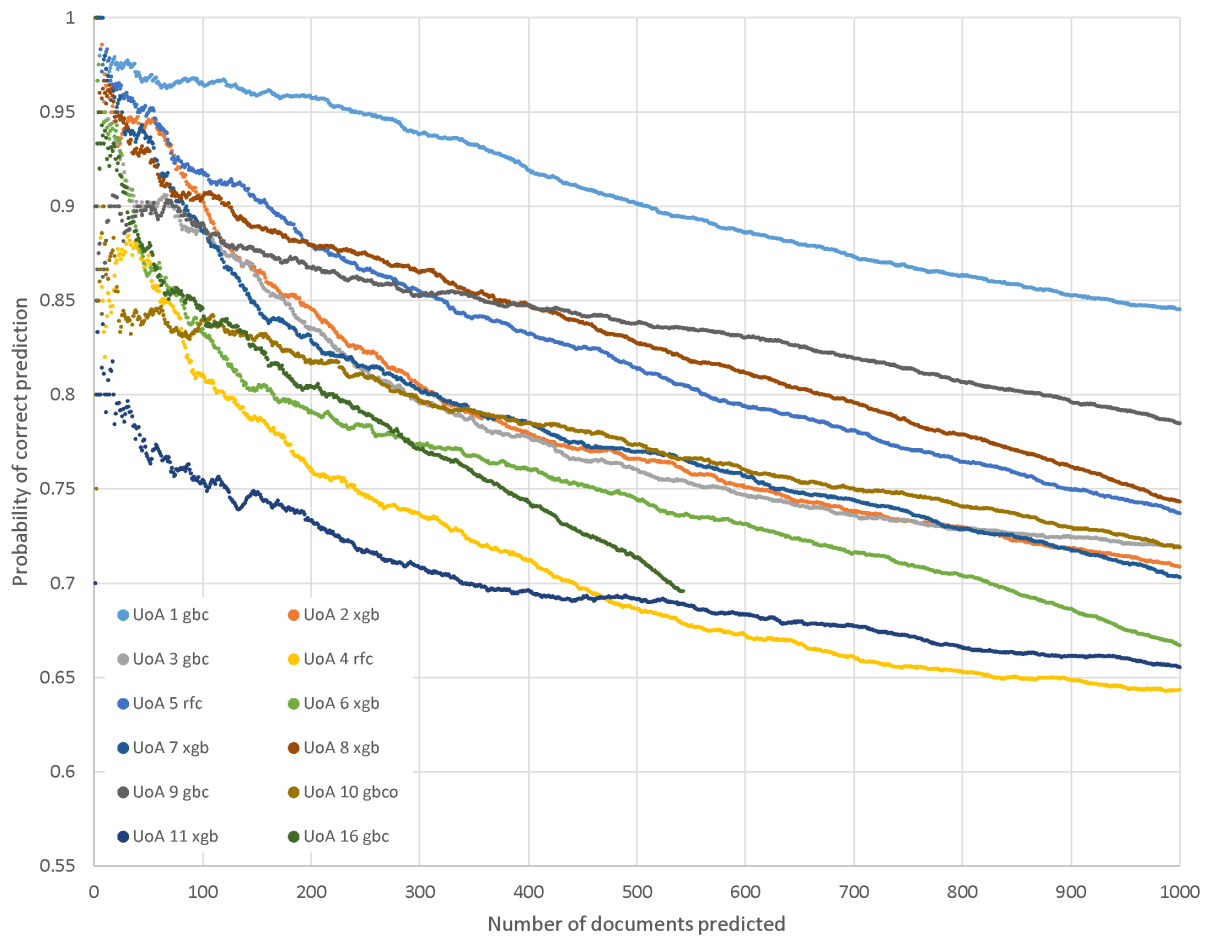

Figure 4.2.2. As above but zoomed in on the first 1000 articles.

If a smaller training set is used then lower numbers of high accuracy documents will be found, especially for the highest accuracy levels (Figure 4.2.3, Figure 4.2.4).

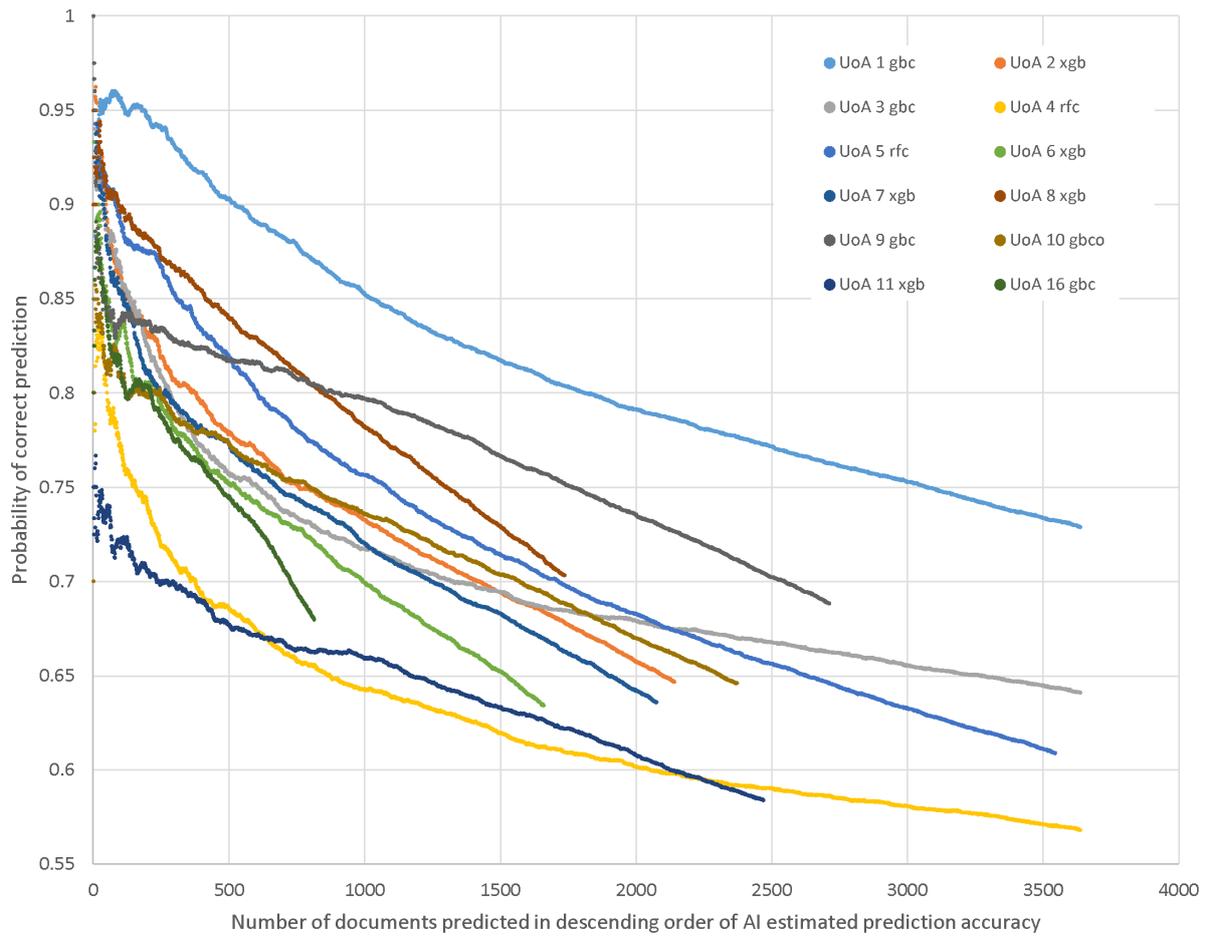

Figure 4.2.3. As above but trained on 25% of the data.

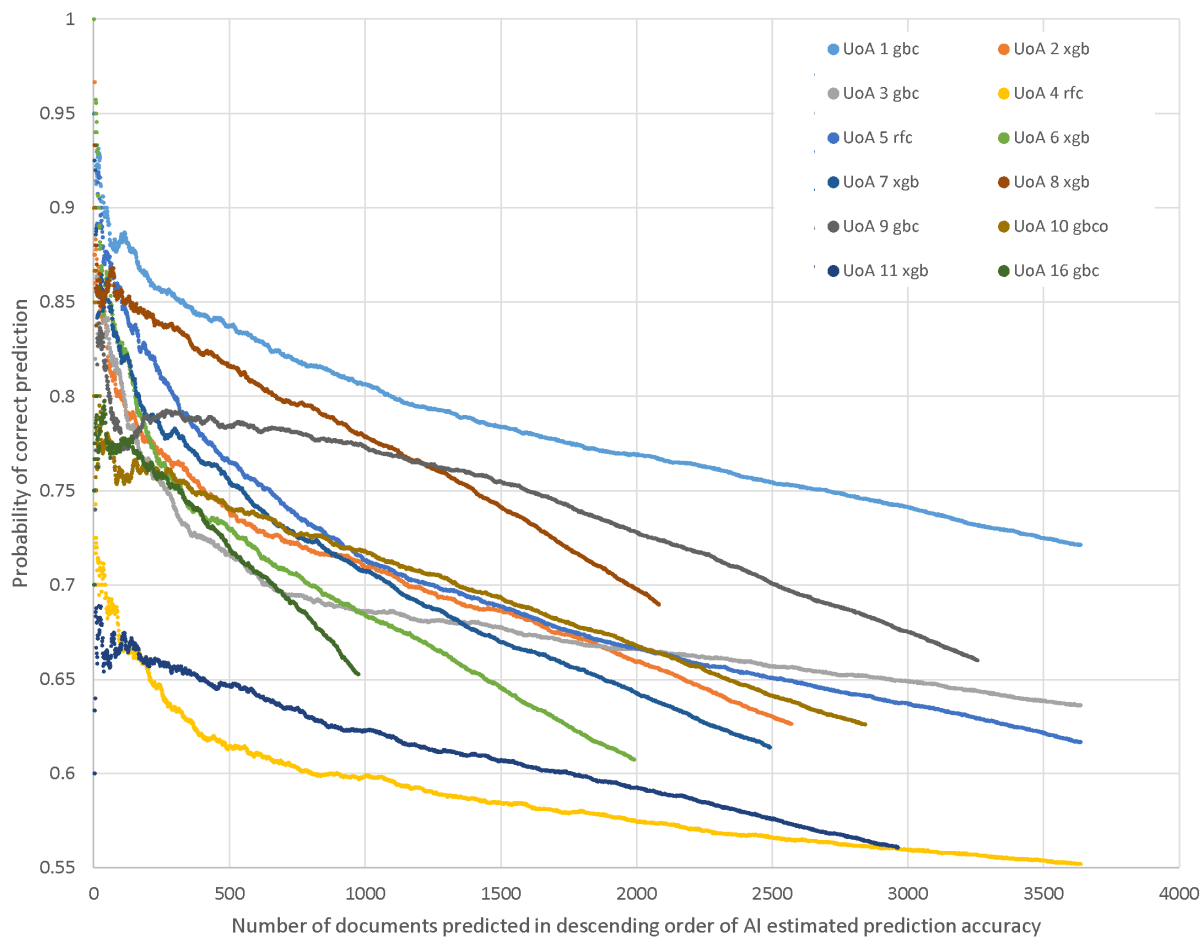

Figure 4.2.4. As above but trained on 10% of the data.

Modifying the above procedure slightly, an alternative strategy for using article-level AI prediction probabilities might be to train the AI model on a smaller amount of human-coded data, then if the accuracy profile is similar, there will be a larger set of articles that can be predicted with the AI at a given threshold. This gives fewer articles which can be predicted at a high level of accuracy when the overall accuracy is lower, however.

      Tables 4.2.1- 4.2.5 summarise the number of articles that could be predicted by the best machine learning method for 2014-18 assuming a minimum accuracy threshold for the AI classified set of 80%, 85%, 90% and 95%. At the 85% level suggested above, xgb tends to be the best method for the UoAs with the highest prediction accuracy. **Classifying individual UoAs produces more high accuracy predictions than classifying entire main panels (although this is not true for 95%) and this would give 2,879 machine learning predicted articles from UoA 1-11 and 16.** Increasing the threshold to 90% more than halves the predicted number to 1385 and at the maximum threshold, halved again to 582 (or 593 if incorporating Main Panel A instead of UoAs 1-6 separately). These accuracy levels refer only to the predicted AI scores. The overall accuracy levels will be much higher after taking into account the majority of articles being human classified.

Table 4.2.1. The number of articles that can be classified by machine learning with an accuracy of **80%** by UoA and Main Panel. The results are for articles 2014-18 with 1000 features and are averages across 10 iterations.

| UoA or Panel | Human classified | Human % | 80% accuracy articles | Top method |
|---|---|---|---|---|
| **1:Clinical Medicine** | 1818 | 25% | 1653 | gbc |
| **2:Public Health, Health Services and Primary Care** | 713 | 25% | 349 | xgb |
| **3:Allied Health Professions, Dentistry, Nurs. Pharm.** | 3481 | 50% | 293 | gbc |
| **4:Psychology, Psychiatry and Neuroscience** | 2922 | 50% | 117 | rfc |
| **5:Biological Sciences** | 1182 | 25% | 568 | rfc |
| **6:Agriculture, Food and Veterinary Sciences** | 553 | 25% | 224 | xgb |
| **7:Earth Systems and Environmental Sciences** | 1384 | 50% | 325 | gbc |
| **8:Chemistry** | 578 | 25% | 789 | gbco |
| **9:Physics** | 904 | 25% | 850 | gbc |
| **10:Mathematical Sciences** | 1579 | 50% | 299 | gbco |
| **11:Computer Science and Informatics** | 1646 | 50% | 63 | xgb |
| **12:Engineering** | 6255 | 50% | 158 | xgb |
| **13:Architecture, Built Environment and Planning** | 424 | 25% | 6 | xgbo |
| **14:Geography and Environmental Studies** | 1158 | 50% | 17 | gbc |
| **15:Archaeology** | 185 | 50% | 11 | mnbo |
| **16:Economics and Econometrics** | 541 | 50% | 212 | gbc |
| **17:Business and Management Studies** | 3767 | 50% | 41 | gbc |
| **18:Law** | 583 | 50% | 4 | xgbo |
| **19:Politics and International Studies** | 159 | 10% | 6 | mnb |
| **20:Social Work and Social Policy** | 511 | 25% | 10 | rfc |
| **21:Sociology** | 474 | 50% | 3 | xgbo |
| **22:Anthropology and Development Studies** | 61 | 10% | 4 | bnbo |
| **23:Education** | 520 | 25% | 4 | rfc |
| **24:Sport and Exercise Sciences, Leisure and Tourism** | 923 | 50% | 10 | gbc |
| **25:Area Studies** | 151 | 50% | 7 | logo |
| **26:Modern Languages and Linguistics** | 63 | 10% | 8 | rfc |
| **27:English Language and Literature** | 106 | 25% | 10 | xgbo |
| **28:History** | 145 | 25% | 4 | rfc |
| **30:Philosophy** | 42 | 10% | 5 | rfco |
| **31:Theology and Religious Studies** | 53 | 50% | 7 | rfc |
| **32:Art and Design: History, Practice and Theory** | 332 | 50% | 5 | xgb |
| **33:Music, Drama, Dance, Performing, Film, Screen** | 87 | 25% | 3 | bnb |
| **34:Communication, Cultural, Media, Library, Info** | 291 | 50% | 6 | rfco |
| **Main Panel A** | 6723 | 25% | 2401 | gbc |
| **Main Panel B** | 6714 | 25% | 1391 | gbc |
| **Main Panel C** | 11380 | 50% | 150 | gbc |
| **Main Panel D** | 409 | 10% | 5 | xgbo |

Table 4.2.2. The number of articles that can be classified by machine learning with an accuracy of **85%** by UoA and Main Panel. The results are for articles 2014-18 with 1000 features and are averages across 10 iterations.

| UoA or Panel | Human classified | Human % | 85% accuracy articles | Top method |
|---|---|---|---|---|
| **1:Clinical Medicine** | 1818 | 25% | 952 | gbc |
| **2:Public Health, Health Services and Primary Care** | 1427 | 50% | 181 | xgb |
| **3:Allied Health Professions, Dentistry, Nurs. Pharm.** | 3481 | 50% | 163 | gbc |
| **4:Psychology, Psychiatry and Neuroscience** | 2922 | 50% | 66 | rfc |
| **5:Biological Sciences** | 2364 | 50% | 308 | rfc |
| **6:Agriculture, Food and Veterinary Sciences** | 1106 | 50% | 86 | xgb |
| **7:Earth Systems and Environmental Sciences** | 1384 | 50% | 142 | xgb |
| **8:Chemistry** | 578 | 25% | 402 | xgb |
| **9:Physics** | 1808 | 50% | 362 | gbc |
| **10:Mathematical Sciences** | 1579 | 50% | 86 | gbco |
| **11:Computer Science and Informatics** | 1646 | 50% | 29 | xgb |
| **12:Engineering** | 6255 | 50% | 60 | gbc |
| **13:Architecture, Built Environment and Planning** | 169 | 10% | 4 | mnb |
| **14:Geography and Environmental Studies** | 1158 | 50% | 8 | gbc |
| **15:Archaeology** | 92 | 25% | 6 | rfco |
| **16:Economics and Econometrics** | 541 | 50% | 102 | gbc |
| **17:Business and Management Studies** | 3767 | 50% | 11 | gbc |
| **18:Law** | 583 | 50% | 3 | xgbo |
| **19:Politics and International Studies** | 797 | 50% | 4 | gbco |
| **20:Social Work and Social Policy** | 1022 | 50% | 5 | xgbo |
| **21:Sociology** | 474 | 50% | 3 | xgbo |
| **22:Anthropology and Development Studies** | 61 | 10% | 2 | bnbo |
| **23:Education** | 520 | 25% | 2 | gbc |
| **24:Sport and Exercise Sciences, Leisure and Tourism** | 923 | 50% | 6 | rfco |
| **25:Area Studies** | 151 | 50% | 4 | xgb |
| **26:Modern Languages and Linguistics** | 63 | 10% | 4 | rfc |
| **27:English Language and Literature** | 106 | 25% | 5 | xgbo |
| **28:History** | 291 | 50% | 3 | gbco |
| **30:Philosophy** | 42 | 10% | 3 | rfco |
| **31:Theology and Religious Studies** | 53 | 50% | 5 | rfc |
| **32:Art and Design: History, Practice and Theory** | 66 | 10% | 3 | bnbo |
| **33:Music, Drama, Dance, Performing, Film, Screen** | 87 | 25% | 3 | bnb |
| **34:Communication, Cultural, Media, Library, Info** | 291 | 50% | 4 | rfco |
| **Main Panel A** | 6723 | 25% | 1534 | gbc |
| **Main Panel B** | 6714 | 25% | 779 | gbc |
| **Main Panel C** | 11380 | 50% | 79 | gbc |
| **Main Panel D** | 409 | 10% | 4 | xgbo |

Table 4.2.3. The number of articles that can be classified by machine learning with an accuracy of **90%** by UoA and Main Panel. The results are for articles 2014-18 with 1000 features and are averages across 10 iterations.

| UoA or Panel | Human classified | Human % | 90% accuracy articles | Top method |
|---|---|---|---|---|
| **1:Clinical Medicine** | 1818 | 25% | 479 | gbc |
| **2:Public Health, Health Services and Primary Care** | 1427 | 50% | 97 | xgb |
| **3:Allied Health Professions, Dentistry, Nurs. Pharm.** | 1740 | 25% | 108 | bnbo |
| **4:Psychology, Psychiatry and Neuroscience** | 2922 | 50% | 28 | rfc |
| **5:Biological Sciences** | 2364 | 50% | 154 | rfc |
| **6:Agriculture, Food and Veterinary Sciences** | 1106 | 50% | 43 | xgbo |
| **7:Earth Systems and Environmental Sciences** | 1384 | 50% | 87 | xgb |
| **8:Chemistry** | 578 | 25% | 148 | xgb |
| **9:Physics** | 1808 | 50% | 134 | xgbo |
| **10:Mathematical Sciences** | 1579 | 50% | 44 | xgbo |
| **11:Computer Science and Informatics** | 1646 | 50% | 15 | gbc |
| **12:Engineering** | 6255 | 50% | 33 | gbc |
| **13:Architecture, Built Environment and Planning** | 169 | 10% | 2 | mnb |
| **14:Geography and Environmental Studies** | 1158 | 50% | 5 | xgbo |
| **15:Archaeology** | 185 | 50% | 5 | log |
| **16:Economics and Econometrics** | 541 | 50% | 48 | gbc |
| **17:Business and Management Studies** | 3767 | 50% | 7 | gbc |
| **18:Law** | 583 | 50% | 2 | xgbo |
| **19:Politics and International Studies** | 797 | 50% | 3 | gbco |
| **20:Social Work and Social Policy** | 511 | 25% | 3 | bnbo |
| **21:Sociology** | 474 | 50% | 2 | xgbo |
| **22:Anthropology and Development Studies** | 61 | 10% | 2 | bnbo |
| **23:Education** | 520 | 25% | 2 | gbc |
| **24:Sport and Exercise Sciences, Leisure and Tourism** | 923 | 50% | 5 | rfco |
| **25:Area Studies** | 151 | 50% | 4 | xgb |
| **26:Modern Languages and Linguistics** | 63 | 10% | 3 | xgbo |
| **27:English Language and Literature** | 106 | 25% | 4 | xgbo |
| **28:History** | 291 | 50% | 2 | mnb |
| **30:Philosophy** | 42 | 10% | 3 | rfco |
| **31:Theology and Religious Studies** | 53 | 50% | 3 | gbc |
| **32:Art and Design: History, Practice and Theory** | 66 | 10% | 2 | rfco |
| **33:Music, Drama, Dance, Performing, Film, Screen** | 87 | 25% | 2 | bnbo |
| **34:Communication, Cultural, Media, Library, Info** | 291 | 50% | 4 | rfco |
| **Main Panel A** | 6723 | 25% | 957 | gbc |
| **Main Panel B** | 13428 | 50% | 400 | gbc |
| **Main Panel C** | 11380 | 50% | 39 | gbc |
| **Main Panel D** | 409 | 10% | 3 | xgbo |

Table 4.2.4. The number of articles that can be classified by machine learning with an accuracy of **95%** by UoA and Main Panel. The results are for articles 2014-18 with 1000 features and are averages across 10 iterations.

| UoA or Panel | Human classified | Human % | 95% accuracy articles | Top method |
|---|---:|---:|---:|---|
| **1:Clinical Medicine** | 3637 | 50% | 239 | gbc |
| **2:Public Health, Health Services and Primary Care** | 1427 | 50% | 46 | xgb |
| **3:Allied Health Professions, Dentistry, Nurs. Pharm.** | 1740 | 25% | 61 | bnbo |
| **4:Psychology, Psychiatry and Neuroscience** | 1461 | 25% | 11 | rfc |
| **5:Biological Sciences** | 2364 | 50% | 46 | rfc |
| **6:Agriculture, Food and Veterinary Sciences** | 553 | 25% | 17 | xgb |
| **7:Earth Systems and Environmental Sciences** | 1384 | 50% | 31 | xgb |
| **8:Chemistry** | 578 | 25% | 40 | xgbo |
| **9:Physics** | 1808 | 50% | 40 | xgb |
| **10:Mathematical Sciences** | 1579 | 50% | 22 | xgbo |
| **11:Computer Science and Informatics** | 1646 | 50% | 9 | gbc |
| **12:Engineering** | 6255 | 50% | 12 | gbc |
| **13:Architecture, Built Environment and Planning** | 169 | 10% | 2 | mnb |
| **14:Geography and Environmental Studies** | 1158 | 50% | 3 | gbc |
| **15:Archaeology** | 185 | 50% | 3 | mnbo |
| **16:Economics and Econometrics** | 541 | 50% | 19 | gbc |
| **17:Business and Management Studies** | 1883 | 25% | 4 | gbco |
| **18:Law** | 583 | 50% | 1 | xgb |
| **19:Politics and International Studies** | 398 | 25% | 2 | gbco |
| **20:Social Work and Social Policy** | 511 | 25% | 2 | rfc |
| **21:Sociology** | 237 | 25% | 2 | logo |
| **22:Anthropology and Development Studies** | 61 | 10% | 2 | bnbo |
| **23:Education** | 520 | 25% | 2 | gbc |
| **24:Sport and Exercise Sciences, Leisure and Tourism** | 923 | 50% | 4 | rfco |
| **25:Area Studies** | 151 | 50% | 2 | xgb |
| **26:Modern Languages and Linguistics** | 63 | 10% | 3 | xgbo |
| **27:English Language and Literature** | 106 | 25% | 2 | xgbo |
| **28:History** | 145 | 25% | 2 | rfc |
| **30:Philosophy** | 42 | 10% | 3 | rfco |
| **31:Theology and Religious Studies** | 53 | 50% | 2 | gbc |
| **32:Art and Design: History, Practice and Theory** | 66 | 10% | 2 | bnbo |
| **33:Music, Drama, Dance, Performing, Film, Screen** | 87 | 25% | 2 | gbco |
| **34:Communication, Cultural, Media, Library, Info** | 291 | 50% | 2 | bnbo |
| **Main Panel A** | 13447 | 50% | 431 | gbc |
| **Main Panel B** | 13428 | 50% | 142 | gbc |
| **Main Panel C** | 11380 | 50% | 14 | gbc |
| **Main Panel D** | 409 | 10% | 2 | xgbo |

The above results for accuracy harness the actual prediction probability, as calculated cumulatively from the data. The AI-estimated prediction probabilities used to rank the articles for the calculations have variable accuracy. They tend to be optimistic an inaccurate for small datasets but pessimistic and relatively accurate for the larger, more accurate datasets from which the majority of AI predictions in the tables above would be drawn. Thus, in practice, the AI-estimated prediction probabilities can be relied on for the more accurate UoAs, which is a conservative approach that would allow slightly fewer AI predictions than shown in the tables.

## *4.3 Strategy 3: Active learning*

As suggested by Dr Petr Knoth of the Open University, active learning during machine learning may increase the proportion of documents that can be predicted with AI at any given level of accuracy. This was adopted for Strategy 3. Although there are many ways of implementing active learning the approach used below was chosen to give informative test results.

To test Strategy 3, the articles were dynamically split into nine batches of 10% for human coding. In the first state, a random 10% of the texts was selected, with their provisional (human) REF scores. In each of the eight subsequent stages, an additional 10% of the articles were selected for provisional REF scores, but this 10% was chosen to be the articles with the lowest AI prediction probabilities from the previous stage. The idea here is that sub-panel members could perform reviews in batches of 10%, stopping when the accuracy for the remaining articles was high enough. For example, if the accuracy level was agreed to be 90% and this was reached after six batches of 10% (i.e., 60% of the articles had been classified by subpanel members) then the remaining 40% could be classified by the AI. Alternatively, if the desired accuracy level was not reached after 90% of the articles had been classified by sub-panel members, then they would also classify the remaining 10%, avoiding AI predictions altogether. Thus, this procedure provides a guaranteed minimum level of accuracy for the AI component but no guaranteed time saving.

Whilst 10% for each batch is suggested here, alternative sizes (e.g., 20%, 25%, or varying size) could also be tried.

This batch active learning strategy has one definite advantage and one possible advantage compared to Strategy 1 and Strategy 2. The definite advantage is that the human coders are directed to score the articles that the machine is least confident about, thereby not "wasting time" scoring articles that could be machine predictable with a high degree of confidence. The possible advantage is that algorithm may benefit more by being fed the scores for borderline cases than from being fed the scores from clear cases, thereby improving its accuracy faster than could be accounted for by the removal of the difficult cases from its test set.

### 4.3.1 Overall active learning machine learning accuracy on the 2014-18 dataset

Accuracy levels for the standard and ordinal variants of the three most promising machine learning methods on the 2014-18 articles, excluding 0 scores and articles with abstracts shorter than 500 characters are illustrated below. Accuracy levels would be slightly lower for 2014-20 articles or if the abstract length requirement were to be dropped.

Unsurprisingly, active learning steadily increases the accuracy of the classifications in all cases (Figure 4.3.1.1). The machine learning method with the highest accuracy at 90% classification was usually xgb (6 UoAs) or gbc (5 UoAs), although xgbo (1 UoA) also appeared as a top method. The worst of the six main AI methods at 90% were always rfc and rfco, confirming that these methods are surprisingly poor at identifying prediction probabilities. None of the UoAs reached 95% accuracy but two reached 90% and seven achieved 85%.

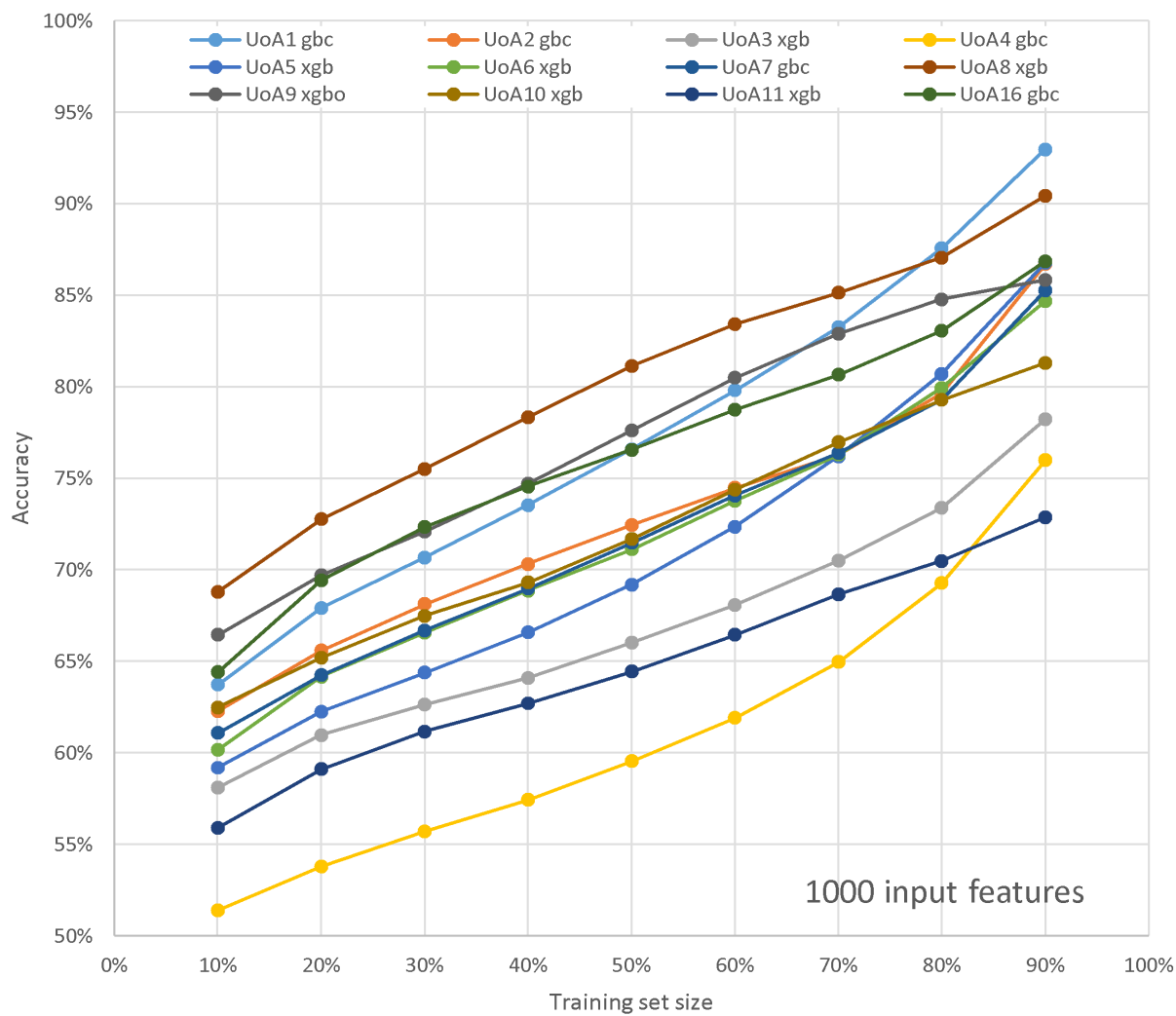

Figure 4.3.1.1. Active learning on UoAs 1-11, 16 showing the results for the machine learning method with the highest accuracy at 90% and 1000 input features. Results are the average of 40 independent full active learning trials.

Taking 85% as the proposed minimum AI accuracy to use the results, AI predictions could be made for the remaining articles in the seven cases when this level is reached, giving a total saving of 3,688 scores (Table 4.3.1.1). This is 809 (28%) higher than the 2,879 predicted at 85% for Strategy 2 (prediction by probability), so strategy 3 is superior to strategy 2 in terms of peer review time saving. Some additional savings would also be possible by applying strategy 2 to the UoAs that do not meet the threshold after 90% of articles had been classified.
      Overfitting (optimistic accuracy predictions) are possible here because the best of six methods is reported. If the second best method was used instead, then 284 less articles would be predicted by the AI (UoA 7 would fall below the threshold).

Table 4.3.1.1. The Number of articles that can be predicted at an accuracy above 85% using active learning in UoAs 1-11,16. Overall accuracy includes the human scored texts for eligible and ineligible articles.

| UoA | Human scores | Human scores % | AI accuracy | Overall accuracy | AI predicted articles |
|---|---|---|---|---|---|
| **1:Clinical Medicine** | 5816 | 80% | 87.6% | 98.4% | 1458 gbc |
| **2:Public Health, Health Services and Primary Care** | 2565 | 90% | 86.7% | 99.2% | 290 gbc |
| **3:Allied Health Professions, Dentistry, Nurs. Pharm.** | 6962 | 100% | | 100.0% | 0 |
| **4:Psychology, Psychiatry and Neuroscience** | 5845 | 100% | | 100.0% | 0 |
| **5:Biological Sciences** | 4248 | 90% | 86.8% | 99.2% | 480 xgb |
| **6:Agriculture, Food and Veterinary Sciences** | 2212 | 100% | | 100.0% | 0 |
| **7:Earth Systems and Environmental Sciences** | 2484 | 90% | 85.3% | 99.1% | 284 gbc |
| **8:Chemistry** | 1617 | 70% | 85.1% | 97.2% | 697 xgb |
| **9:Physics** | 3249 | 90% | 85.9% | 99.1% | 368 xgbo |
| **10:Mathematical Sciences** | 3159 | 100% | | 100.0% | 0 |
| **11:Computer Science and Informatics** | 3292 | 100% | | 100.0% | 0 |
| **16:Economics and Econometrics** | 972 | 90% | 86.9% | 99.2% | 111 gbc |
| **Total** | | | | | **3688** |

Increasing the number of input features to 2000 (Figure 4.3.1.2) or 5000 (Figure 4.3.1.3) changes the active learning results very little. Possibly due to normal statistical fluctuations due to random factors, the number of AI predicted articles for 2000 input feature is lower at 2911 and for 5000 input features is higher at 3911. In the latter case, the increase is due to UoA 6 crossing the 85% threshold (its accuracy at 90% training set size increasing 0.6% from 84.7% to 85.3%).

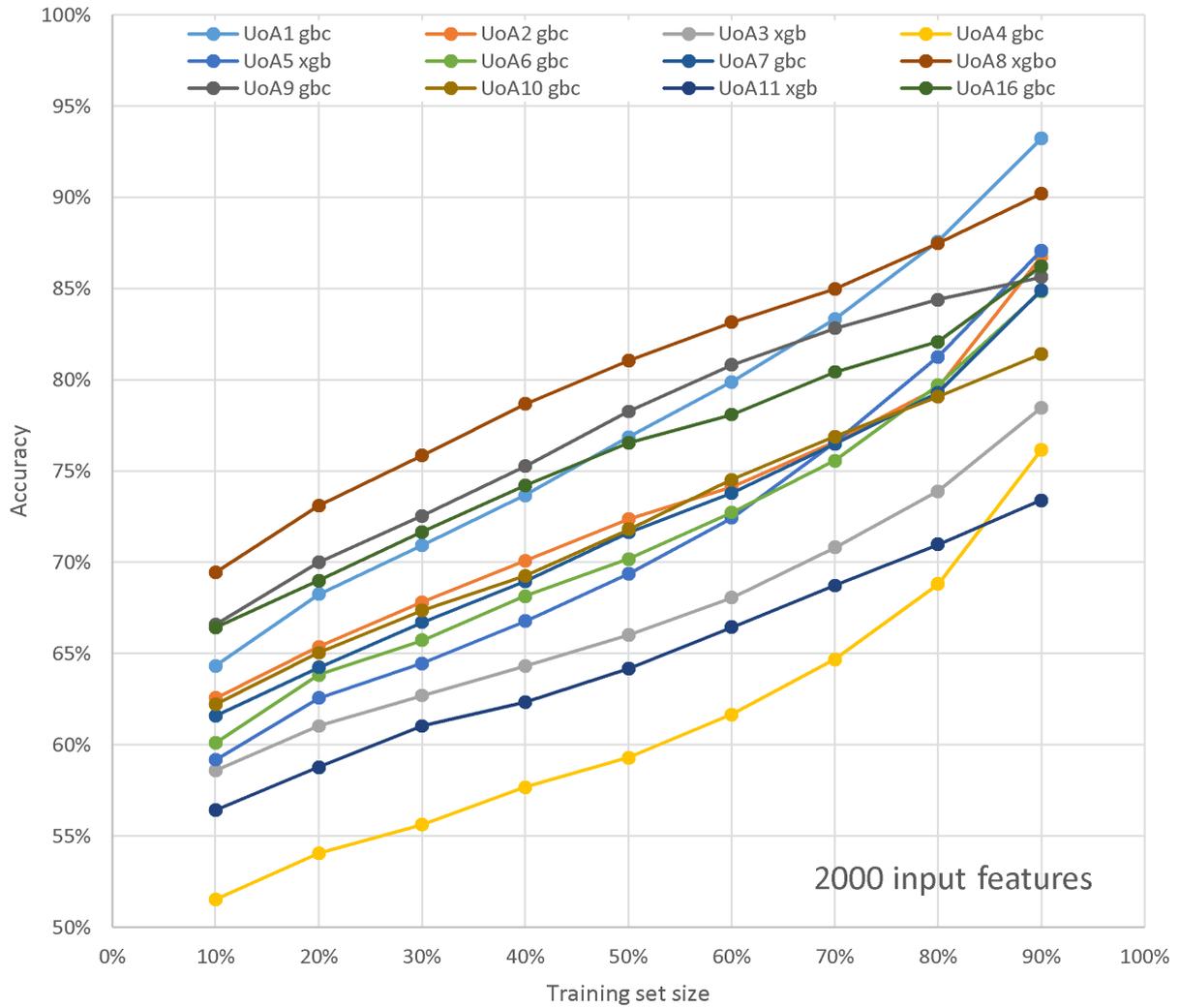

Figure 4.3.1.2. Active learning on UoAs 1-11, 16 showing the results for the machine learning method with the highest accuracy at 90% and **2000** input features. Results are the average of 10 independent full active learning trials.

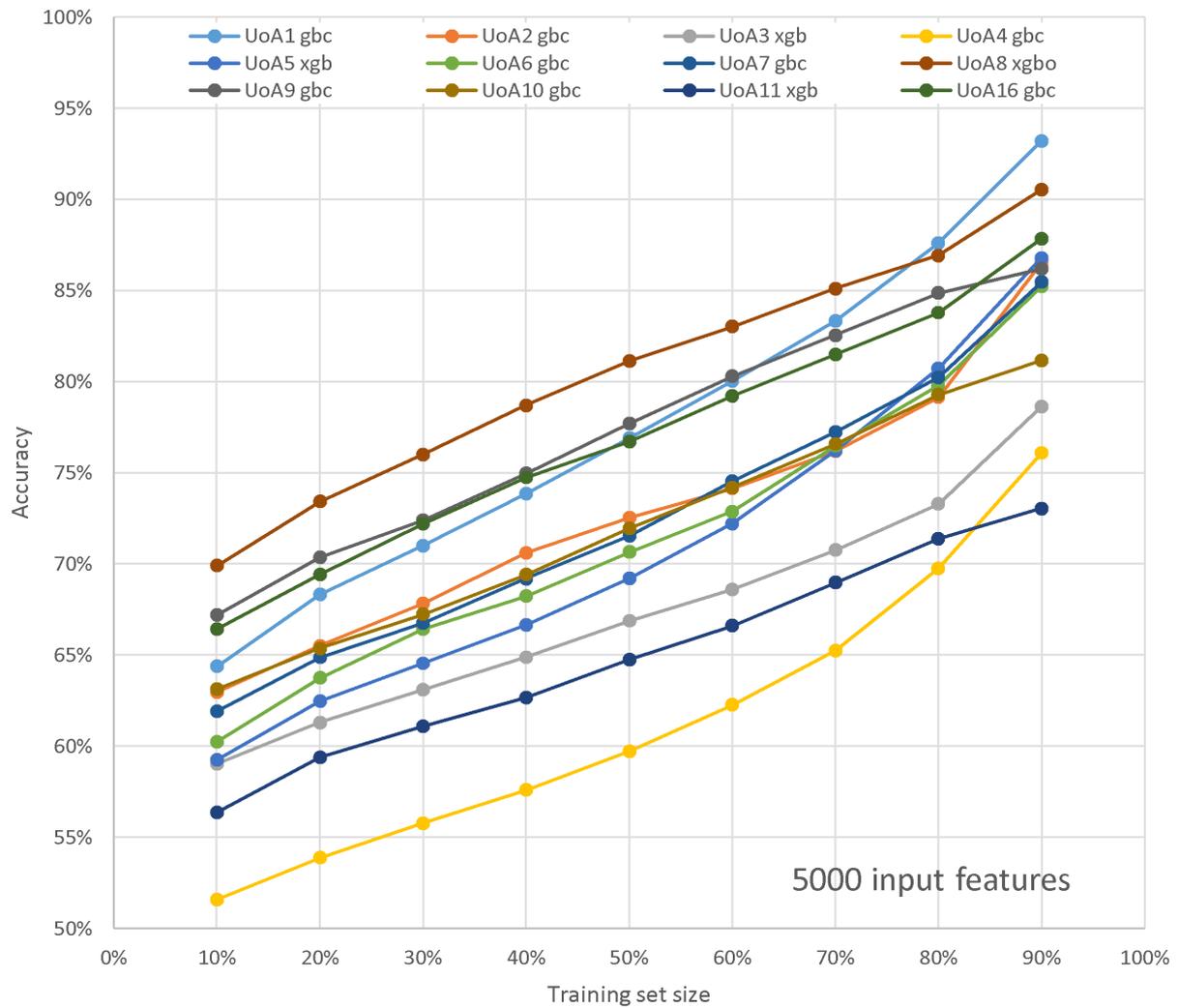

Figure 4.3.1.3. Active learning on UoAs 1-11, 16 showing the results for the machine learning method with the highest accuracy at 90% and **5000** input features. Results are the average of 10 independent full active learning trials.

### 4.3.2 Institutional score shifts

Institutional score shifts from AI predictions are generally smaller for active learning due to the higher accuracy achieved (Figure 4.3.2.1). The score shifts on all eligible articles (right hand axis) are lower due to the smaller proportion of articles predicted with AI. The error bars are relatively large compared to the accuracy for smaller submissions (including compared to Figure 4.1.2.1) because of the smaller number of articles predicted by the AI in the active learning case. Since the eligible articles constitute 62.6% of all REF journal articles, the overall score shifts are about two thirds of the figure on the right-hand y-axis.

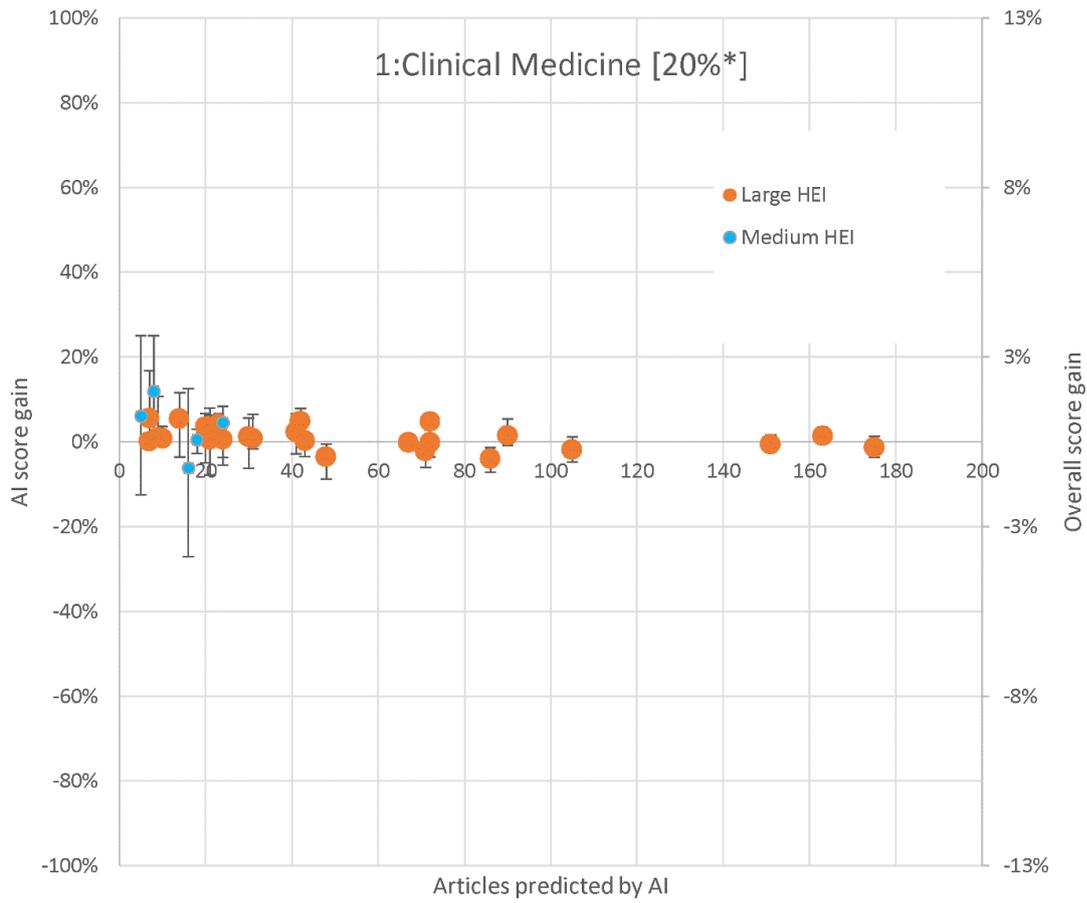

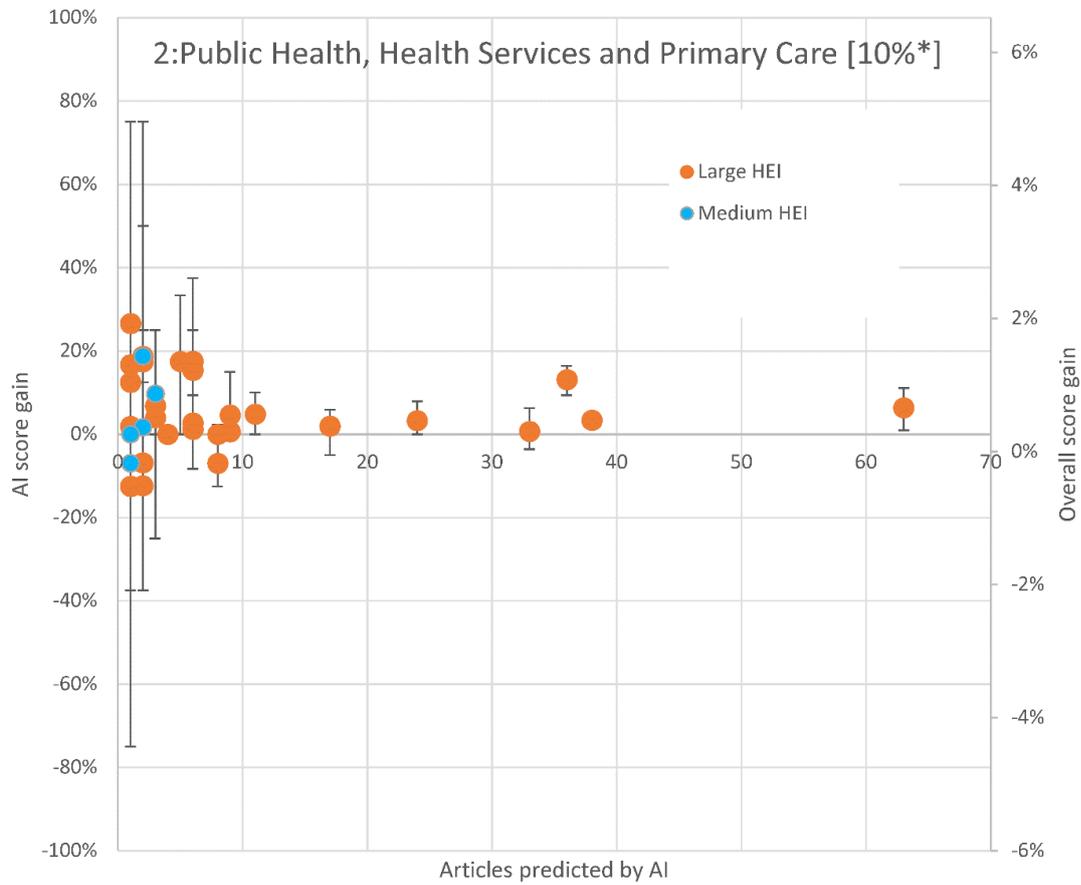

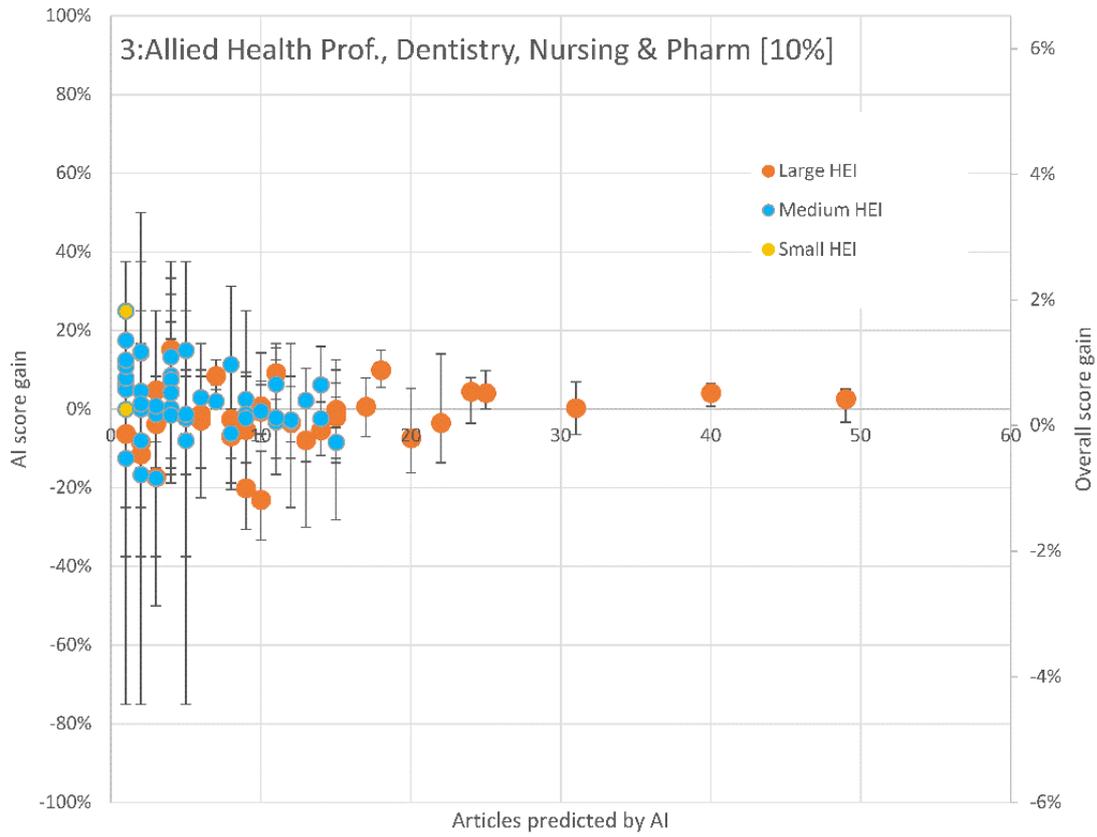

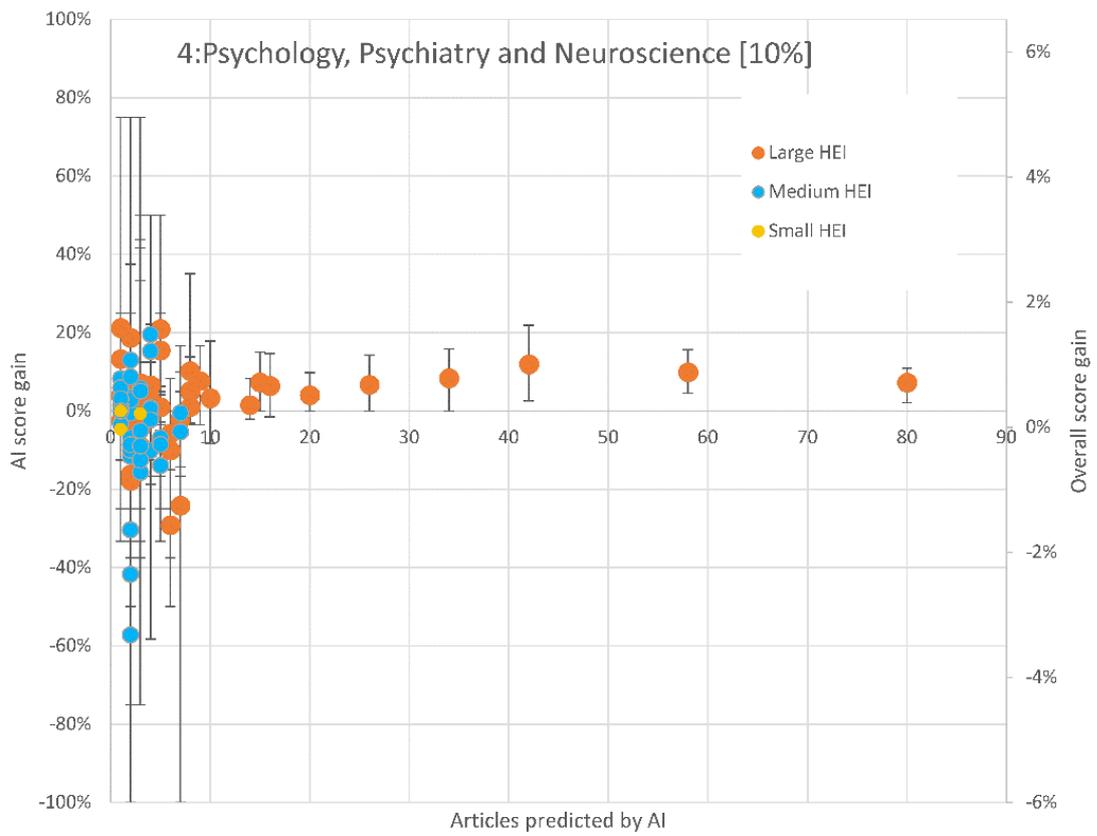

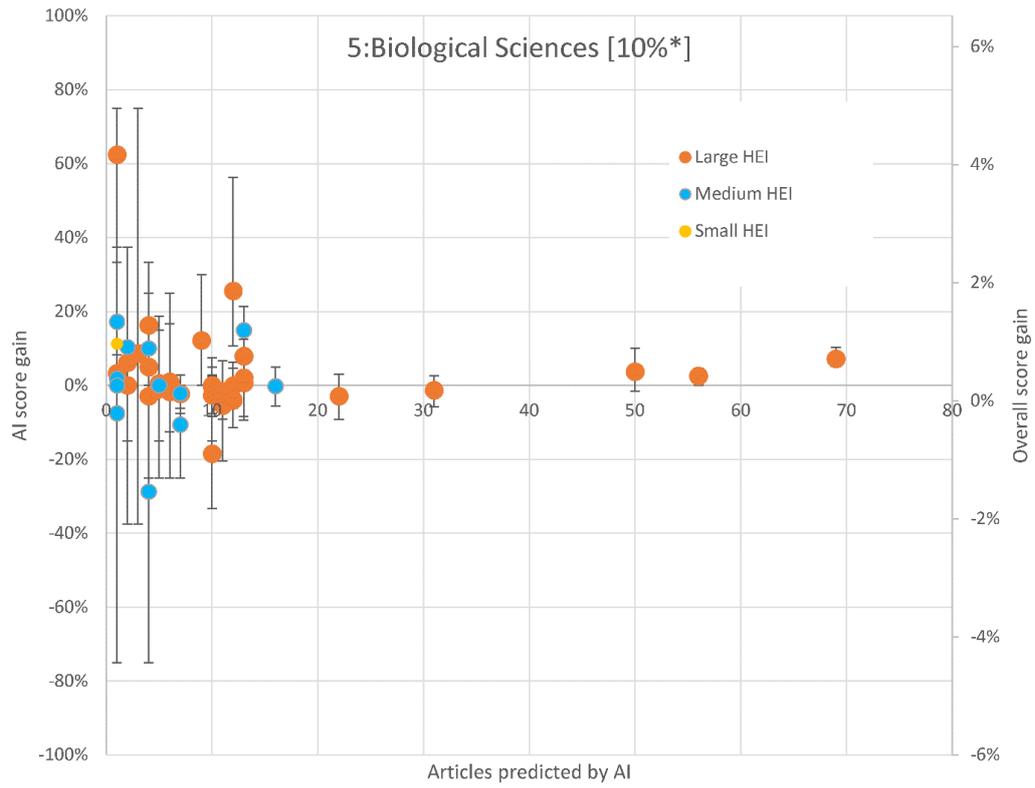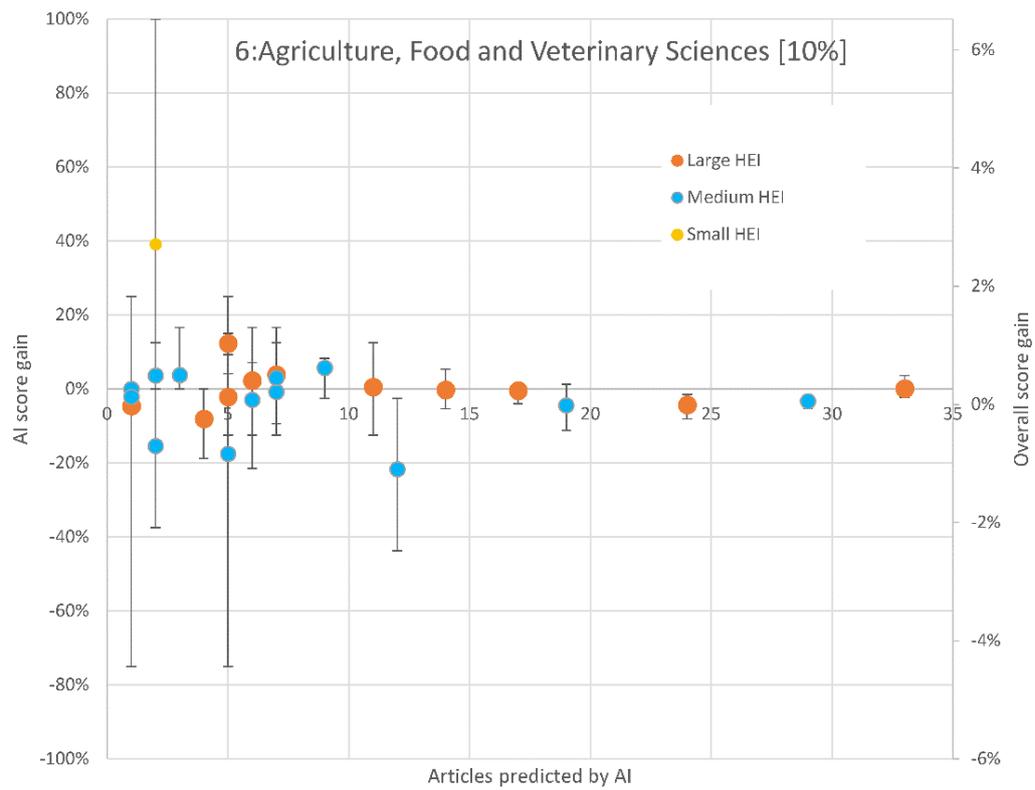

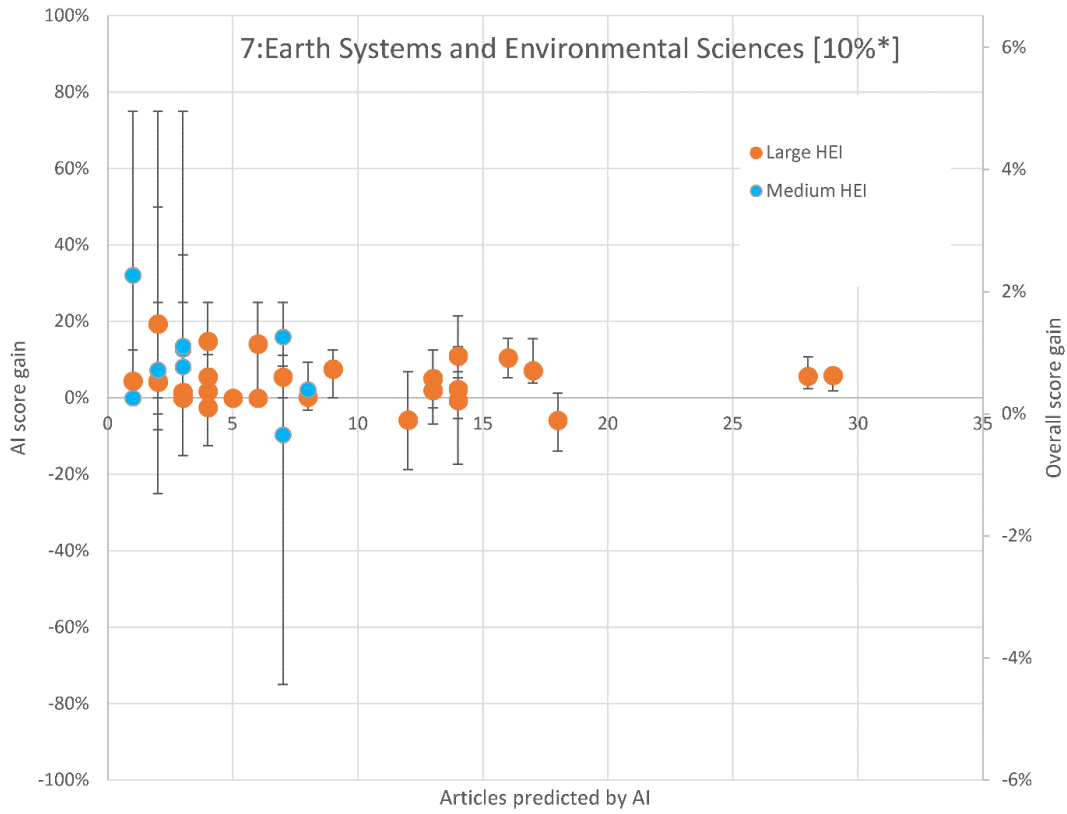
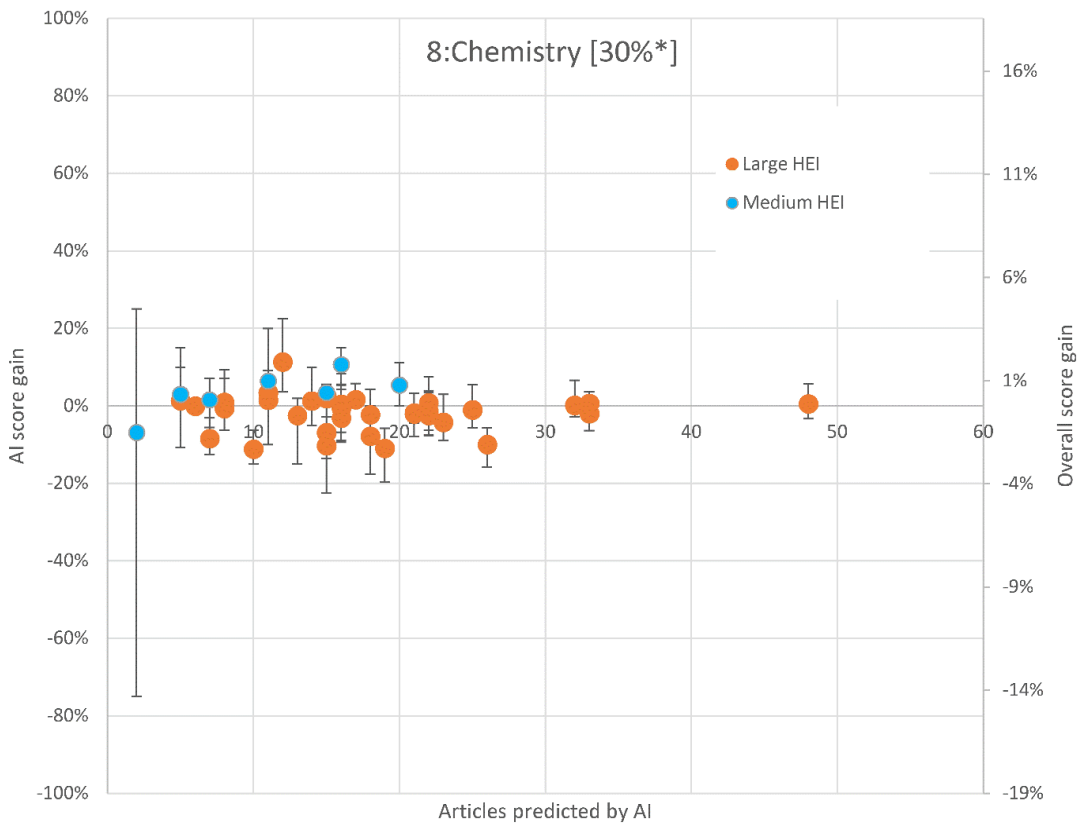

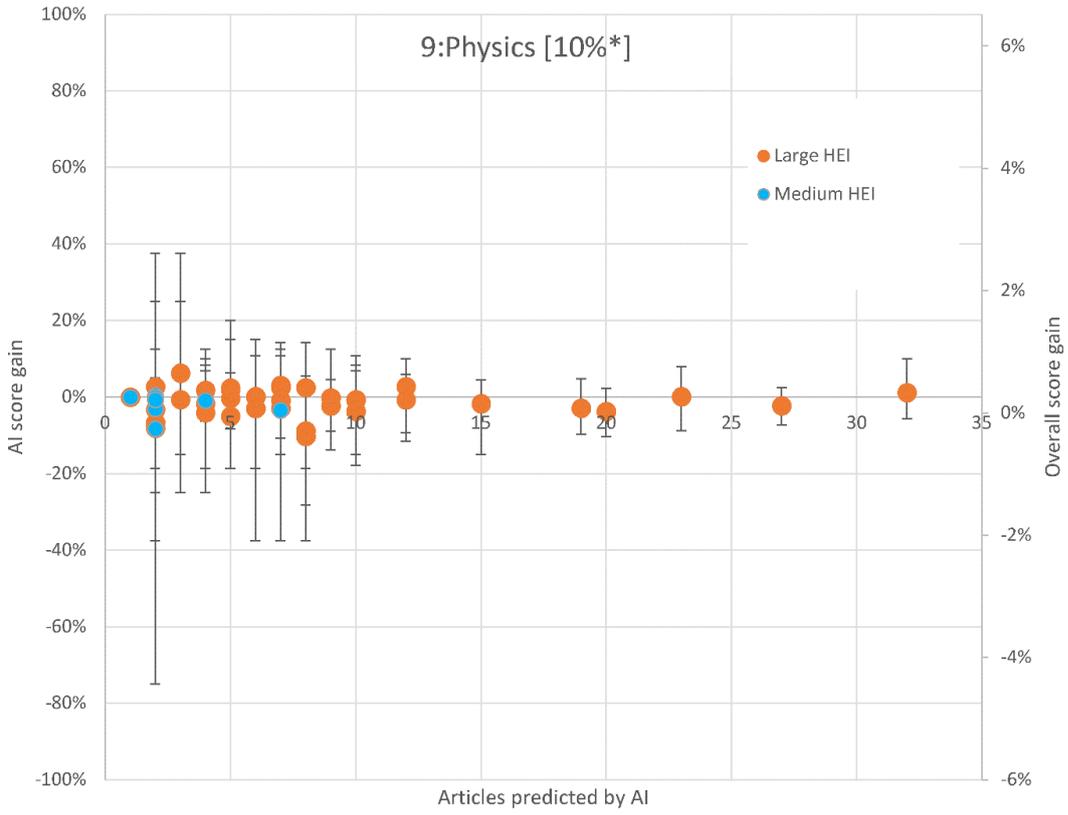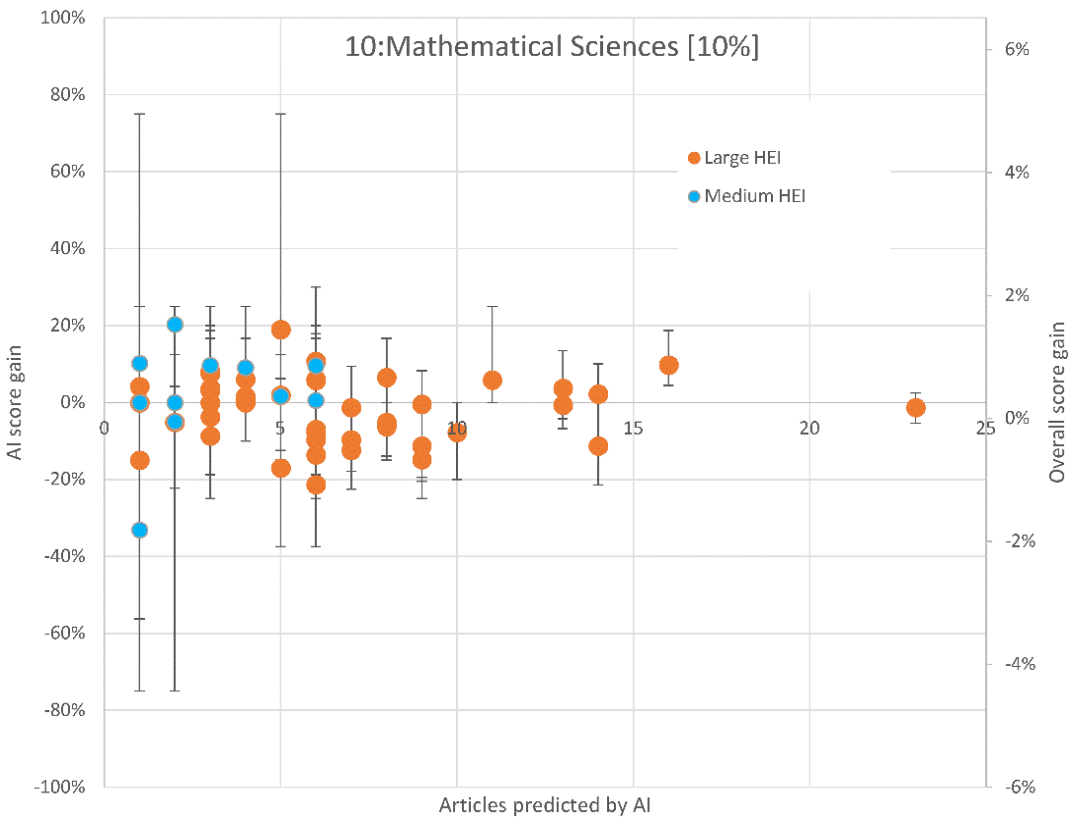

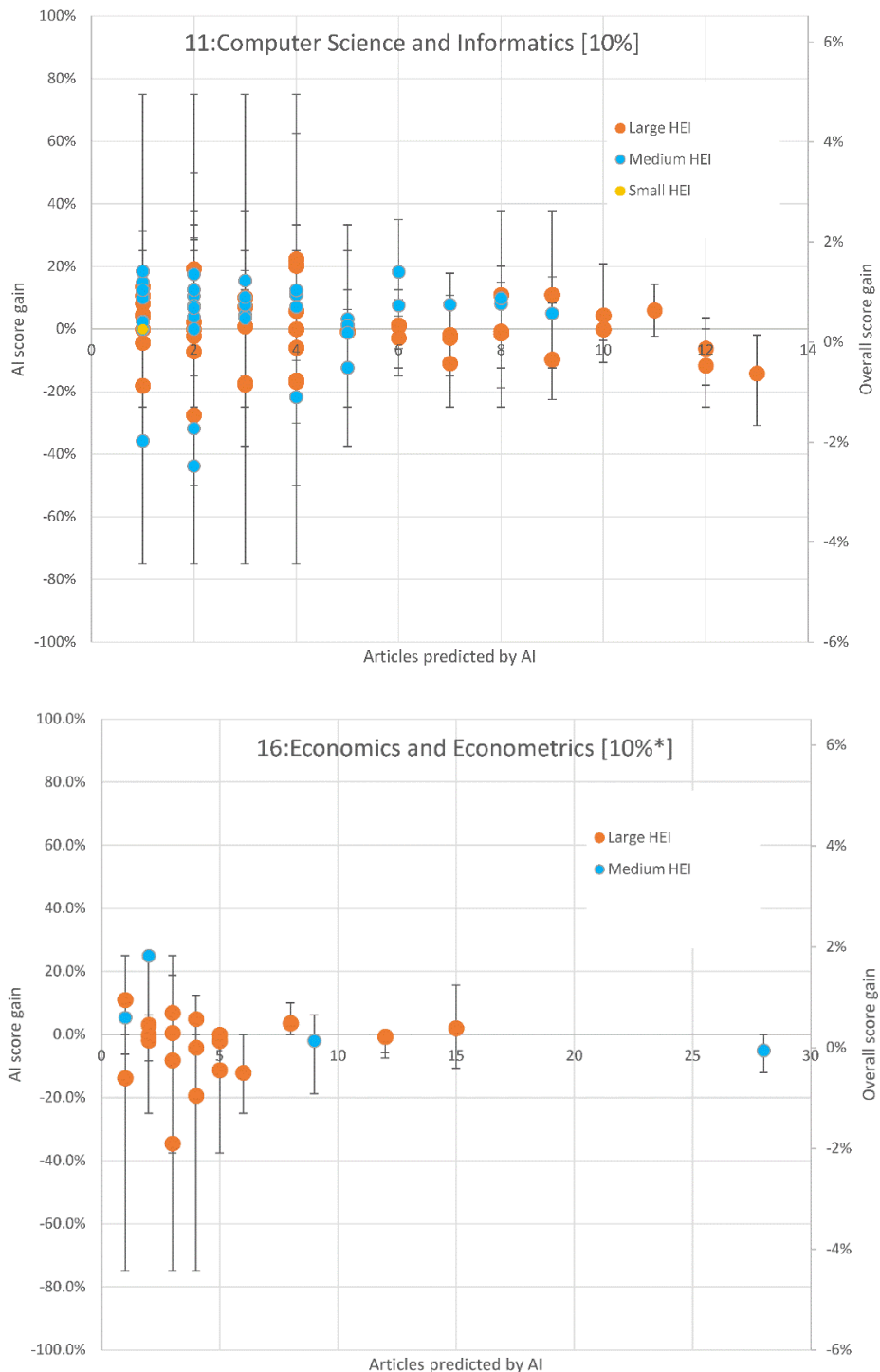

Figure 4.3.2.1. The average REF AI institutional score gain on UoA 1:Clinical Medicine to UoA 16: Economics and Econometric for the most accurate machine learning method with active learning, stopping at 85% accuracy on the 2014-18 data and **bibliometric + journal + text inputs, after excluding articles with shorter than 500 character abstracts**. AI score gain is a financial calculation (4*=100% funding, 3*=25% funding, 0-2*=0% funding). The x axis records the number of articles with predicted scores in one of the iterations. The right-hand axis shows the overall score gain for all REF journal articles. Error bars indicate the highest and lowest values from 10 iterations. Captions indicate the proportion of journal articles predicted, starred (and the figures at full size) if the 85% accuracy active learning threshold is met.

For the UoAs using active learning, the largest average overall score shift for the biggest 5 submissions is 1.9%, with 2.6% in the worst case out of the ten iterations of the machine learning (Chemistry) (Table 4.3.2.1). Overall, the score shifts for smaller submissions can be larger due to less statistical averaging of errors. These can be 3.9% on average (Biological Sciences) or 14% in the worst case out of the ten iterations of the machine learning (Chemistry). Thus, the score shifts introduced tend to be minor except for small submissions. There may need to be special consideration given to small submissions to prevent this.

Table 4.3.2.1. Maximum average AI score shifts for five largest HEI submissions and for all submissions. The same information for the largest AI score shifts rather than the average score shifts. Overall figures include all human coded journal articles.

| UoA or Panel | Human scores % | Max HEI av. score shift (overall) | Max top 5 HEIs av. score shift (overall) | Max HEI largest score shift (overall) | Max top 5 HEIs largest score shift (overall) |
|---|---|---|---|---|---|
| **1:Clinical Medicine** | 80% | 12% (1.5%) | 1.9% (0.2%) | 27% (3.4%) | 5.4% (0.7%) |
| **2:Public Health, H. Services & Primary Care** | 90% | 27% (1.7%) | 13% (0.8%) | 75% (4.7%) | 16% (1.0%) |
| **3:Allied Health Prof., Dentist Nurs Pharm** | 100% | | | | |
| **4:Psychology, Psychiatry & Neuroscience** | 100% | | | | |
| **5:Biological Sciences** | 90% | 63% **(3.9%)** | 7.3% (0.5%) | 75% (4.7%) | 10% (0.6%) |
| **6:Agriculture, Food & Veterinary Sciences** | 100% | | | | |
| **7:Earth Systems & Environmental Sciences** | 90% | 32% (2.0%) | 11% (0.7%) | 75% (4.7%) | 16% (1.0%) |
| **8:Chemistry** | 70% | 11% (2.1%) | 10% **(1.9%)** | 75% **(14%)** | 14% **(2.6%)** |
| **9:Physics** | 90% | 10% (0.6%) | 3.7% (0.2%) | 75% (4.7%) | 10% (0.6%) |
| **10:Mathematical Sciences** | 100% | | | | |
| **11:Computer Science & Informatics** | 100% | | | | |
| **16:Economics and Econometrics** | 90% | 35% (2.2%) | 5.1% (0.3%%) | 75% (4.7%) | 19% (1.2%) |

### 4.3.3  Systematic score shifts by HEI size, submission size and submission quality

The systematic score shifts in this section tend to be in the same direction as for Strategy 1, but weaker. In many fields there is a moderate tendency for smaller HEIs, smaller submissions, and lower scoring submissions to have larger AI prediction gains (Figure 4.3.3.1). In 10 of the 12 UoAs 1-1,16, smaller HEIs had an AI advantage, with the strength of the correlation with size usually being moderate (-0.2 to 0). Thus, smaller institutions would gains slightly from active learning AI Strategy 3, and larger institutions would lose.

There is a moderate tendency for lower scoring UoAs to gain from the AI predictions, with the correlation strength being between -0.45 and 0.15, for 11 of the UoAs in Figure X. Thus, replacing human scores with AI predictions using Strategy 1 would result in a moderate score shift in favour of weaker submissions. Again, other factors being equal, an article with an AI prediction error is more likely to be lose if it is from a high scoring submission. The correlations are based on small sample sizes, however, so other factors may be at work (e.g., 95% confidence intervals (Fisher transformation) for almost all correlations would include 0).

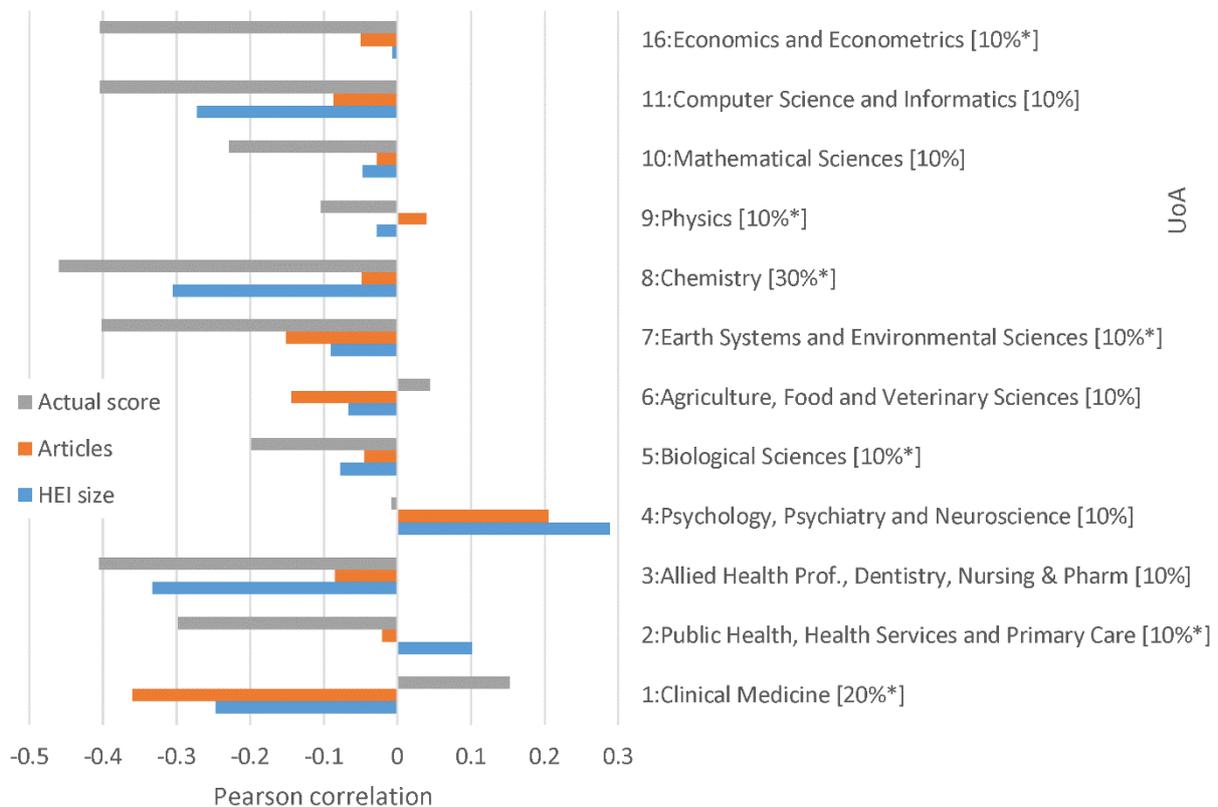

Figure 4.3.3.1. Pearson correlations between institutional size (number of articles submitted to REF) or submission size (number of articles submitted to UoA) or average institutional REF score for the UoA and average REF AI institutional score gain on UoA 1:Clinical Medicine to UoA 16: Economics and Econometric for the most accurate machine learning method with active learning, stopping at 85% accuracy on the 2014-18 data and **bibliometric + journal + text inputs, after excluding articles with shorter than 500 character abstracts**. Captions indicate the proportion of journal articles predicted, starred if the 85% accuracy active learning threshold is met.

### 4.3.4 Gender and Early Career Researcher (ECR) status score shifts

Switching from human REF scores to AI predictions does not systematically work in favour or against ECRs in any of UoAs 1-11,16, with the minor exception of UoA 2, where there is an average 4% relative loss for ECRs (Figure 4.3.4.1). Since only 10% of eligible articles are predicted in UoA 2, the net loss for ECRs is only 0.3% overall. Thus, the implications for ECRs are minor.

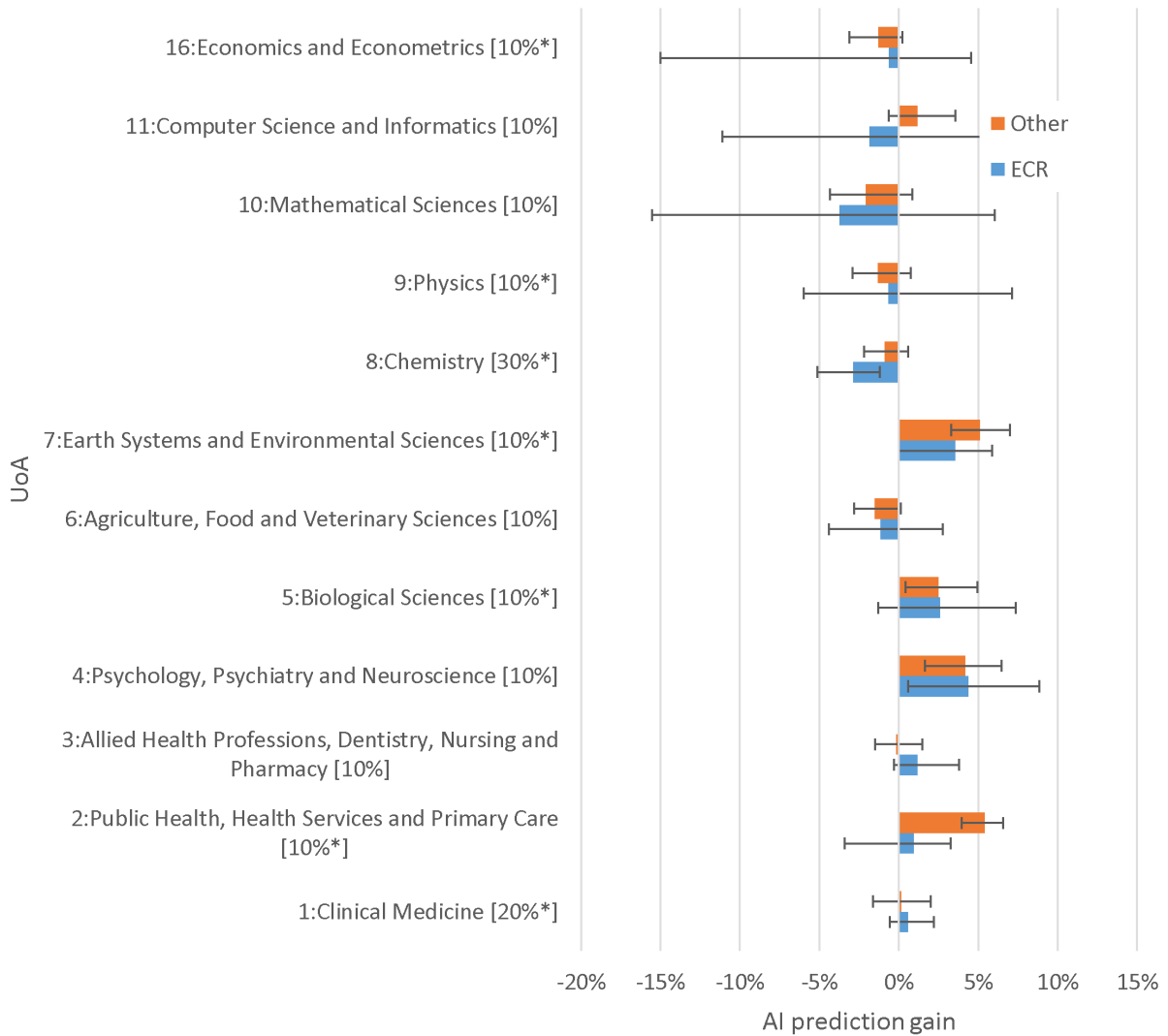

Figure 4.3.4.1. Average REF score AI prediction gains (AI score subtract reviewer score) for ECRs and experienced researchers in the twelve most predictable UoAs for 2014-18 data and the most accurate machine learning method with active learning, stopping when accuracy is 85% or **90%** of the 2014-18 data and **bibliometric + journal + text inputs** ten times. REF score is a financial calculation (4*=100% funding, 3*=25% funding, 0-2*=0% funding). Error bars show the highest and lowest value from ten separate sets of AI predictions. Captions indicate the proportion of journal articles predicted, starred if the 85% accuracy active learning threshold is met.

Switching from human REF scores to AI predictions with active learning does not systematically work in favour or against male or female first authored articles in UoAs 1-11, 16 (Figure 4.3.4.2), but there is an average advantage for males in some and females in others. The biggest advantage of any UoA qualifying for active learning (i.e., starred in the figure) is a 5% male advantage (0.4% overall with human classifications), which is small.

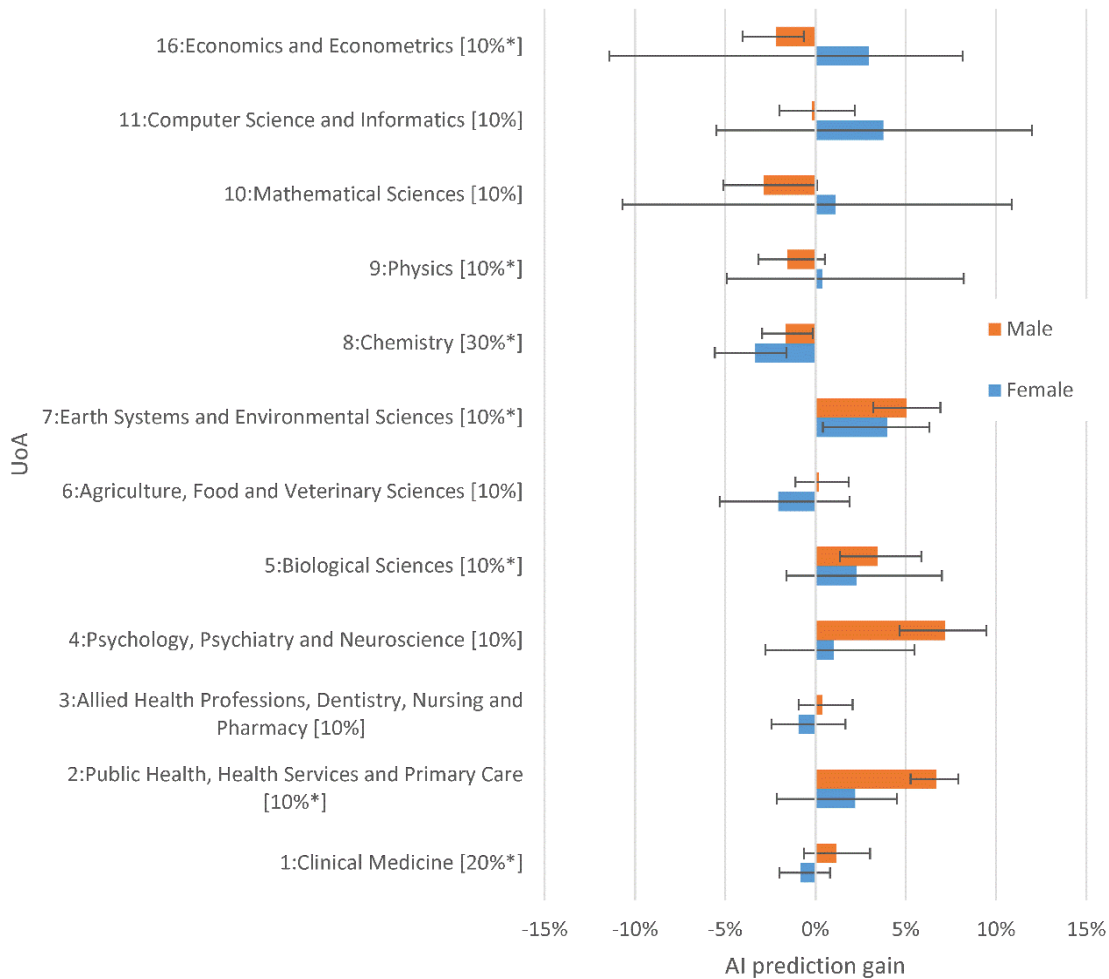

Figure 4.3.4.2. Average REF score AI prediction gains for male and female first authored articles in the twelve most predictable UoAs for 2014-18 data and the most accurate machine learning method with active learning, stopping when accuracy is 85% or **90%** of the 2014-18 data and **bibliometric + journal + text inputs** ten times. REF score is a financial calculation (4*=100% funding, 3*=25% funding, 0-2*=0% funding). Error bars show the highest and lowest value from ten separate sets of AI predictions. Captions indicate the proportion of journal articles predicted, starred if the 85% accuracy active learning threshold is met.

### 4.3.5 Interdisciplinarity score shift tests

Active learning AI score predictions did not systematically favour outputs flagged as interdisciplinary or non-interdisciplinary research overall or by a large margin in any UoA (Figure 4.3.5.1).

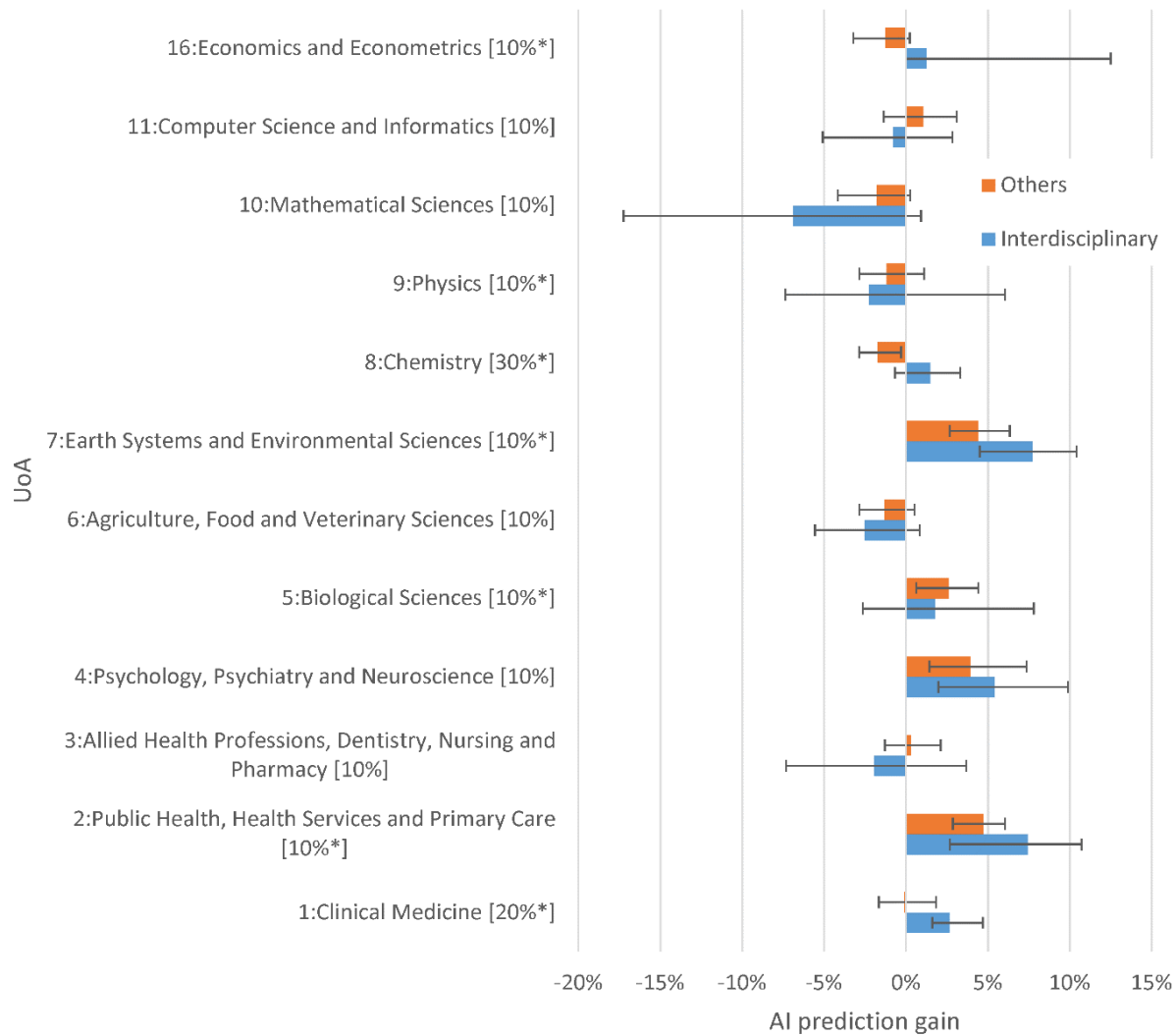

Figure 4.3.5.1. Average REF score AI prediction gains for articles flagged as interdisciplinary or not in the twelve most predictable UoAs for 2014-18 data and the most accurate machine learning method with active learning, stopping when accuracy is 85% or **90%** of the 2014-18 data and **bibliometric + journal + text inputs** ten times. REF score is a financial calculation (4*=100% funding, 3*=25% funding, 0-2*=0% funding). Error bars show the highest and lowest value from ten separate sets of AI predictions.

# 5 Analysis of machine learning results

## 5.1 Reasons why some article scores are difficult to predict with AI

Five hundred articles from the five most predictable UoAs that had been incorrectly predicted by the AI but had a high AI estimated probability of a correct prediction were individually examined to identify possible reasons for the discrepancies. The purpose was to gain insights into relatively intractable issues that may evade any kind or AI solution or that, alternatively, might be predictable with different methods. These articles were extracted from an earlier version of the input set, which did not include length information.

This was a difficult task for us, lacking the subject expertise for these fields, and there were few cases where we were confident about why a discrepancy had occurred. For example, we saw cases of large-scale multi-centre research into life threatening conditions that had been published in prestigious journals and were quite highly cited but had been scored only 3* by relevant sub-panels. In such cases we might speculate that the sub-panel members spotted problems or limitations that the journal reviewers had not or took a different perspective on the value of the findings (e.g., if they were negative, routine, or had been superseded by other approaches).

The following are suggested reasons for discrepancies between AI predictions and sub-panel provisional scores. The length consideration was addressed by adding page counts to the input sets (as reported in this document: preliminary AI results without length are not reported). The other reasons for discrepancies could not be translated into new inputs.

- **Incorrect article metadata**: One article matched a different article in Scopus. In this case the Scopus record had an incorrect DOI either due to an historical error in Scopus (the live Scopus record was checked and found to be correct) or due to an error in University of Wolverhampton software processing the Scopus data.
- **Article length**
  - **Short articles**: Sub-panel members might give lower scores to short articles despite them being highly cited, if they tend to reflect preliminary results or less content overall. Although article length is a useful AI input in theory, many articles are online only with the number of pages, even if also available in PDF, not recorded in bibliometric databases. From the page information recorded in the REF database, 13% of the articles did not have a first page number and were therefore probably online only. For this reason, the bibliometric information used in the AI system (page counts) is incomplete and UoA median lengths were used when page numbers were missing. Moreover, since font sizes and page sizes differ (e.g., A4, A5) page counts are crude indicators of article length. Word counts and character counts in the full texts experiments did not provide a more accurate system, however. The underlying problem seems to be that some prestigious journals require short articles, often with extensive supplementary materials files containing most of the methods. Thus, a short article might be prestigious, preliminary, or with little content, depending on the journal. The problem is therefore that there are no currently available inputs that reliably quantify the size of the content of an article.
  - **Long articles**: The same issue arose for some long articles. Some were perhaps insufficiently condensed research accounts published in perhaps lower quality journals, whereas other articles were almost books in journal form. Thus, again, long articles might have been penalised or rewarded by reviewers, depending on the nature of the content. This could not be quantified by the inputs.
- **Unusual contribution types**:
  - **Valued low citation research issues or nonstandard contribution types**: In predictable fields in which citations are a good indicator of quality, there may still be issues or types of contribution that get few citations but are still respected (e.g., medical ethics contributions to medicine). These might be overlooked by the AI if they were rare or if they were common but lacked distinctive keywords that could be leveraged. There does not seem to be a solution to this problem. This issue may be widespread in much of the social sciences, arts and humanities but could affect all fields to some extent.
  - **Highly cited relatively minor contributions**: Some journal articles may describe contributions to the research infrastructure, such as ethical guidelines for publishing, put together by large committees (rather than research teams), attracting many citations, and published in a reputable journal, but not containing what the field might recognise as a substantial research contribution.
    - **Meta-papers**: Articles partly based on reviews or evidence syntheses might be highly cited but might be considered to make relatively minor contributions compared to papers mentioning the same methods but applying them to a study.
- **Negative or expected results**: papers reporting large clinical studies with negative results might be generate high predictions but be downgraded by human assessors for lack of impact. Similarly, large scale studies with valuable findings might be downgraded for being relatively routine in terms of methods, such as if they were follow-up or replication studies.
- **Specialist issues**: A topic might be considered too niche to be given a high score in some fields.
- **Developing country location**: Human assessors might tend to score research lower if it focuses on poorer nations. It is also possible that research in poorer nations had less robust methods due to a lack of resources.
- **Narrow geographic focus**: A paper might be downgraded by assessors for a narrow scope, which might make it less generalisable and more of a case study than a comprehensive evaluation. The AI can't learn rare names so would not learn the narrow focus of any paper.

## *5.2 Discussion of overfitting issues*

The overfitting issue in machine learning is that the accuracy of a system on one dataset may be misleadingly high compared to other datasets to which it is applied. This can easily occur if the training data and test data overlap (not done above), if no development set is used (the preliminary testing on different data performed the partial role of a development set for this report) or there are many options evaluated, with decisions made based on the most accurate option (as done to some extent above). In the context of this report, overfitting is the suggestion that the approaches reported above for REF2021 data would be less accurate for future REFs. Here are some key points to consider.
- The main results comparing all AI algorithms on all UoAs and year ranges investigated only report the accuracy of the best AI algorithm. This accuracy may be a slight overestimate since six algorithms had similar levels of accuracy, so the results cherry-pick the best one (rfc, gbc, xgb, or ordinal variants), even though it may have been the best only by chance.
- Some of the above detailed sections report the results for only the best machine learning method (rfc, rfco, or xgb) for five UoAs 2014-18. For the reasons in the bullet point above, the statistics generated from these may be slightly optimistic.
- The methods above to select the best AI algorithm used 100% of the relevant data (some for training, some for testing), whereas a future practical application would not have 100% human classified data and would therefore be less effective at picking the most accurate AI solution. Using the standard 10-fold cross-validation method, this reduces the amount of data for training the models by 10% (e.g., from 50% to 45%) as well, reducing the accuracy of the models built for testing.

## 5.3 Differentiating between 1* and 2* scores

The main results combined 1* and 2* scores into a single category to reduce the disparity between category sizes and aid the machine learning through more balanced training sets. Predicting all four classes (1* vs. 2* vs. 3* vs. 4*) tends to slightly reduce the overall accuracy and rfc is the most accurate classifier in most cases (Figure 5.3.1). Thus, whilst the three-class problem gives the most accuracy, the four-class problem could reasonably be applied as a second approach to differentiate between the 1* and 2* predictions. There are very few 1* predictions, however. For example, in UoA1 only 3% of the 1* and 2* scores for 2014-18 were 1*.

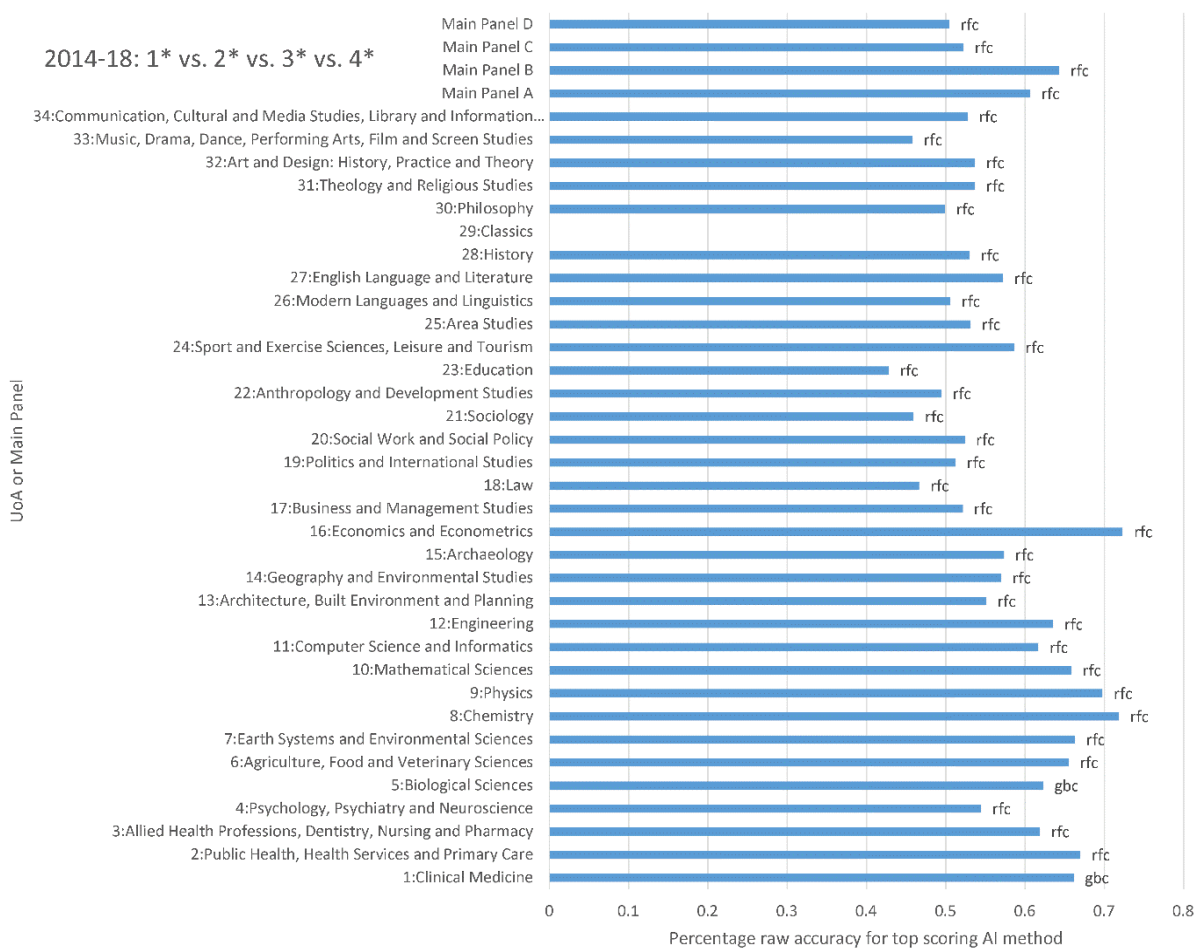

Figure 5.3.1. The percentage accuracy for the most accurate machine learning method predicting for classes (1* vs. 2* vs. 3* vs. 4*), trained on **50%** of the 2014-18 Input Set 3: Bibliometrics, journal impact and text, after excluding articles with shorter than 500-character abstracts **and duplicate articles within each UoA**. No models were built for Classics due to too few articles. Only the three most accurate methods in standard format (not ordinal) were used.

## 6 Reasons why higher accuracy was not achieved

The AI predictions in some cases have the highest accuracy ever reported anywhere in the world for predicting REF score profiles, despite careful checking against overfitting. For example, the Pearson correlation of 0.998 for UoA 1 has not been surpassed before for any field. This is because the input set has been culled from an extensive literature review and new indicators added that are more powerful than any used before (e.g., MNLCS instead of JIF for journals). Nevertheless, the change in scores for individual submissions to a UoA is much more important than this and the article-level accuracies have not approached 100% and could only exceed 72% by using a strategy that systematically excludes hard-to-predict articles. Following on from the analysis of possible reasons for incorrect predictions for individual articles (Section 5.1), there are both article-level and field-level issues, with the latter explaining the lower accuracy for some UoAs.

## 6.1 UoA differences in prediction accuracy and field-level factors

An important reason why the accuracy differed between fields is that some UoAs had too few journal articles to build a model from, either because they were small or because journal articles were a minority of the outputs. Two other generic important factors affect the extent to which scores in a field are fundamentally predictable with the AI strategies used here: the extent to which citations are good indicators of research quality and the extent to which experts agree on what constitutes research quality.

The citation-based inputs were the most powerful in UoAs with higher AI prediction accuracy, so an important limitation on prediction accuracy is the extent to which citations are relevant in a field. Generically, citations primarily reflect the scholarly impact of an article rather than its originality or rigour (Aksnes et al., 2019), but they do not reflect scholarly impact at all in some fields.

The case for using citations as evidence of scholarly impact is that scholars cite work that has influenced them when producing new studies (Merton, 1973). This applies to hierarchical fields, such as the natural sciences. In other fields, core books might be by far the most influential (e.g., Marx's *Capital*, Merton's *Sociology of Science,* Knorr Cetina's *Epistemic cultures: How the sciences make knowledge,* Eddo-Lodge's *Why I'm no longer talking to white people about race*). Or, as in areas of the arts and humanities, citations may reflect sources, criticism, and relatively random inspirations (e.g., Foster & Ford, 2003) rather than generally influential work. In many fields, background and perfunctory citations may dominate important sources of influence and there may be great flexibility in which articles to cite, with social factors dominating the choice (Figure 6.1.1).

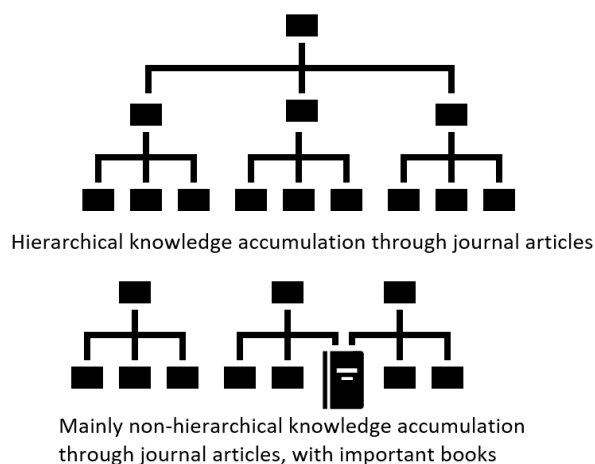

Figure 6.1.1. Hierarchical and non-hierarchical knowledge organisation of fields. Boxes indicate journal articles and lines underneath them indicate their references.

A second UoA-specific factor is that there are substantial disciplinary differences in the extent to which specialists agree on the quality of work. Whilst substantial disagreements between expert journal reviewers occur routinely in all areas of scholarship, the frequency and depth of disagreement between reviewers has systemic components. The organisational models of Whitley (2000) are helpful to understand this, and three of them are used here (Figure 6.6.2). Like all models they are wrong/oversimplifications, but in this context they are useful. They tend towards caricature and the distinctions between disciplines are not as sharp as they suggest (Trowler et al., 2012), especially in the current era of widespread interdisciplinarity and standardised assessment practices. Nevertheless, they point to important underlying factors that affect AI prediction accuracy.

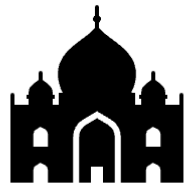

Conceptually integrated bureaucracy – broad agreement on core aspects of research

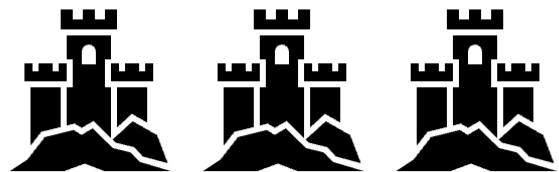

Polycentric oligarchy – competing claims for research quality

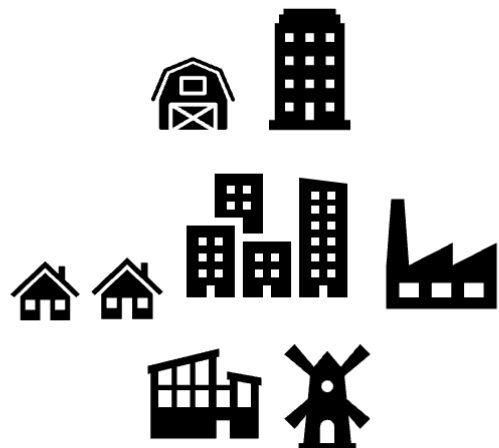

Fragmented adhocracy – diverse research objects and uncertainty about quality

Figure 6.1.2. Three types of research organisation (Whitley, 2000).

Fields that are *conceptually integrated bureaucracies*, such as high energy physics, have a high degree of agreement about what to research, how to research it, and what constitutes good quality research. This has evolved partly out of necessity to control access to expensive equipment. In such fields, quality judgements between REF reviewers should be relatively consistent, even if this is not the perspective of individual scholars based on conflicting referee reports. Conceptually integrated bureaucracies are perfect for AI because there is general agreement on all core aspects of the field.

At the opposite extreme, *polycentric oligarchy* fields host competing paradigms where each school regards the contributions of other schools as invalid or weak (e.g., qualitative vs.

quantitative; Keynesian vs. free market economics). Quality scores for articles in UoAs dominated by polycentric oligarchies depend greatly on which paradigms are represented by the panel members, if not all are. An article may not fare well when evaluated by the other side in a paradigm war (Munoz-Najar Galvez, et al., 2020) and the Italian version of the REF has procedures to deal with this in the social sciences and humanities (Bonaccorsi, 2018). Polycentric oligarchies are difficult for AI because quality is strongly disputed, unless only one paradigm is represented. In this case the AI would learn and repeat the prejudices of the assessors' paradigm.

In between these, *fragmented adhocracies* have little agreement about what to study, how to study it and the quality of the results. In such fields, all quality judgements are highly subjective and depend on whether there are any panel members with interests relevant to the article reviewed. Most of the arts and humanities and some social sciences are probably like this, including library and information science. For example, no panel members had the expertise to fairly assess some outputs with rare topics (e.g., scientometrics). Fragmentation can mean that panel members struggle to evaluate robustness, originality, and significance for research outside their narrow specialism. They may not recognise originality because it is of a type that is not relevant in their specialism (e.g., methods originality dominating one specialism, research object originality in another). For example, our own research includes articles introducing new web-based sources of evidence for scientometric indicators. We believe that an expert evaluator would regard this as original because most of the field relies on traditional indicators, but we fear that a non-expert evaluator from our library and information management field would regard these as not original because they do a similar type of thing (introduce a new indicator) or, even worse, regard them as pedestrian because there are already many articles about research indicators (many journals are full of them). In this context, there is not a correct answer. Others have observed that in some fields, subjectivity is unavoidable, and all quality evaluations are inherently unfair. Thus, in the latter two cases, there is essentially not an agreed quality score for articles in the field, so it would be impossible for an AI system to accurately predict something that does not exist. Fragmented adhocracies are difficult for AI because fragmented fields have more and weaker patterns and the quality of the articles is inherently uncertain.

## *6.2 Generic problems for prediction accuracy and article-level factors*

AI prediction accuracy is limited by partial patterns in the relationship between the inputs and research quality and by anomalies in the data, as found by the analysis of incorrect predictions. General rules, such as that extremely highly cited articles tend to be 4* in some fields, are more difficult to detect by AI when there are exceptions. This is because the AI must detect the pattern despite the exceptions, which means that less common pattens will not be detected. For complete accuracy, the AI then needs to learn to recognise exceptions. This is technically impossible for AI because exceptions are rare and therefore do not have a pattern. To give an extreme example, if there is one exception to a pattern then if it is in the AI training set then it could conceivably be detected (although one is not a pattern) but then it would not occur in the set of articles to be predicted. Conversely, if the exception is only in the prediction set then the AI would need to be clairvoyant to detect it.

A second generic issue is that patterns can be partial and difficult to detect as a result, given the limited number of human-scored articles available to detect the pattern. This is an aspect of the well-known *curse of dimensionality*: huge numbers of article scores are needed to detect patterns from the high dimensional inputs. For example, the system probably has not always been able to detect that some articles are short because they are short form limited contributions, such as letters or commentaries, whilst others are short because they are in prestigious journals with length restrictions.

The field delimitations used to normalise the citation-based inputs (up to 330 Scopus narrow fields) is mainly journal-based and so this is a limitation. A more accurate article-based system might make the citation-based inputs more powerful.

Most importantly, the AI system can only access surface level characteristics of an article from its title, abstract and keywords, as well as a range of supporting bibliometric information (normalised citations, authorship team properties, journal impact). REF scores are allocated by experienced field experts that have already ingested a mass of knowledge about field norms, field progress and societal values that is not available to the AI system (Figure 6.2.1). This, in addition to reading and comprehending article full texts, gives them a substantial advantage because quality evaluation is complex and multi-faceted (Langfeldt et al., 2020).

An effective evaluation would have to judge the many different aspects of rigour that could only be possible with a close reading of the full text combined with specialist methods knowledge, including from experience of learning about and conducting similar types of research and understanding key methods issues. The AI state of the art for rigour is limited to detecting numerical inconsistencies in common statistical tests (Nuijten & Polanin, 2020), and the ability to automatically detect high level rigour issues is currently inconceivable. For example, perhaps a construction engineering experiment had used a type of oven that previous research had shown to be sub-optimal for bricks with organic trace matter: an expert might recognise it but all current AI would not, even if fed the full texts of all relevant articles. An expert would also need to draw upon their disciplinary expertise to evaluate the originality/novelty claims of an article: the AI system might read the claim in the abstract but could not evaluate it. Finally, the expert would need to draw upon societal and field knowledge to evaluate the significance of an article, whilst the AI could only guess at scholarly significance from citation and journal information.

Even an experienced academic from a different field would be likely to have a low rate of agreement with panel members (e.g., the 54% inter-UoA agreement rate suggested above, although this is at least partly caused using different criteria). It seems almost certain that in most or all UoAs, some panel members would be incapable of effectively assessing outputs in their UoA from other specialisms (e.g., neuroscience vs. social psychology in UoA 4). An AI system would need a way of acquiring specialist knowledge from multiple areas to compete with panel members on accuracy.

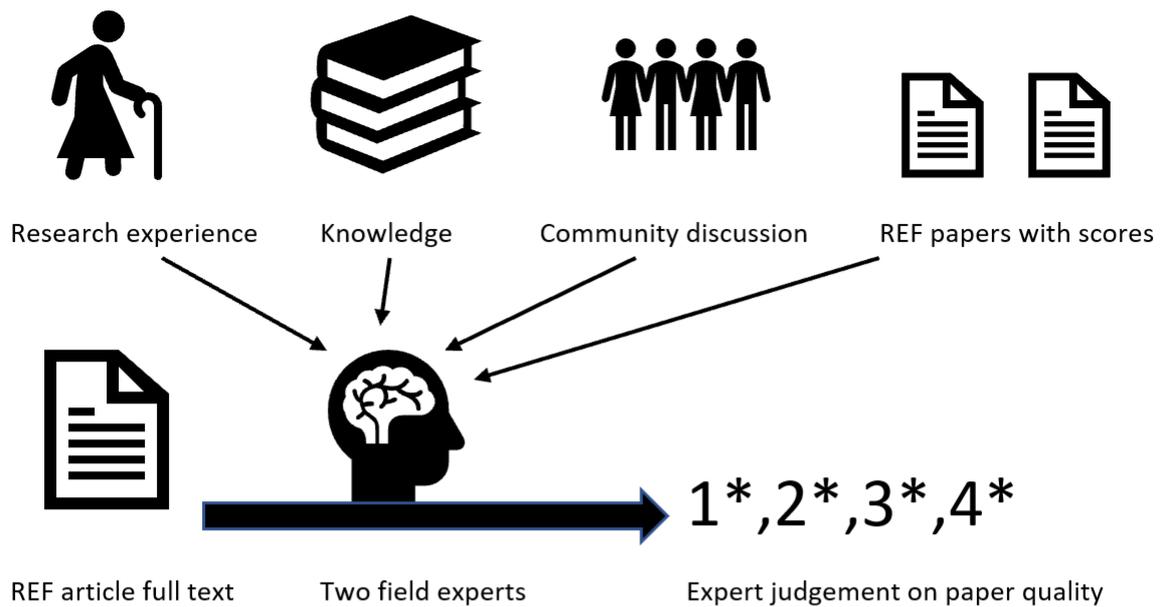

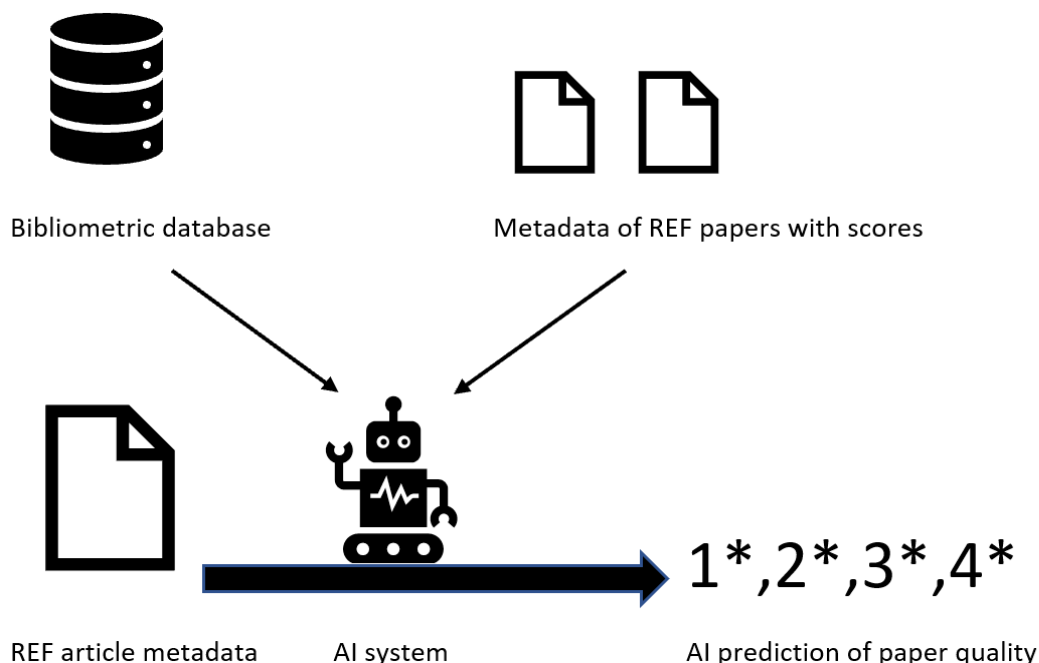

Figure 6.2.1. A REF panel member (top) has decades of experience in which she has read lots of papers and books, learned from discussions with and feedback from community members. She has also read the full text of every REF submission allocated to her and discussed quality issues about them in detail with one other member of staff before making a final decision. The AI system (bottom) only has access to scores and metadata about the articles and bibliometric information so cannot judge rigour, can only guess at originality and can only partially assess significance in the best case.

## 6.3 Future strategies for improved accuracy

As suggested by the above, to achieve a substantially higher level of accuracy at REF output score prediction, an AI system would need to learn the knowledge of an experienced field expert. Fundamentally, this means it needs access to a much larger set of richer data to learn enough patterns. This would need to comprise the following, which are currently impossible but are likely to become more practical during the next decade.
- Full text of all articles in a clean format to be able to check for methodological rigour.
- Full text of all scholarly articles in the world (or a substantial fraction) to be able to check for originality.
- Quality scores or acceptable alternatives (e.g., journal-based proxies) for a large set of articles.
- Detailed funding information for articles.
- Millions of peer review reports from open peer review sites to be able to learn which aspects of an article may be criticised or praised for originality, significance, and rigour.
- More fine-grained metadata to detect different article types more effectively.
- Deeper theoretical models of disciplinary operationalisations of research quality to enable the generation of more powerful input indicators.

In summary, because of the complex and multifaceted nature of research quality, AI may essentially need to become as knowledgeable as a scientist about their field to make effective judgements. This would entail a radically more powerful system that is capable of effectively ingesting not only a substantial proportion of the disciplinary literature but also other communication artefacts, such as peer review reports and external evaluations (for societal impact). Thus, the goal is to make an effective AI scholar that, almost as a by-product, would be able to make reasonably effective quality judgements. This seems like a distant goal now.

# 7 Responsible metrics and the four proposed strategies

Ethical issues can be revisited in the light of the proposed strategies and the performance of the system. As argued above, all systems seem to be compatible with DORA despite using journals as system inputs because (a) a relatively robust journal impact metric is used, and (b) journal-level inputs are not used exclusively but are employed alongside numerous article-level inputs. In addition, this report includes clear evidence that journals containing articles with the same quality score are the exception rather than the rule and the proposed approaches all use human classification for most articles. The four strategies for using AI systems can be evaluated for responsible use by discussing how well they fit two well-known sets of recommendations for research metrics.

## 7.1 Metric tide principles

The Metric Tide responsible use of metrics for research assessment recommendations are as follows (Wilsdon et al., 2015a).
- **Robustness**: basing metrics on the best possible data in terms of accuracy and scope.
- **Humility**: recognising that quantitative evaluation should support – but not supplant ‐ qualitative, expert assessment.
- **Transparency**: keeping data collection and analytical processes open and transparent, so that those being evaluated can test and verify the results.
- **Diversity**: accounting for variation by field, and using a range of indicators to reflect and support a plurality of research and researcher career paths across the system.
- **Reflexivity**: recognising and anticipating the systemic and potential effects of indicators, and updating them in response.

All the systems here use carefully curated sets of inputs, so meet the **robustness** criteria unless important inputs have been overlooked. The **transparency** issue has been discussed above. In broad terms the data collection and processing is transparent in the sense that the methods can be declared in advance. Nevertheless, those evaluated could not verify the algorithmic results afterwards because they probably will not be told which articles were scored by AI and certainly will not be told the score of any articles. This is not a drawback of the AI approach in the sense that the same is true for the current REF. The issue of **diversity** is met by using a range of inputs and only scoring journal articles with AI so that, for example, books, software, and performances are not penalised.

**Reflexivity** is a more substantial issue. Strategies 2-4 seem unlikely to generate systemic effects since a small minority of articles would be affected by the AI predictions, so it does not seem sensible for institutions to prioritise inputs that are likely to statistically associate with higher REF scores (e.g., number of authors, journal impact) rather than quality directly. Nevertheless, research quality is difficult to assess and there may be a temptation to look for simplistic alternatives that are easy to identify and are "glorified" by potential inclusion in the AI system. This problem would be much more substantial in relevant UoAs if strategy 1 were to be adopted because there would be a realistic chance that articles would be assessed by the AI.

## 7.2  Leiden manifesto

The Leiden Manifesto lists the following ten principles (quoted from: Hicks et al., 2015).
1. *Quantitative evaluation should support qualitative, expert assessment*. Strategies 1-3 violate this rule, at least for the articles predicted. This seems reasonable, however, as long as the degree of error is known, measurable and (in general) accepted by those evaluated. These criteria are not usually applicable to bibliometrics, but apply in the current case, if the sector adopts one of the proposed strategies.
2. *Measure performance against the research missions of the institution, group or researcher*. This does not apply to the output component of the REF, although it applies to other components.
3. *Protect excellence in locally relevant research*. The analysis of the incorrect score predictions suggested that **locally relevant research may be disadvantaged by the AI score predictions**, especially if they name a specific geographic region.
4. *Keep data collection and analytical processes open, transparent and simple*. The input data and cleaning processes can be open and transparent, but **the machine learning methods are inherently complex**, as are the human decision-making processes that they would replace or supplement.
5. *Allow those evaluated to verify data and analysis*. As above, this is not allowed for REF scores currently and would not be allowed for the AI predictions.
6. *Account for variation by field in publication and citation practices*. This is built into the system with field normalisation of the citation-related indicators and 34 different field-based AI systems
7. *Base assessment of individual researchers on a qualitative judgement of their portfolio*. This is not relevant to the REF.
8. *Avoid misplaced concreteness and false precision*. This is satisfied by the three or four point scale outputs.
9. *Recognize the systemic effects of assessment and indicators*. This has been discussed above in terms of reflexivity for the Metric Tide.
10. *Scrutinize indicators regularly and update them*. This can be achieved and the AI systems automatically update themselves for each new input set.

## 7.3 Disabled academics

The idea to partly replace human scores with AI scores has potential implications for disabled academics because of the reliance upon standard formats. A disabled academic may tailor their outputs to fit their disability and may be penalised for producing outputs that do not match the standard format of an excellent article, as identified by the AI. For example, a visually impaired academic might write shorter articles and be penalised through the length input into the AI and an academic with communication problems may prefer to work alone and be penalised with low collaboration inputs. These same issues apply to human REF scores, so an ideal solution might be to allow academics with disabilities to declare how their disability influence their outputs. This could not be an AI input, but disabled academics might be given the option to opt out of AI classifications if this is implemented.

# 8 Sector engagement

Focus groups and presentations at the steering group for this project and the UK Forum for Responsible Research Metrics were used to gain feedback from the sector on the use of AI in research assessment based on the results of this project. These events were run primarily in May and June after the initial report had been finalised. Brief summaries were shared with the participants in a format agreed with the steering group. The purpose of all events was to explain the main findings and gain critical feedback from the sector about the elements of the recommendations and their implications that are acceptable or problematic. Care was taken to gain a wide range of inputs because overall agreement is unlikely. The focus groups were run online to reduce the barriers to participation, given that gaining a wide range of perspectives is a key goal.

## 8.1 Suggestions from participants

**Suggestions for improvement**
- Consider incorporating corresponding author information (x2). In chemistry, the corresponding author is often a PhD student and the career of the supervisor, normally the corresponding author, is more relevant than that of the first author.
- Incorporate discipline codes assigned by sub-panels, perhaps splitting separately by discipline within UoAs.
- Many papers in high impact journals in some fields are short with extensive supplementary materials. Incorporating a way to effectively identify this could improve accuracy in relevant UoAs.
- Base AI predictions on the 13-point scale used internally by subpanels, possibly including all scores, not just the final agreed score. Computing apparently used software to compensate for reviewer strictness and this may need to be taken into account. This information might also be used to calculate inter-assessor agreement rates
- Allow subpanels to use the predictions to replace the third reviewer for UoAs that had three for some or all outputs.

**Reasons blocking higher accuracy**
- UoAs can contain diverse but related fields, with different relationships between article quality and the AI system inputs. For example, UoA4 includes psychology, neuroscience, and psychiatry and "While there is a reasonable degree of overlap between the three with high levels of cross-disciplinary collaborations in some areas," the quality criteria [] are somewhat distinct." The AI prediction accuracy may be higher for neuroscience and psychiatry than for psychology and this would show at the institutional level since different types of HEI often specialise in different areas.
- Some specialisms contain many high-citation, low value papers which renders citation-based indicators unhelpful for the UoA as a whole.
- Some papers might be extremely high quality but published in a journal without a high reputation/JIF because it is appropriate for their target audience, or they legitimately have a national focus. These would be disadvantaged by AI.
- Fields without journal hierarchies will be harder to predict with AI.
- Some decisions require a very close reading of a full-text, for example to decide whether an article is a genuine systematic review (grade 1* to 4*) or just a review (grade 0).
- Panel members may simultaneously assess a set of outputs from the same institution with borderline quality and randomly round some up and others down to give the correct overall score profile, even though individual output scores are inaccurate. No AI system could cope with this and it reduces article-level accuracy for all AI systems. Other UoAs deliberately assessed articles in a random order to minimise the chance of institutional bias, so would not have done this.

**Uses of the system**
- Run AI in parallel with peer review for next REF.
- Consider replacing one of the two human evaluators with AI predictions in some UoAs. Or allow subpanels to make this choice as part of strategy 4. But do not reduce the total number of assessors because breadth is important.
- The AI might identify borderline outputs that need extra human evaluation.
- AI may be useful for interdisciplinary outputs as an alternative to other means of evaluation . It might be used instead of cross-referral (two independent suggestions). Cross-referrals were seen as problematic in REF2021 because they may have caused delays, caused unexpected extra work and may have been rejected; however, at least one UoA had no problems with interdisciplinary work.
- Offer the system to institutions to help them select outputs (a controversial suggestion), because this is time consuming.
- Use the system to help with score calibration within panels.

**Choice of strategy**

Two focus group members recommended strategy 3 and one recommended that the AI is not used. The remainder were in favour of a version of 4. In general, panel chairs (a small sample) seemed to be most strongly against strategies 2 and 3 (multiple stages) due to the extra complications of AI prediction delays.
- Strategy 3 is impractical since REF panel members have specific expertise and the system-selected 10% at each stage might give an unbalanced set.
- Strategy 3 is highly undesirable because allocating outputs to review is extremely time consuming and anything that interferes with it would cost more time than it would save.
- Strategy 3 will not work because early scores needed substantial calibration, for example because the extreme scores were rarely used in at least two UoAs (on the extended scale used during the deliberations).

- An advantage of Strategy 3 (perhaps also 2 and 4) is that the assessors may be more confident in the AI if they interact with it.
- Strategy 2/3 give too little time saving to be worth the extra complexity, including because they seem likely to primarily classify articles that a field expert would be able to quickly score accurately anyway.
- Strategies 2-3 will make allocation more complex, which would not be worth the time saving. Sub-panel members lose most of their summer evaluating outputs and must be given their allocation in one go at the start so that they can manage their time and other commitments (including personal). Complexity and deadlines (e.g., Strategy 3) will greatly slow the process, which is restricted at times to the pace of the slowest member.
- The choice of strategies 1-3 would partly depend on whether senior figures (e.g., vice chancellors) have an appetite for risk.
- Strategy 4 is risk averse and provides continuity. Strategy 4 could also function partly as a pilot testing phase.
- Since the AI predictions would not be available at the start for Strategy 4, this would either create unfairness between outputs, especially if the results were used to replace and not supplement the bibliometrics. Delaying decisions on borderline outputs before the AI predictions became available would create significant extra work (panel chair).
- Only a higher accuracy system could be considered for strategies 1-3. 85% accuracy seems too low, but there were no suggestions of a level that would be acceptable. One UoA which reported a "surprisingly high" degree of agreement between subpanel members, indicated 90% or even 95% might be needed for acceptability. Higher accuracy is especially needed for UoAs with small submissions where one or two errors could have a large influence on rankings. Institutions are risk averse and will not accept the possible ranking changes due to AI.
- Some of those preferring strategy 4 suggested a parallel exercise to test the AI further.
- The extra administration burden for strategy 4 would not be worth the added value.

**Systemic effects of AI**
- Institutions might use AI to predict scores to help them decide what to submit. Some may try to game the system (e.g., by preferring to submit hyperauthored articles), although this occurs already. For example, an article being in a good journal or highly cited would give senior research managers confidence that it was good enough to submit. AI might encourage wider bad behaviour such as unnecessarily inviting international authors. In fields with many outputs per year (e.g., Chemistry, Physics), this may primarily influence which outputs are submitted rather than which outputs are produced.
- Including citations as an input might provide an incentive to self-cite. The same for the other inputs, such as team size. Universities might seize on anything that they can control that might help attract a better score in the next REF.
- Perverse incentives might apply even for strategies 2, 3 and even 4, since some people may pick up on the mildest signals and emphasise them, such as at the institutional level. Any type of mild favouring of citation impact is to be discouraged.
- A strength of the REF is that it disrupts traditional hierarchies by allowing teams to recognise excellent research in unexpected places. Incorporating bibliometrics might work against this by having a conservative influence and therefore be anti-innovation.

- Similarly to the above, a system that has predictable inputs (bibliometrics, author teams) will have a conservative influence and tend to reproduce existing hierarchies. This will reinforce the already strong citation-related incentives in some fields (e.g., business, economics) and promote an unhealthy culture. A strength of the REF is its ability to allow innovation in unexpected places.
- Conservatism in terms of journals and journal impact will work against innovation by pushing people away from publishing innovative work in new journals.
- Including JIF-like inputs cuts across the repeated anti-JIF message in the REF to sub-panel members and might represent a philosophical shift. The sector is moving away from journal reliance and including publication information in any way represents a step in the wrong direction and goes against the spirit of DORA.
- Allocating scores without analysing the content of an article goes against the spirit of research evaluation.
- AI solutions work against human expertise, which are at the heart of the REF rationale and this goes away from EU moves against metrics.
- AI strategies applied differently by UoA potentially cause wider issues that need time to be thought through.

**Other considerations and points**
- REF2021 panel members had extensive and repeated anti-bias training and warnings, so differences with AI predictions are unlikely to reflect bias correction, except at a minor level.
- Some subpanel members might only need 10 minutes to score an output, so the overall cost of peer review would be much lower than hypothesised in this document if this is common.
- Consistency was important in REF2021. It might be important to have a uniform process across UoAs so that none feel that they are treated differently. Conversely, more people thought that it is fine to have opt-in AI along the lines of the current bibliometrics procedures.
- A separate procedure for small HEIs because of their high variability is unacceptable in principle.
- Rules will be needed to limit when the predictions can be viewed for strategy 4, to prevent evaluators from being overly influenced by them.
- The AI should be built into the REF system before it starts so that there are no delays.
- If REF becomes more frequent, then the AI will be less useful because the most recent two years are ignored.
- Institutions might consider submitting to a non-AI UoA if there is a choice.
- Including citation data as an important input in the system represents a radical shift from the current focus on relying on content and this change needs to be communicated effectively.
- The best work is done in books in some UoAs so AI for journal articles seems irrelevant. The REF might need a more radical overhaul so AI predictions may be redundant.
- Impact assessment takes a small fraction of the time that output assessment does, so is a more time-efficient exercise.

**Future system modifications**
- Systematic collection of associated data access statements and associated code in future REFs might provide additional inputs.

- Get feedback from panel members on articles where they disagree with the AI score.
- Is the 1* grade too rare to be meaningful in the REF anymore?
- Consider whether big team science should be evaluated differently, particularly for 100+ author papers, which is currently controversial and a change in procedures for this would need to be reflected in AI system changes.

# 9 Summary/issues for consideration

The main results are also available in the executive summary at the start of this document.

## 9.1 Strategies for using AI in research evaluation

Based on the above results and taking 85% as the proposed minimum accuracy to use the results, Strategy 1 is unacceptable for all UoAs. With Strategy 3 (active learning), AI predictions could be made for articles in the seven UoAs when 85% accuracy is reached, giving a total saving of 3,688. This is 809 (28%) higher than the 2,879 predicted at 85% for Strategy 2 (prediction by probability), so Strategy 3 is technically superior to Strategy 2 in terms of peer review time saving at a theoretically acceptable level of accuracy.

1. **Strategy 1: Classical machine learning**: Sub-panel members classify 50% of the eligible journal articles published 2014-18 in UoAs 1,2,6-10,16 with 65%-72% accuracy and use AI predictions for the rest. This would result in score shift between institutions within UoAs of up to 7% overall for larger submissions and 8% overall for smaller submissions. In the least inaccurate case, 12,639 articles from UoAs 1,2,6-10,16 can be predicted with 65%-72% accuracy (effectively 9.3% of all REF2021 articles, taking duplicates into account).
2. **Strategy 2: Identifying high prediction accuracy articles with prediction by probability**: Sub-panel members classify 50% of the journal articles for the most predictable fields and identify a highly predictable subset of the remainder to classify with AI for medium and large institutions, then human classification for the remainder. This would change the overall scores little but would only reduce the number of articles to be predicted by a small amount: 2,879 articles in total at 85% accuracy for the predicted set (effectively 2.1% of all REF2021 articles).
3. **Strategy 3: Active learning**: Sub-panel members classify 10% of the outputs for all UoAs, then the AI identifies an additional 10% of difficult to classify articles, sending them to the sub-panel members to classify, repeating this until all articles have been classified by sub-panel members or the remaining articles can be classified with AI above a threshold (e.g., 85% accurate). This would allow 3,688 articles to be predicted by the AI at 85% accuracy for the predicted set (effectively 2.7% of all REF2021 articles). The saving is 30% in UoA 8, 20% in UoA 1, and 10% in UoAs 2,5,7,9,16.
4. **Strategy 4**: Strategy 2 is run when at least 50% of the articles have been scored by a sub-panel but the predictions and estimated prediction accuracies are given to the sub-panels to decide how to use them. The following are the most likely options.
5. **Strategy 4a: AI informing or cross-checking human scores**: The AI predictions are used as evidence to inform or cross-check peer judgements in selected UoAs. This might replace the bibliometric information currently used and would carry more weight. The AI predictions might even be taken from the models built for REF2021 and used to provide initial predictions for some UoAs to inform expert judgement. The prediction accuracies would be lower, however, due to the time differences.

6. **Strategy 4b: AI replacing one human reviewer**: The AI predictions replace a third human reviewer (possibly the second if there are only two), keeping the other human reviewer(s), who would have the power to override the AI. The human reviewer might decide how much time to spend on an article based on how sure the AI was about the conclusion.
7. **Strategy 4c: End of assessment mopping up**: The AI predictions and their associated probabilities are given to sub-panel reviewers to use how they see fit towards the end of the process. They can then be used to mop up problematic cases and as a sanity check. Access to predictions from models built by other UoAs would allow a degree of automated interdisciplinary input, which may help with this issue that was problematic for some or all UoAs in REF2021.

## 9.2 Advantages of AI in research evaluation

- **Time saving/cost efficiency**. Nationally, the evaluation phase consumes much of a year of the lives of around a thousand senior UK researchers. For example, subpanel 1 (out of 34) has 34 members and 6 assessors, most of whom are professors. In 2020/21 it generated "a substantial workload for individual members, especially in reviewing outputs."[13] Assuming that 1000 panel members each spend a quarter of a year assessing outputs then this is 250 person-years in total. Given that journal articles comprise 82% of all REF outputs, each percentage point of journal articles classified with AI in the ref would give a 2 person-year time saving (although costing AI development time). Costing each year at £80,000, this translates into a saving of £160,000 per percentage point of journal articles classified by AI. Administrative costs for AI might run to £200,000, which would need to be subtracted. These figures (other than 1000 panel members) are completely without evidence, so this is a wild estimate. On this basis, suggestion 1 above breaks even and the other two are less efficient.
- **Increased objectivity/reduced subjectivity**: Although AI can learn bias (e.g., institutional, gender) and may be biased against some types of research (e.g., humanities-oriented contributions to medicine), AI would at least apply the same rules to outputs from all submissions in the same UoA. For example, the same output from multiple institutions in the same UoA would always get the same AI score. When the AI introduces a score shift, this could be reducing human bias/inaccuracy rather than introducing AI bias/inaccuracy.
- **Potential for future improvement**: If any strategy is implemented then experience with it may suggest avenues for future improvement. It also sets an incentive (a real-world application) and a basis from which AI researchers can research prediction strategies. It would be a useful investigation for possible evaluation of AI in research evaluation that might open the way to future more accurate versions, especially if public peer review text can be harnessed in the future and structured full text becomes available.

## 9.3 Disadvantages of AI in research evaluation

- **New sources of bias**: AI might be biased against some minority types of research that score badly on traditional field indicators (e.g., humanities-oriented contributions to medicine), disadvantaging institutions specialising in these. It is biased against high scoring submissions and larger HEIs in the same way that bibliometrics probably are, and Strategy 4 may reduce these biases compared to bibliometrics since the AI predictions are more accurate.
- **Perverse incentives**: AI might lead institutions to emphasise the aspects of articles included in the AI inputs, such as journal impact, article citations, or authorship teams, skewing the results or the operation of science. This would be mitigated by the AI not relying on a small set of inputs and using a learning strategy that does not guarantee that higher scores translate into higher predictions (e.g., it might penalise humanities-oriented contributions in high impact journals or with many authors). This incentive would presumably be mild for strategies 2-4 where the overwhelming majority of articles would be scored by human evaluators rather than the AI. Nevertheless, the inclusion of journal inputs might complicate the current clarity of the DORA message.
- **Increased system complexity**: Including an extra stage in the evaluation procedure changes it from a conceptually simple system (human peer review) to a more complex mix of human and AI, which may reduce its understandability. There may also be substantial practical difficulties with the proposed two stage review process where sub-panel members review half, the AI predicts the most predictable articles from the remainder, and the sub-panel members, in a second stage, predict the remaining articles.
- **Possibly reduced system credibility**: The REF may lose some credibility if examples are found where AI predictions are clearly wrong (even if they statistically tend to be mostly correct, a more subtle point). This may lead some to lose trust in the system.
- **Accuracy may change over time**: Changes in science may reduce the percentage of outputs for which the AI is accurate. For example, increased publishing in megajournals may reduce the effectiveness of journal-level inputs in the AI and increased publication of outputs other than journal articles would have the same effect.

---

[13] https://www.ref.ac.uk/media/1009/annex_a_ref_2017_03.pdf

## 10 Recommendations

Despite an appetite amongst panel members to reduce the considerable burden of the REF on panel members, AI should only be used to support peer review and not replace it. Because of the feedback from the focus groups, strategies 1 and 2 are not wanted and strategy 3 is mostly not wanted. The three versions of Strategy 4 have support within some UoAs but also opposition from other UoAs on reasonable grounds. It is therefore not possible to recommend a strategy that will be acceptable to all UoAs for which the AI provides the highest accuracy. We think that the best strategy is using Strategy 4 as additional to bibliometrics, allowing UoAs to use the AI predictions and probabilities as they deem appropriate. In some UoAs, 4c fills an identified gap in the system, 4a can add robustness to the results and 4b may provide minor time saving. Nevertheless, on balance, the inclusion of journal factors and gameable AI inputs is unhealthy for academia and, combined with the extra administrative burden, the minor advantages given by AI and the lack of time savings are insufficient to justify its use, so we recommend that it is not adopted. Instead, we recommend that it is used for pilot testing only in the next REF with a view to further development for future use, perhaps with effective full text processing. If there is an appetite in the sector to adopt it anyway then we recommend Strategy 4. In more detail for both Strategy 4 and Strategy 5 (pilot testing):
- Peer review is at the heart of REF and AI systems cannot yet replace human judgements. They can currently only exploit shallow attributes of articles to guess

- their quality and are not capable of assessing any meaningful aspects of originality, robustness and significance. They are not accurate enough to replace expert scores, they would encourage conservative behaviour, such as targeting high impact journals, and they would encourage gaming, such as gift authorship or citation cartels. AI predictions should not therefore replace peer review scores, or reduce the number of peer reviewers within a sub-panel.
- *Strategy 4 only*: The AI predictions are not ready to replace the current use of bibliometrics (partly because they are not available at the start). Instead, sub-panels should be given the option to consult the AI predictions when at least half of the outputs have been reliably scored. The predictions and probabilities can be used as the UoA decides, as such as mopping up final disagreements that the bibliometrics do not help with, a second opinion on difficult interdisciplinary outputs that have not been successfully cross-referred, and as a sanity check to look for anomalies. Alternatively, sub-panel members may choose to examine the AI predictions to evaluate their potential for their UoA rather than to support decisions about outputs.
- *Strategy 4 only*: The AI system should predict scores for REF journal articles and make the predictions and the prediction probabilities available to the subpanels that opt to receive them near the end of the assessment period. They should complement but not replace the bibliometrics, which should continue to be made available throughout the assessment period. The overall prediction accuracy for the UoA should also be presented for context.
- AI models should be built separately for each UoA for all except the two most recent years as a combined set. The models should make predictions on all articles from all UoAs (except the two most recent years) in case other UoAs want to use the predictions on interdisciplinary research submitted to them.
- *Strategy 4 only*: AI predictions should be hidden from evaluators until (i) at least 50% of the articles have been evaluated (ii) enough norm-referencing of scores within a sub-panel has occurred that the sub-panel scores are close to the eventual level of accuracy and (iii) the subpanel is towards the end of the assessment period and dealing with remaining difficult cases. *For Strategy 5*, the outputs should be only used for pilot testing.
- The inputs to the system are as specified in this document (the maximum set). *Strategy 4 only*: Subpanels should be given the option of considering the corresponding author to be the most important position rather than the first author (e.g., in Chemistry they probably designed the proposal and funded the work).
- New AI models should be trained from these inputs with provisional scores from the next REF rather than using AI models built from provisional REF2021 scores because a 60% increase in 4* journal articles between REF2014 and REF2021, combined with changes in the journal publishing system mean that the REF2021 AI systems may not predict reliably for the next REF.
- *Strategy 4 only*: In the longer term, Strategy 4 may replace the bibliometrics and perform a similar role, helping to resolve difficult cases. Panel members should be asked for their attitudes towards this after seeing the AI predictions for their UoA and completing the next REF.
- Although the main results in this report are based evaluating 50% of the outputs, the system should use all available scores because this will improve accuracy.

- *Strategy 4 only*: The importance of ignoring JIFs and journals during evaluations should be continued and emphasised, with the journal component of the AI system inputs being allowed as the sole exception, and explicitly explained as making a very minor contribution to REF decision making (secondary even to bibliometrics). This would ideally encourage assessors that favour JIFs to completely ignore them, leaving them to the AI. Similarly, as for bibliometrics currently, the importance of directly evaluating the quality of articles using disciplinary expertise and reading them should continue to be emphasised.
- The AI system is complex and should be seamlessly built into the REF computer system at least a year in advance so that it can be tested and does not cause delays. The python code for building the AI models is available to help.
- The tender process for the bibliometric information supplier should include a requirement to calculate and display the AI predictions from the bibliometric and other information, as specified above. Consider also adding the requirement for an effective article level classification scheme to the bibliometric tender to help panel chairs allocate outputs to reviewers. [This recommendation is also in the literature review.]
- During the next REF, UKRI should make plans to save sub-panel assigned disciplinary classifications and fine-grained scores on the extended scale because these will be valuable additional inputs for future AI experiments.
- During the next REF, institutions should be encouraged to self-archive versions of their articles that are suitable for text mining to support future, more powerful AI. Ideally, they would be in standard XML format (e.g., https://jats.nlm.nih.gov/publishing/) but in practical terms a plain text version or a watermark-free PDF would be helpful. [This recommendation is also in the literature review.]
- A deeper understanding of how experts make peer review judgements in different fields is needed together with innovative ideas for developing AI to exploit this understanding. Future research is needed to address these challenges. Possible sources of this include pilot studies with volunteers on REF-like tasks, where participants explain their decisions, and existing feedback from institutional mock REF exercises. The first would need careful framing if run by UKRI to avoid pressurising participants and the second would need GDPR clearance, such as through agreement by output authors and reviewers individually. This exercise could also be watchful for decision-making criteria that may be biased in terms of equality, diversity and inclusion (EDI).

# 11 Acknowledgements

Thank you to members of the steering group for comments on earlier drafts: Andy Hepburn (Research England), Steven Hill (Research England), Petr Knoth (Open University), Duncan Shermer, (Research England), and Jennifer Stergiou (University of Northumbria and ARMA Chair). Thank you also to the anonymous UoA panel members who agreed to participate in focus groups to discuss the provisional results. Finally, thank you also to Petr Knoth, Maria Tarasiuk and Matteo Cancellieri (http://bsdtag.kmi.open.ac.uk/) for supplying the full text of 59,194 REF-submitted articles from the CORE (https://core.ac.uk/) repository of open access papers (Knoth & Zdrahal, 2012). This study was funded by Research England, Scottish Funding Council, Higher Education Funding Council for Wales, and Department for the Economy, Northern Ireland as part of the Future Research Assessment Programme

(https://www.jisc.ac.uk/future-research-assessment-programme). The content is solely the responsibility of the authors and does not necessarily represent the official views of the funders.

# 13 Supplementary materials

The following supplementary files are available separately in arXiv.org and are summarised here http://cybermetrics.wlv.ac.uk/TechnologyAssistedResearchAssessment.html. All after the first are self-contained supplementary analysis of the REF2021 provisional journal article quality scores.

## 13.1 Main outputs on AI automation in the REF

1. *Literature review*: Reviews research related to possible AI automation of various REF tasks. Makes a list of separate recommendations for the future REF in terms of tasks that could be partly automated.
2. *Statistical analysis: factors able to predict REF scores*: Compares the relative strengths of the initially proposed inputs for machine learning. This was used to help select the inputs for the AI experiments in the main report.
3. *Predicting article quality scores with machine learning*: This summarises the AI findings of the main report above but sets the results in a wider research context.

## 13.2 Additional analyses investigating aspects of REF scoring

4. *Do bibliometrics introduce gender, institutional or interdisciplinary biases into research evaluations?* Shows that bibliometrics may introduce biases against high quality departments when used as indicators of research quality. **Implications**: REF sub-panels using bibliometrics should be warned about the slight bibliometric bias against high scoring departments.
5. *Is big team research fair in national research assessments? The case of the UK Research Excellence Framework 2021*. Shows that highly collaborative articles probably do not skew REF results, except possibly in a few UoAs. **Implications**: Supports maintaining the status quo in terms of allowing collaborative articles to count at full value for each author.
6. *In which fields are citations indicators of research quality?* Identifies the UoAs in which citations can reasonably be used as quality indicators and shows that there is no UoA with a citation threshold for 4* research. **Implications**: Supports continued use of bibliometrics and gives evidence to UoAs about the extent to which citations agree with quality scores in their area.
7. *In which fields is journal impact an indicator of article quality?* Identifies the UoAs in which journal citation rates can reasonably be used as quality indicators and shows that the journal alone never determines the quality of an article. **Implications**: Gives evidence that can be used to support the UKRI DORA commitment and to convince sub-panel members that JIFs are never substitutes for reading articles.
8. *Does the perceived quality of interdisciplinary research vary between fields?* Identifies a partial hierarchy of quality judgement strictness for interdisciplinary research submitted to multiple UoAs. **Implications**: Raises an issue for discussion about the extent to which quality standards are equivalent between UoAs.

9. *Can qualitative research be world-leading? Terms in article titles, abstracts, and keywords associating with high or low quality*. Identifies writing styles, methods and topics within UoAs that associate with higher and lower quality research as well as likely reasons for some of them. **Implications**: Raises an issue for discussion about whether some types of research are appropriately valued in REF scoring.

## 13.3 Additional analyses for wider UKRI policy

10. *Are co-authored articles higher quality in all fields? A science-wide analysis*. Identifies when collaboration gives added value to REF articles. **Implications**: Information to UKRI research funders about the fields in which collaboration might need to be supported or encouraged more than others.
11. *Are internationally co-authored journal articles higher quality? The UK case 2014-2020*. Identifies the UoAs and countries for which international collaboration associates with a quality advantage. **Implications**: Information to UKRI research funders about the fields in which international collaboration might need to be supported or encouraged more than others.
12. *Do altmetric scores reflect article quality?* Assesses the extent to which altmetrics from Altmetric.com can reflect research quality, as scored by REF panel members, showing that Tweeter counts are stronger than previously thought as research quality indicators. Whilst Mendeley readers are the strongest research quality indicator, they are slightly less strong than citation counts. Also shows that field normalised citation metrics can be worse than raw citation counts as research quality indicators for individual research fields. **Implications**: Altmetrics can be used with more confidence than before as early indicators of research quality, although their power varies substantially between fields.
13. *Is research funding always beneficial? A cross-disciplinary analysis of UK research 2014-20.* Gives evidence that research declaring funding sources tends to attract higher REF scores in all UoAs. **Implications**: Current sector-wide incentives to seek funding for research do not appear to be detrimental to research overall in any discipline.